%% file: master.tex
\documentclass[11pt,a4paper]{article}
\pdfoutput=1 % if your are submitting a pdflatex (i.e. if you have  images in pdf, png or jpg format) 

\usepackage{jinstpub} % for details on the use of the package, please
\usepackage[colorinlistoftodos]{todonotes}

\usepackage{graphicx}
\usepackage{amssymb}
\usepackage{wasysym}
\usepackage{rotating}
\usepackage{array}
\usepackage{lineno}

\newcommand{\Fig}[1]{Figure~\ref{#1}}
\newcommand{\Equ}[1]{Eq.~\ref{#1}}
\newcommand{\Sec}[1]{Section~\ref{#1}}
\newcommand{\Tab}[1]{Table~\ref{#1}}
\newcommand{\GEVcc}{\mbox{$\mathrm{GeV}/{{\it c}^2}$}}
\newcommand{\GEVc}{\mbox{$\mathrm{GeV}/{{\it c}}$}}
\newcommand{\GEV}{\mbox{$\mathrm{GeV}$}}
\newcommand{\MEVcc}{\mbox{$\mathrm{MeV}/{{\it c}^2}$}}
\newcommand{\MEVc}{\mbox{$\mathrm{MeV}/{\it c}$}}

\newcommand{\PPI}{\mbox{$\pi^{+}$}}
\newcommand{\MPI}{\mbox{$\pi^{-}$}}
\newcommand{\PMP}{\mbox{$\pi^{\pm}$}}
\newcommand{\PIo}{\mbox{$\pi^{0}$}}
\newcommand{\MPL}{\mbox{$\mu^{+}$}}
\newcommand{\KPL}{\mbox{$K^{+}$}}

\newcommand{\KL}{\mbox{$K_{L}$}}
\newcommand{\KPM}{\mbox{$K^{\pm}$}}
\newcommand{\KTAU}{\mbox{$\KPL \rightarrow \PPI \MPI \PPI$}}
\newcommand{\KTAUo}{\mbox{$\KPL \rightarrow \PPI \PIo \PIo$}}
\newcommand{\KTP}{\mbox{$\KPL \rightarrow \PPI \PIo$}}
\newcommand{\KMN}{\mbox{$\KPL \rightarrow \MPL \nu$}}
\newcommand{\PNN}{\mbox{$K \rightarrow \pi \nu \bar\nu$}}
\newcommand{\PNNP}{\mbox{$\KPL \rightarrow \PPI \nu \bar\nu$}}
\newcommand{\PNNL}{\mbox{$\KL \rightarrow \PIo \nu \bar\nu$}}
\newcommand{\Pll}{\mbox{$K^+\to\pi^+\ell^+\ell^-$}}
\newcommand{\PGG}{\mbox{$\PIo \rightarrow \gamma \gamma$}}
\newcommand{\ZVT}{\mbox{${\rm Z}_{\rm vertex}$}}

\begin{document}
\title{The beam and detector of the NA62 experiment at CERN}
\collaboration {The NA62 collaboration}
\abstract{ NA62 is a fixed-target experiment at the CERN SPS dedicated to measurements of rare kaon decays. Such measurements, like the branching  fraction of the $\PNNP$ decay, have the potential to bring significant insights into new physics processes when comparison is made with precise theoretical predictions. For this purpose, innovative techniques have been developed, in particular, in the domain of low-mass tracking devices. Detector construction spanned several years from 2009 to 2014. The collaboration started detector commissioning in 2014 and will collect data until the end of 2018. The beam line and detector components are described together with their early performance obtained from 2014 and 2015 data.}

March 27, 2017\\
\maketitle
\flushbottom
%%%%     to get authors list
\clearpage
\section*{The NA62 Collaboration\renewcommand{\thefootnote}{\fnsymbol{footnote}}\footnotemark[1]\renewcommand{\thefootnote}{\arabic{footnote}}}\label{sec:auth}
\input{authors.tex}
%%%%%% end authors list
\clearpage

%\begin{linenumbers}
\section{Introduction}\label{sec:intro}
\input{Sec1_v4.tex}
\section{Design and analysis principles}\label{sec:design}
\input{Sec2_v4.tex}
%\end{linenumbers}
%\resetlinenumber
%\begin{linenumbers}
\section{High intensity kaon beam line}\label{sec:K12-beam}
\input{Sec3-beam_v2.tex}
\clearpage
%\end{linenumbers}
%\resetlinenumber
%\begin{linenumbers}
\section{Kaon tagger (KTAG)}\label{sec:KTAG}
\input{Sec4-KTAG_v3.tex}
%\end{linenumbers}
%\resetlinenumber
%\begin{linenumbers}
\section{Beam spectrometer (GTK)}\label{sec:GTK}
\input{Sec5-GTK_v3.tex}
%\end{linenumbers}
%\resetlinenumber
%\begin{linenumbers}
\section{Charged Anti-coincidence detector (CHANTI)}\label{sec:CHANTI}
\input{Sec6-Chanti_v3.tex}
%\end{linenumbers}
%\resetlinenumber
%\begin{linenumbers}
\section{Straw spectrometer (STRAW)}\label{sec:STRAWSpec}
\input{Sec7-STRAW_v3.tex}
%\end{linenumbers}
%\resetlinenumber
%\begin{linenumbers}
\section{Photon-Veto system (PV) }\label{sec:PV}
\input{Sec8-PV_3.tex}
%\end{linenumbers}
%\resetlinenumber
%\begin{linenumbers}
\section{Ring Imaging Cherenkov counter (RICH)}\label{sec:RICH}
\input{Sec9-RICH_v2.tex}
%\end{linenumbers}
%\resetlinenumber
%\begin{linenumbers}
\section{Charged Particle Hodoscopes}\label{sec:CHOD}
\input{Sec10-CHOD_v2.tex}
%\end{linenumbers}
%\resetlinenumber
%\begin{linenumbers}
\section{Muon Veto system (MUV)}\label{sec:MUV}
\input{Sec11-MUV12_v5.tex}
\input{Sec11-MUV3_v2.tex}
%\end{linenumbers}
%\resetlinenumber
%\begin{linenumbers}
\section {Additional veto detectors} \label{sec:EXTRA}
\input{Sec12-Vetoes_v4.tex}
%\end{linenumbers}
%\resetlinenumber
%\begin{linenumbers}
\section{Trigger and Data Acquisition system (TDAQ)}\label{sec:TDAQ}
\input{Sec13_TDAQ_v3.tex}

%\end{linenumbers}
%\resetlinenumber
%\begin{linenumbers}
\section{Online system, Data handling and Control}\label{sec:online}
\input{Sec14_PCfarm_v2.tex}
\input{Sec14_RunC_v3.tex}
\input{Sec14_DCS_v3.tex}

%\end{linenumbers}
%\resetlinenumber
%\begin{linenumbers}
\section{Performance validation}\label{sec:valid}
\input{Sec15-perf2015_v3.tex}
%\end{linenumbers}
%\resetlinenumber
%\begin{linenumbers}
\section{Conclusion}
\input{conclusion_v1.tex}
%\end{linenumbers}
%\resetlinenumber
%\begin{linenumbers}
\section*{Acknowledgements}
\input{acknow.tex}

%\end{linenumbers}

\end{document}

%% file: authors.tex
%%%%
{\bf Universit\'e Catholique de Louvain, Louvain-la-Neuve, Belgium}\\
E.~Cortina Gil, E.~Mart\'in Albarr\'an\footnotemark[1], E.~Minucci, G.~N\"ussle\footnotemark[2], 
S.~Padolski\footnotemark[3], P.~Petrov, 
N.~Szilasi, B.~Velghe\footnotemark[4]\\[2mm]
%%%%%%%
{\bf University of Sofia ``St. Kl. Ohridski'', Sofia, Bulgaria}\\
G.~Georgiev, V.~Kozhuharov\footnotemark[5], L.~Litov \\[2mm]
%%%%%%%
{\bf Charles University, Prague, Czech Republic}\\
T.~Husek, K.~Kampf, M.~Zamkovsky \\[2mm]
%%%%%
{\bf Institut f\"ur Physik and PRISMA Cluster of Excellence, Universit\"at  Mainz, Mainz, Germany}\\
R.~Aliberti, K.~H.~Geib, G.~Khoriauli, K.~Kleinknecht, J.~Kunze, 
D.~Lomidze\footnotemark[6], 
R.~Marchevski, L.~Peruzzo, M.~Vormstein, R.~Wanke, A.~Winhart\footnotemark[7] \\[2mm]
%%%%%%
{\bf INFN, Sezione di Ferrara, Ferrara, Italy}\\
M.~Bolognesi, V.~Carassiti, S.~Chiozzi, A.~Cotta Ramusino, A.~Gianoli, R.~Malaguti \\[2mm]
{\bf Dipartimento di Fisica e Scienze della Terra dell'Universit\`a  e INFN, Sezione di Ferrara, Ferrara, Italy}\\
P.~Dalpiaz,  M.~Fiorini,  E.~Gamberini, I.~Neri,  A.~Norton, F.~Petrucci, M.~Statera, H.~Wahl \\[2mm]
%%%%%%
{\bf  INFN, Sezione di Firenze, Sesto Fiorentino, Italy}\\
F.~Bucci,  R.~Ciaranfi,  M.~Lenti, F.~Maletta, R.~Volpe \\[2mm]
{\bf Dipartimento di Fisica e Astronomia dell'Universit\`a e INFN, Sezione di Firenze, Sesto Fiorentino, Italy}\\
A.~Bizzeti\footnotemark[8], A.~Cassese\footnotemark[9], E.~Iacopini \\[2mm]
%%%%%%
{\bf Laboratori Nazionali di Frascati dell'INFN, Frascati, Italy}\\
A.~Antonelli, E.~Capitolo, C.~Capoccia, A.~Cecchetti, G.~Corradi, V.~Fascianelli, 
F.~Gonnella\footnotemark[7], G.~Lamanna\footnotemark[10], R.~Lenci, G.~Mannocchi, S.~Martellotti, M.~Moulson, 
C.~Paglia, M.~Raggi\footnotemark[11], V.~Russo, M.~Santoni, T.~Spadaro, D.~Tagnani, S.~Valeri, T.~Vassilieva \\[2mm]
{\bf INFN, Sezione di Napoli, Napoli, Italy}\\
F.~Cassese,  L. Roscilli \\[2mm]
{\bf Dipartimento di Fisica ``Ettore Pancini'' e INFN, Sezione di Napoli, Napoli, Italy} \\
F.~Ambrosino, T.~Capussela, D.~Di Filippo\footnotemark[12], P.~Massarotti, M.~Mirra,
M.~Napolitano, G. Saracino  \\[2mm]
%%%%%%%
{\bf INFN, Sezione di Perugia, Perugia, Italy}\\
M.~Barbanera\footnotemark[13], P.~Cenci, B.~Checcucci, V.~Duk,  L.~Farnesini, 
E.~Gersabeck\footnotemark[14], M.~Lupi\footnotemark[15], A.~Papi, M.~Pepe, M.~Piccini, G.~Scolieri  \\[2mm]
%Dipartimento di Fisca dell'Universit\`a e 
{\bf Dipartimento di Fisica e Geologia dell'Università e INFN, Sezione di Perugia,  Perugia, Italy} \\
D.~Aisa, G.~Anzivino, M.~Bizzarri, C.~Campeggi, E.~Imbergamo, A.~Piluso, C.~Santoni\footnotemark[16] \\[2mm]
%%%%%
{\bf INFN, Sezione di Pisa, Pisa, Italy}\\
L.~Berretta, S.~Bianucci, A.~Burato, C.~Cerri, R.~Fantechi,  S.~Galeotti\renewcommand{\thefootnote}{\fnsymbol{footnote}}\footnotemark[2]\renewcommand{\thefootnote}{\arabic{footnote}}, 
G.~Magazzu', M.~Minuti,  A.~Orsini, G.~Petragnani, L.~Pontisso, F.~Raffaelli,  F.~Spinella \\[2mm]
{\bf Scuola Normale Superiore e INFN, Sezione di Pisa, Pisa, Italy} \\
G.~Collazuol\footnotemark[17],  I.~Mannelli  \\[2mm]
{\bf Dipartimento di Fisica dell'Universit\`a e INFN, Sezione di Pisa, Pisa, Italy}\\
C.~Avanzini, F.~Costantini, L.~Di~Lella, N.~Doble, M.~Giorgi, S.~Giudici,  E.~Pedreschi,  
R.~Piandani, G. Pierazzini\renewcommand{\thefootnote}{\fnsymbol{footnote}}\footnotemark[2]\renewcommand{\thefootnote}{\arabic{footnote}}, 
J.~Pinzino, M.~Sozzi, L.~Zaccarelli \\[2mm]
%%%%%%
{\bf  INFN,  Sezione di Roma, Roma, Italy} \\
A.~Biagioni, E.~Leonardi, A.~Lonardo, P.~Valente, P.~Vicini  \\[2mm]
{\bf Dipartimento di Fisica,  Sapienza Universit\`a  e INFN, Sezione di Roma, Roma, Italy} \\
G.~D'Agostini \\[2mm]
%%%%%
{\bf  INFN, Sezione di Roma ``Tor vergata'', Roma, Italy}\\
R.~Ammendola, V.~Bonaiuto\footnotemark[18], N.~De~Simone\footnotemark[15], 
L.~Federici\footnotemark[19], A.~Fucci, G.~Paoluzzi, A.~Salamon, G.~Salina, 
F.~Sargeni\footnotemark[19] \\[2mm]
%%%%%%
{\bf INFN, Sezione di Torino, Torino, Italy}\\
C.~Biino, G.~Dellacasa, S.~Garbolino\footnotemark[20], F.~Marchetto, S.~Martoiu\footnotemark[21],  G.~Mazza, A.~Rivetti  \\[2mm]
 {\bf Dipartimento di Fisica Sperimentale dell'Universit\`a e INFN, Sezione di Torino, Torino, Italy}\\
 R.~Arcidiacono\footnotemark[22], B.~Bloch-Devaux,   M.~Boretto, 
L.~Iacobuzio\footnotemark[7], E.~Menichetti, D.~Soldi \\[2mm]
 %%%%%%
{\bf Universidad Aut\'onoma de San Luis Potos\'i, San Luis Potosí, Mexico}\\
J.~Engelfried, N.~Estrada-Tristan\\[2mm]
%%%%%
{\bf Horia Hulubei National Institute of Physics and Nuclear Engineering, Bucharest-Magurele, Romania}\\
A.~M.~Bragadireanu, O.~E.~Hutanu \\[2mm]
%%%%%%
{\bf Joint Institute for Nuclear Research, Dubna, Russia}\\
N.~Azorskiy,  V.~Elsha, T.~Enik, V.~Falaleev, L.~Glonti, Y.~Gusakov, S.~Kakurin, V.~Kekelidze,
S.~Kilchakovskaya, E.~Kislov, A.~Kolesnikov, D.~Madigozhin, M.~Misheva\footnotemark[23], 
S.~Movchan, I.~Polenkevich, Y.~Potrebenikov, V.~Samsonov, S.~Shkarovskiy, S.~Sotnikov,
L.~Tarasova, M.~Zaytseva, A.~Zinchenko\renewcommand{\thefootnote}{\fnsymbol{footnote}}\footnotemark[2]\renewcommand{\thefootnote}{\arabic{footnote}}\\[2mm]
%%%%%%
{\bf Institute for Nuclear Research -- Russian Academy of Science, Moscow, Russia}\\
V.~Bolotov\renewcommand{\thefootnote}{\fnsymbol{footnote}}\footnotemark[2]\renewcommand{\thefootnote}{\arabic{footnote}}, 
S.~Fedotov, E.~Gushin,  A.~Khotjantsev, A.~Khudyakov, A.~Kleimenova, 
Yu.~Kudenko\footnotemark[24], A.~Shaikhiev\\[2mm]
%%%%%% 
{\bf Institute for High Energy Physics -- State Research Center of Russian Federation, Protvino, Russia}\\
A.~Gorin,  S.~Kholodenko, V.~Kurshetsov, V.~Obraztsov, A.~Ostankov, V.~Rykalin, V.~Semenov, V.~Sugonyaev, O.~Yushchenko
 \\[2mm]
 %%%%%
{\bf Comenius University, Bratislava, Slovakia}\\
L.~Bician, T.~Blazek, V.~Cerny, M.~Koval\footnotemark[15], R.~Lietava\footnotemark[7] \\[2mm]
%%%%%
\clearpage
{\bf CERN, European Organization for Nuclear Research, Geneva, Switzerland}\\
G.~Aglieri~Rinella, J.~Arroyo~Garcia, S.~Balev\renewcommand{\thefootnote}{\fnsymbol{footnote}}\footnotemark[2]\renewcommand{\thefootnote}{\arabic{footnote}}, 
M.~Battistin, J.~Bendotti, F.~Bergsma, S.~Bonacini, F.~Butin, A.~Ceccucci,  P.~Chiggiato, 
H.~Danielsson, J.~Degrange, N.~Dixon, 
B.~Döbrich, P.~Farthouat,  L.~Gatignon, P.~Golonka, S.~Girod, A.~Goncalves~Martins~De~Oliveira, R.~Guida, F.~Hahn, E.~Harrouch, M.~Hatch, P.~Jarron\footnotemark[25], O.~Jamet, 
B.~Jenninger,  J.~Kaplon,
A.~Kluge, G.~Lehmann-Miotto, P.~Lichard, G.~Maire, A.~Mapelli, J.~Morant, M.~Morel, J.~No\"el,
M.~Noy, V.~Palladino\footnotemark[26], A.~Pardons, F.~Perez-Gomez, L.~Perktold, M.~Perrin-Terrin\footnotemark[27], P.~Petagna, K.~Poltorak, 
P.~Riedler, G.~Romagnoli, G.~Ruggiero\footnotemark[28], T.~Rutter, J.~Rouet, V.~Ryjov, A.~Saputi, 
T.~Schneider, G.~Stefanini, C.~Theis, S.~Tiuraniemi, %\todo {Tiuraniemi new affi?}, 
F.~Vareia Rodriguez, S.~Venditti\footnotemark[29], M.~Vergain, H.~Vincke,  P.~Wertelaers \\[2mm]
%%%%%%
{\bf University of Birmingham,  Birmingham, United Kingdom}\\
M.~B.~Brunetti, S.~Edwards, E.~Goudzovski, B.~Hallgren, M.~Krivda, C.~Lazzeroni, N.~Lurkin, D.~Munday, 
F.~Newson\footnotemark[30], C.~Parkinson, S.~Pyatt, A.~Romano, X.~Serghi, A.~Sergi,  R.~Staley, 
A.~Sturgess  \\[2mm]
%%%%%%%
{\bf University of Bristol, Bristol, United Kingdom}\\
H.~Heath, R.~Page \\[2mm]
%%%%%%%
{\bf University of Glasgow, Glasgow, United Kingdom}\\
B.~Angelucci\footnotemark[29], D.~Britton, D.~Protopopescu, I.~Skillicorn \\[2mm]
%%%%%%%
{\bf University of Liverpool, Liverpool, United Kingdom}\\
P.~Cooke, J.~B.~Dainton, J.~R.~Fry, L.~Fulton, D.~Hutchcroft, E.~Jones, T.~Jones, 
K.~Massri\footnotemark[15],
E.~Maurice\footnotemark[31], K.~McCormick, P.~Sutcliffe, B.~Wrona \\[2mm]
%%%%%
{\bf George Mason University, Fairfax, U.S.A.}\\
A.~Conovaloff, P.~Cooper, D.~Coward\footnotemark[32], P.~Rubin \\[2mm]
%%%%%%
{\bf University of California Merced,  Merced, U.S.A.}\\
R.~Winston \\[2mm]
%%%%%%
\setcounter{footnote}{0}
\renewcommand{\thefootnote}{\fnsymbol{footnote}}
\footnotetext[1]{Corresponding author: NA62 Editorial Board, email: na62eb@cern.ch}
\footnotetext[2]{Deceased}
\renewcommand{\thefootnote}{\arabic{footnote}}
% Martin
\footnotetext[1]{Now at iC-Haus GmbH, Bodenheim, Germany}
% Nüssle
\footnotetext[2]{Now at AVO/ADAM S.A., Geneva, Switzerland}
%Padolski
\footnotetext[3]{Now at Brookhaven National Laboratory, Upton, U.S.A.}
%Velghe
\footnotetext[4]{Now at TRIUMF, Vancouver, Canada}
% Venelin
\footnotetext[5]{Also at Laboratori Nazionali di Frascati dell'INFN, Italy}
%Kunze
%\footnotetext[4]{Now at ...,Germany}
% Lomidze
\footnotetext[6]{Now at Universit\"at Hamburg, Hamburg, Germany}
% Gonnella, Winhart, Iacobuzio
\footnotetext[7]{Now at University of Birmingham, United Kingdom}
% Bizzeti
\footnotetext[8]{Also at Universit\`a di Modena e Reggio Emilia, Modena, Italy}
% Cassese
\footnotetext[9]{Now at Altran, Italy}
%Lamanna
\footnotetext[10]{Now at Universit\`a di Pisa e NFN, Sezione di Pisa, Pisa,Italy}
% Raggi
\footnotetext[11]{Now at Sapienza Universit\`a di Roma, Roma, Italy}
%Di Filippo
\footnotetext[12]{Now at Info Solution S.p.A., Milano, Italy}
% Barbanera
\footnotetext[13]{Now at INFN, Sezione di Pisa, Pisa, Italy}
%Gersabeck
\footnotetext[14]{Now at Ruprecht-Karls-Universit\"at Heidelberg, Germany}
% Lupi, DeSimone, Massri
\footnotetext[15]{Now at CERN, Geneva, Switzerland}
% Santoni 
\footnotetext[16]{Now at Imagination Technologies, Kings Langley, United Kingdom}
%Collazuaol
\footnotetext[17]{Now at Dipartimento di Fisica dell'Universit\`a e NFN,  Sezione di Padova, Padova, Italy}
%Bonaiuto
\footnotetext[18]{Also at Dipartimento di Ingegneria Industriale dell'Universit\`a di Roma ``Tor Vergata", Roma, Italy}
% Federici,Sargeni
\footnotetext[19]{Also at Dipartimento di Ingegneria Elettronica dell'Universit\`a di Roma ``Tor Vergata", Roma, Italy}
% Garbolino
\footnotetext[20]{Also at  Waters Corporation, Milford (MA), U.S.A.}
%Martoiu
\footnotetext[21]{Now at IFIN-HH, Bucharest, Romania}
%Arcidiacono
\footnotetext[22]{Also at Universit\`a degli Studi del Piemonte Orientale, Vercelli, Italy}
%Misheva
\footnotetext[23]{Now at Institute of Nuclear Research and Nuclear Energy of Bulgarian Academy of Science (INRNE--BAS), Sofia, Bulgaria}
%Kudenko
\footnotetext[24]{Also at National Research Nuclear University (MEPhI), Moscow, and 
Institute of Physics and Technology, Moscow, Russia}
% Jarron
\footnotetext[25]{Now at Universit\`a degli Studi di Torino, Torino, Italy}
% Palladino
\footnotetext[26]{Now at Imperial College, London, United Kingdom}
%Perrin-Terrin
\footnotetext[27]{Now at Universit\'e Catholique de Louvain, Louvain-La-Neuve, Belgium}
%Ruggiero
\footnotetext[28]{Now at University of Liverpool, Liverpool, United Kingdom}
%Venditti, Angelucci
\footnotetext[29]{Now at CAEN, Viareggio, Italy}
% newson
\footnotetext[30]{Now at Tessella, Burton upon Trent, United Kingdom}
%Maurice
\footnotetext[31]{Now at Laboratoire Leprince Ringuet, Palaiseau, France} 
%Coward
\footnotetext[32]{Also at SLAC, Menlo Park (CA), U.S.A.}
%%%%%%

%% file: Sec1_v4.tex
\subsection{Physics motivation}
Investigation of quark mixing and CP violation in $K$ and $B$ meson decays has been one of the most active areas of high-energy physics in past decades. The present knowledge
of the Cabibbo-Kobayashi-Maskawa quark-mixing matrix~\cite{Cabibbo:1963yz,Kobayashi:1973fv}
is based on both precise experimental measurements  and theoretical calculations.

In the Standard Model (SM), charged weak currents are responsible for the transitions between quarks. The properties of these transitions have significant implications. For instance, 
the conservation of probability leads to the cancellation of flavour-changing neutral currents (FCNC) at tree level, 
known as the GIM mechanism~\cite{Glashow:1970gm}. This suppression makes the observation of FCNC a sensitive test of the SM: any deviation from predictions would give a clear sign of physics beyond the SM.

Within the SM, one can determine fundamental parameters such as quark masses, mixing parameters and phenomenological quantities (e.g. decay constants and form factors)  allowing the interpretation of the
observed properties of hadrons in terms of
the fundamental quark constituents. In another approach, one can fix the SM parameters using theoretical and phenomenological determinations to make firm predictions and look for deviations in the data. The latter approach works particularly well for kaons because lattice QCD and Chiral Perturbation Theory are excellent tools in this domain.
While the energy frontier is limited by the reach of the Large Hadron Collider (LHC) in terms of centre-of-mass energy, no such limitation exists in principle for rare decays, making them a highly valued complementary approach in the search for new phenomena.

The theoretical \cite{Buras:2015qea} and experimental \cite{Artamonov:2009sz,Ahn:2009gb} 
 status of the most interesting rare kaon decays is displayed in Table~\ref{Tab:Kpinn}. The gap between theory and experiment is striking: 
the main goal of the NA62 experiment~\cite{Anelli:2005ju} is to match the 10\% theory precision for the $\PNNP$ decay rate. 

\begin{table}[hb]
 \setlength{\tabcolsep}{3ex}
 \caption{Expected and measured branching fractions of $\PNN$  decays.}
 \label{Tab:Kpinn}
 \vspace{1ex}
 \centering
 \begin{tabular*}{0.90\textwidth}{@{\extracolsep{\fill} }  p{4.5cm} l l }
 \hline\hline
 \hspace{1ex}Decay mode & \multicolumn{2}{c}{Branching ratio ($\times 10^{10}$)} \\
 &    Theory \cite{Buras:2015qea} & Experiment \cite{Artamonov:2009sz,Ahn:2009gb}  \\
 \hline
 \hspace{0ex} $\PNNP$  & $0.84\pm 0.10$& $1.73^{+1.15}_{-1.05}$\\
 \hspace{0ex} $\PNNL$  & $0.34\pm 0.06$& $< 260$ (90\% CL)\\
 \hline\hline
 \end{tabular*}
 \end{table}
 
\subsection{Performance requirements} \label{ssec:req}
\begin{figure}[ht]
\begin{minipage}{0.5\linewidth}
\begin{center}
\includegraphics[scale=0.40]{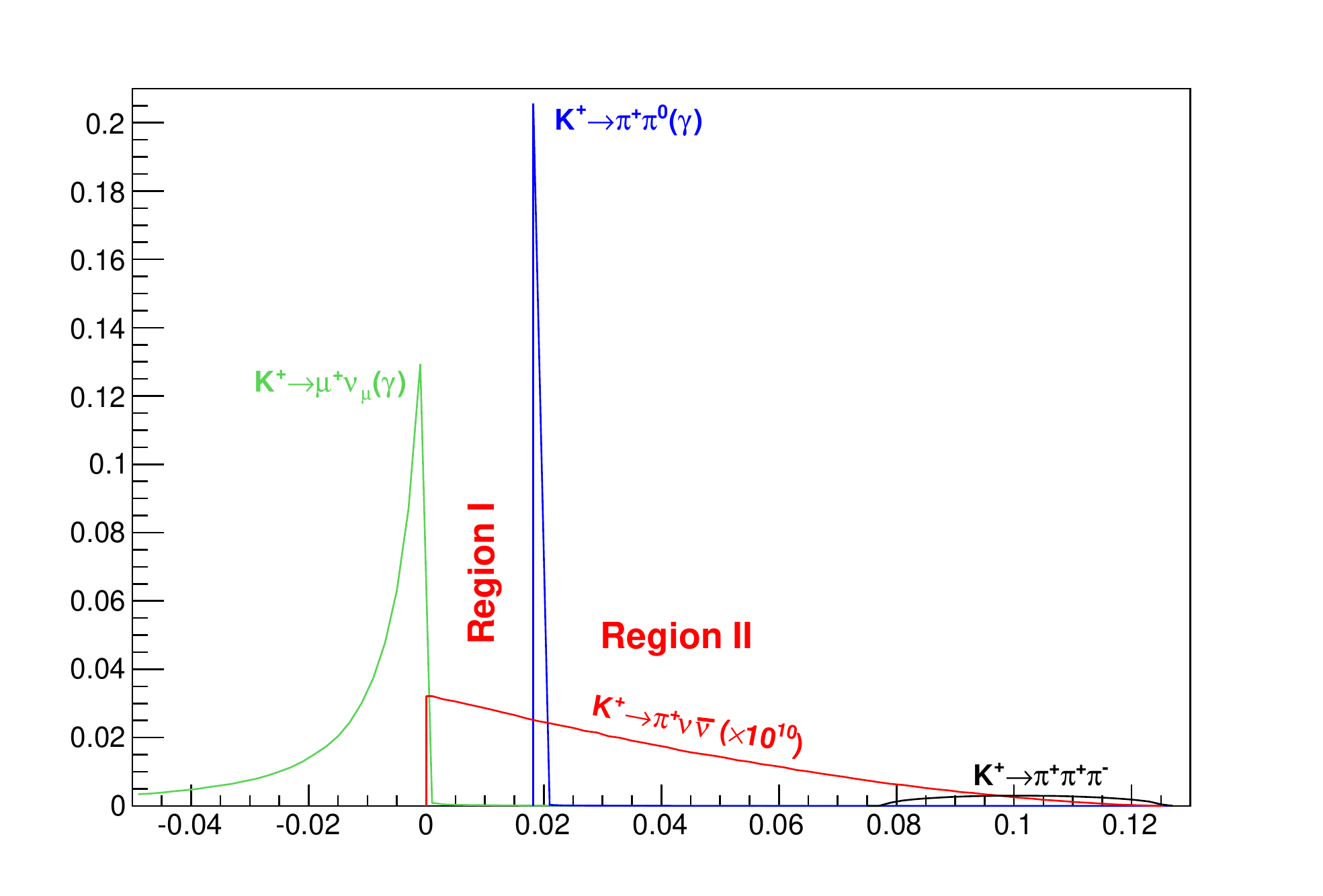}
\put(-95,0){$ m^{2}_{\rm miss}$ [GeV$^2 / c^4$]}
\put(-232,60){\rotatebox{90}{$1 / \Gamma_{\rm tot}$ ~d$\Gamma$ / d$m^{2}_{\rm miss}$}}
\end{center}
\end{minipage}
\begin{minipage}{0.5\linewidth}
\begin{center}
\includegraphics[scale=0.40]{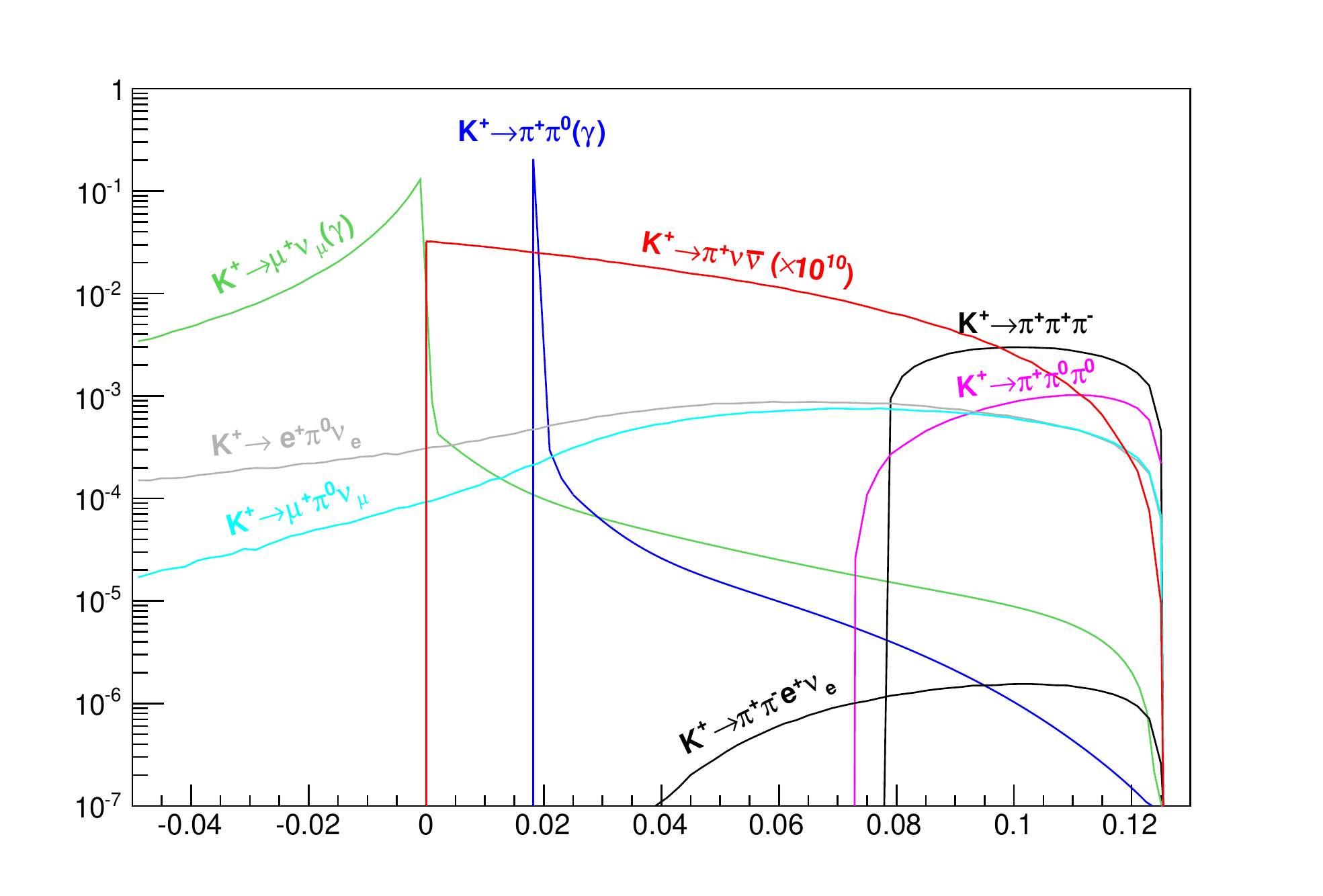}
\put(-95,0){$ m^{2}_{\rm miss}$ [GeV$^2 / c^4$]}
\put(-232,60){\rotatebox{90}{$1 / \Gamma_{\rm tot}$ ~d$\Gamma$ / d$m^{2}_{\rm miss}$}}
\end{center}
\end{minipage}
\caption{\label{fig:mmiss2} Distribution of the $m^{2}_{\rm miss}$ variable for  kaon decay modes with the largest contribution (left, linear scale);  all decay modes (right, log scale).}
\end{figure}

Collecting $\sim$100 $\PNNP$  events (based on the assumption of the SM branching ratio) and assuming a 10\% detector acceptance  means an exposure to $10^{13}$
kaon decays and achieving a rejection factor of $10^{12}$ for the other decay modes.
The signature of the $\PNNP$ signal is very simple: one incoming kaon decays to a single-detected charged-pion track, and is weakly constrained compared to all other kaon decays.
Kinematic rejection of the most abundant kaon decay modes is obtained by selecting two restricted 
regions (Region I and Region II, as shown in \Fig{fig:mmiss2}-left) of the $m^{2}_{\rm miss}$ distribution defined as:
\begin{equation}\label{eq:mmis2}
  m^{2}_{\rm miss} = (P_{K} - P_{\pi})^2 ,
\end{equation} 
where $P_{K}$ is the 4-momentum of the parent particle, assumed to be a kaon and $P_{\pi}$ is the 4-momentum of the decay particle, assumed to be a pion.  

The control of the resolution of the $\KTP$ peak  at the level  of 0.001 $(\GEVcc)^2$  is mandatory to 
obtain  a kinematic  rejection factor of $\cal{O}$$(10^4 - 10^5 )$  as seen in \Fig{fig:mmiss2}.
To achieve this resolution, the  momentum and angle of the kaon must be measured with a precision of $\sim$0.2\% and 0.016 mrad, respectively, while the momentum  and angular resolutions for the 
downstream track  must be better than 1\% and 0.060~mrad, respectively.

Charged-particle identification  (to reject muons with respect to pions)  and photon rejection  (to
 reject  $\PIo$ decays to photons and radiative decays), of $\cal{O}$$(10^7 )$  and 
 $\cal{O}$$(10^8  )$, respectively,  should complement the background rejection in  Region I and Region II.
  
Precise time matching between the incident kaon and downstream track  at  the 100~ps level, is essential to reject accidental coincidences when operating at an intense hadron beam flux (\Sec{sec:design}).

%% file: Sec2_v4.tex
\subsection{Choice of detector layout}
The choice of the decay-in-flight technique is motivated  by the possibility of obtaining an integrated flux of  $\cal{O}$$(10^{13} )$ kaon decays over a few years of data-taking with a signal acceptance 
of about 10\%, leading to the collection of  about 100 SM events in the $\PNNP$ channel.

The 400 $\GEVc$ proton beam from the CERN SPS accelerator  enables the production of a 
75~$\GEVc$ secondary kaon beam.
The advantage of using a high-energy proton beam is the reduction of non-kaon-related accidental background due to the higher kaon production cross section.
The disadvantage of high-energy protons and, consequently, of a high-energy secondary beam,
is that pions and protons cannot be separated efficiently from kaons. 
 During the SPS spill of 3~s effective duration, the particle rate in the NA62 positive secondary  hadron beam is 750~MHz, of which about 6\% is from $K^+$, leading to 5~MHz of $K^+$ decays in the 65 m long decay region.
Therefore the upstream detectors that measure the momentum and the direction of the incoming kaons are exposed to a particle flux about 16 times larger than the kaon flux.  Note that 75\% of the kaons do not decay before hitting the beam dump at the end of the beam line .
The \textbf{kaon beam line} properties are described in \Sec{sec:K12-beam}.
 
\begin{figure}[ht]
\includegraphics[width=1.0\linewidth]{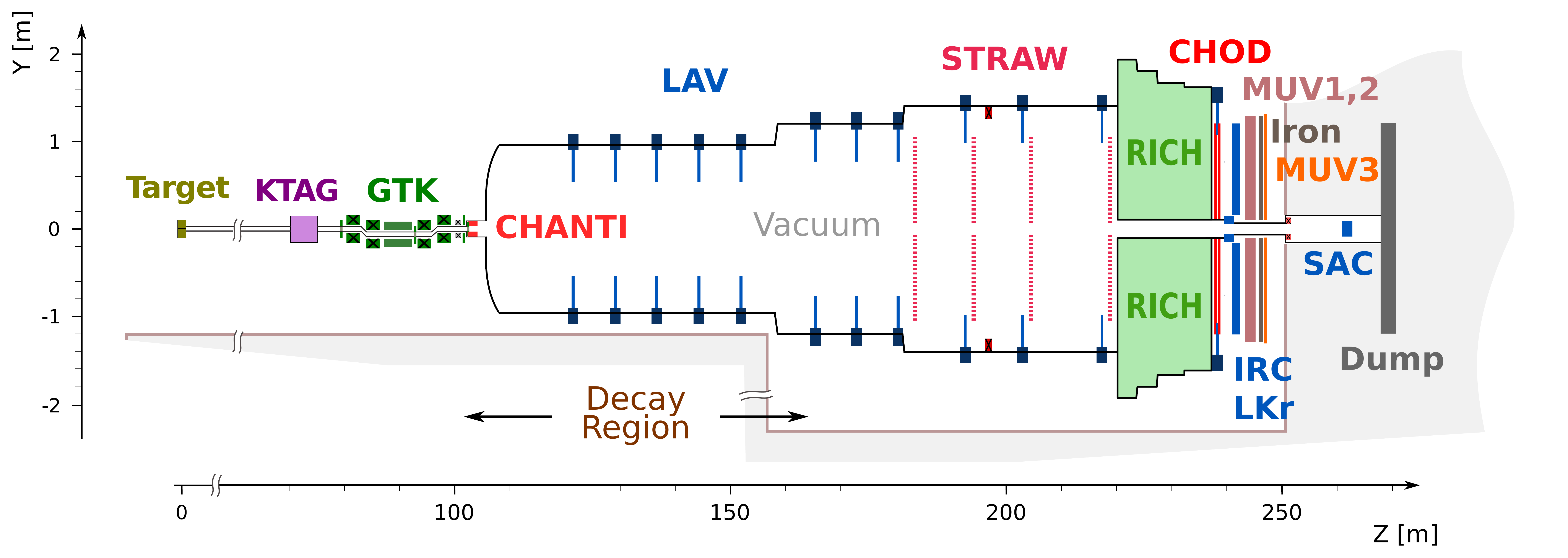}
 \caption{\label{fig:detectorview1}
 Schematic vertical  section through the NA62 experimental setup. The main elements for the detection of 
 the  $\KPL$ decay products are located along a 150~m long region starting 121~m downstream of
the kaon production target. Useful $\KPL$ decays are detected in a 65~m long 
decay region. Most detectors have an approximately cylindrical shape around the beam axis.
An evacuated passage surrounding the  beam trajectory allows the intense (750 MHz) flux of un-decayed 
beam particles to pass through without interacting with detector material before reaching the dump.}
\end{figure}

The scale and reference system for the experimental layout are displayed  in 
\Fig{fig:detectorview1}.  The beam line defines the Z axis with its origin at the kaon production target and beam particles travelling in the positive direction, the Y axis points vertically up, and the X axis is horizontal and directed to form a right-handed coordinate system.

Following breaks in the vacuum to accommodate some of the beam line elements, the beam spectrometer GTK and all detectors surrounding the decay region as well as
 the spectrometer detecting the final-state particles are placed in vacuum 
to avoid interactions and scattering of the beam and to obtain improved resolution for measured kinematic quantities.
The time matching between the incoming kaon and the outgoing charged particle is  essential to keep 
the correct association probability greater than 99\% and must be kept under control at the level of 100--150~ps.
Kaon identification is provided by a CEDAR differential Cherenkov  counter equipped with 
a photon detection system \textbf{KTAG}. \Sec{sec:KTAG} gives detailed information  on the design and performance of the KTAG .

The beam spectrometer \textbf{GTK} consists of three silicon pixel  stations providing momentum 
and direction measurements of the incoming kaons. A low-mass tracking system is 
essential  to minimize inelastic scattering of beam particles in the detector material that could  mimic an 
isolated outgoing charged particle from a decay. Information on the  
design and performance of the beam tracker is found in \Sec{sec:GTK}. The guard ring detector 
\textbf{CHANTI} (for CHarged Anti-coincidence detector), installed downstream of the GTK to detect inelastic scattering interactions in the 
last station, is described in \Sec{sec:CHANTI}.

 Downstream of the decay region, the \textbf{STRAW} tracker measures 
 the trajectories and momenta of the charged products of $\KPL$ decays.
 To minimize multiple scattering, the straw chambers, which are constructed of ultra-light material, are installed inside the vacuum. The tracker consists of four chambers and a large-aperture dipole magnet (MNP33) providing 
a momentum kick to charged particles of $270~\MEVc$ in the horizontal plane. A description of the spectrometer design and  performance is given in \Sec{sec:STRAWSpec}.

 A system of photon-veto detectors provides hermetic coverage for photons produced in the decay region and propagating at angles up to about 50~mrad with respect to the detector axis.
  These detectors (\Sec{sec:PV}) include, from larger to smaller angular coverage:
  \begin{itemize}
  \item a series of 12 annular Large-Angle photon Veto detectors (\textbf {LAV)};
  \item the Liquid Krypton electromagnetic calorimeter (\textbf{LKr});
  \item the Intermediate-Ring (\textbf{IRC}) and Small-Angle (\textbf{SAC}) calorimeters.
  \end{itemize}

The Ring Imaging CHerenkov detector \textbf{RICH} situated downstream of the last straw chamber includes a 17~m long radiator volume filled with neon gas at one atmosphere and ensures the separation of electrons, muons, pions, and kaons. 
Details of its design and performance are given in \Sec{sec:RICH}.

The RICH detector is followed by a system of hodoscope counters \textbf{CHOD} 
constructed from scintillator slabs and tiles (\Sec{sec:CHOD}). The counters provide a time resolution of $\sim$150~ps, which is precise enough to define the reference trigger time for other detectors.

Two hadronic calorimeters (MUV1, MUV2) and a plane of scintillating tiles (MUV3) behind 80~cm of iron form a pion/muon identification system, as explained in \Sec{sec:MUV}.

Additional counters installed at optimized locations provide hermetic coverage for charged particles produced in multi-track kaon decays; they are described in \Sec{sec:EXTRA}.

The detectors are operated and interconnected by a high-performance trigger and data acquisition system (\textbf{TDAQ}, \Sec{sec:TDAQ}). The \textbf{Online} Data handling, Control  and safety aspects are documented in \Sec{sec:online}.

Lastly, \Sec{sec:valid} focuses on the 
validation of performances obtained using data recorded in 2014--2015.

\subsection{Data samples for performance studies} \label{ssec:samples}
Specifically  selected data samples are used to 
evaluate the detector performance. Two types of beam configurations are employed for this purpose: 
1) standard ``kaon runs'', using the beam line as described in \Sec{sec:K12-beam}, 
and 2) special ``muon runs'',  where the slits in the collimators of the first beam line achromat (TAX1,2) 
are closed to dump the secondary beam and to generate a relatively pure flux of muons illuminating almost uniformly the downstream detectors.

The analysis of kaon run data requires further event selection to identify and exploit particular decay modes, like $\KTAU$, $\KMN$ and $\KTP$.

\paragraph{Three-track event samples:}
The $\KTAU$ decay offers a tool to fine tune the calibration of the straw spectrometer by comparing the reconstructed three-pion mass
with the Particle Data Group (PDG) \cite{PDG2014} kaon mass, to measure the average kaon direction and momentum, to study the properties of the beam and to calibrate the GTK beam spectrometer.  
The selection of $\KTAU$ decays requires at least three tracks reconstructed in the straw spectrometer.
A least-squares vertex fit including the effect of the residual magnetic field measured in the decay region defines the position of the origin of the three tracks (decay vertex). Cuts on the longitudinal position of the decay vertex, typically between 105 and 180 m from the target, 
and on the fit quality are used to select the final sample.   

\paragraph{Single-track event samples:}
Single track decays,  as in $\KMN$ and $\KTP$ modes, are used to study particle identification and photon rejection.
 These events are selected by requiring a single track reconstructed in the straw spectrometer  and a CHOD signal geometrically compatible to within 50~mm of the track extrapolation.
 The CHOD  signal is the time reference for the charged particle in the final state. The squared missing mass is the main  kinematic variable in the analysis. It is  defined as      
\begin{equation}\label{eq:mmis}
  m^{2}_{\rm miss} = (P_K - P_{\rm particle})^2 ,
\end{equation} 
where $P_{K}$ denotes the 4-momentum of the parent particle assumed to be a kaon and 
$P_{\rm particle}$ the 4-momentum of the decay particle assumed to be a muon or a pion, depending on the selection.  $P_{K}$ is obtained from the average kaon momentum and direction, as measured from $\KTAU$. 

The closest distance of approach ($CDA$) between 
the track reconstructed in the STRAW spectrometer and the nominal kaon direction defines the position of the kaon decay vertex. Cuts on $CDA$ and the longitudinal vertex coordinate  ($\ZVT$) are employed to select kaon decays, thus rejecting the beam-halo contribution.
Studies that use data samples with very low background contamination can also use
the event-by-event kaon momentum and direction measured by the GTK tracker. In this case the $CDA$ between 
the tracks measured in the STRAW spectrometer and in the GTK tracker is used together with timing information to match the initial- and final-state tracks. Typical cuts are $CDA<25 (10)$~mm and 
$125(105)< \ZVT <165$~m from the target  for events reconstructed without (with) the GTK. 

To select $\KMN$ decays, the muon mass is assigned  to $P_{\rm particle}$ in \Equ{eq:mmis} and a cut 
around the square of the corresponding neutrino mass is used to  select the sample kinematically. 
A typical cut compatible with the kinematic resolution is $|m^{2}_{\rm miss}|<0.01 ~(\GEVcc)^2$. 
Depending on the type of study performed, additional conditions are required to clean the selected 
muon sample further: 
KTAG signals compatible with a kaon in time with the CHOD; no signals compatible with photons in LKr, LAV, IRC and SAC; presence of hits in MUV3 to study MUV1,2 performance; calorimetric conditions selecting minimum ionizing particles to study MUV3 response; combined information from MUV3 and calorimeters to study muon identification in RICH. 

To select $\KTP$ decay candidates, the charged-pion mass hypothesis is used in  \Equ{eq:mmis} 
together with the condition $0.012<m^{2}_{\rm miss}<0.025 ~(\GEVcc)^{2}$. 
Additional requirements may include a   signal from the KTAG in time with $\PPI$; no MUV3 hits in time 
with the  $\PPI$; use of the photon detectors to select different $\PGG$ topologies; calorimetric 
conditions in LKr, MUV1, and MUV2 for pion identification to study, for example, the RICH 
performance. 

\paragraph{Calorimeter-selected event samples:} 
Another strategy to select $\KTP$ samples makes use of the LKr electromagnetic calorimeter to 
reconstruct the $\PIo$ without employing the straw spectrometer for $\PPI$ reconstruction. This selection is used to study the straw spectrometer performance.
In this case, events with at least two clusters 
of energy deposition in the LKr calorimeter with energies above 3 GeV are selected. The longitudinal coordinate of the decay vertex ($\ZVT$) is computed from the cluster energies and positions
under the hypothesis that they are produced by photons from the decay of a $\PIo$ . Events are selected if exactly one pair of clusters satisfies the condition  
$115< \ZVT <170$~m from the target. The position of the vertex allows the 4-momentum of the 
$\PIo$ ($P_{\pi^0}$) to be computed, and the expected momentum of the $\PPI$ is derived using the average kaon momentum and direction measured from $\KTAU$. 
A LKr cluster  and a CHOD signal  associated with the expected position of the $\PPI$ impact point 
are required  to clean up the sample further. A cut on the squared missing mass defined by \Equ{eq:mmis}  (where $P_{\rm particle} = P_{\pi^0}$)
around the squared mass of the charged pion is the final condition and reduces the
background below the percent level. 

%% file: Sec3-beam_v2.tex
\begin{figure}[ht]
\begin{center}
\includegraphics[width=0.8\linewidth]{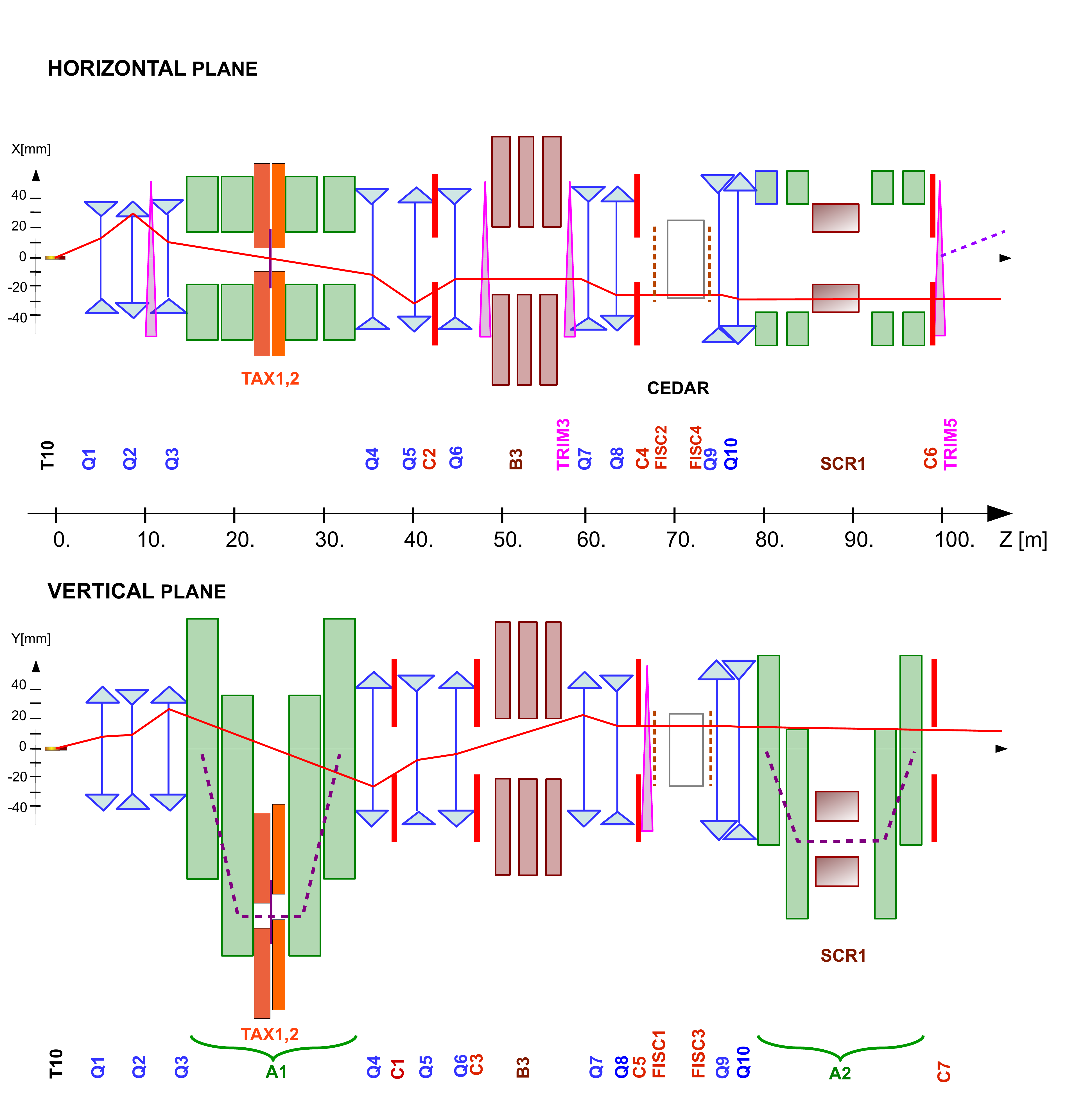} 
\end{center}
\caption{\label{fig:K12beam-schematic} Schematic layout and optics of the high-intensity $\KPL$  beam from the T10 target to the entrance of the decay region. In each view, the solid line corresponds to  the trajectory of a particle leaving the target from the centre at nominal momentum and at the angle indicated. The dashed line indicates the trajectory of an initially on-axis 75 $\GEVc$ momentum particle .
}
\end{figure}

The primary proton beam is extracted at 400~$\GEVc$ from the CERN SPS
accelerator and directed via the P42 beam line to the T10 target (400 mm long, 2 mm diameter beryllium) located in a  tunnel connecting the SPS to the underground experimental hall
\cite{Doble:1977}. 
A ``straight-line'' layout, joining the  
target station T10 to the centre of the existing LKr calorimeter, has been adopted for the positively charged secondary beam.
This secondary, high-intensity hadron beam (K12) is derived from the T10 target at a  central momentum of $+75~\GEVc$, chosen both in sign and magnitude to maximize the  fraction of kaons with respect to the flux of incident protons and to other hadrons in the beam. 
A front-end momentum selection and a downstream momentum measurement are then each performed by an ``achromat'' consisting of four dipole magnets. The first pair of magnets in each achromat produces a parallel displacement of the beam, while the second pair returns it to the undeviated axis. 
The advantages of such a design are that it saves longitudinal space, is favourable for sweeping away muon background and keeps open possible future options of providing either two simultaneous, oppositely charged  $\KPM$ beams  
(as for the NA48/2 experiment  \cite{na482:ag}) or a neutral $\KL$ beam (as for the NA48 experiment  \cite{NA48:2007}).
The K12 beam line  has a length of 101.3 m up to the final collimators (C6, C7).  \Fig{fig:K12beam-schematic} schematically shows the beam optics, calculated using the 
 program {\tt TRANSPORT} \cite{Brown:1980}. 

\subsection{K12 Beam line layout}\label{ssec:beam-layout}

The T10 target is immediately followed by a 950~mm long, water-cooled, copper collimator,
offering a choice of bores of different apertures; a 15~mm diameter hole
is generally selected to transmit the desired secondary particles. The
first active elements of the high-intensity beam are a triplet of
radiation-hard, small-aperture, quadrupole magnets (Q1, Q2, Q3),
which collect a large solid angle acceptance ($\pm$ 2.7~mrad horizontally and  $\pm$1.5~mrad 
vertically) at 75 $\GEVc$ central momentum (\Fig{fig:K12beam-schematic}). Shortly
downstream follows a front-end achromat (A1) to select the beam of 75~$\GEVc$  
with a 1\% rms momentum bite.
The achromat consists of four vertically-deflecting dipole magnets. The first two  magnets produce a parallel downward displacement of the beam by 110~mm, while the following two magnets return the beam onto the original axis. In between, the beam passes through a set of
graduated holes in two motorized and water-cooled beam-dump units, TAX1 and TAX2, to make the momentum selection 
whilst absorbing the remaining primary proton beam and unwanted secondary particles (\Fig{fig:K12beam-schematic}). 

At a double (horizontal and vertical) focus between TAX1 and TAX2, a ``radiator'' consisting of an arrangement of tungsten plates with a choice of  thickness up to 5~mm $(1.3~X_{0})$ is introduced into the beam. It is optimized to cause positrons to lose sufficient energy by Bremsstrahlung for them to be subsequently rejected, whilst minimizing the loss of hadrons by scattering.

A following triplet of quadrupoles (Q4, Q5, Q6) serves to refocus the beam in the vertical plane and to render it parallel with limited width in the horizontal plane. The drift-space between these quadrupoles  is occupied by  two collimators (C1, C2), which redefine the vertical and horizontal acceptance of the transmitted beam.  A subsequent collimator (C3) redefines the beam at the second focus in the vertical plane. 
At this point the positrons that have been degraded in momentum by the radiator between TAX1 and TAX2 are sufficiently separated from the hadron beam to be absorbed in the C3 collimator.

The beam  then passes  through a  40~mm diameter, almost field-free bore, in iron plates which are inserted between the poles of  three 2 m long dipole magnets (B3).  The vertical magnetic field in the iron surrounding the beam serves to sweep aside muons of both signs, whilst the deviation of the beam due to the small stray-field inside the bore is cancelled by two steering dipoles (TRIM 2 and TRIM 3 before and after B3).

A differential Cherenkov counter (CEDAR) \cite{Bovet:1982} equipped with 8 new arrays of photodetectors (KTAG)  serves to identify the $\KPL$ in the beam (\Sec{sec:KTAG}). This requires the beam to be rendered parallel, for which purpose the CEDAR is preceded  by two quadrupoles  (Q7, Q8), as well as  by horizontal and vertical cleaning collimators (C4, C5) to absorb particles in the tails of the beam.  

\begin{figure}[ht]
\includegraphics[width=1\linewidth]{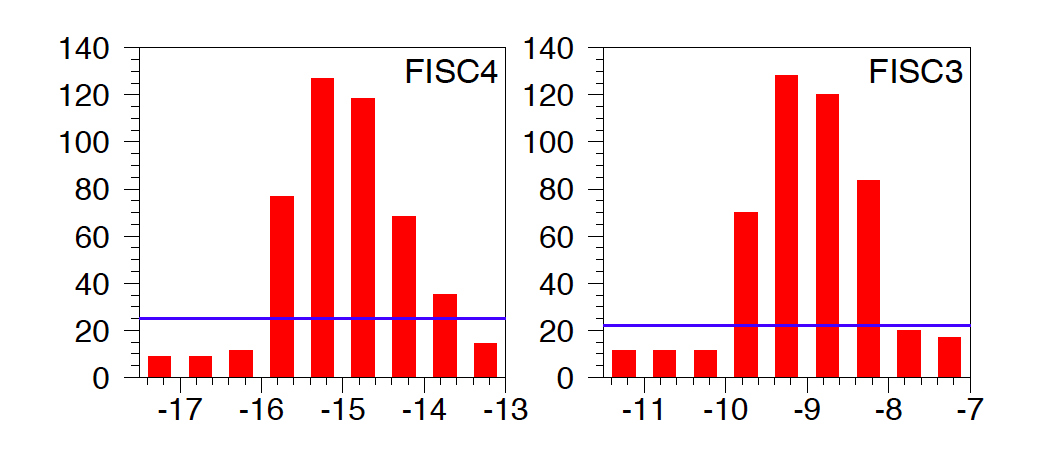}
\put(-258,5){ X [mm]}
\put(-372,168){ mean  $-14.8$ mm}
\put(-372,158){ rms 0.5 mm}
\put(-420,78){\rotatebox{90}{counts [arbirary unit]}}
\put(-62,5){ Y [mm]}
\put(-178,168){ mean  $-8.9$ mm}
\put(-178,158){ rms 0.5 mm}
\caption{\label{fig:FISC34}Profiles of particles in the horizontal (FISC4) and vertical (FISC3) plane of 
the FISC filament counter downstream of the CEDAR that have also traversed the corresponding upstream FISC counter. The mean and rms values shown correspond to the bins in the peak region (coincidence rate above the threshold line), the rms widths of 
$\Delta{\rm X} = \Delta{\rm Y }= 0.5$ mm correspond to  angular divergences of 0.07~mrad. }
\end{figure}
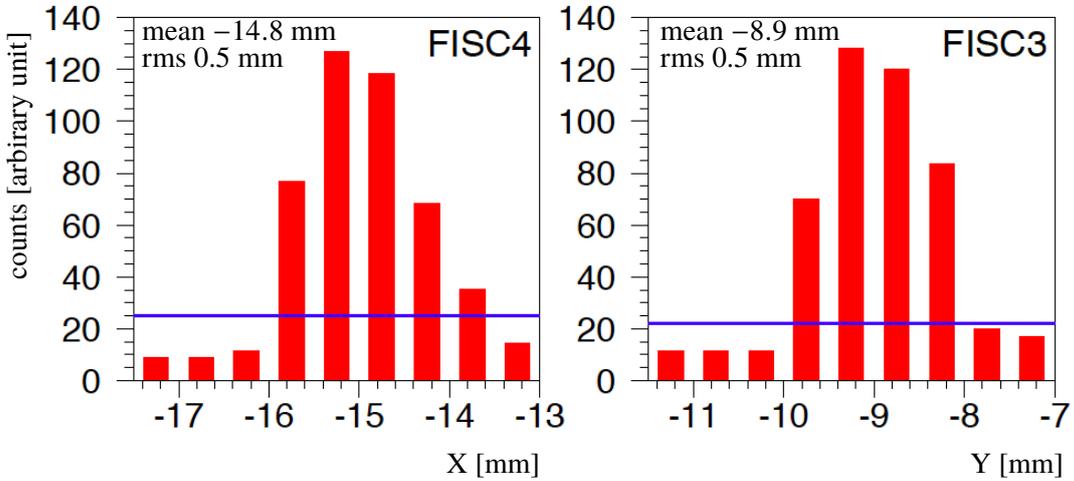

Two pairs (vertical and horizontal) of filament scintillator counters (FISC 1, 3 and FISC 2, 4) are installed up- and downstream of the CEDAR. When connected in coincidence respectively, they permit the mean divergence of the beam to be measured and tuned to zero and the remaining, intrinsic divergence to be verified in each plane (\Fig{fig:FISC34}). 

\begin{figure}[ht]
\begin{center}
\includegraphics[width=0.9\linewidth]{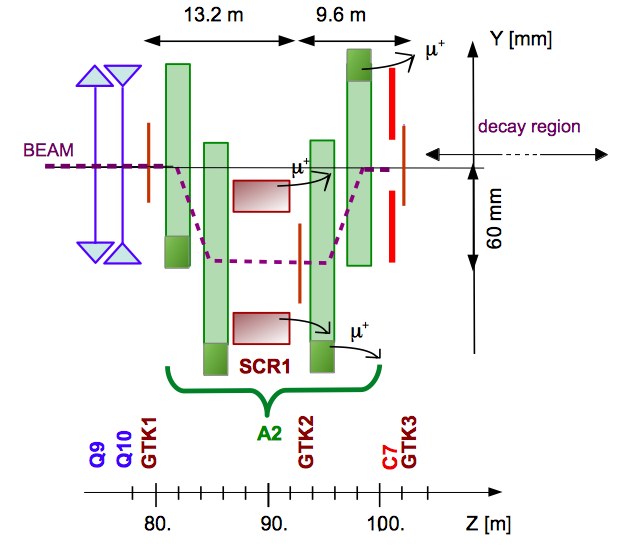}
\end{center}
\caption{\label{fig:achromat2} Schematic layout of the beam tracking and momentum measurement in the second achromat (A2). The beam is deflected vertically by 60~mm and returned to its nominal direction after the momentum measurement. Muons are swept away by the scraper SCR1 and the return yokes of the last two C-shaped magnets of the achromat (dark shaded areas).} 
\end{figure}

Following the CEDAR, a doublet of relatively weakly-focusing quadrupoles (Q9, Q10)
match the beam through the tracking and momentum-measurement stage, shown schematically in  
\Fig{fig:achromat2}, and determine the beam divergence and size through the apertures of the 
downstream detectors. 
The beam tracking system GTK (\Sec{sec:GTK}) consists of three stations, each composed of silicon pixel detectors installed in the beam vacuum. The stations are arranged so that the space between GTK1 
and GTK3 is occupied by a second achromat (A2), composed of four, vertically-deflecting,
C-shaped dipole magnets. The return yokes of the third and fourth
magnets, as well as a toroidally-magnetized iron collimator SCR1, defocus
 muons which leave the beam in the momentum-dispersed section between the 
second and third magnets of the achromat (\Fig{fig:achromat2}). GTK2 is located in the
same section, just after the magnetic collimator SCR1, where the $+75~\GEVc$  beam has a parallel, 
downward displacement of $\Delta{\rm Y } = -60$~mm and hence a dispersion of 0.6~mm per percent $\Delta$p/p.  GTK 3, located at 102.4 m from the target, marks the entrance plane at the beginning of the decay region.
The cleaning collimators (C6 ,C7)  preceding GTK3 are intended to intercept background outside the
beam acceptance.

\begin{figure}[ht]
\includegraphics[width=1.\linewidth]{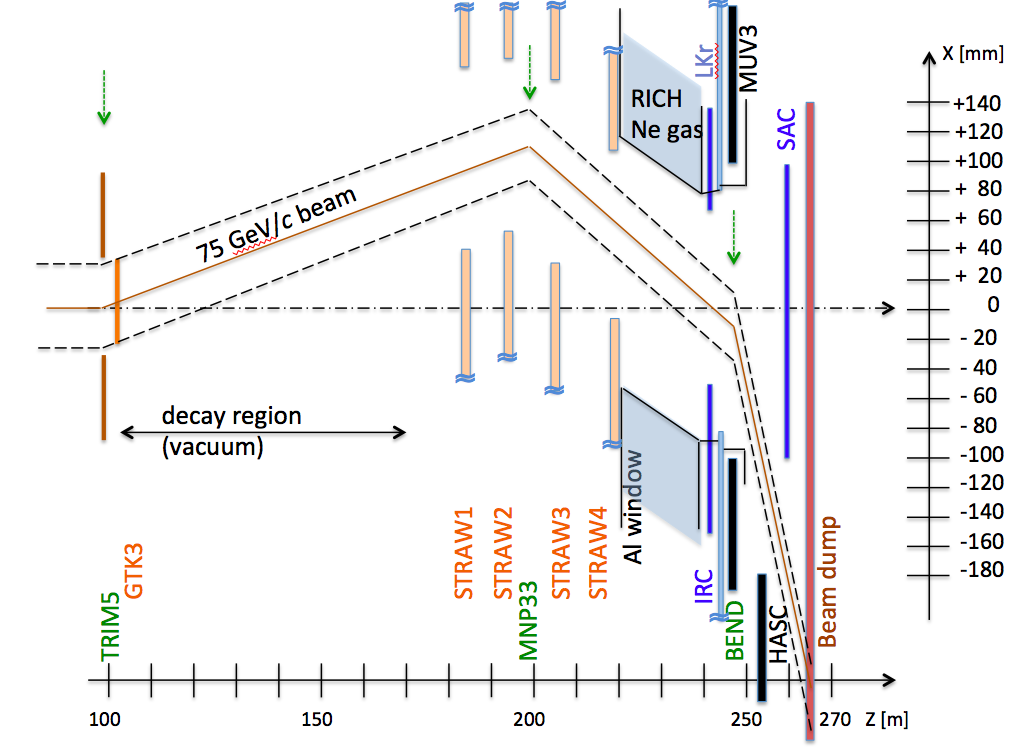}
\caption{\label{fig:K12beam-downstream} Close view of the  beam line layout through the decay region and the detectors in the (X, Z) plane. At the entrance of the decay region the beam has a horizontal angle of $+1.2$ mrad, the beam is then deflected in the spectrometer magnet MNP33 by $-3.6$ mrad to 
match the central aperture of the LKr calorimeter. 
After MUV3,  the dipole magnet BEND deflects 
charged particles associated with the beam away from the SAC and sends them  into the beam dump.  The dashed lines correspond to the two sigma width of the beam profile.
The vertical arrows indicate the bending centre of each magnet. Note the different scales along the two axes.}
\end{figure}

In addition, a horizontal steering magnet (TRIM 5) is used to deflect the beam 
 towards positive X  by an angle of $+1.2$ mrad. This angle is adjusted so that 
 the subsequent $-3.6$ mrad deflection 
 towards negative X, due to the downstream spectrometer magnet 
 MNP33, directs the beam back through the central aperture of the LKr calorimeter and subsequent 
 detectors, as shown schematically in \Fig{fig:K12beam-downstream}.

The decay region is contained in the first 60~m of a large 117~m long tank, starting  
102.4~m downstream of the target. This tank is evacuated to a residual 
pressure of $\sim 10^{-6}$ mbar using up to seven cryo-pumps. The tank hosts 11 LAV detectors  and the four spectrometer STRAW chambers,  and consists of 19 cylindrical sections made of steel or stainless steel. The vessel diameter increases from 1.92 m in the first section after  GTK3 to 2.4 m in the middle section and to 2.8 m in the spectrometer region. 
Every second component (LAV, STRAW or vacuum vessel element) has an extendible flange (telescopic extension of the tube) with a sliding vacuum seal. When retracted, these flanges provide enough longitudinal space to allow the removal of one of the detector elements. The vacuum tank  is closed off at its downstream end by a thin aluminium window 
(2~mm thickness), separating the tank from the neon gas of the 17~m long  RICH counter. 
The flange around a hole in the centre of the window is attached to a thin-walled aluminium beam 
tube  of inner diameter 168~mm, displaced to X $= +34$ mm and converging to the reference
axis at an angle of $-1.8$ mrad to follow the trajectory of the beam, which is thus transported in 
vacuum throughout the length of the detectors (\Fig{fig:K12beam-downstream}).

The magnetic spectrometer includes two pairs of straw tracking chambers (\Sec{sec:STRAWSpec}),
on either side of the large aperture dipole magnet (MNP33). The chambers cover the full acceptance 
outside a 118~mm diameter material-free passage around the beam path. 
The dipole magnet provides a horizontal momentum  kick of 270 $\MEVc$ deflecting the 75 $\GEVc$
beam by $-3.6$ mrad, so as to converge to, and then cross the undeviated axis at a point
2.8 m downstream of the centre of the LKr calorimeter (\Fig{fig:K12beam-downstream}).

\begin{figure}[ht]
\includegraphics[width=1\linewidth]{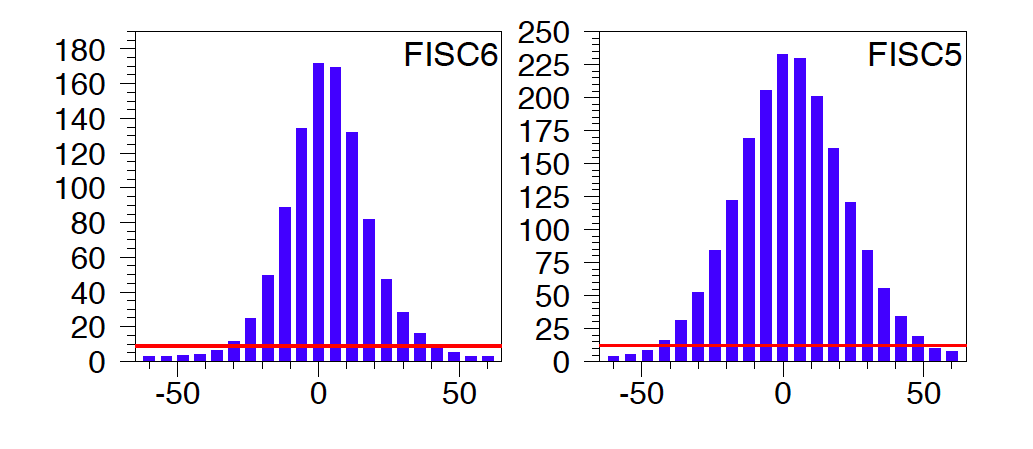}
\put(-258,5){ X [mm]}
\put(-374,163){ mean  2.8~mm}
\put(-374,153){ rms 13.4~mm}
\put(-420,78){\rotatebox{90}{counts [arbirary unit]}}
\put(-62,5){ Y [mm]}
\put(-180,163){ mean  3.0~mm}
\put(-180,153){ rms 18.5~mm}
\caption{\label{fig:FISC56}Beam profile in the filament scintillator counters located downstream of the LKr calorimeter in the horizontal (FISC6)  and vertical (FISC5)  plane. The mean and rms values correspond to the bins above the threshold line.}
\end{figure}
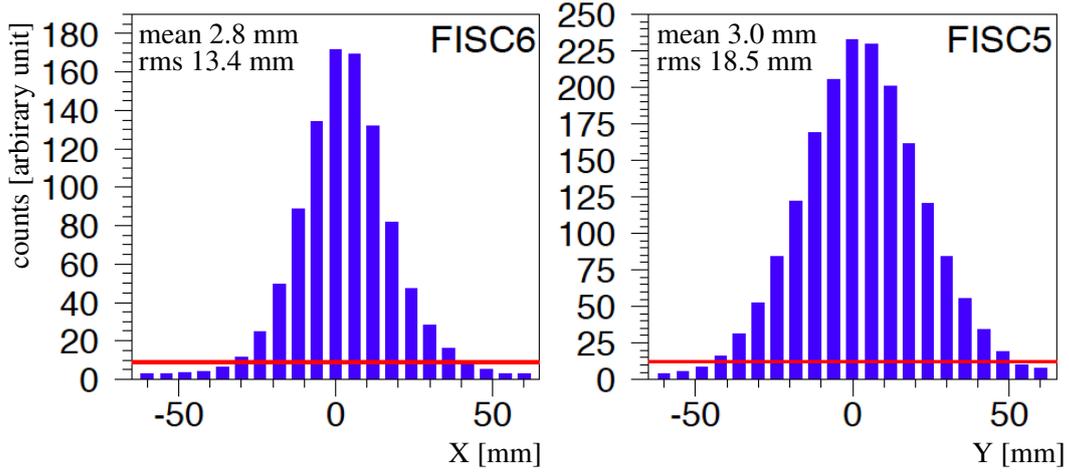

Close to this crossing point, a pair of larger filament scintillator counters (FISC 5, 6), installed in vacuum, allows the beam to be observed and steered correctly (\Fig{fig:FISC56}). The beam is finally deflected towards negative X through a
further angle of $-13.2$ mrad by a dipole magnet (BEND), so as to clear the SAC (\Sec{ssec:SAV}), 
located inside the beam vacuum vessel, 11.8~m further downstream. The beam is finally absorbed in a beam dump composed of iron surrounded by concrete at a sufficient 
distance behind the detector to diminish the effects of back-splash. To monitor the profile and intensity of 
the beam, a wire chamber with analogue readout and an ionization chamber are located in the space between the vacuum exit window and the beam dump.
A more detailed description of the beam can be found in \cite{NA62:2010}.

\subsection{Beam parameters and particle rates}\label{ssec:beam-parameters}
The principal characteristics of the high-intensity $\KPL$ beam are listed in \Tab{tab:K12-parameter}, 
where the effective solid angle and momentum acceptance, as well as the beam sizes and divergences, 
are calculated using the ray-tracing program {\tt TURTLE} \cite{Brown:1974}. 
The design particle composition of the 75 $\GEVc$ beam is obtained by interpolation of data at 60 and  120 $\GEVc$ according to the empirical formula in \cite{Atherton:1980}. 

The muons accompanying a secondary beam contribute a major part of the single-particle flux to
which the detectors outside the beam are exposed. The transport and
decay to $\mu^\pm  \nu $ of a wide spectrum of $ \PMP$ and $\KPM$ originating in the target
have been simulated using the program {\tt HALO} \cite{Iselin:1982}. This program tracks the
parent particles and their decay muons inside the beam apertures and the ``halo'' muons leaving the 
apertures through the vacuum tubes, magnet yokes and shielding surrounding the beam. Results of such calculations are given in \Tab{tab:K12-muonrates} which provide a fair approximation to the observed muon rates.

The intrinsic divergence of the parallel beam, as tuned for the CEDAR, is obtained
for the horizontal (vertical) plane by the width of the distribution of counts in the
0.2~mm-wide scintillator filament of FISC 4 (FISC 3) downstream of the
CEDAR, connected in coincidence with the similar scintillator of FISC 2
(FISC 1), located 7.14 m upstream. The count distributions are plotted in \Fig{fig:FISC34}
where the derived angular divergences are consistent with the nominal values. The profiles recorded 
on the pair of larger filament scintillator counters (FISC 6, 5), located near the crossing point of the 
beam with the reference axis at the exit of the LKr calorimeter, are shown in \Fig{fig:FISC56}.  It should be noted that these profiles 
are obtained with $N_2$ (at 1.75 bar) in the CEDAR. The width of the beam is expected to decrease by $9\%$  in each plane, if the CEDAR is operated with $H_2$. 

\begin{table}[h]
\centering
\caption{Principal parameters of the high intensity K12 beam according to design and at nominal intensity.
\label{tab:K12-parameter}}
\vspace{2ex}
\begin{tabular*}{1.0\textwidth}{llc}
\hline 
\hline
\multicolumn{2}{l}{SPS spill length (s) / SPS cycle time (s)} & 0.3 \\
\multicolumn{2}{l}{Effective SPS duty cycle (s/s)} & $\sim$ 0.2   \\
\multicolumn{2}{l}{SPS protons per pulse of 4.8 s} & $3.3 \times 10^{12}$ \\
\multicolumn{2}{l}{Instantaneous proton rate per effective s of spill \hspace*{4ex}} & $1.1 \times 10^{12}$ \\
\hline
 Beam acceptance &  Horizontal & $\pm$ 2.7~mrad \\
                 &  Vertical & $\pm$ 1.5~mrad \\
\hline
                                 & mean momentum & 75 $\GEVc$  \\
$\KPL$ momentum: & \hspace{0pt}Effective $\Delta$p/p      & 1.6\% \\
                                 & \hspace{0pt}$\Delta$p/p  rms &  1.0\% \\
\hline
Divergence at CEDAR (rms) &  Horizontal    & 0.07~mrad \\
                                                &  Vertical   & 0.07~mrad \\
\hline
Divergence at GTK (rms) &  Horizontal  &  0.11~mrad \\
                                           &  Vertical    &  0.11~mrad  \\
\hline
Beam size at GTK (rms)   &  Horizontal   & $\pm$ 26.4~mm \\
					  &  Vertical & $\pm$ 12.0~mm \\
\hline
Fiducial decay length: &  &  60~m \\
 $\Delta$ Z = fiducial length/decay length & & 0.107 \\
Decay fraction: &  $(1-e^{-\Delta z}$)  &0.101 \\
\hline
                                              &    $p$       &   173 MHz 	  (23\%) 	\\
                                              &   $\KPL$   &  45 MHz 		 (6\%) 	\\
Instantaneous beam rates    &   $\PPI$    &  525 MHz  	   (70\%) 	\\
                                             &   $\mu^+$  &  $\sim$ 5 MHz    $ (< 1\%)$ \\
                                             &    Total       & 750 MHz   (100\%) 	\\
\hline
\multicolumn{2}{l}{Total beam flux per pulse} & $2.25 \times 10^{9}$ \\
\multicolumn{2}{l}{Estimated $\KPL$ decays per year} & $5 \times 10^{12}$ \\
\hline 
\hline
\end{tabular*}
\end{table}

\renewcommand{\arraystretch}{1.2}
\begin{table}[ht]
\centering
\caption{Estimated instantaneous muon rates from pion and kaon decays that traverse the detectors at nominal beam intensity.}
\label{tab:K12-muonrates}
\vspace{2ex}
\begin{tabular*}{1.0\textwidth}{lcccc}
\hline
\hline
 & &  Total & Max. Intensity &  Dose \\
Detector & Area &  Rate & per effective s &  Rate \\
& [cm\textsuperscript{2}] & [MHz] & [kHz/cm\textsuperscript{2}] & [Gy/year]\\
\hline
CHANTI  & & & & \\
\hspace{12pt} ($4.8 <\lvert$X$\rvert <15$;  $3.3 < \lvert$Y$\rvert <1 5$ cm) & 838 & 1.5 & $\sim$ 70& $\sim$ 35 \\
\hline
Large Angle Veto & & & & \\
\hspace{12pt} LAV1 ($53.6 <$  r $< 90.6 $ cm) & $1.7 \times 10^{4}$ & 1.7 & $\sim$ 0.4 & $\sim$ 0.2 \\
\hspace{12pt} LAV1-12 total & $26.1 \times 10^{4}$ & 11.2 & &  \\
\hspace{12pt} LAV1-12 OR & $26.1 \times 10^{4}$ & 4.0  & &  \\
\hline
Straw Chamber 4 & & & & \\
\hspace{12pt} Full plane ($5.9 <  \lvert$X$\rvert$,$\lvert$Y$\rvert < 105$ cm) & $4.2 \times 10^{4}$ & 7.9 & & \\
\hspace{12pt} Nearest straw vertical & 210 & 0.6 & $\sim$ 60 & $\sim$ 30 \\
\hspace{12pt} Nearest straw horizontal  & 210 &0.3 & $\sim$ 10 & $\sim$ 5 \\
\hline
RICH & & & & \\
\hspace{12pt} Mirrors ($9 <$  r $< 120$  cm) & $4.5 \times 10^{4}$ & 9.7 & $\sim$ 50 & $\sim$ 25 \\
\hline
IRC & & & & \\
\hspace{12pt}($6 < $ r $ < 14.5$ cm) & 547 & 7.6 & $\sim$ 70 & $\sim$ 35 \\
\hline
LKr & & & & \\
\hspace{12pt}($12 < $ r $< 120$ cm) & $4.5 \times 10^{4}$ & 8.4 &  & \\
\hspace{12pt}($12 < $ r $< 16$ cm) & 352 & 1.2 & $\sim$ 30 & $\sim$ 15 \\
\hline
\hline
\end{tabular*}
\end{table}
\renewcommand{\arraystretch}{1.0}

%% file: Sec4-KTAG_v3.tex
\subsection{Design and construction}

Kaons make up a minority (6\%) of the K12 beam and are identified by the KTAG detector. Cherenkov light is produced in the gaseous radiator volume of a CERN W-type CEDAR, a  differential Cherenkov counter with achromatic ring focusing designed in the late 1970s to discriminate kaons, pions and protons in unseparated charged-particle beams extracted from the CERN SPS~\cite{Bovet:1982}. 
In the NA62 configuration, the CEDAR with its gas volume of 0.94~${\rm m}^3$ has been filled with nitrogen ($N_2$) at 1.75~bar at room temperature. This represents, with the CEDAR windows, a total of $3.5\times 10^{-2}~X_0$ of material in the path of the beam. Alternatively, the NA62 CEDAR can be filled with hydrogen ($H_2$) at 3.9~bar which reduces the material thickness to $7\times 10^{-3}~X_0$ and decreases the beam emittance by about 9\% in each plane.

\begin{figure}[ht]
\begin{center}
\includegraphics[width=1.0\linewidth]{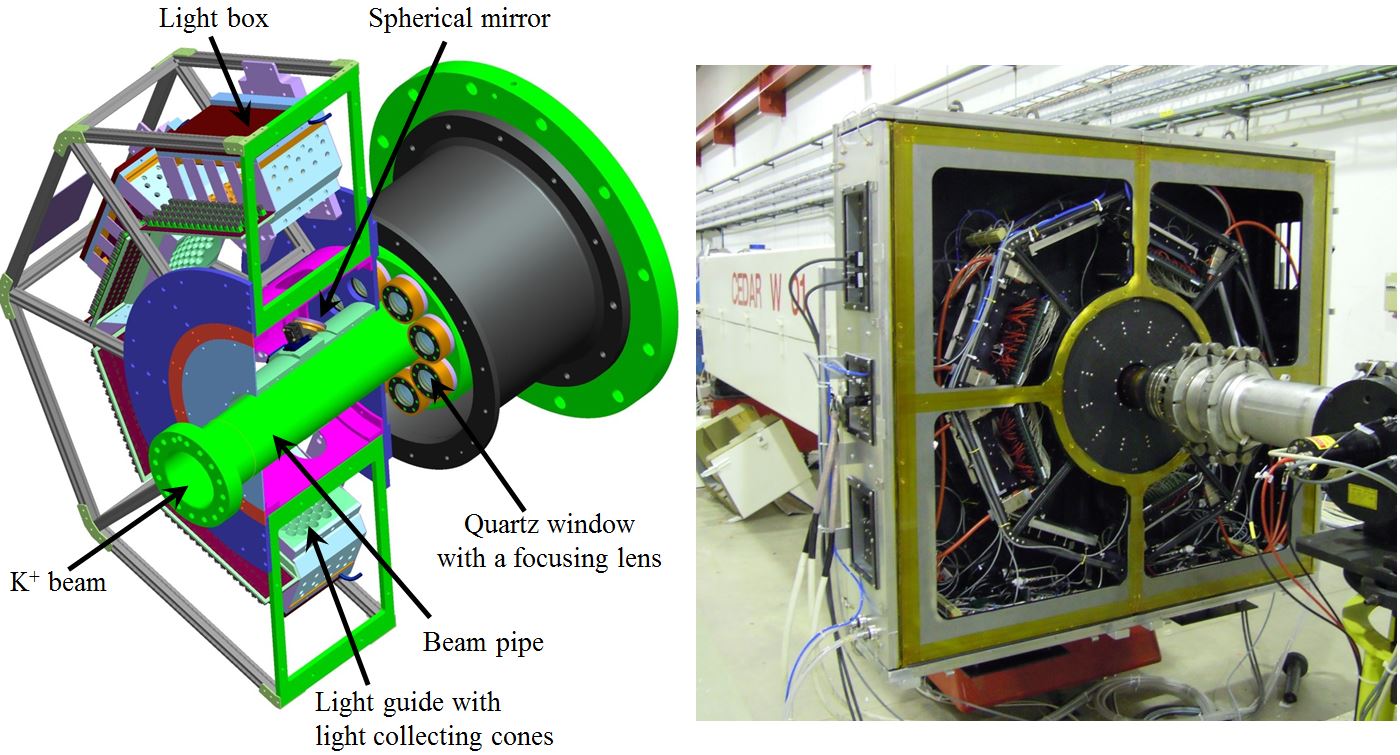}
\end{center}
\caption{\label{fig:KTAG-Picture1c} Left: Drawing of the upstream part of the CEDAR and the KTAG. Right: KTAG and CEDAR in the NA62 beam line during a test run in 2012, with four of the eight sectors equipped.}
\end{figure}

The CEDAR gas volume and optics are suitable for use in NA62, but the original photodetectors and readout electronics are not capable of sustaining the nominal 45~MHz kaon rate in the NA62 beam line, nor of providing timing resolution at the required level of 100~ps. The KTAG detector, which includes new photon detection and readout systems, has been developed to meet these requirements. Details of the CEDAR internal optics can be found in~\cite{Bovet:1982}, while the KTAG optics and mechanics developed for the NA62 experiment are shown in Figure~\ref{fig:KTAG-Picture1c}. 

The radiator gas is kept at room temperature, and its pressure can be varied between vacuum and 5~bar. 
The CEDAR detector design requires the angular beam divergence to be below 100~$\mu$rad in each plane, and the angular alignment of the detector and beam axis to be of a similar accuracy.
The pressure in the CEDAR is chosen so that only light from the desired particle type passes through an annular diaphragm of fixed central radius and varying radial aperture. The light is focused to exit the vessel through eight quartz windows and then focused onto eight spherical mirrors. The mirrors reflect the light  radially outwards into eight light boxes (referred to as sectors) located in an insulated, cooled Faraday enclosure flushed with $N_2$ gas and equipped with environmental monitors. 
The entrance to each box is a light guide consisting of a matrix of 64~closely spaced conical sections of 15~mm (4~mm) outer (inner) radius cut into a 17~mm thick aluminium plate of spherical section with centre of curvature at the virtual focus of the Cherenkov light. The interior of each cone is lined with aluminized Mylar, and 48~Hamamatsu\textsuperscript{TM} photomultipliers (PM, 32~of R9880 type and 16~of R7400 type) are set into the outer curved surface of each light guide, matching the cones precisely; 
16 peripheral cones in each sector are not instrumented.

Differential signals from the anode and the last dynode of each photomultiplier are read into front-end boards consisting of a mother board with 64 analogue differential inputs and outputs, an embedded local monitor board (ELMB) for remote control and services, and 8 mezzanine cards each with an 8-channel 
NINO ASIC~\cite{NA62:2010}. The low-voltage differential signal (LVDS) outputs feed into 128-channel TDC boards, enabling the times of leading and trailing edges of the signal to be measured, and thereby permitting slewing corrections to be implemented offline. The TDC boards are used as daughter boards for TEL62 boards, for a maximum of 512 channels readout in each TEL62 board 
(\Sec{sec:TDAQ}.2).  
Additional splitter boards are installed between front-end and TDC boards to equalize the data rate in each TEL62. Further details of the detector optimization, construction, mechanics, readout and electronics can be found in \cite{Goudzovski:2015}.

\boldmath
\subsection{Performance with $N_2$ radiator gas in 2015}
\unboldmath

\begin{figure}
\begin{center}
\includegraphics[width=0.90\linewidth]{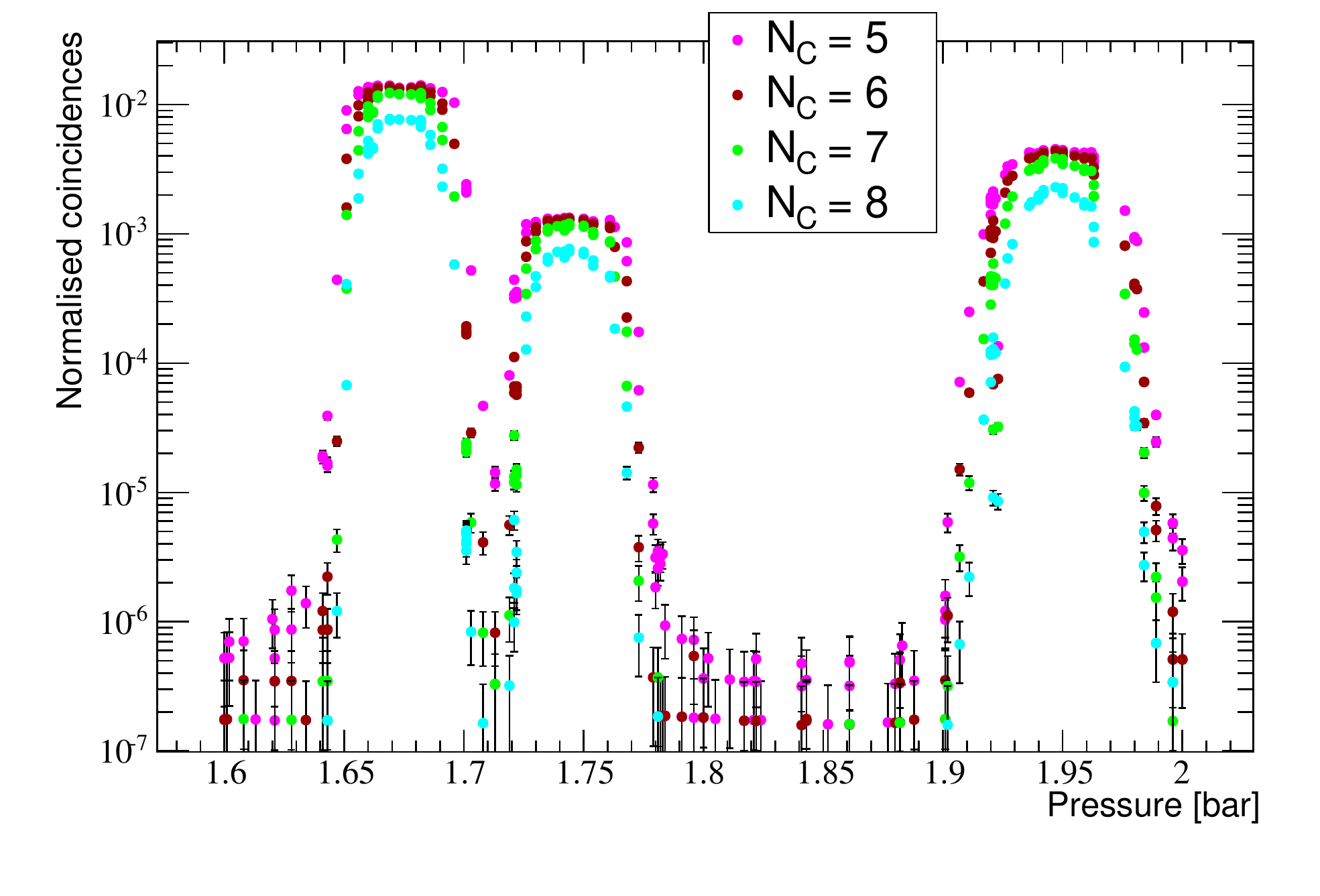}
\put(-79,150){\Large \boldmath $p$}
\put(-228,125){\Large \boldmath $\KPL$}
\put(-275,170){\Large \boldmath $\PPI$}
\end{center}
\vspace{-10mm}
\caption{\label{fig:KTAG-Picture2} Results of a CEDAR pressure scan with $N_2$ radiator gas at the nominal diaphragm aperture of 1.5~mm. For each minimum number of KTAG sectors required in coincidence $N_{\rm C}$ (ranging from 5 to 8), the normalised number of events satisfying the requirement is plotted versus the $N_2$ pressure. Well separated pion, kaon and proton peaks are visible.}
\end{figure}

Alignment of the CEDAR optical axis to the beam axis is achieved by minimizing the left-right and top-bottom asymmetries of the measured light distribution in the KTAG sectors. The orientation of the optical axis is adjusted  
for a set of decreasing diaphragm apertures, until the desired aperture is reached. Following the alignment procedure, the $N_2$ pressure is optimized to obtain maximum light yield. The discriminatory power of the KTAG for identifying $\PPI$, $\KPL$ and protons is shown in \Fig{fig:KTAG-Picture2} in terms of the number of coincidences of signals in the sectors plotted as a function of the $N_2$ gas pressure. Clear separation of the $\PPI$, $\KPL$ and proton peaks is obtained by requiring a coincidence of signals in at least 5~sectors. At the pressure of 1.75~bar optimal for $\KPL$ identification, the mean number of PM signals per kaon is~20.

\begin{figure}
\begin{minipage}{0.5\textwidth}
\centering
\includegraphics[width=1.\linewidth]{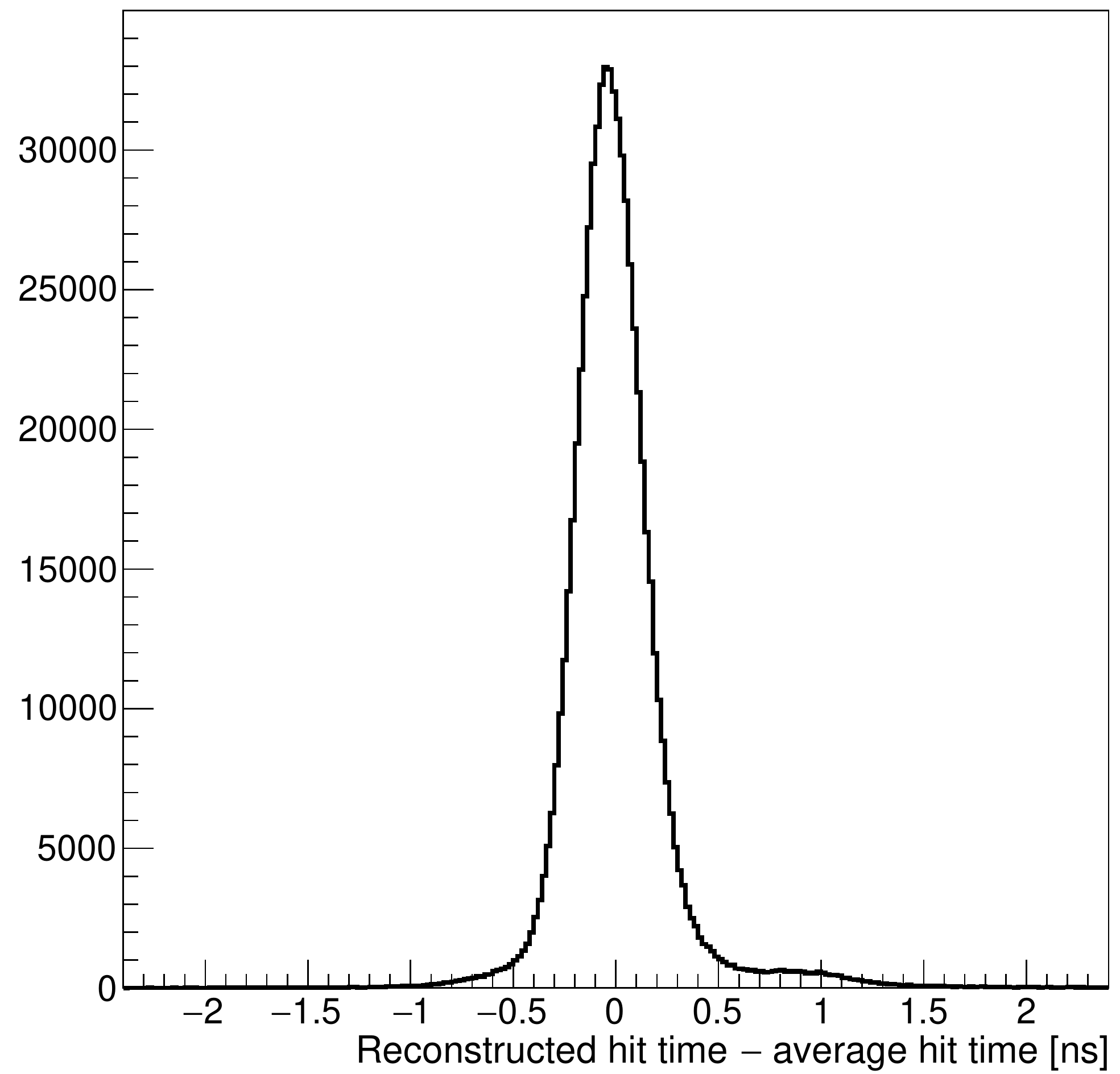}
\end{minipage}
\begin{minipage}{0.5\textwidth}
\centering
\includegraphics[width=1.\linewidth]{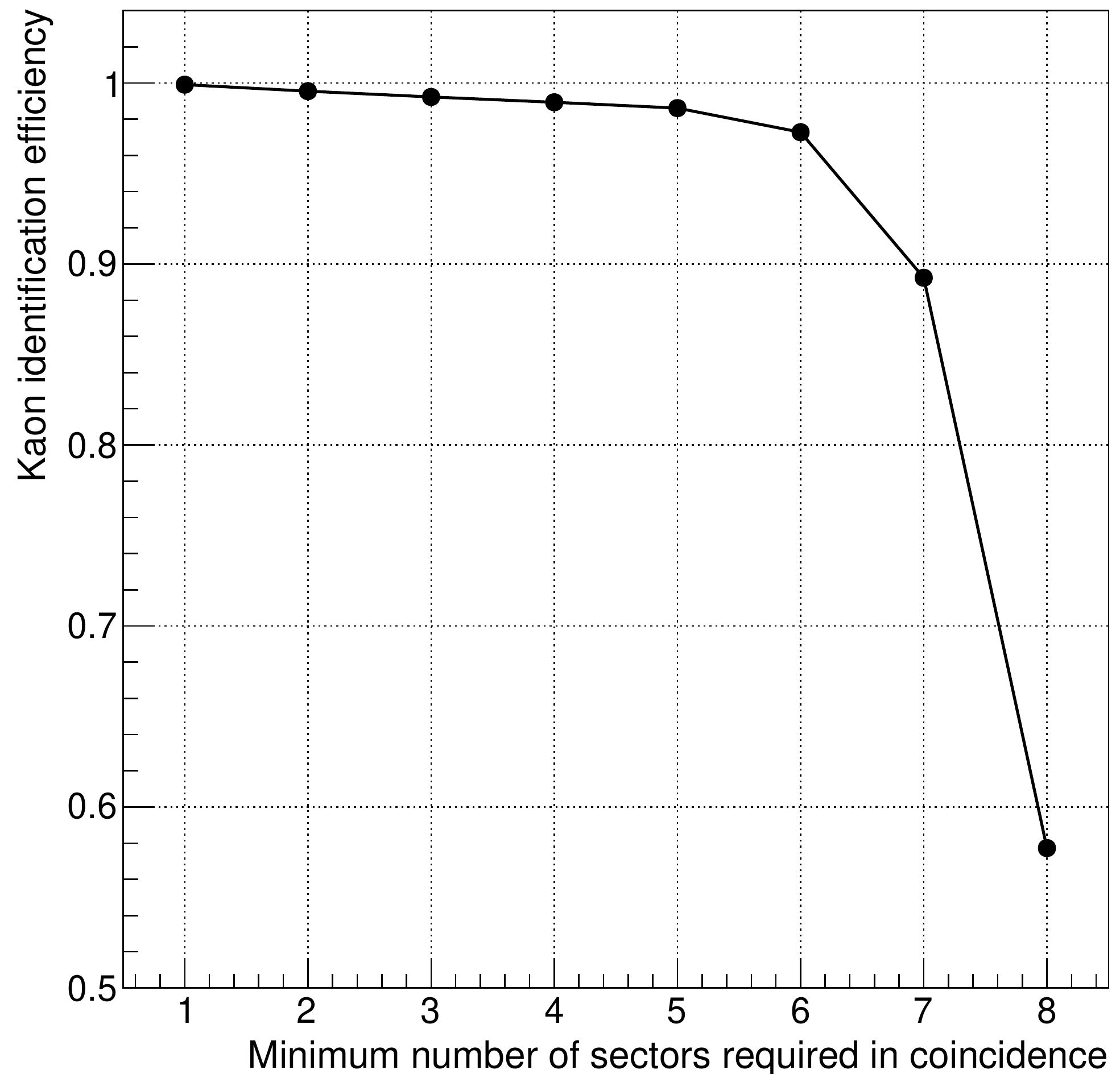}%
\end{minipage}
\caption{\label{fig:KTAG-Picture3} KTAG performance measured with the 2015 data at the nominal 1.5~mm diaphragm aperture with $N_2$ radiator in the CEDAR. Left: Time resolution of a typical PM signal; a Gaussian fit to the central peak gives a resolution of 160~ps, while the rms of the distribution, including late signals due to the elastic scattering of the photoelectron in the first dynode, is 300~ps; this leads to kaon time resolution of 70~ps. Right: Kaon identification efficiency as a function of the minimum number of sectors required in coincidence.}
\end{figure}

The KTAG time resolution for the individual PM signals is measured from the difference of hit time in a PM and the average time of all signals produced by a beam particle. With the time offset and slewing corrections implemented, the rms time resolution in a single channel is 300~ps (\Fig{fig:KTAG-Picture3}-left). With 20~PM signals detected on average per beam kaon (with the average rate of 2.3~MHz/channel at the nominal 45~MHz kaon rate), kaon time resolution of 70~ps is achieved.

The KTAG kaon identification efficiency measured with reconstructed $\KTP$ decays is found to exceed 98\% when requiring Cherenkov light in coincidence in at least 5~sectors (\Fig{fig:KTAG-Picture3}-right). For the same coincidence requirement, the probability of misidentifying a pion as a kaon while operating at the kaon pressure is estimated from the pressure scan data to be ${\cal O}(10^{-4})$. This estimate does not account for misidentification due to pile up, which dominates at the beam intensity of $2.25 \times 10^9 $ particles per 3 s effective pulse.

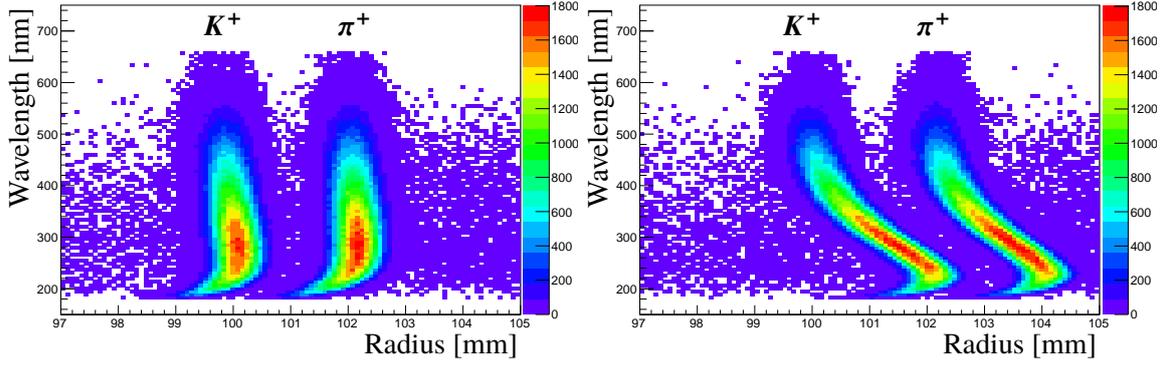
\begin{figure}
\begin{minipage}{0.5\textwidth}
\centering
\includegraphics[width=1.0\textwidth]{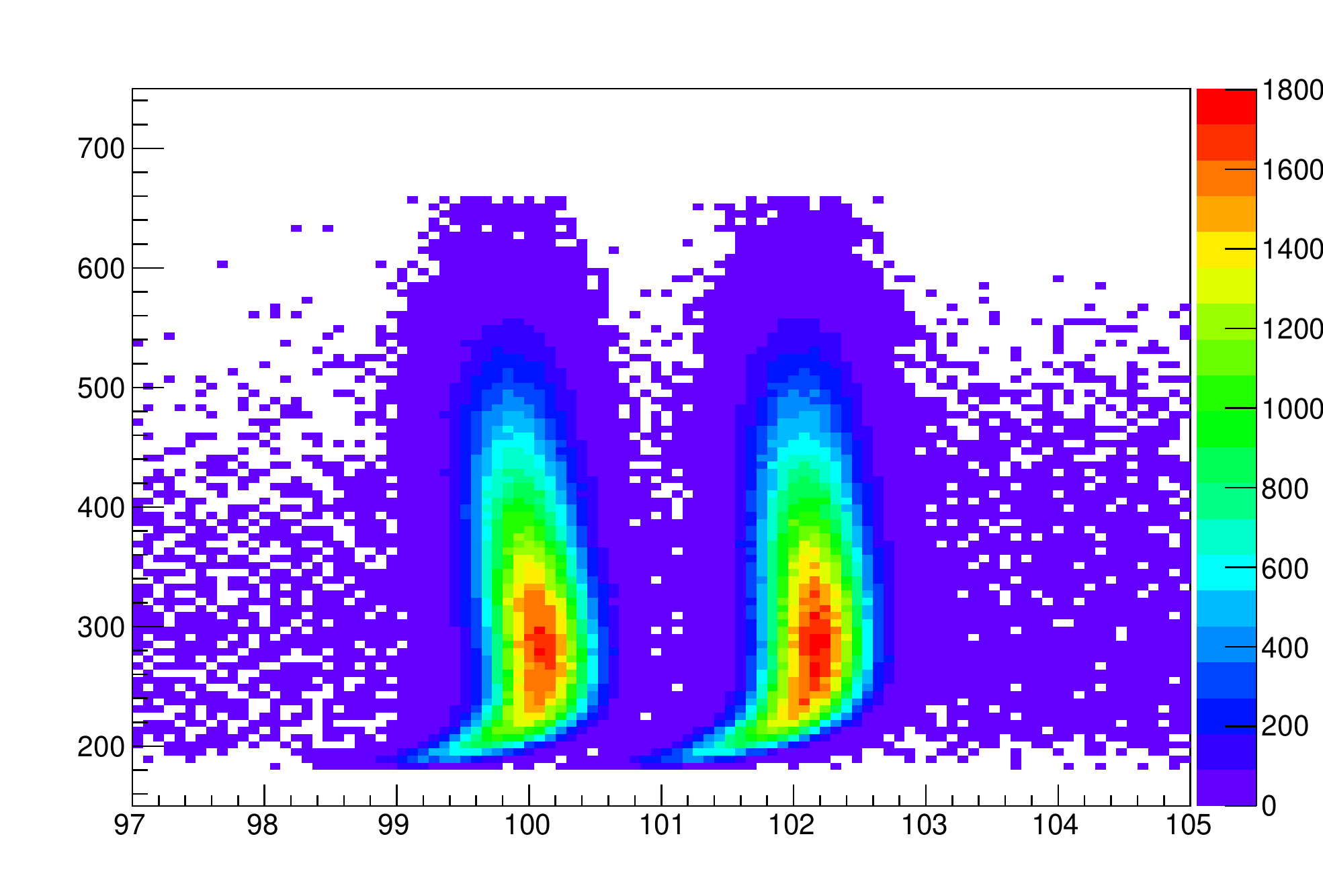}%
\put(-140,120){\boldmath $\KPL$}
\put(-90,120){\boldmath $\PPI$}
\put(-80,0){Radius [mm]}
\put(-213,55){\rotatebox{90}{Wavelength [nm]}}
\end{minipage}
\hfill
\begin{minipage}{0.5\textwidth}
\centering
\includegraphics[width=1.0\textwidth]{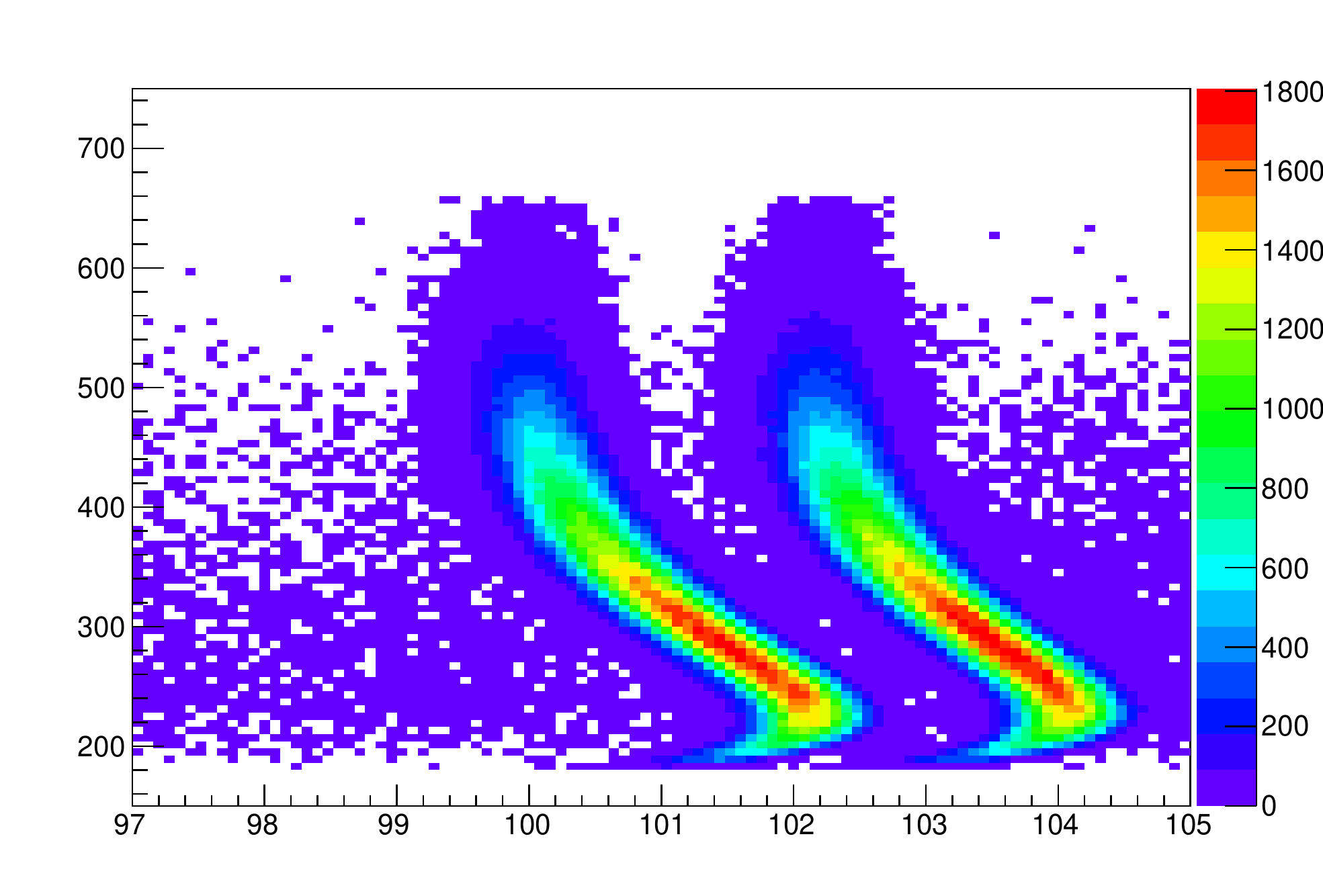}%
\put(-140,120){\boldmath $\KPL$}
\put(-90,120){\boldmath $\PPI$}
\put(-80,0){Radius [mm]}
\put(-213,55){\rotatebox{90}{Wavelength [nm]}}
\end{minipage}
\caption{Simulated distributions of wavelength vs radius at the CEDAR diaphragm plane for Cherenkov photons produced by the $\KPL$ and $\PPI$ beam for $N_2$ (left) and $H_2$ (right) radiator, weighted with the quantum efficiency~\cite{Goudzovski:2015}.}
\label{fig:Diaphragm}
\end{figure}

Regarding the choice of the operating gas for the CEDAR, it is important to note that the W-type CEDAR was originally designed for a $N_2$ radiator, where the optics are fully corrected for dispersion of the Cherenkov light and all wavelengths form narrow cones for each particle species.
In order to reduce the background due to scattering in the gas, NA62 has kept the option of filling the CEDAR with $H_2$ gas, where the optics do not correct the chromatic dispersion as for $N_2$. This causes a broadening of  the Cherenkov peaks making the separation between kaons and pions more difficult (Figure~\ref{fig:Diaphragm}). When $H_2$ is used the light yield is expected to be about 30\% lower, depending on the fine-tuning of the pressure and diaphragm opening. This will, in turn, decrease the detection efficiency and worsen the time resolution.

%% file: Sec5-GTK_v3.tex
\begin{figure}[ht]
\begin{center}
\includegraphics[width=0.96\linewidth]{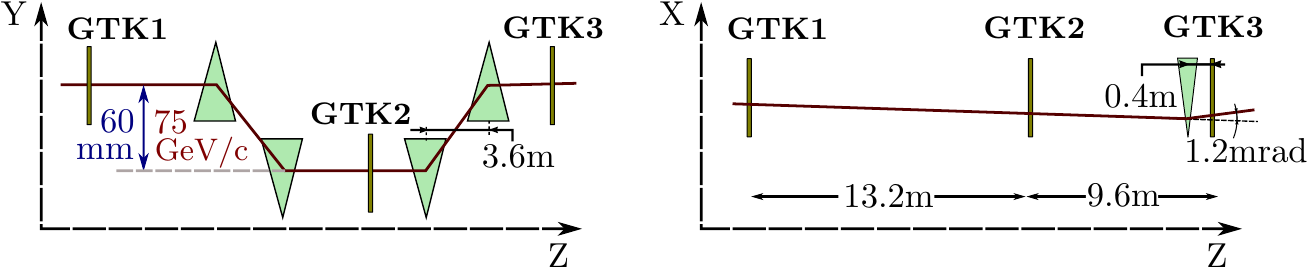}
\end{center} 
\caption{\label{fig:GTK-Layout} Schematic layout of the GTK stations within the achromat in the vertical and horizontal views.}
\end{figure}

The beam spectrometer provides precise measurements of momentum, time and direction of the incoming beam particles. The spectrometer is located inside the vacuum pipe immediately upstream of  the $\KPL$  decay region and is composed of three similar stations installed around four dipole magnets arranged as an achromat (\Fig{fig:GTK-Layout} ). The particle momentum can be derived from the vertical displacement of the trajectory in the second station. 

As discussed in \Sec{ssec:req}, the GTK has been designed to measure the momentum of the 75~$\GEVc$ beam particles to 0.2\% precision and their direction, dX/dZ and dY/dZ, at the exit of the achromat to $16~\mu$rad precision. The high beam rate (750~MHz and up to 1.5~MHz/mm$^2$ around the detector centre) requires a hit time resolution better than 200~ps \cite{Aglieri:2013}. The material budget for each of the three stations was chosen to be less than 0.5\% $X_0 $, corresponding to about $500~\mu$m of silicon. Finally, the detector has to sustain a high level of radiation. The last three requirements involved significant design efforts.

\begin{figure}[ht]
\begin{minipage}{0.5\linewidth}
\begin{center}
\includegraphics[width=1.\linewidth]{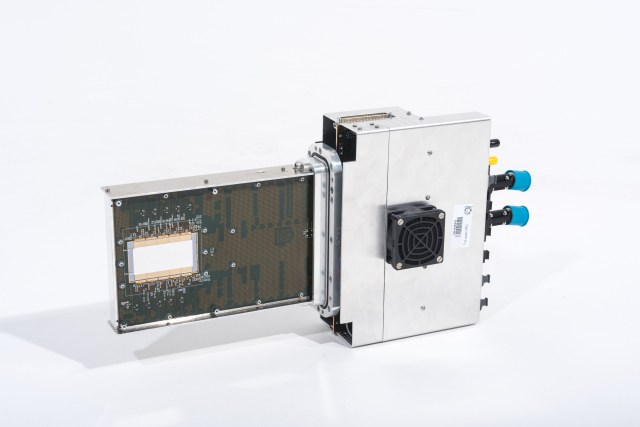} 
\end{center}
\end{minipage}
\begin{minipage}{0.5\linewidth}
\begin{center}
\includegraphics[width=1.\linewidth]{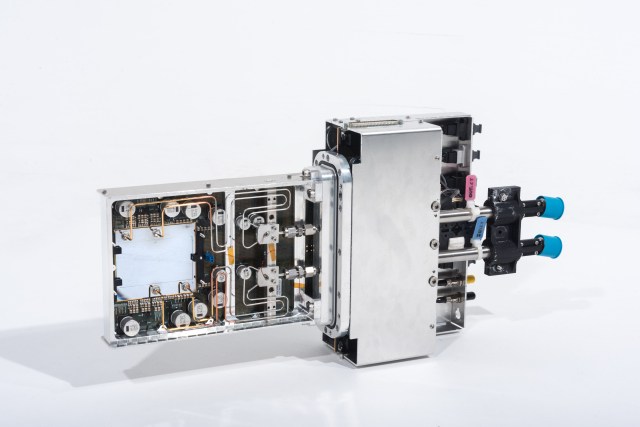} 
\end{center}
\end{minipage}
\caption{\label{fig:GTK_photo} Picture of an assembled station looking at the sensor side (left) and at the cooling side (right).}
\end{figure}

\begin{figure}[ht]
\begin{center}
\includegraphics[width=0.8\linewidth]{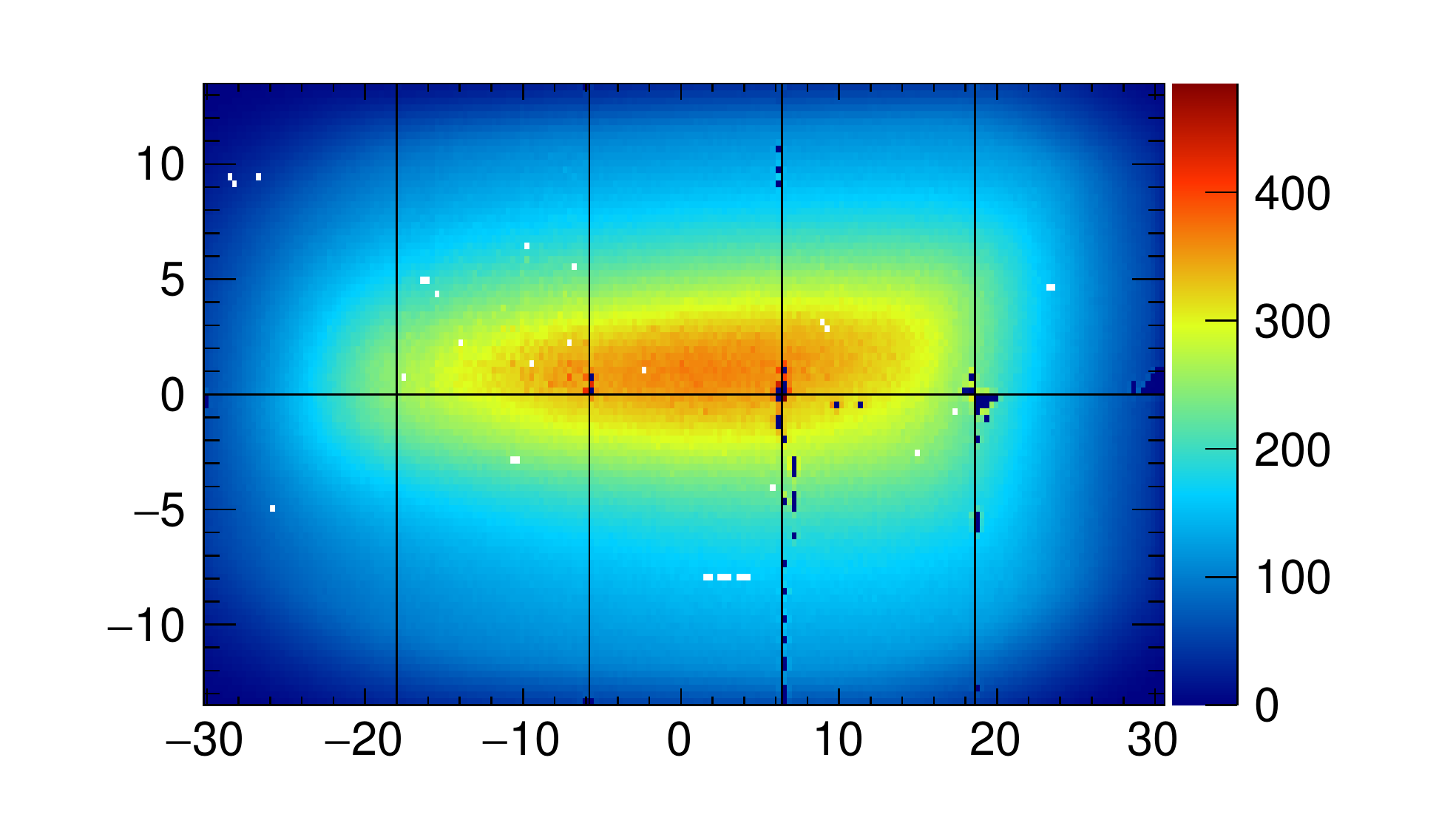}
\put(-100,5){X [mm]}
\put(-330,140){\rotatebox{90}{Y [mm]}}
\put(-20,90){\rotatebox{90}{Hit rate [kHz/${\rm mm}^2$]}}
\end{center}
\vspace{-5mm}
\caption{\label{fig:GTK-Picture2}  Illumination of the GTK1 station  operated  in the beam at about 30\% of the nominal intensity. The grid lines correspond to  the chip boundaries.}
\end{figure}

Each station (\Fig{fig:GTK_photo}) is a hybrid silicon detector consisting of 18 000 pixels of $300~\times ~300~\mu{\rm m}^2$ area each, arranged in a matrix of 200 $\times$ 90 elements corresponding to a total area of 62.8 $\times$ 27~${\rm mm}^2$ (\Tab{tab:GTK-table1}). The matrix is read out by application-specific integrated circuits (ASIC) arranged in two rows of five chips  (\Fig{fig:GTK-Picture2}), with each chip serving 40 $\times$ 45~pixels.

\setlength{\tabcolsep}{6pt}
 \begin{table}
 \setlength{\tabcolsep}{3ex}
 \renewcommand{\arraystretch}{1.1}
 \caption{Gigatracker  detector dimensions.}
 \label{tab:GTK-table1}
 \vspace{2ex}
 \centering
  \begin{tabular*}{0.8\textwidth}{@{\extracolsep{\fill} }  l c }
 \hline\hline
 \\[-2.5ex]
\textbf{Full Detector:} &  \\
\hspace{30pt}  Full size & 60.8 $\times$ 28.2 mm\textsuperscript{2} \\
\hspace{30pt}  Active area & 60.8 $\times$ 27 mm\textsuperscript{2} \\
\hspace{30pt}  Number of pixels & 18 000\\
\hspace{30pt}  Number of read-out chips  & 10\\
\hline
\textbf{Read Out Chip (ROC): }     &  \\
\hspace{30pt}  Full size      &  12 $\times$ 19.5 mm\textsuperscript{2}\ \\
\hspace{30pt}  Pixel matrix & 12 $\times$ 13.5 mm\textsuperscript{2} \\
\hspace{30pt}  Number of pixels read & 1 800\\
\hspace{30pt}  Pixel size & 300 $\times$ 300 $\mu$m\textsuperscript{2}\\
\hline
\textbf{Thickness:} &  \\
\hspace{30pt}  Sensor & 200 $\mu$m  \\
\hspace{30pt}  Read-out chips & 100 $\mu$m \\
\hspace{30pt}  Cooling plate &  210 $\mu$m\\
hspace{30pt}  Total thickness in active area & 510 $\mu$m\\
 \hline\hline
 \end{tabular*}
 \end{table}

Reducing the multiple coulomb scattering has two main advantages: it improves the angular resolution and reduces the background produced by interactions in the last station, where the products are not swept away by the magnetic field.

At the position of GTK2, the beam has a vertical spatial dispersion of 0.6~mm per percent of $\Delta$p/p. In order to obtain a momentum resolution of 0.2\% and accounting for the effect of multiple Coulomb scattering, a pixel size of $300 \times 300~\mu{\rm m}^2$ was considered appropriate. In these conditions, the contributions to the angular resolution from geometrical parameters and multiple Coulomb scattering are similar.

Finally, the expected radiation level damages the silicon structure in the sensor introducing defects that trap charges reducing the signal yield and consequently degrading the time resolution. The effect of the radiation damage can be mitigated by a combination of cooling the detectors, increasing the bias voltage, and replacing the damaged assemblies. 

\subsection{Design and construction}\label{ssec:GTK-detector}
The sensors have been made from 200 $\mu$m thick wafers of $\geq$ 3 k$\Omega$cm
resistivity silicon. 
The most probable signal yield is approximately 15000 electron-hole pairs (equivalent to 2.4 fC) for a 
beam particle crossing the silicon, which is sufficient to discriminate the physical signal
 from electronics noise, which is 0.04 fC. Read-out chips (ROC) were thinned to 
 100 $\mu$m before bonding them to the sensors with  10 -- 15 $\mu$m high SnAg bumps. A planar 
 matrix of p-in-n pixel diodes has been chosen as the base configuration, although the front-end
amplifier is bi-polar and n-in-p structures can also be handled.

The detectors are located in a harsh radiation environment. At nominal beam intensity the detectors 
are exposed to a fluence corresponding to $4 \times 10^{14}$ one-MeV neutron equivalent 
cm$^{-2}$ in one year (200 days) of data taking. In order to minimize ageing effects due 
to radiation damage, the detectors are operated at approximately $-15~^\circ$C.
 It is expected that they can be operated continuously for more than 100 days without any significant performance degradation \cite{Ziock:1993} under these conditions. The detector mechanics has been designed such that detectors can be replaced rapidly, for example, during one of the regular short accelerator stops.

The detector achieves a single hit time resolution better than 200 ps if
the sensor is operated in an over-depleted regime with a bias voltage of
300~V or higher \cite{Aglieri:2015}. To operate the sensor 
at this high bias voltage, a multi-guard ring structure was implemented to establish a gradual 
voltage drop between backside potential and the sensitive region.

The ASIC, called TDCPix, has been designed in 130~nm CMOS technology. The time-over-threshold
 (ToT) is measured, allowing for an efficient correction for the time-walk.
To satisfy the stringent requests concerning time resolution, the 
amplifier, with peaking time of 5~ns, and the ToT discriminator were both located within the 
$300 \times 300 ~\mu{\rm m}^{2}$ pixel area.  Furthermore, the pixel area includes 
a 5-bit DAC to trim the discriminator threshold and a configuration register \cite{Aglieri:2011}. The 
signals are  then transmitted to the end-of-column (EoC) region for the time measurement.

For each signal associated to a beam-particle crossing the sensor, the
times of the rising and trailing edges are measured with a delay-locked-loop  based (DLL) TDC. The clock frequency of 320.632~MHz, derived from the NA62 main clock frequency, is divided into a 5-bit
fine-time corresponding to a TDC time bin of 97~ps. To limit
the number of DLLs and TDCs, a pair of columns shares the same DLL and five
pixels share the same TDC, resulting in 20 DLLs per chip and 9 two-channel TDCs per
column. In the EoC region, a hit-arbiter maximizes the throughput whenever two or more pixels sharing the same TDC are hit. The hits are stored in a FIFO ready to be transmitted to the offline read-out.

The expected maximum hit rate in a chip exceeds $10^{8}$~particles/s. Considering the hit word size 
of 48 bits, the total bit-load to be transmitted is estimated to be close to 6 Gbit/s. The chip is split in 
quarters (10 columns), each served by a 3.2~Gbit/s serial transmitter.

The radiation environment can affect the chip function by inducing single event upsets (SEU). 
To mitigate the problems due to bit-flip the logic was triplicated and a majority decision is taken.

The dissipated power is not uniform over the detector, being
approximately 0.4 W/cm\textsuperscript{2} in the sensitive area and
3.2 W/cm$^{2}$ in the whole  EoC region (7.2~cm$^2$), where most of the digital
functions are located; hence both areas require active cooling \cite{Romagnoli:2015qxs}. In this
context, the aim of the cooling is twofold: a) removing swiftly the heat produced by the
ASIC electronics and transferring it outside the vacuum pipe and b) keeping the sensor at a stable
 low temperature.
The low temperature helps to mitigate the loss in gain caused by radiation damage.

\begin{figure}[ht]
\begin{center}
\includegraphics[width=0.7\textwidth]{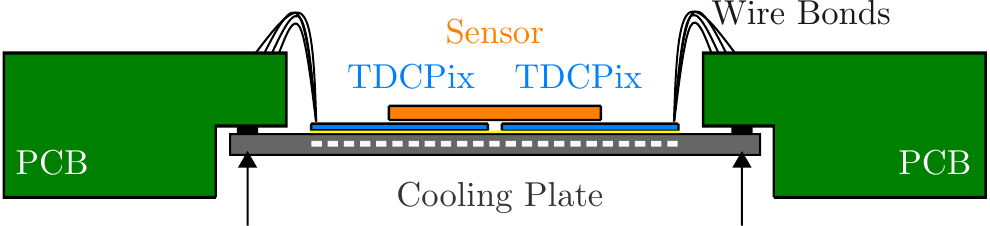}
\end{center}
\caption{\label{fig:GTK-Picture3} Sketch of the detector after assembly. The 
different components from top to bottom are: sensor, TDCPix and cooling plate.}
\end{figure}

The design chosen is quite challenging: a cooled liquid (single
phase C$_{6}$F$_{14}$)  flows through 150~parallel micro-channels which act as heat exchangers. The
micro-channels are etched on a silicon wafer, 140~$\mu$m thick,
by Deep Reactive Ion Etching. The cover of the channels is
obtained by bonding a flat silicon wafer, 70~$\mu$m thick, to the channel
wafer. The channels have a cross section of
$200 \times 70  ~\mu{\rm m}^{2}$ and they are separated by 200~$\mu$m walls.
To limit the pressure drop necessary to flow the liquid,  the micro-channels are split into two groups with independent inlet and outlet tubes. 
The plate is then glued with double-face adhesive tape to the back-side of the read-out
chip \cite{Nuessle:2013}. The sketch in Figure \ref{fig:GTK-Picture3} represents the cooling concept: the heat developed by the TDCPix is efficiently removed by the cooling plate,
which is kept at low temperature by the liquid. Consequently, the temperature of the silicon 
sensor is linked to that of the read-out chip.

The cooling fluid is supplied by a cooling plant common to the three GTK stations.
The liquid is cooled by an externally
controlled chiller and the circulation is forced through the micro-channels at a pressure close
to 3 bar with a flow of 2 g/s/station. The temperature at the chiller
output can be controlled over the range $-25 $ to 0~$^{\circ}$C. In case of vacuum failure or 
overheating, adequate interlocks protect the system from damage.

\subsection{Performance in 2015}\label{ssec:GTK-Results}

During the 2015 beam period, the three stations were installed and equipped with detectors, and all the infrastructure including vacuum, mechanics, cooling, and data acquisition,  were in place and operational.
GTK data, synchronized to the NA62 L0 trigger (\Sec{ssec:GTK-readout}),
were collected  and the detector was partially commissioned. 
The detectors were operated at 0~$^{\circ}$C temperature by circulating C$_{6}$F$_{14}$ at 2~g/s, which allowed a pressure drop of around 3.2~bar across the cooling plates. Pixel-signal thresholds were adjusted and pixel-to-pixel thresholds equalized with a test-bench procedure performed before installation. The thresholds were set to 0.7~fC, corresponding to 30\% of the most probable charge of a minimum ionizing particle.
During the data taking, the bias voltages were set and scanned in the range between 200 and 300~V.
All the GTK read-out cards were installed, the optical-fibre communication to the TDCPix was established and the data read-out was proven to be fully functional and reliable.

The first operation of the GTK system 
confirmed that the mechanics, the cooling and electronics were properly working, although further calibration and data analysis are needed to exploit the full potential of the detector.  
\begin{figure}[ht]
\begin{center}
\includegraphics[width=0.9\linewidth]{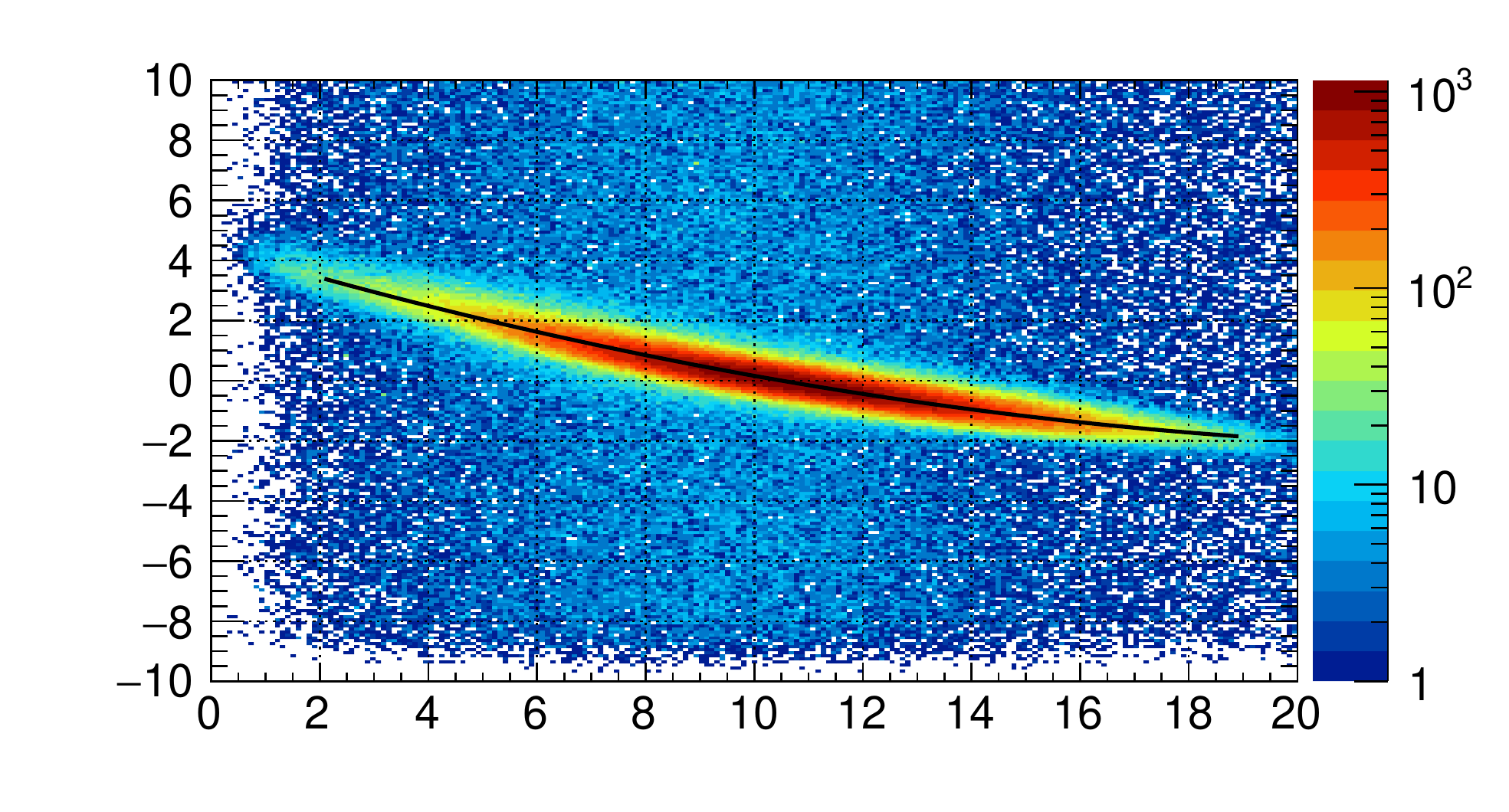}
\put(-90,5){ToT [ns]}
\put(-370,110){\rotatebox{90}{$ {\rm t}_{\rm GTK} - {\rm t}_{\rm KTAG}$ [ns]}}
\end{center}
\vspace{-7mm}
\caption{\label{fig:GTK-Picture9} Time-walk as a function of time-over-threshold (ToT). The line corresponds to the result of a quadratic fit of the most probable value of the time-walk.}
\end{figure}

\begin{figure}[ht]
\begin{minipage}{0.5\linewidth}
\begin{center}
\includegraphics[scale=0.42]{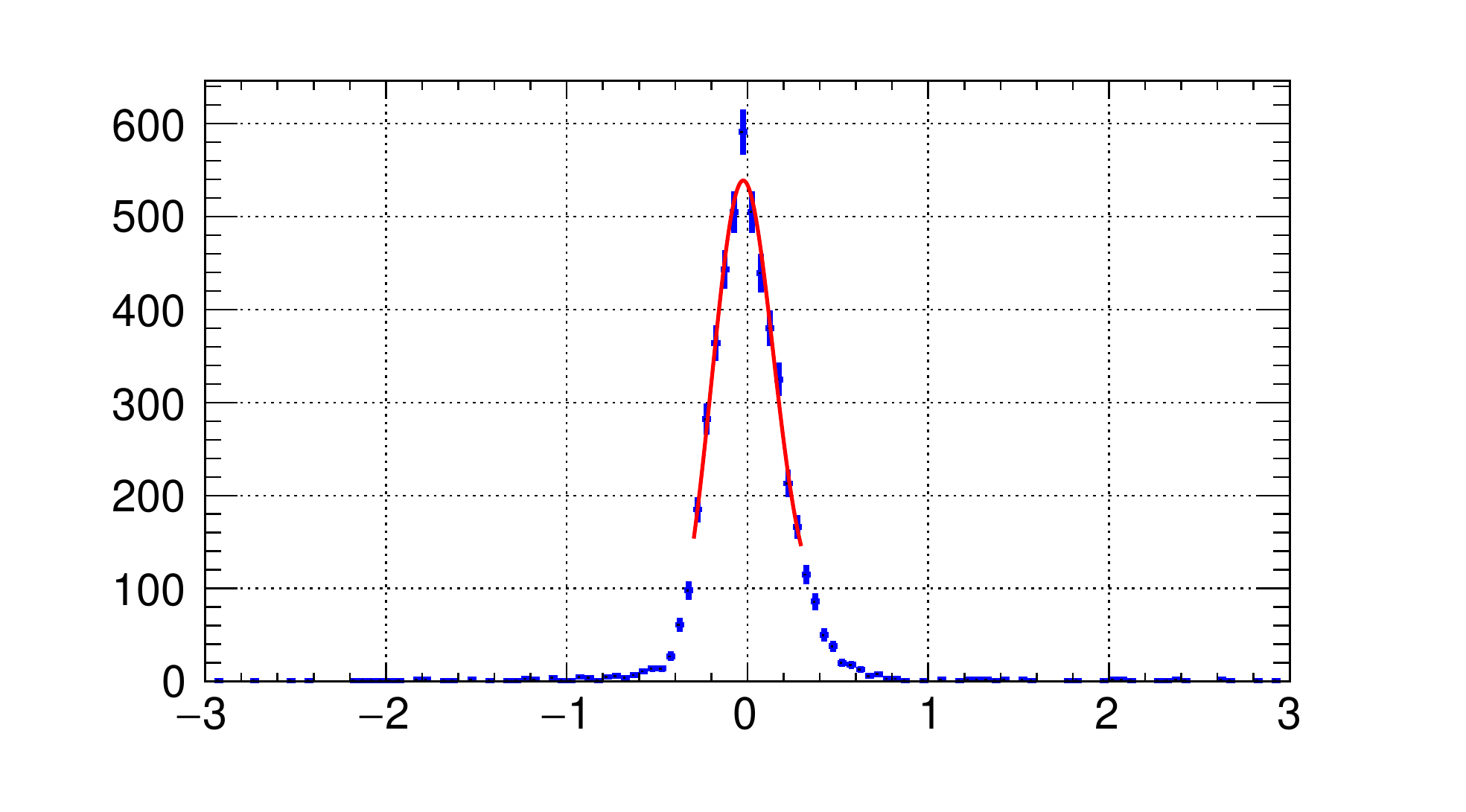}
\end{center}
\end{minipage}
\put(-70,40){Single pixel}
\put(-77,24){$\sigma_{\rm{pixel}}$ = 149 ps}
\put(-125,-66){${\rm t}_{\rm hit} - {\rm t}_{\rm KTAG}$ [ns]}
\put(-210,-10){\rotatebox{90}{count / 50 ps}}
\begin{minipage}{0.5\linewidth}
\begin{center}
\includegraphics[scale=0.42]{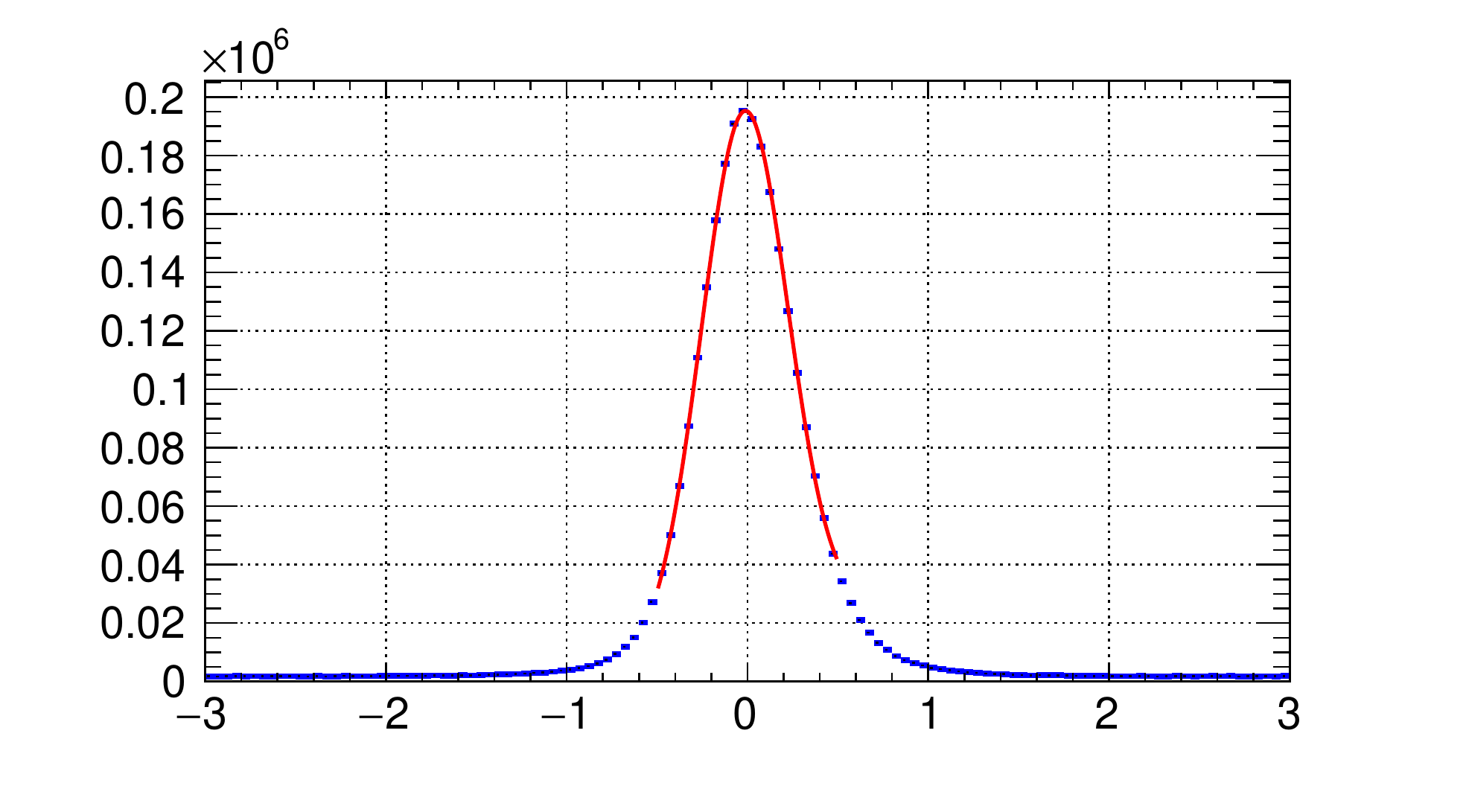}
\end{center}
\end{minipage}
\put(-65,40){GTK1}
\put(-77,24){$\sigma_{\rm GTK}$ = 225 ps}
\put(-125,-66){${\rm t}_{\rm hit} - {\rm t}_{\rm KTAG}$ [ns]}
\put(-210,-10){\rotatebox{90}{count / 50 ps}}
\caption{\label{fig:GTK-Picture7} Time resolution of a single pixel (left) and GTK1 station (right) operated at a bias voltage of 300 V. The quoted resolution  is obtained subtracting the KTAG 70 ps contribution.
}
\end{figure}
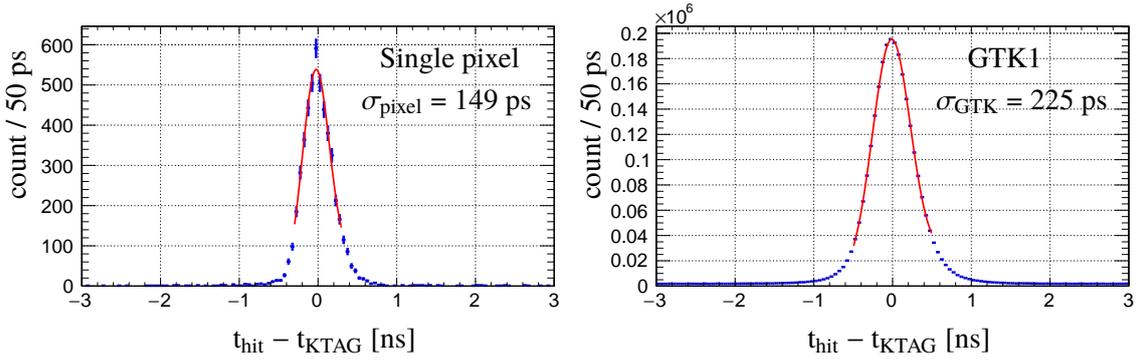

\paragraph{Time Resolution:}
The time of a hit in a pixel is reconstructed by applying two corrections.
First, a time offset correction is evaluated, which depends primarily on the station and the chip and, at second order, on the  column and row of the pixel. This correction is straightforward and is derived by comparing the GTK hit time to other detectors providing a  precise timing, such as the NA48-CHOD or KTAG. 
Secondly, a time-walk correction takes into account  Landau fluctuations of the signal. Figure \ref{fig:GTK-Picture9} shows the time-walk as a function of the ToT.  
The corrected hit-time is obtained from the raw time and the fitted time-walk correction as a function of the ToT value.

The time resolution is obtained by comparing the corrected hit time to the KTAG
time, which has a resolution of  70~ps. When selecting a sample of hits from a
single pixel the time resolution, after unfolding the KTAG contribution, is found
to be 150~ps (\Fig{fig:GTK-Picture7}-left). This value confirms the results obtained 
with the prototype detector when operated at a similar bias voltage \cite{Aglieri:2015}. Resolutions 
75~ps larger are obtained when all the pixels of a station are included in the sample
(\Fig{fig:GTK-Picture7}-right).
This increase may be due to remaining systematic uncertainties in the GTK time offsets.

\paragraph{Momentum and angular Resolution:}
From a sample of 
$\KTAU$ decays, fully reconstructed with the STRAW spectrometer, kaons are traced 
back to the GTK stations and the alignment offsets in the transverse plane of each 
individual station are derived. The momentum calibration is obtained
from the same sample of 
$\KTAU$ decays  by equalizing the average kaon momentum measured with the GTK
to the one reconstructed with the STRAW spectrometer.

The impact of the GTK performance on the kinematic reconstruction is quantified 
by the improved width (rms) of the $ m^{2}_{\rm miss} $ distribution  (\Equ{eq:mmis2}) as a function of the pion momentum for a selected sample of 
$\KTP$ decays,  where $P_K$ is measured with the GTK spectrometer and $P_\pi$ is measured with the STRAW spectrometer.
Including precise GTK information instead of nominal beam values decreases the width  of the 
distribution,  as can be seen in \Fig{fig:datakine}  (\Sec{sec:valid}), 
 in agreement with the value expected from the GTK specifications (\Sec{ssec:req}).

%% file: Sec6-Chanti_v3.tex
\begin{figure}[htbp!]
\begin{center}
\includegraphics[width=0.95\linewidth]{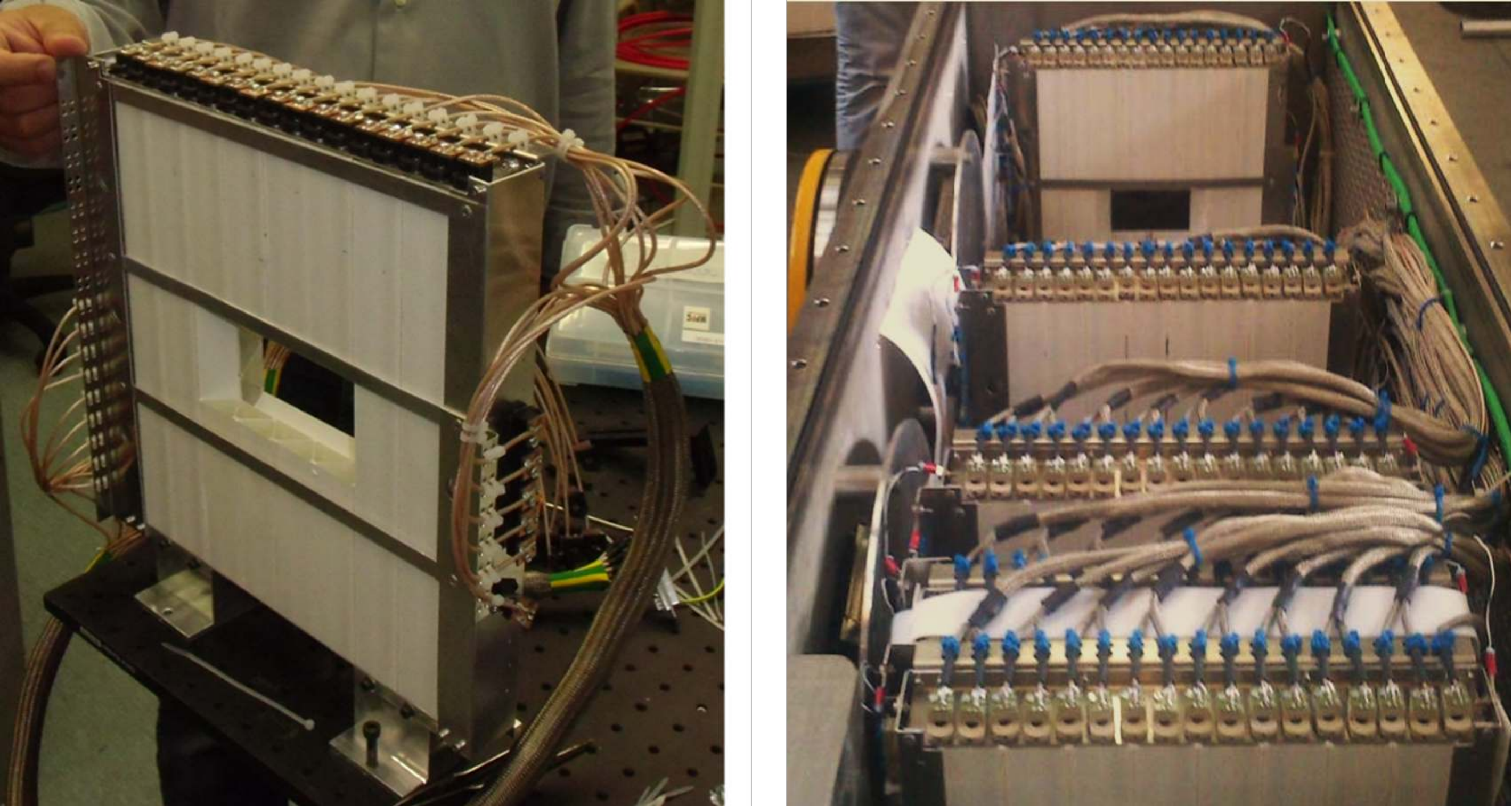}
\end{center}
\caption{\label{fig:CHANTI-Picture3}Left: A CHANTI station. Right: Upstream part of the CHANTI vacuum vessel, equipped with the first five stations. The photograph was taken during the assembly.}
\end{figure}

The CHANTI detector provides rejection for background from inelastic interactions of the beam with the most downstream GTK station, GTK3. Particles resulting from these interactions can enter the acceptance of the detector, creating  background for $\PNNP$  
events. The inelastic interactions on GTK3 are not, however, the only source of activity in the CHANTI, which detects also the muon halo close to the beam and a fraction of the charged particles generated upstream of GTK3.
The CHANTI is composed of six square hodoscope stations $300 \times 300$ mm$^2$ in cross section with a $95 \times 65$ mm$^2$ hole in the centre to leave room for the beam (\Fig{fig:CHANTI-Picture3}-left). The first station is placed 28 mm downstream of GTK3, and the distance between each station and the next one approximately doubles for successive stations, so that the angular region between 49 mrad and 1.34~rad is covered hermetically for particles generated on GTK3. 
GTK3 and all CHANTI stations are located inside the same vacuum vessel.
\Fig{fig:CHANTI-Picture3}-right shows a picture of the open vacuum vessel during installation. The stations  are made of scintillator bars of triangular cross section read out with fast wavelength-shifting (WLS) fibres coupled to silicon photomultipliers (SiPMs). 

\begin{figure}[ht]
\begin{minipage}{0.5\linewidth}
\begin{center}
\includegraphics[width=0.85\linewidth]{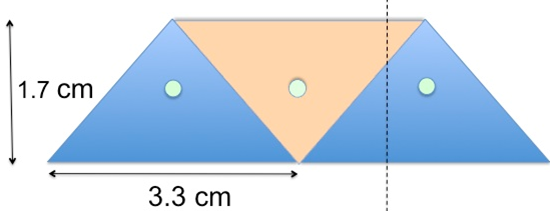}
\end{center}
\end{minipage}
\begin{minipage}{0.5\linewidth}
\begin{center}
\includegraphics[width=0.85\linewidth]{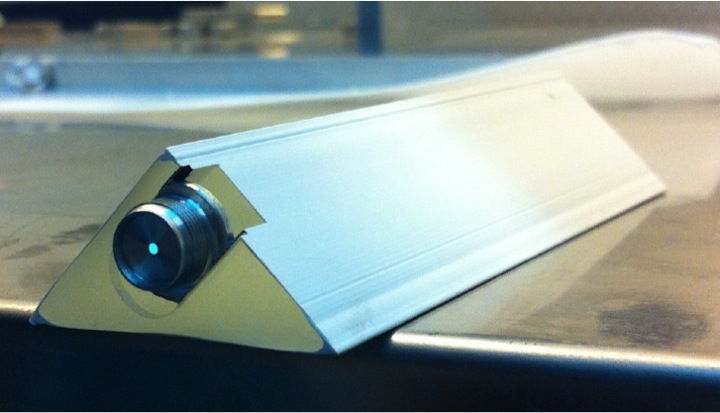}
\end{center}
\end{minipage}
\caption{\label{fig:CHANTI-Picture1} Left: Arrangement of the bars to form a plane. Right: A triangular bar, constituting the building block of the CHANTI detector. The WLS fibre is visible in the centre. }
\end{figure}

Each station has two readout planes, with the bars oriented vertically and horizontally to form X and Y 
views. For each view, the triangular bars are arranged into a plane 
as shown in \Fig{fig:CHANTI-Picture1}-left; particles incident from the front of the detector generally 
traverse two bars.  Each of the six stations consists of 48 bars,  adding to a total of 288 bars.
A more detailed description of the CHANTI detector may be found in \cite{Ambrosino:2016}.

\subsection{Design and construction}\label{ssec:CHANTI:Detector}

The CHANTI polystyrene-based scintillator bars
were extruded complete with the hole for the WLS fibre and a thin, co-extruded layer of TiO$_2$, which provides diffuse reflectivity along the lateral surfaces.

To construct the CHANTI stations, the scintillator bars were first cut
to three different lengths: 30 cm, 11.75 cm, and 10.25 cm. One side of each bar was then machined to create the seat for the connector that couples the fibres to the SiPM. Bicron BCF-92, multi-cladded, 1 mm diameter WLS fibres were used for light readout; the fibres were mirrored at the end opposite to the SiPM by Al sputtering under vacuum, cut to the desired length, and glued to the connectors. 
The free space between the fibre and the scintillator was filled with
an optically transparent silicone-elastomer glue. A low outgassing structural glue was used to fix the connector to the scintillator.

Each individual bar was tested with cosmic rays prior to assembly of the CHANTI stations. This test is crucial because individual bars cannot be replaced after gluing. Including spares, about 300 bars were built and individually tested over a period of about six hours in a climate-controlled chamber. Fewer than 6\% of the bars failed the test. A normally incident cosmic ray crossing two bars forming a plane releases about 70 photoelectrons (pe) shared among the two bars (\Fig{fig:CHANTI-Picture1}-left); the detector can thus be operated with thresholds of 5--6~pe per channel with negligible efficiency losses.

The SiPMs (Hamamatsu MPPC-S10362-13-050-C)  have an active area of 1.3 $\times$ 1.3 mm$^2$ and a pixel size of 50 $\times ~50 ~\mu$m$^2$, with a total of 667 pixels.
The current-voltage curve for each SiPM was measured at fixed temperature in a climate-controlled chamber to obtain the breakdown voltage $V_{\rm bd}$ at a working temperature of 25 $^{\circ}$C ($V_{\rm bd}$ is typically 70~V). 
To determine the temperature dependence of $V_{\rm bd}$, the test was repeated at different temperatures over the range 5--35 $^\circ$C for 25\% of the SiPMs. Good agreement was obtained with the value for the temperature coefficient quoted by Hamamatsu (50~mV/K).
The typical dark count rate at 25 $^{\circ}$C for a threshold of 0.5 pe was found to be approximately 800~kHz, again in agreement with manufacturer specifications.

The materials used to construct the CHANTI stations were chosen to be compatible with operation in vacuum.
The outgassing rate was measured to be less than $3\times10^{-5}$ mbar litre per second per station. The vacuum vessel hosting both GTK3 and the six CHANTI stations is made of two rectangular stainless-steel chambers connected by a tube to form a single, 2165 mm long vacuum volume. Each chamber is equipped with flanges and 16-channel, D-sub 37-pin feedthrough connectors for signal and high-voltage.

\subsection{Front-End electronics}\label{ssec:CHANTI-Electronics}

The SiPMs require a reverse bias voltage of about 70~V and produce fast signals  with rise times  below 1~ns, amplitudes of few $\mu$V/pe on 50 $\Omega$ load and 
gains of about $7\times10^5$. The bias-voltage stability, however, must be kept under control at the per mille level for satisfactory gain stability. Moreover, since the typical signal is relatively small, it must be amplified before it can be discriminated and/or digitized. Therefore, a custom, all-in-one front-end board (CHANTI-FE) was designed and is used for CHANTI. 

The signals from each of the six stations are grouped on three vacuum feedthrough connectors 
and input to nine custom-designed, VME 9U CHANTI-FE boards. Each CHANTI-FE board can accept up to 32 channels. For each channel, the board sets the SiPM bias voltage, monitors the bias-voltage current, and amplifies the signal. 
Each board can also monitor up to four PT100 temperature probes placed on the detector; this allows the bias voltage to be adjusted to maintain constant SiPM gain in response to temperature variations. 
Other important characteristics of the CHANTI-FE board include
\begin{itemize}
\item bias-voltage setting with 10 mV precision, stable with temperature and time;
\item current readout with nA precision in the range  $0 - 2500$ nA; 
\item fast amplification by a factor of 25 of the analogue signal on a 50$~\Omega$ load;
\item adjustable  settings and thresholds via CANOpen standard communication.
 \end{itemize}

The analogue outputs from the CHANTI-FE boards are converted to LVDS signals by a  LAV-FE board (\Sec{ssec:lav-fee}) modified to handle the dynamic range of the CHANTI signals. The boards generate an output signal equal in duration to the time during which the input signal is above threshold (time-over-threshold, ToT). Each input channel is compared to two programmable thresholds (high and low), so that the 32 physical input channels are mapped to 64 logical output channels. The LVDS signals are digitized into leading and trailing edge times by the standard TDCB+TEL62 readout system (\Sec{sec:TDAQ}).

The thresholds are calibrated for each electronic channel as described in \cite{Ambrosino:2016}. This procedure enables the threshold to be set at the desired  number of photoelectrons for each channel.

\subsection{Performance in 2015 \label{ssec:CHANTI-Performance}}
\begin{figure}
\begin{minipage}{0.5\linewidth}
\begin{center}
\includegraphics[width=1.\linewidth]{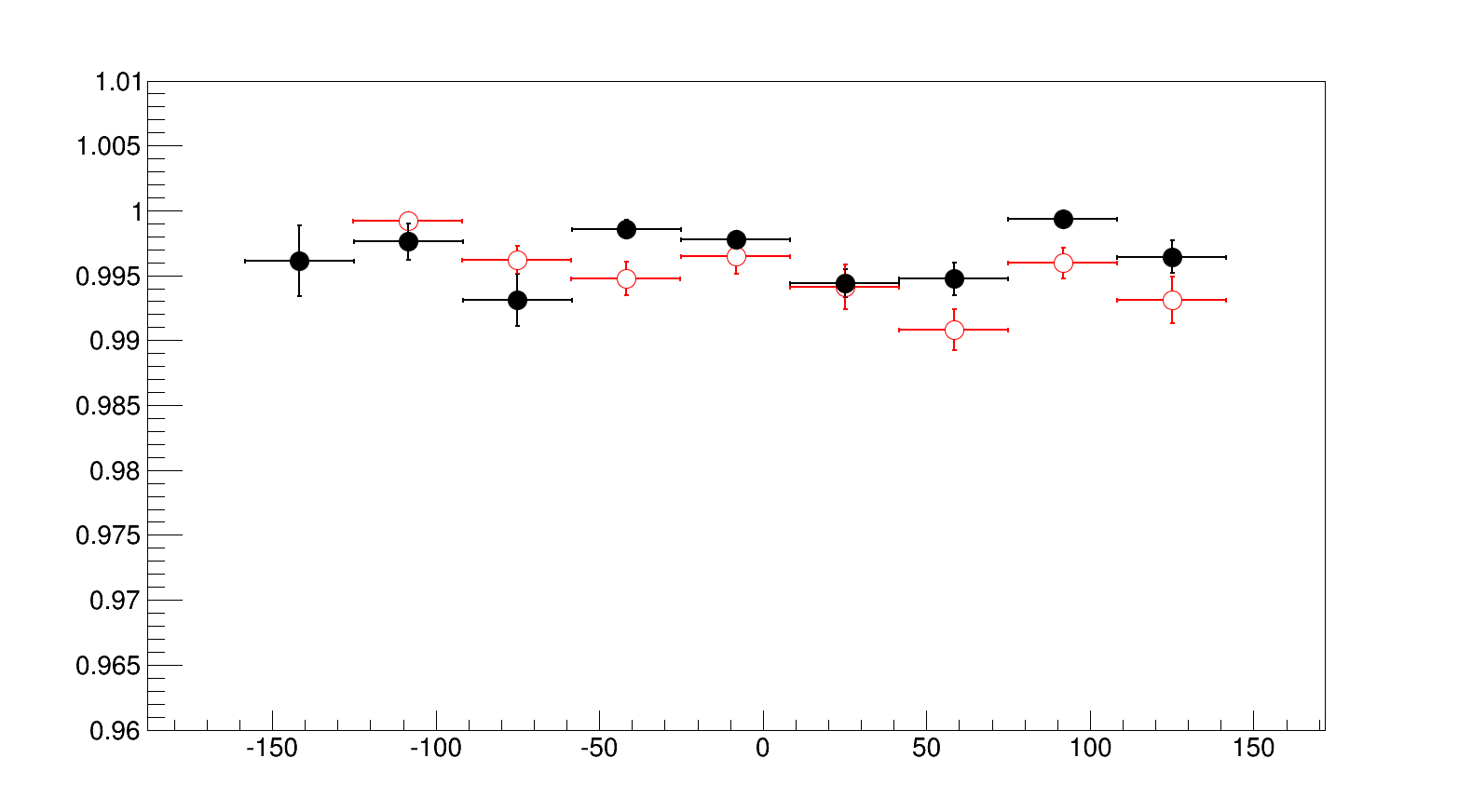}
\put(-80,-5){X or Y [mm]}
\put(-215,65){\rotatebox{90}{Efficiency}}
\end{center}
\end{minipage}
\begin{minipage}{0.5\linewidth}
\begin{center}
\includegraphics[width=1.11\linewidth]{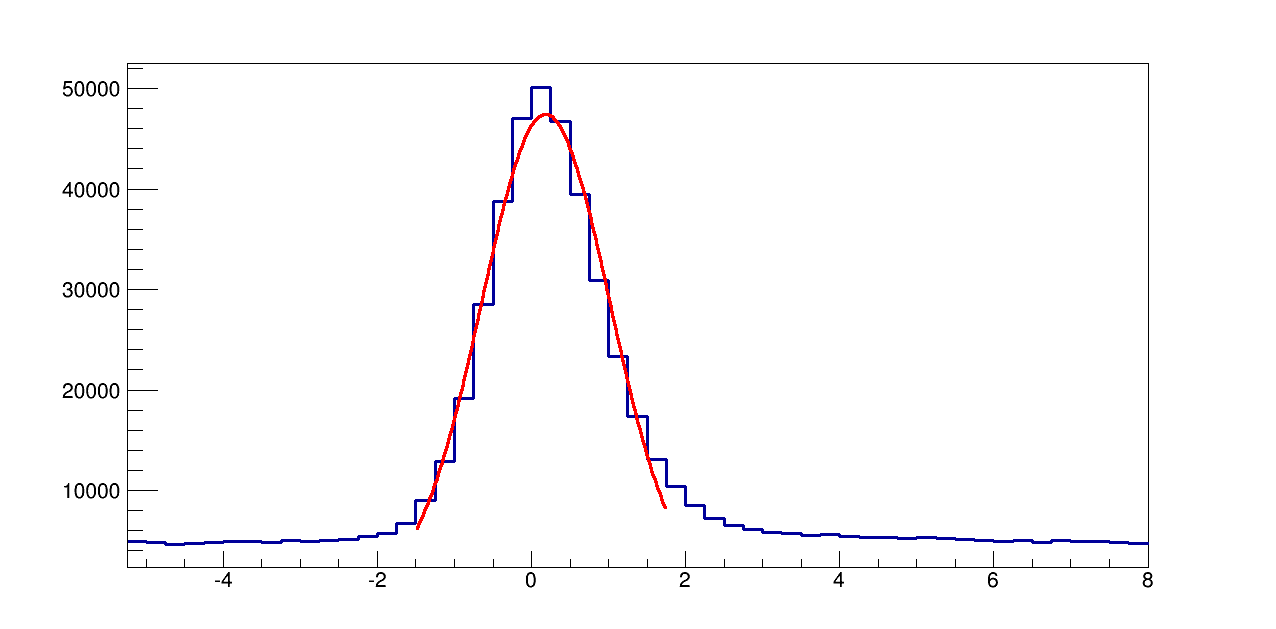}
\put(-108,70){$\sigma_{\rm CHANTI}$  = 830 ps}
\put(-115,-5){${\rm t}_{\rm CHANTI} - {\rm t}_{\rm KTAG}$  [ns]}
\end{center}
\end{minipage}
\caption{\label{fig:CHANTI-Picture4}Left: Measured detection efficiency in a CHANTI station for a sample of muons. Black full (red open) dots refer to X (Y) view, respectively.  Right: Difference between the CHANTI candidate time (after corrections for channel-by-channel time offsets and slewing) and the time measured in the KTAG. The quoted resolution  is obtained after subtraction of the KTAG 70~ps contribution.}
\end{figure}

The CHANTI was installed and commissioned in 2014 and fully exploited during the 2014 and 2015 data taking periods. 
The efficiency and spatial resolution were measured using muon runs (\Sec{ssec:samples}), which provide a clean sample of straight and penetrating minimum ionizing particles (MIPs). At a threshold of about 50 mV, the efficiencies for all views of all stations were measured to be uniform and above 99\%  (\Fig{fig:CHANTI-Picture4}-left), in good agreement with prior laboratory tests on prototype detectors with cosmic rays. 

Given the geometry of the detector, a straight MIP normally incident on a station shares its energy release between two adjacent bars in a way inversely proportional to the impact parameter of the track with respect to the bar centres. 
The position of a hit in a station is determined from the hit positions in each bar weighted by the energy release in each.
The charge in each bar is not measured directly, but is evaluated as a function of the ToT.
The resolution is found to be about 2.5 mm, which is better by nearly a factor of two than the intrinsic resolution from the pitch of the fibres (16.5~mm/$\sqrt{12}$). The relative alignment of the six CHANTI stations was checked with muon halo tracks; the deviations from the nominal positions were measured to be below 0.5~mm, which matches the design specifications.

\begin{figure}[htbp!]
\vspace{-58mm}
\begin{center}
\includegraphics[scale=0.6]{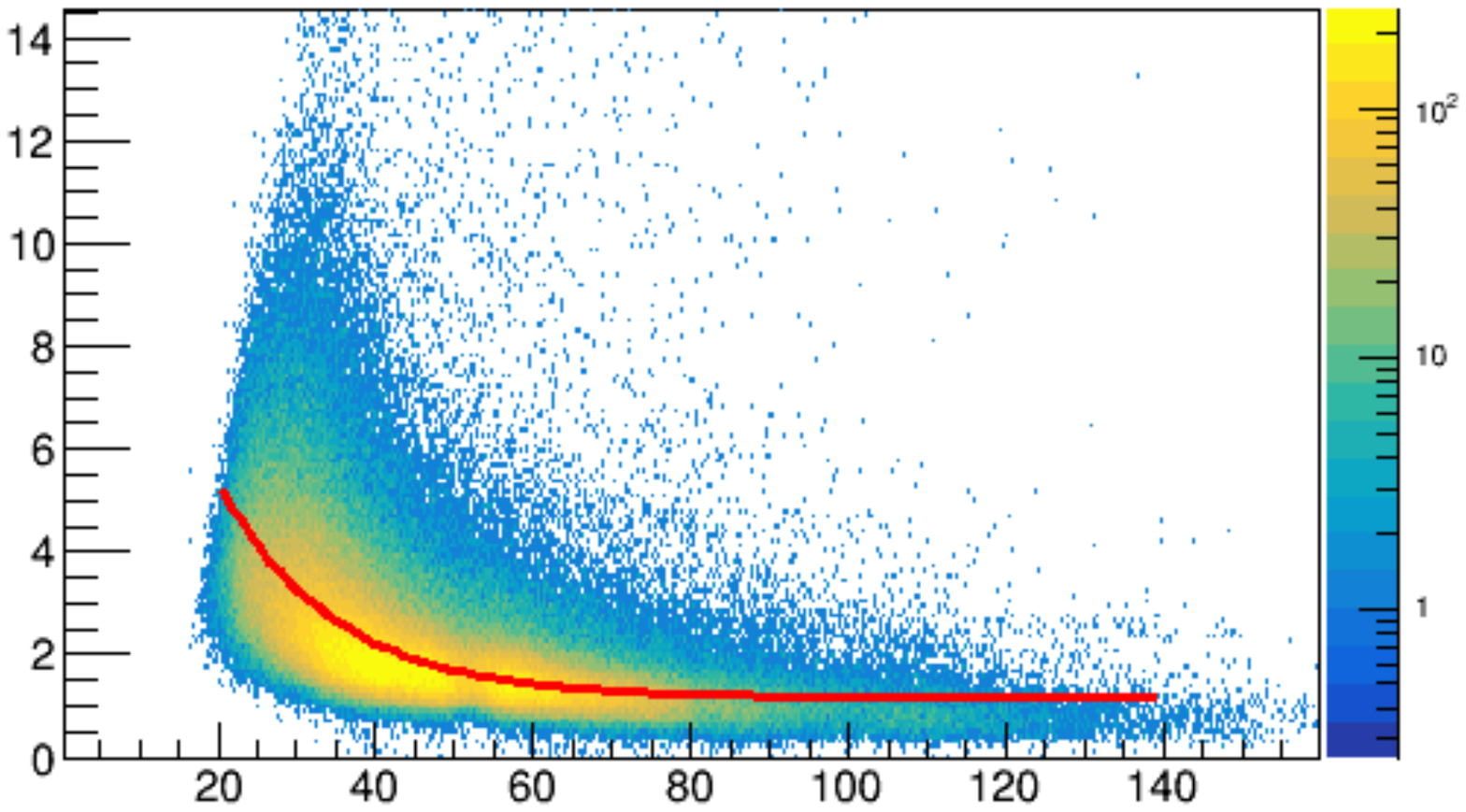} 
\put(-99,149){ToT [ns]}   
\put(-335,215){\rotatebox{90}{Slewing correction [ns]}}
\end{center}
\vspace{-53mm}
\caption{\label{fig:CHANTI-Picture7}Slewing correction, evaluated for signal with both thresholds crossed, as a function of low threshold time-over-threshold (ToT).  The line corresponds to the average correction parameterization.}
\end{figure}

The LAV-FE boards provide two digital outputs for each channel, which correspond to two different programmable threshold levels. This feature allows corrections for slewing to be calculated by linear extrapolation when both thresholds are crossed. An average slewing correction calculated from the ToT is used when only the lower threshold is crossed; the slewing correction dependence on the low threshold ToT is shown in Figure \ref{fig:CHANTI-Picture7}.

After slewing corrections, the typical single hit time resolution in the CHANTI is found to be 1.14~ns.
A CHANTI signal is usually made of more than one hit (typically two hits per view) that are close in space and time. The time resolution for CHANTI signals with respect to the reference time from the KTAG is found to be about 830~ps after subtraction of the 70~ps KTAG contribution.
The distribution of the time difference  between the CHANTI candidate and the 
KTAG candidate closest in time is shown in \Fig{fig:CHANTI-Picture4}-right.
The overall rate on the CHANTI, when requiring at least one hit in both X and Y views, is expected to be about 7.2 MHz at nominal beam intensity with the kaon inelastic interactions contributing only 11\% of the total rate. The measured time resolution is therefore adequate to keep the random veto probability to an acceptable level of a few percent.
A detailed study of the CHANTI random veto probability has been performed on a sample of 2015 data recorded at 40\% nominal beam intensity \cite{MirraPhD:2016}. 
It has shown that,  if a 6~ns wide time window is used to veto events, about 1.8\% of genuine $\KPL$ decays with one track in the final state are vetoed.

The tracking capability of the CHANTI is helpful in distinguishing beam inelastic interactions on GTK3 
from events with charged particles generated upstream. This is done by selecting events where at least one track segment, formed from CHANTI hits in the different stations,  extrapolates back to the GTK3 region in the XY plane, which enables the selection of a sample of inelastic-interaction events that are useful for background studies.

%% file: Sec7-STRAW_v3.tex
%%%%%%%%%%%%%%%%%%%%%
One of the major challenges of the NA62 experiment is to reduce the background from 
the most abundant kaon decays and accidental tracks (\Sec{sec:intro}). 
The required kinematic rejection depends critically on precision low-mass systems that track the incoming kaon and the final state particles.   
The straw spectrometer measures the trajectories and the momenta of the charged particles produced in the kaon decay. It extends over a length of 35 m along the beam line, starting  $\sim$ 20~m after the decay region. It consists of four straw chambers and a large aperture dipole magnet (MNP33) providing an integrated field of 0.9~Tm  (\Fig{fig:detectorview1}).
To minimize multiple scattering the chambers are built of light material and are installed inside the vacuum tank. The total amount of material in the spectrometer corresponds to 1.8\% $X_0$.

\subsection{Design and construction}\label{ssec:STRAW-Design}

\begin{figure}[h]
\begin{minipage}{0.4\linewidth}
\begin{center}
\includegraphics[scale=0.6]{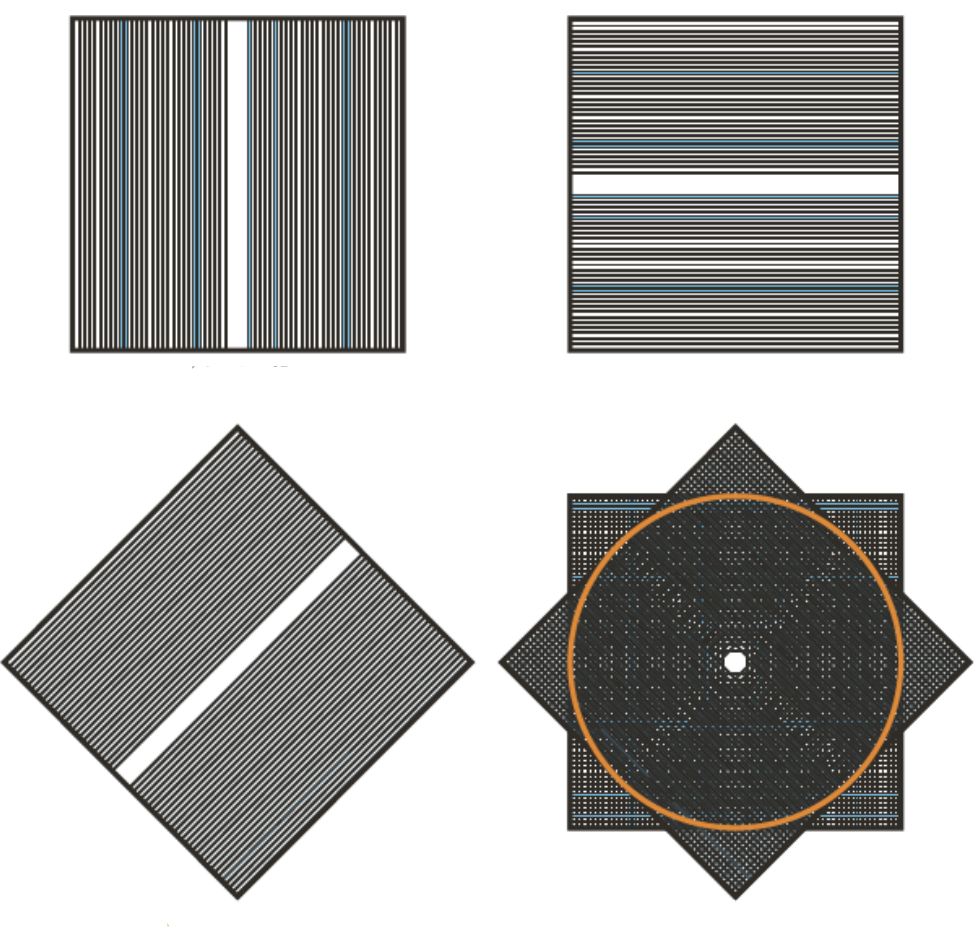}
\end{center}
\end{minipage}
\color{black}
\put(-160,13){X view}
\put(-55,13){Y view}
\put(-160,-100){U view}
\put(-71,-100){\vector(1,0){57}}
\put(-18,-100){\vector(-1,0){60}}
\put(-55,-120){{\diameter 2.1 m }}
\begin{minipage}{0.5\linewidth}
\begin{center}
\includegraphics[scale=0.22]{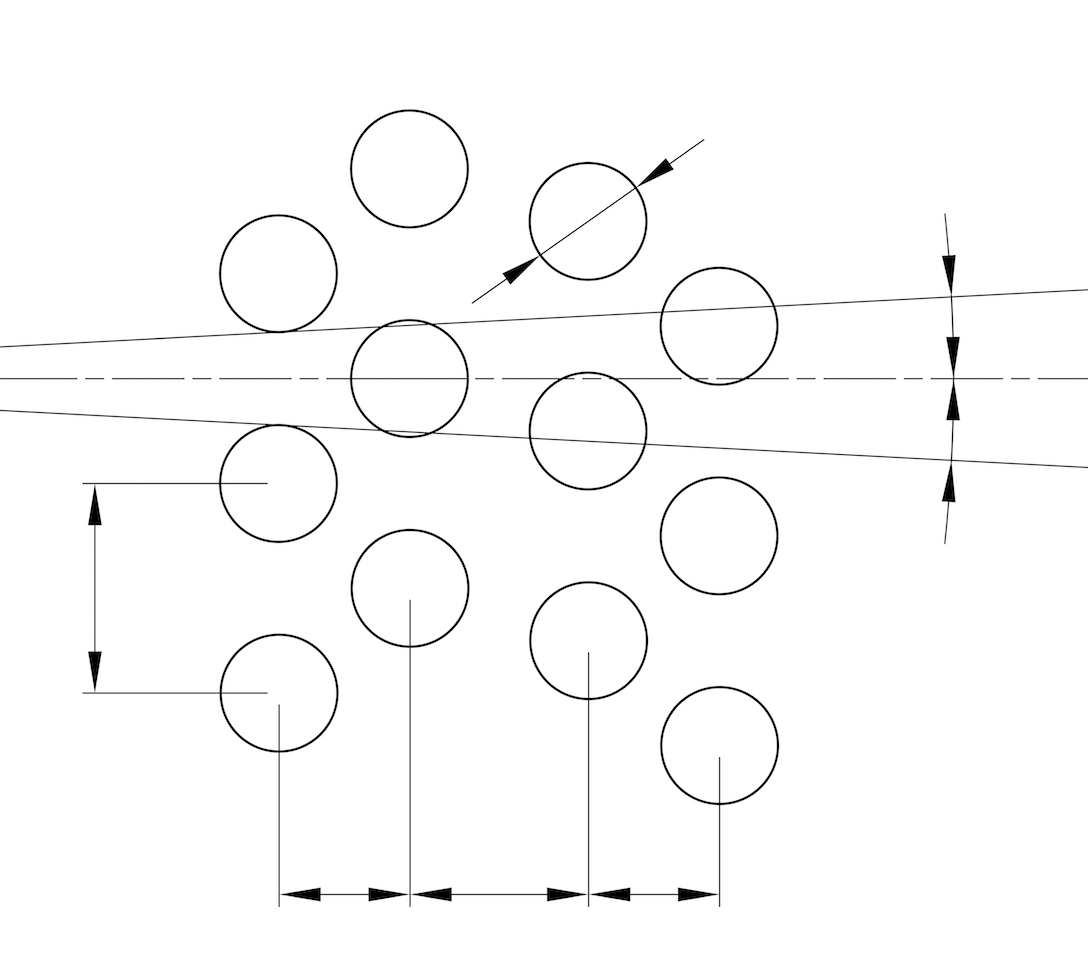}
\end{center}
\end{minipage}
\put(-105,90){{\diameter 9.82~mm }}
\put(-20,32){{3 $^\circ$}}
\put(-20,12){{3 $^\circ$}}
\put(-215,-20){{17.6 mm}}
\put(-158,-120){{11~mm \ 15~mm \ 11~mm}}
\caption { \label {fig:straw-layout}Left: One straw chamber is composed of four views (X, Y, U, V) and each view measures one coordinate. Near the middle of each view a few straws are left out forming 
a free passage for the beam. 
Right: 
The straw geometry is based on two double layers per view with sufficient overlap to guarantee at least two  straw crossings per view and per track, 
as needed to solve the left-right ambiguity.  
The $\pm ~3^\circ$  angle corresponds to the angular range of tracks produced in kaon decays and 
detected within the geometrical acceptance of the spectrometer. }  
\vspace{6mm}
\begin{minipage}{1.0\linewidth}
\begin{center}
\includegraphics[width=0.9\linewidth]{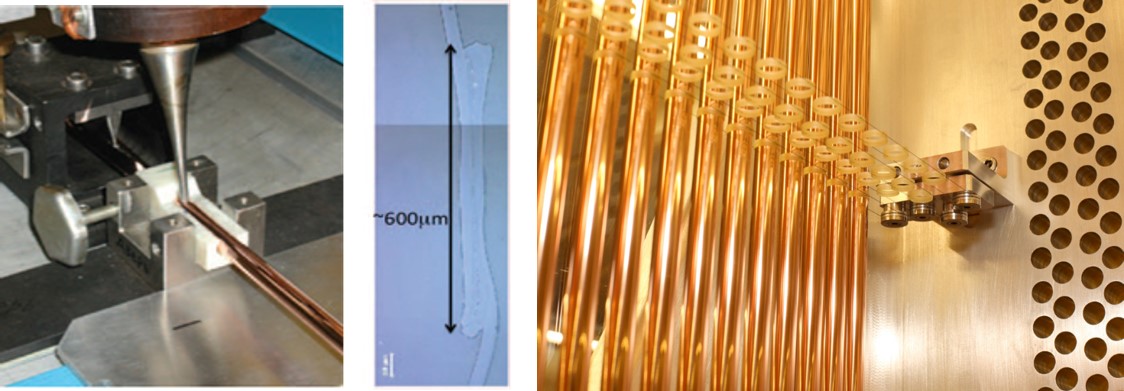}
\end{center}
\caption{\label{fig:STRAW-Picture3}Left: Close view of the straw welding machine with the sonotrode (ultrasonic welding head) in the centre. Middle: Microscope image showing the cross section of the weld. Right:  Close view of the spacer arrangement: one layer of the vertical view is already mounted; in front are three spacers mounted, waiting for the next set of straws to be inserted. The straw is supported in two positions along its  length. }
\end{minipage}
\end{figure}

The module design is optimized to minimize multiple scattering and to give uniform space
resolution over the full active area.  Each straw chamber is composed of two modules. One module contains two views measuring X ($0^\circ$), Y ($90^\circ$)
and the other module contains the  U ($-45^\circ$) and V ($+45^\circ$) views (\Fig{fig:straw-layout}-left).
The chamber active area is a circle of 2.1~m outer diameter centred on the longitudinal Z axis.
Each view has a gap of about 12 cm without straws near the centre, such that, after overlaying the four views, an octagon shaped hole of 6~cm apothem is created for the beam passage.  As the beam has an angle of 
$+1.2$ mrad and $-3.6$ mrad in the horizontal plane, upstream and downstream of the magnet, respectively  (\Fig{fig:K12beam-downstream}),
this hole is not centred on the Z axis, but has offsets (along the X direction) 
for each chamber (\Tab{tab:Straw-table1}).  
High detection redundancy is provided through a  straw arrangement with four layers per view, which guarantees at least two hits per view, i.e. 8 --12 hits per track per chamber (\Fig{fig:straw-layout}-right). Due to the 12 cm  gap without straws in each view, the number of hits per chamber is not evenly distributed over the detector surface.

\begin{table}[ht]
\setlength{\tabcolsep}{2ex}
\caption{Straw spectrometer characteristics.}
\label{tab:Straw-table1}
\vspace{2ex}
\centering
\begin{tabular*}{0.8\textwidth}{@{\extracolsep{\fill} }  l c c c}
\hline \hline
Number of chambers: &  \multicolumn{3}{c}{4}    \\
Number of views / chamber & \multicolumn{3}{c}{4}   \\
Order of views along beam direction & \multicolumn{3}{c}{U, V, X, Y }   \\
Number of straws per view & \multicolumn{3}{c}{448}   \\
Beam hole size (octagon apothem) & \multicolumn{3}{c}{6 cm}    \\
\hline
Beam Hole Offset & X & Y  & Z \\ 
\hspace {20pt} Chamber 1 & 101.2 mm & 0 mm & 183 m  \\
\hspace {20pt} Chamber 2 & 114.4 mm & 0 mm & 193 m \\
\hspace {20pt}  MNP33      &                 &         &  198 m \\
\hspace {20pt} Chamber 3 &   92.4 mm & 0 mm & 204 m \\
\hspace {20pt} Chamber 4 &   52.8 mm & 0 mm &  218 m \\
\hline
\hline
\end{tabular*}
\end{table}

Each chamber contains 1792 straws of 9.82~mm diameter and  2160~mm length.  The gas inside the straws is  a mixture  of  70\% Ar and 30\% CO$_2$ at atmospheric pressure. The straws are operated in the vacuum of the decay tank. 
They are sufficiently separated from each other  to allow the 
 flexible straws to increase their diameter when the decay tank is under vacuum. 

Each straw is made from 36 $\mu$m thick polyethylene terephthalate (PET),
coated with 50 nm of copper and 20 nm of gold on the inside. The straw is fabricated by ultrasonic 
welding with the seam along the straw axis (\Fig{fig:STRAW-Picture3}-left). 
The gold-plated tungsten anode wires are 30 $\mu$m in diameter. They are tensioned at 80 g without supports and crimped at both ends of the straw.

Dedicated tooling was developed to guarantee straight straws positioned to high precision.
The tooling applies a fixed pre-tension of 1.5 kg to each straw prior to the gluing of the straw ends into the aluminium frames, minimizing the need for supporting material.
The straw tension decreases the deflection of the straws under the force of gravity. This effect is particularly important for the horizontal straws and the straws at 45~degrees, because the anode wires have to be well centred inside.  
However, the straw tension alone is not sufficient to keep the straw deflection within the required  limits. 
For this reason two light ``spacers'' consisting of two tungsten wires (0.1~mm in diameter) under tension with precision-glued ULTEM\textsuperscript{TM} rings support the straw in two positions along its length (\Fig{fig:STRAW-Picture3}-right). The rings are glued to 
the wires under tension using dedicated tooling with a relative precision between rings
of $20~\mu$m, and the tension in the two wires is 1.6 kg each. 
Each spacer is  
adjustable in position and tension (i.e. position and precise distance between adjacent rings). The spacers are installed and positioned accurately  prior to the straw installation. To align the rings in the spacer with the holes in the aluminium
frame, two wires are mounted in two holes in the aluminium frame (at the position of the straws). The wires pass through the corresponding spacer rings and the distance between the wires and the
centre of the rings is measured. The adjustment screws in the spacer
supports are used to correct the position if needed. 

\begin{figure}[h]
\begin{minipage}{1.0\linewidth}
\centering
\includegraphics[width=0.58\linewidth]{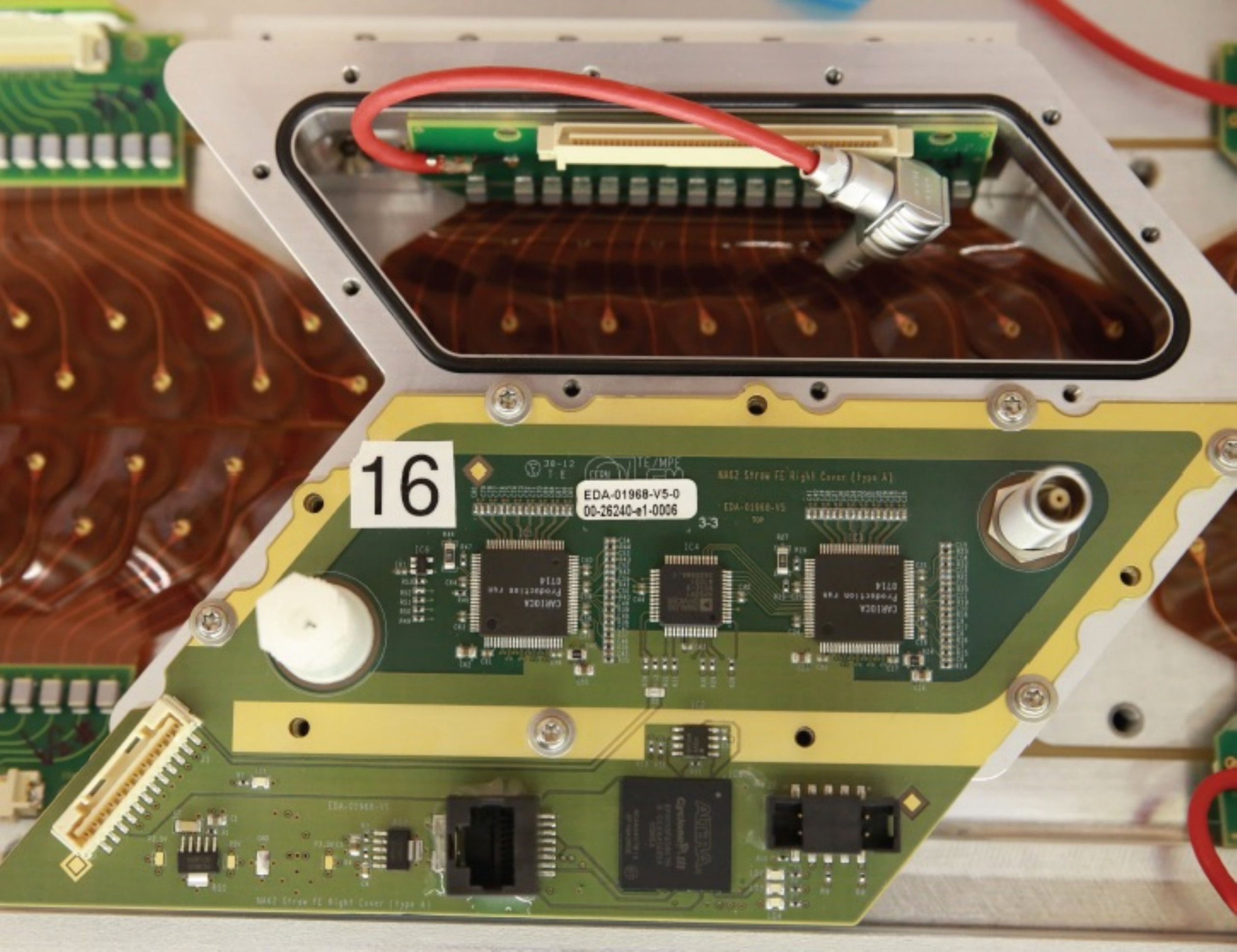}
\end{minipage}
\put(-425,110){Gold plated crimp-tubes connect the signal wires to the web}
\put(-85,90){Dismountable}
\put(-80,80){manifolds}
\put(-85,5){Two 8-channel}
\put(-80,-5){ASICs}
\put(-85,-60){FPGA hosting}
\put(-80,-70){the TDC}
%Add arrows
\color{red}
\thicklines
\put(-408,105){\vector(1,-1){72}}
\put(-85,80){\vector(-3,-2){45}}
\put(-88,0){\vector(-4,-1){70}}
\put(-85,-65){\vector(-1,0){120}}
\color{black}
\caption{\label{fig:STRAW-Picture5} Picture of the straw-end connectivity and the gas manifold. A
flex-rigid circuit board (web) connects the high-voltage to the signal wires and transmits the
signal to the front-end cover. The cover also houses the high-voltage connector and the feedthrough for the gas.}
\end{figure}
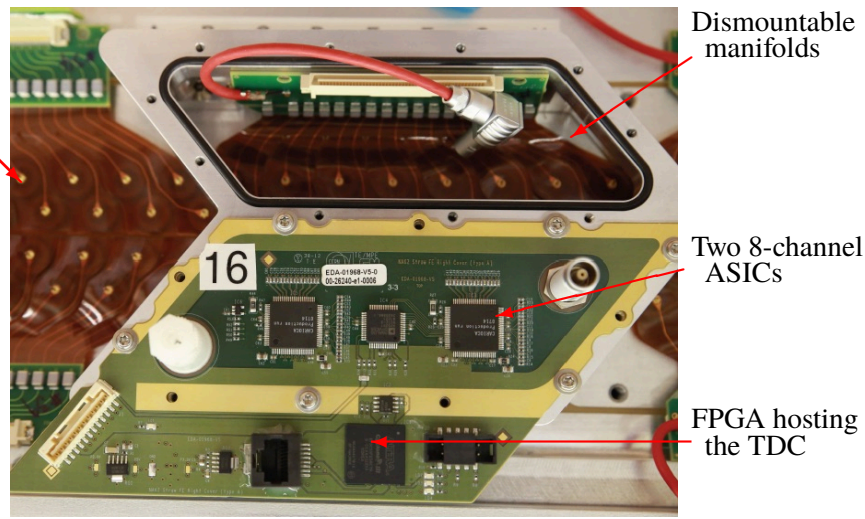

A flex-rigid circuit (web) makes the connection of the anode signal wires 
to the front-end cover (\Fig{fig:STRAW-Picture5}). The front-end cover, which has a modularity of 16 straws, closes the active gas volume. 
It  is designed to resist the force of a pressure difference of $\sim$ 1 bar in case of the 
unlikely event of a broken straw. In case a straw breaks, the gas system detects a pressure drop
and closes both the inlet and the outlet within a few milliseconds.

\subsection{Calibration and operation}\label{ssec:STRAW-calib}
\begin{figure}[ht]
\begin{center}
\includegraphics[width=0.9\linewidth]{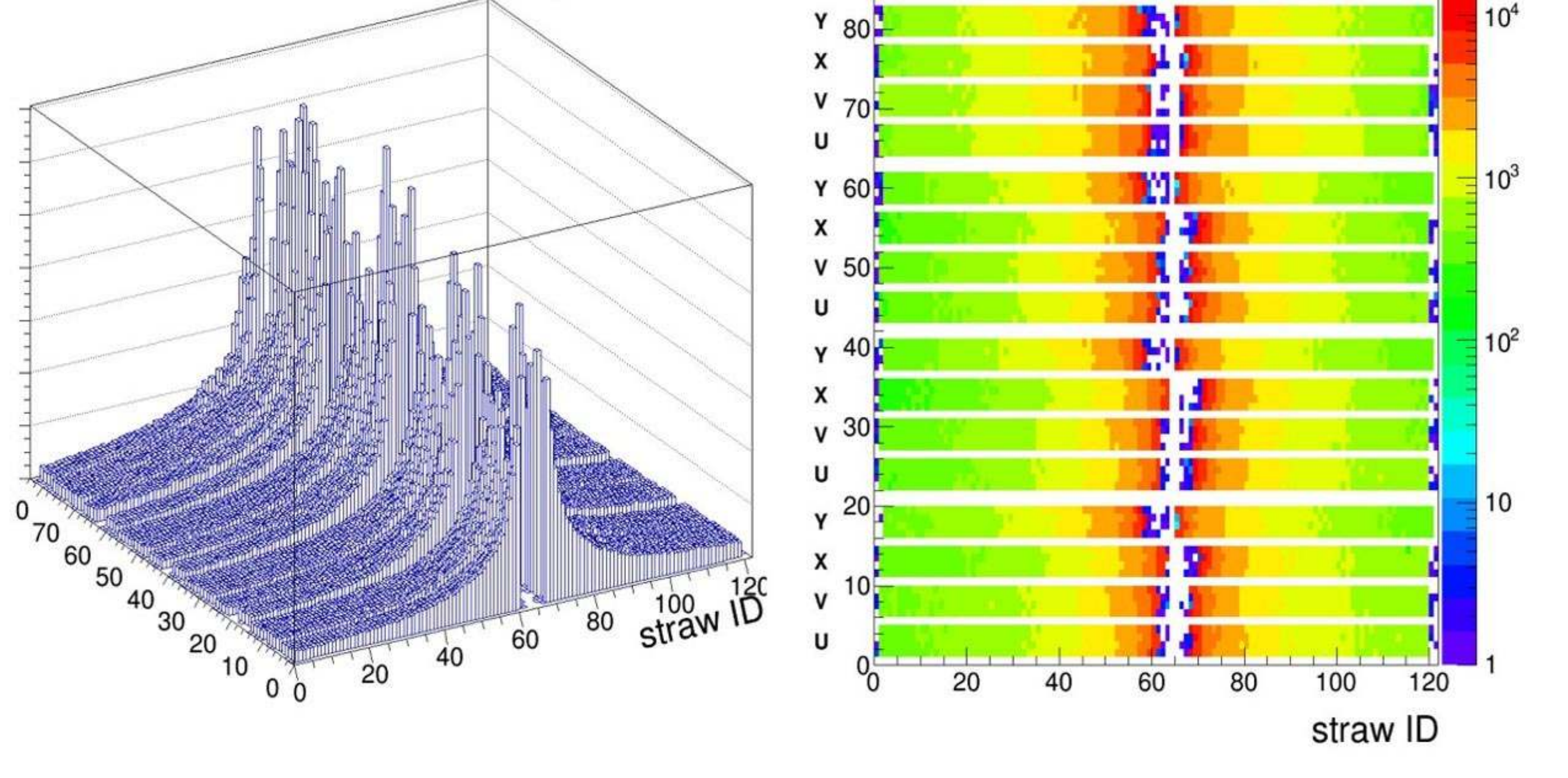}
\put(-410,70){\rotatebox{90}{count/straw [arbitrary unit]}}
\end{center}
 \caption{\label{fig:Straw-Picture9} {Online plot showing the hit activity straw by straw in all four chambers.  The histogram shows triggered data with the expected profile. The requirement for no material to interact with the beam means no straws in the beam region, as shown by the gap in the plots. 
 The entries  with very low hit rates (dark blue) are not equipped with straws and count only electronic noise. }}
\end{figure}

The four straw chambers were fully operational in 2015 including the custom front-end electronics and data acquisition system (described in \Sec{ssec:Straw-readout}).   An example of the straw activity is shown in \Fig{fig:Straw-Picture9}. 
 
Two types of data sets are available from the straw readout, allowing for a number of checks and monitoring tasks:
\begin{itemize}
\item data sent through the back-end interface to a local PC containing all hits in the straws during a given time period. Due to the 
high data rate, only a subset of the total data is sent through this channel.
\item triggered data sent to the PC farm, which is the normal data acquisition stream. 
\end{itemize}

The maximum drift time in the straws is approximately 140 ns and both leading and trailing edges of the signal are read out. The leading edge is used to measure the drift time which gives the lateral position of a track crossing a straw using the radial distance (R) to the wire-drift time R-t correlation. 

\begin{figure}[ht]
\begin{center}
\includegraphics[width=0.9\linewidth]{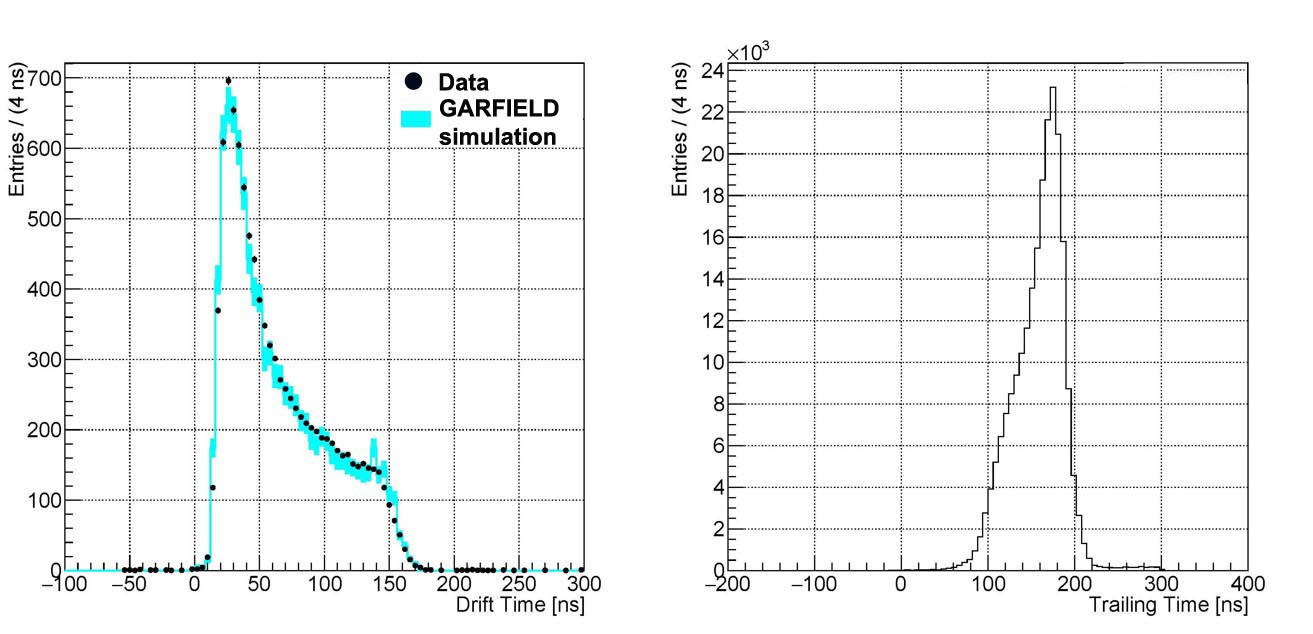}
\end{center}
 \caption{\label{fig:Straw-Picture8} {Left: Measured and simulated drift time spectra (leading edge). The dots represent data and the coloured line the simulation obtained using {\tt GARFIELD} \cite{Veenhof:3000}. Right: Measured trailing edge time distribution.}}
\end{figure}

Because the maximum drift time in a cylindrical tube is the same for all tubes,
the trailing edge gives a measurement of the absolute time of a hit and can be used to aggregate hits belonging to the same track (\Fig{fig:Straw-Picture8}).

\begin{figure}[ht]
\begin{minipage}{0.25\linewidth}
\begin{center}
\includegraphics[scale=0.3]{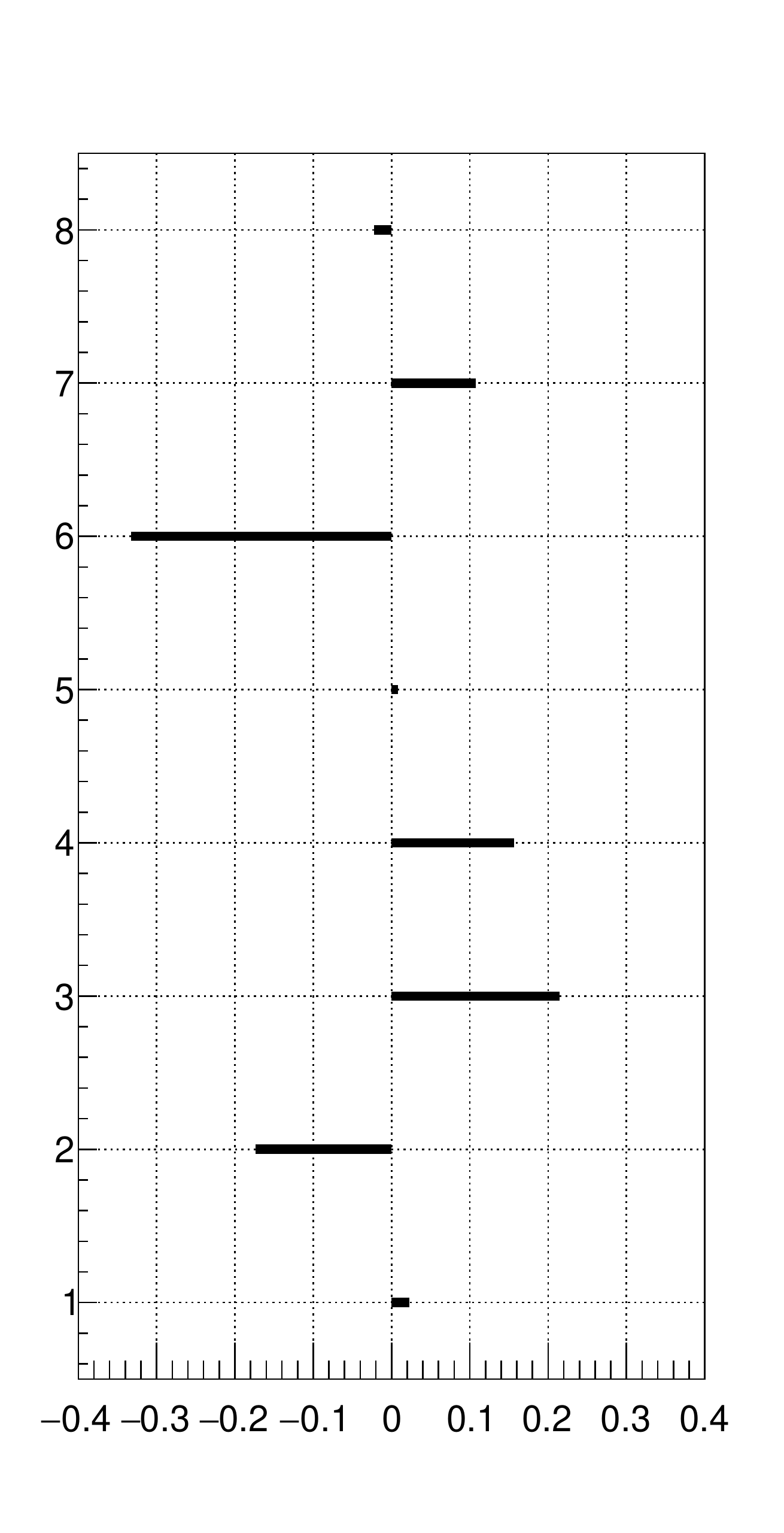}
\put(-88,0){X deviation [mm]}
\end{center}
\end{minipage}
\begin{minipage}{0.25\linewidth}
\begin{center}
\includegraphics[scale=0.3]{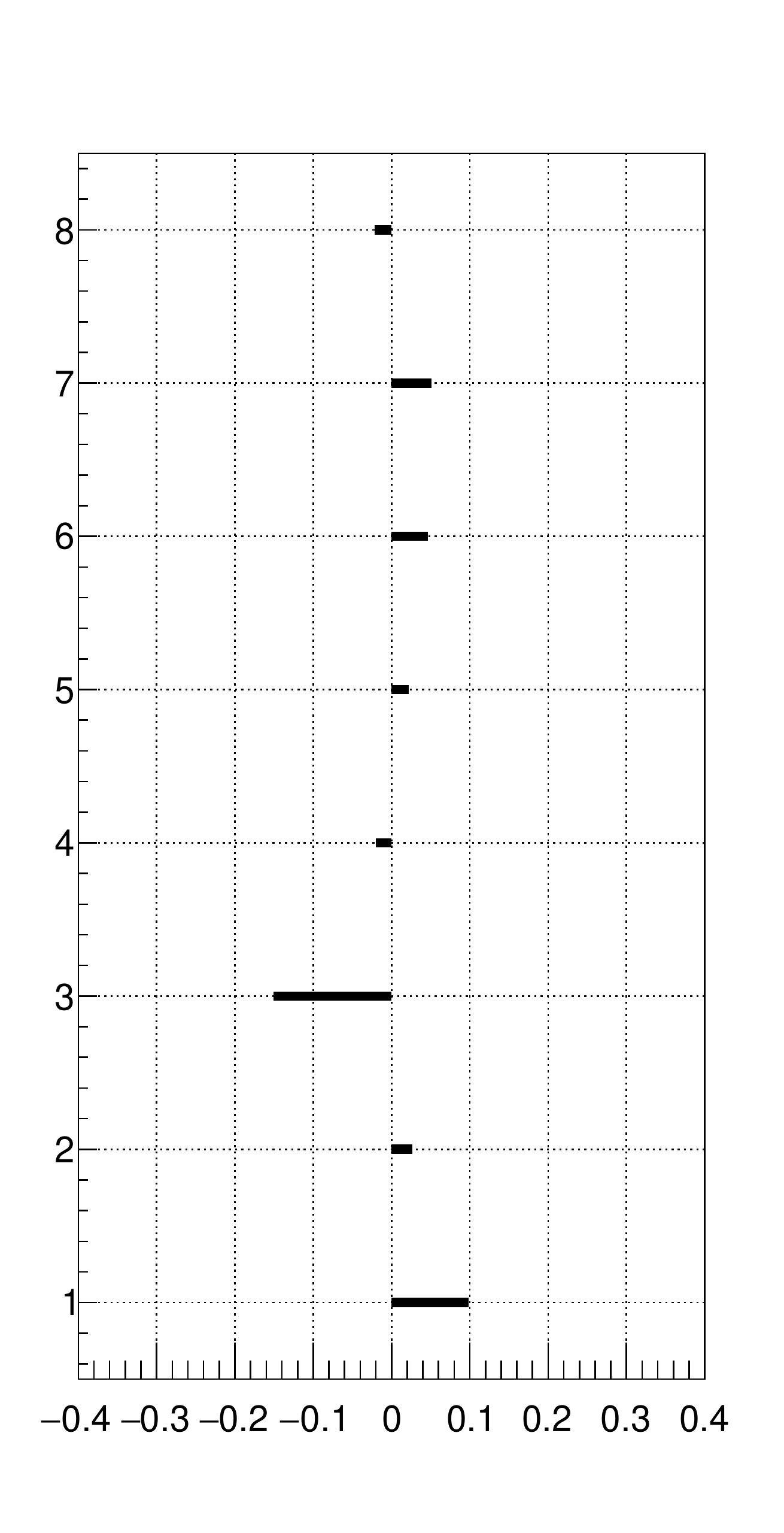}
\put(-88,0){Y deviation [mm]}
\put(-116,125){\rotatebox{90}{Module number}}
\end{center}
\end{minipage}
\begin{minipage}{0.5\linewidth}
\begin{center}
\includegraphics[scale=0.71]{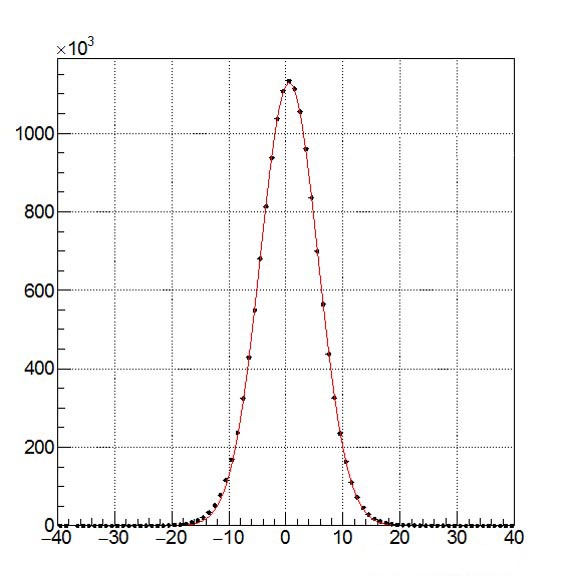}
\put(-102,0){${\rm t}_{\rm track} - {\rm t}_{\rm reference}$ [ns]}
\put(-88,171){$\sigma_{\rm track}$ = 5.1 ns}
\end{center}
 \end{minipage}
\caption{\label{fig:Straw-Picture11} {Left: Measured deviations in X and Y after the positional alignment. The deviation from zero is plotted in mm for each module (numbered from 1 to 8). Right: Track time resolution for tracks having at least 27 straw hits and at least four hits in every chamber.}}
\end{figure}

The straw chambers have been installed to a positional accuracy of $\pm$ 200 $\mu$m and an additional software alignment of each individual sense wire is needed to obtain the design accuracy for the track momentum and direction measurements. In 2015 a preliminary software alignment has been obtained using straight muon tracks  (with the spectrometer magnetic field set to zero)  using an iterative alignment procedure.  The tracks are reconstructed with all but one module   
and the hits in the missing module are compared to the extrapolated positions.  The observed deviations are at the expected level and the procedure converges after 3 or 4 iterations (\Fig{fig:Straw-Picture11}-left). 

Information from the trailing edge of the straw signal is used to calculate the time of the track. For tracks traversing all four chambers, the average number of straws per track is 27 and the track time resolution is found to be 5.1~ns (\Fig{fig:Straw-Picture11}-right).

The NA62 experiment  uses the same magnet (MNP33) as previously used in the NA48 experiment.  The magnet cross-sectional dimension is 4.4 m wide by 4 m high and it is 1.3~m long, with an opening of the iron yokes of 2.4 m vertically and 3.2 m horizontally. The overall weight of the magnet is 105 tonnes. 
The magnet is used in conjunction with a custom made stainless steel vacuum tank section leaving an active fiducial region of 2.37 m diameter.  The dipole magnet provides an integrated vertical field  $ \int B \cdot dl \approx 0.9 $ Tm. The field is mainly vertical pointing toward negative Y coordinate, 
the integral of the other components being smaller by a factor of $10^{-3}$. 

The magnet main properties are summarized in  \Tab{tab:mnp33-parameter}. 

\begin{table*}[h]
\setlength{\tabcolsep}{2ex}
\caption{Main parameters of the magnet for the nominal operating current.}
\label{tab:mnp33-parameter}
\vspace{2ex}
\centering
\begin{tabular*}{0.8\textwidth}{@{\extracolsep{\fill} }  l c  }
\hline \hline
Measured field at the centre of the magnet  &  0.38 T \\
Integrated field  &  0.9 Tm \\
Maximum field close to the poles  &  $< 1.0$ T \\
Transverse momentum kick  & 270 MeV/$c$\\
Effective magnetic length  & 2.13 m \\
Effective aperture diameter  & 2.37 m\\
\hline
\hline
\end{tabular*}
\end{table*}
In the  NA62 layout, the spectrometer  magnet was dismounted and moved to a new position,  
and a new field measurement was performed.

\begin{figure}[ht]
\begin{center}
\includegraphics[width=0.65\linewidth]{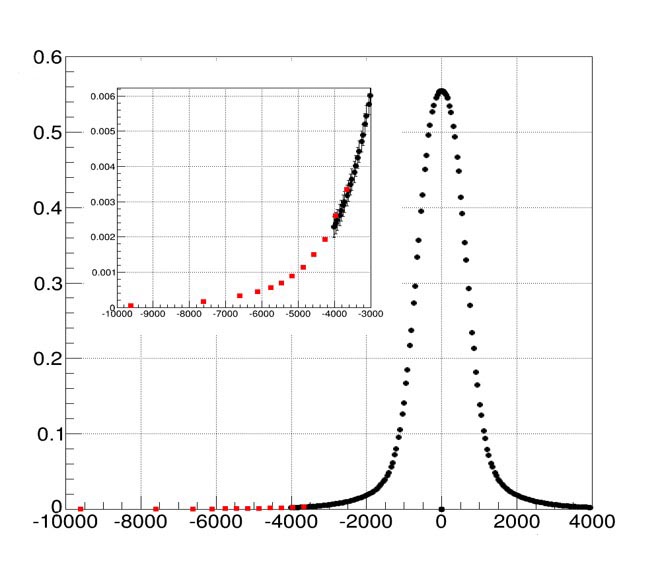}
\put(-50,0){Z [mm]}
\put(-280,190){\rotatebox{90}{B$_{\rm Y}$ [T]}}
\end{center}
 \caption{\label{fig:MNP33-Picture2} {The main magnetic field component B\textsubscript{Y} at (X=500 mm, Y=800 mm) along the Z direction (Z=0 is at the centre of MNP33). The insert shows data from the fringe field measurements superimposed on those of MNP33.}}
\end{figure}

 To ensure complete geometrical coverage, magnetic field measurements within the vacuum region were made covering a cone with half-angle of 20 mrad starting shortly after the GTK3 plane and extending into the fringe field of MNP33 beyond the second straw chamber, a total distance of 90 metres. The magnetic field of the dipole magnet was measured over a grid spanning $\pm$ 4 m across  the centre of the magnet in the Z direction with a small overlap region at the centre, and over the entire aperture of the magnet in the transverse XY plane. Three orthogonal Hall probes were used to measure the X, Y, and Z-components of the magnetic field at a given position \cite{Aleska:2008}.   
The bulk of the  measurements from which the final field map was produced were taken with a spacing of 80~mm in  Z, and a total of approximately 250 000 measurements of  
the 3 magnetic field components  were made. An example  of the  B\textsubscript{Y} component  measurement versus Z, is shown in \Fig{fig:MNP33-Picture2} at some (X ,Y) position.
Full details about these measurements can be found in \cite{magnet}.

\subsection{Performance in 2015}
\begin{figure}[ht]
\begin{minipage}{0.45\linewidth}
\begin{center}
\hspace{-5mm}
\includegraphics[scale=0.38]{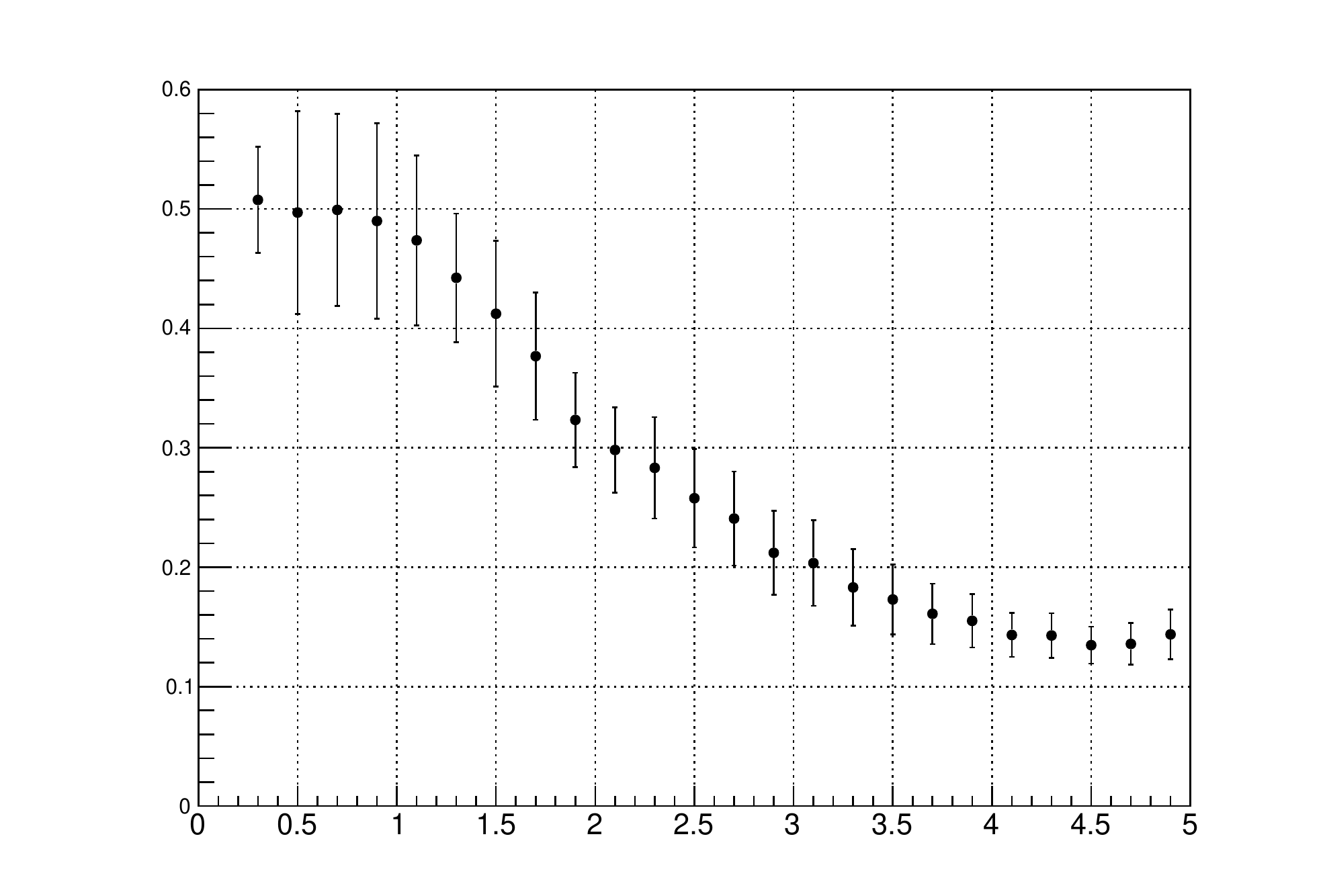}
\put(-165,0){Distance to the anode wire [mm]}
\put(-205,58){\rotatebox{90}{Resolution [mm]}}
\end{center}
\end{minipage}
\begin{minipage}{0.45\linewidth}
\begin{center}
\includegraphics[scale=0.4]{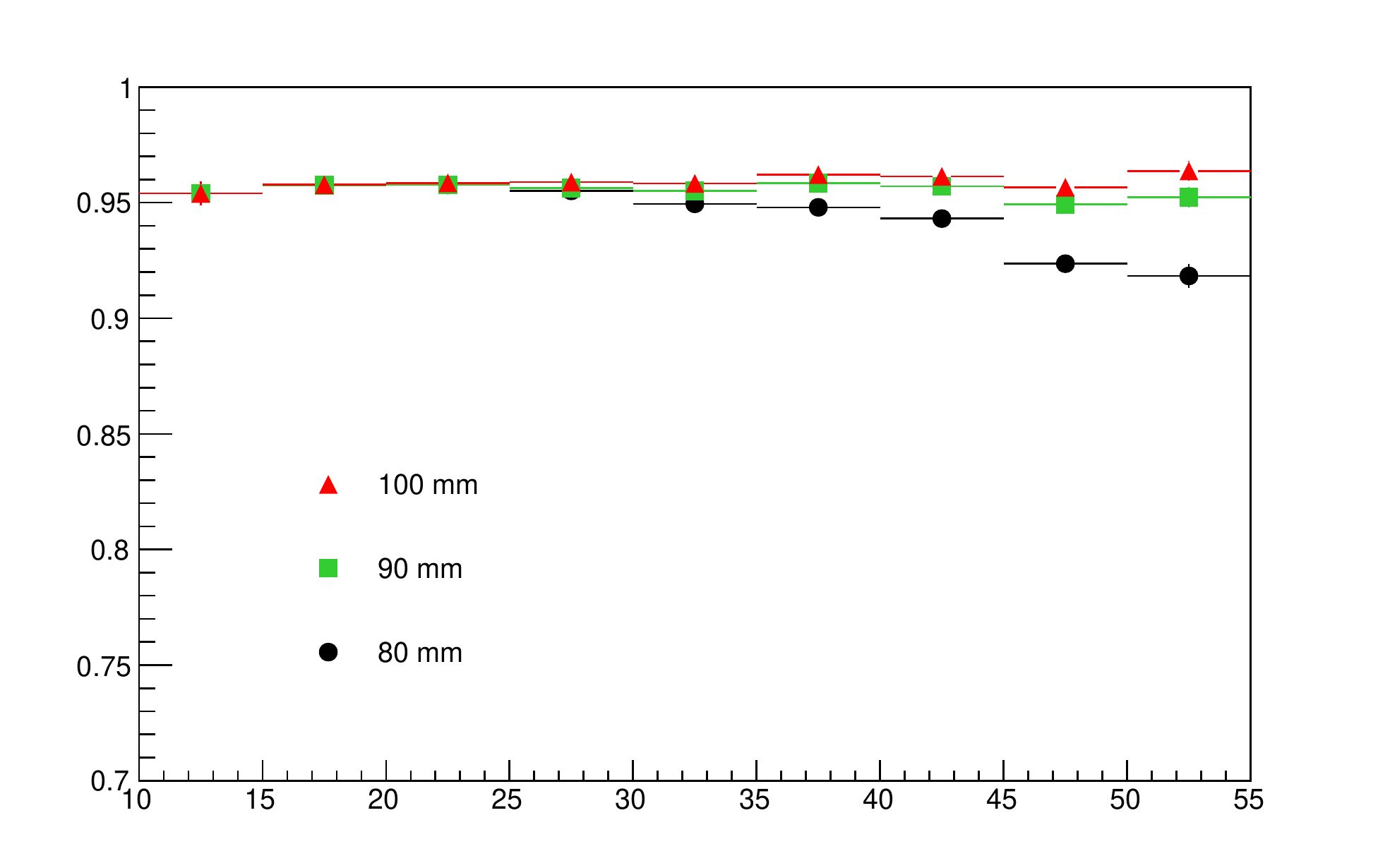}
\put(-137,0){Track momentum [$\GEVc$]}
\put(-225,85){\rotatebox{90}{Efficiency}}
\end{center}
\end{minipage}
\caption{Left: Measured resolution as function of the distance to the anode wire. The drift time was converted to a radial distance using the R-t correlation as obtained using  {\tt GARFIELD}.  
Right: Global track-reconstruction efficiency as a function of momentum for tracks at distances from the beam axis greater than shown in the legend . The efficiency was measured from a sample of reconstructed  $\KTP$  decays from the 2015 data.  }
\label{fig:STRAW-Picture10}
\end{figure}

The R-t correlation  
 is obtained using a {\tt GARFIELD} simulation \cite{Veenhof:3000}
and depends on high voltage, straw geometry and gas properties (temperature, pressure and composition).  \Fig{fig:STRAW-Picture10}-left shows the resulting tracking performance in terms of resolution for fitted tracks at different distances to the anode wire within a straw. The resolution becomes significantly better close to the wall of the straw as expected, where it approaches 130 $\mu$m. The average number of straw hits per track is 27 and only tracks with hits in all four chambers are considered. 

To estimate  the track reconstruction efficiency, a calorimeter-selected sample of $\KTP$
is used (\Sec{ssec:samples}).
The $\PPI$ candidate emitted at  the kaon decay vertex is propagated through the successive chamber planes.  
The track reconstruction efficiency as function of momentum for different radial distances to the beam line at straw chamber 1 is shown in \Fig{fig:STRAW-Picture10}-right. 

\begin{figure}[ht]
\begin{minipage}{0.5\linewidth}
\begin{center}
\includegraphics[width=0.84\linewidth]{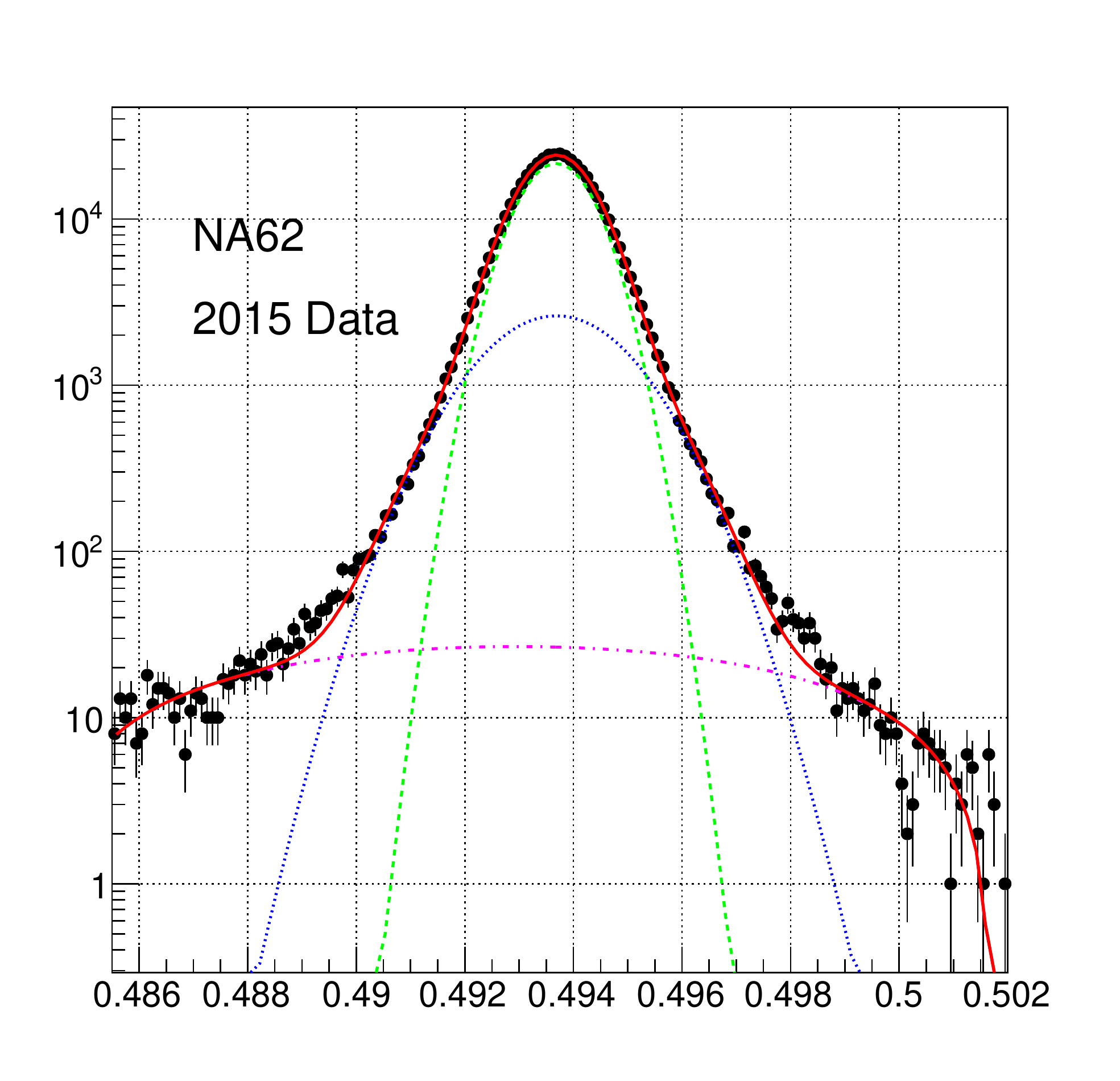}
\put(-115,0){M($\PPI\PPI\MPI$) [$\GEVcc$]}
\end{center}
\end{minipage}
\hspace{-12mm}
\begin{minipage}{0.5\linewidth}
\begin{center}
\includegraphics[width=1.24\linewidth]{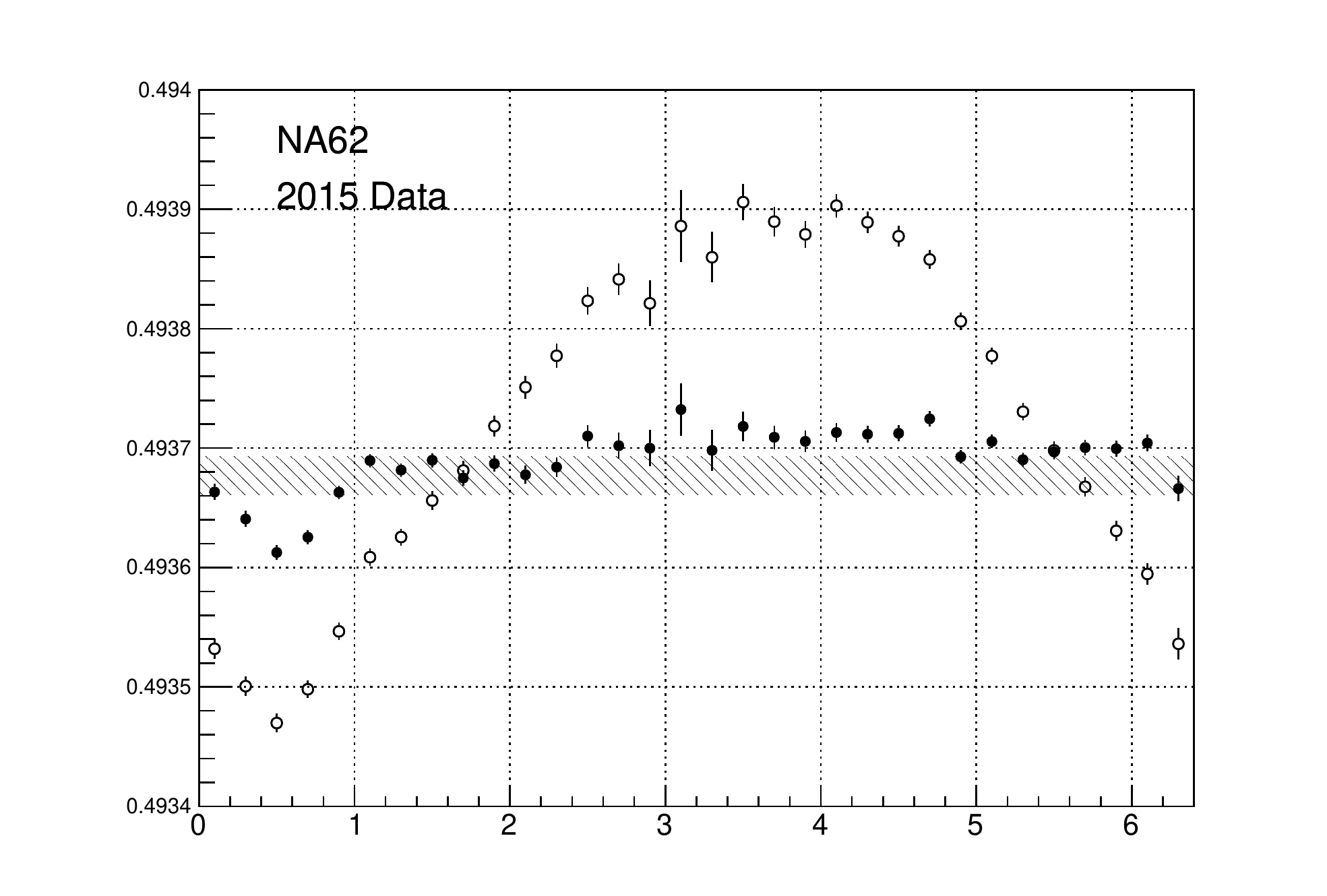}
\put(-80,0){$\phi(\MPI)$ [rad]}
\put(-257,53){\rotatebox{90}{$\langle$ M($\PPI\PPI\MPI$) $\rangle$ [$\GEVcc$]}}
\end{center}
\end{minipage}
 \caption{\label{fig:kmass}Left: Reconstructed three pion mass of selected three-track events. The distribution is fitted by two Gaussian functions (dashed and dotted lines) describing the core and the  tails and a polynomial (dash-dotted line) describing the combinatorial background. Right: Mean value of the reconstructed three pion mass  as a function of the azimuthal angle of the $\pi^-$ track at the decay vertex. Large variations (open circles) are observed if the residual  field is not taken into account when propagating the three tracks backwards to the decay vertex.  Full dots correspond to the same sample with the residual  field taken into account. The shaded band corresponds to the uncertainty on the PDG kaon mass.} 
\end{figure}
 The calibration of  the spectrometer has been studied using reconstructed $\KTAU$ decays (\Sec{ssec:samples}). The reconstructed three pion mass  of \Fig{fig:kmass}-left shows good agreement with the PDG kaon mass \cite{PDG2014} and subsequently the adequate quality of the momentum scale.
The measured resolution is obtained from a two Gaussian fit with rms of  $0.7~\MEVcc$  describing the core and $1.2 ~\MEVcc$  describing the tails. The rms obtained in a single Gaussian fit  is $\sim0.9~\MEVcc$ .
The impact of the residual  field in the decay region can be demonstrated by the variation of the mean reconstructed three pion mass with the azimuthal angle of the $\pi^-$ track at the decay vertex  in the transverse plane to the kaon line of flight (\Fig{fig:kmass}-right).  

 The first evaluation of the  performance of the straw chambers is in agreement with the design specifications (\Sec{ssec:req}) in terms of space point  and track time resolutions.
 The resulting track momentum resolution of the spectrometer  is consistent with:
 \begin{equation}
 \frac{ \sigma({\rm p}) } {\rm p} = 0.30\% \oplus 0.005\% \cdot {\rm p}  ~,
  \end{equation}
 where  p is expressed in $\GEVc$.  The track angular resolution decreases from 60~$\mu$rad at 10 $\GEVc$ to 20~~$\mu$rad at 50 $\GEVc$ momentum. Both resolutions satisfy the performance requirements.

%% file: Sec8-PV_3.tex
Kinematic cuts provide a rejection of $\sim$$10^4$  for the main background $\KTP$ (\Sec{sec:intro}). The additional requirement for the $\PPI$  momentum to be in the range  15 to 35 $\GEVc$
ensures that the photons from the $\PIo$ decay have an energy of at least 40~GeV, leaving  the remaining rejection factor of $10^8$ to the photon-veto system.
Four  different types of calorimeters are used to identify and reject  background kaon decays with photons  in the final state.
Much of the rejection power is provided by the NA48 liquid krypton calorimeter (LKr). 
The other photon-veto detectors were constructed specifically 
for NA62: the large-angle veto (LAV) detectors to detect photons emitted at angles larger than those 
covered by the LKr, and the intermediate-ring calorimeter (IRC)  and small-angle calorimeter (SAC) 
to intercept forward, high-energy photons that would otherwise escape down the beam pipe. 
 The angular coverage of these detector systems 
 and the required photon detection efficiencies  are summarized in \Tab{tab:phveto}.
 
\begin{table}[ht]
 \setlength{\tabcolsep}{3ex}
 \renewcommand{\arraystretch}{1.1}
 \caption{\label{tab:phveto}Characteristics of the NA62 photon-veto detectors.}  \vspace{2ex}
 \centering
 \begin{tabular}{@{}ccl@{}}
 \hline\hline
 \\[-2.5ex]
 Photon-veto detector  & Angular coverage (mrad) & Design inefficiency\\[0.5ex] 
 \hline \\[-1.75ex]
 LAV  & 8.5--50 & $10^{-4}$ for $E_\gamma > 200$~MeV \\
 [1ex] \hline \\[-1.75ex]
  LKr &  1 -- 8.5 & $10^{-3}$ for $E_\gamma > 1$~GeV\\
                 & & $10^{-5}$ for $E_\gamma > 10$~GeV\\
  [1ex] \hline \\[-1.75ex]
 IRC, SAC & 0 --1 & $10^{-4}$ for $E_\gamma > 5$~GeV\\[1ex]
 \hline\hline
 \end{tabular}
 \end{table}

\subsection{Large-Angle Veto system (LAV)} \label{sec:LAV}

The ring-shaped large-angle photon vetoes are placed at 11
positions along the vacuum volume (\Fig{fig:detectorview1}) while the 12\textsuperscript{th} station is located about 3~m upstream of the LKr calorimeter and is operated 
in air. The LAV detectors provide full geometric coverage for photons from decays within the
decay volume emitted at angles from 8.5 to 50~mrad with respect to the Z axis.

Because of the anti-correlation between the energies and emission
angles of photons from background decays such as $\KTP$ in events that survive the fiducial cut 
on the $\PPI$ track, it is not possible for both photons from the $\PIo$ decay to have angles larger than 
the limit of the LAV angular coverage, and in only 0.2\% of these events 
one photon is emitted at an angle greater than this limit. For events in which one of the two photons is not
intercepted, the energy of the escaping photon is below 200~MeV.
Conversely, more than 95\% of photons inside the LAV angular acceptance have energies 
higher than this value. 
To secure the required overall photon rejection power of $10^8$, these 
photons must be detected with an inefficiency of less than $10^{-4}$.
At nominal beam intensity, the rates on the LAV detectors are dominated by the muon halo, which is expected to contribute about 4~MHz to the event rate in the LAV system (\Tab{tab:K12-muonrates}). 
For the accidental rate to be kept to the percent level with a $\pm5\sigma$ coincidence window, the LAVs must have a time resolution of $\sim$1~ns for 1 GeV photons.
In addition, an energy resolution of at least 10\% for 1 GeV photons is desired for precise application of the veto threshold, as well as to facilitate use of the LAVs for the selection of control samples.
Finally, the LAVs must be compatible with operation in vacuum.

\subsubsection{Design and layout} \label{ssec:LAV}
The NA62 LAV detector reuses lead-glass blocks recycled from the OPAL electromagnetic calorimeter barrel \cite{OPAL+91:NIM}, which became available in 2007. Tests of a prototype detector were performed with the electron beam of the Frascati Beam-Test Facility, demonstrating that the technology was suitable for use in NA62 \cite{A+07:veto}.  The inefficiency for the detection of individual tagged electrons was measured to be  $1.2^{+0.9}_{-0.8} \times 10^{-4}$ at 203~MeV and $1.1^{+1.9}_{-0.7}\times 10^{-5}$at 483~MeV.  This choice resulted in significant savings on construction costs.

\begin{figure}
\centering
\includegraphics[width=0.8\linewidth]{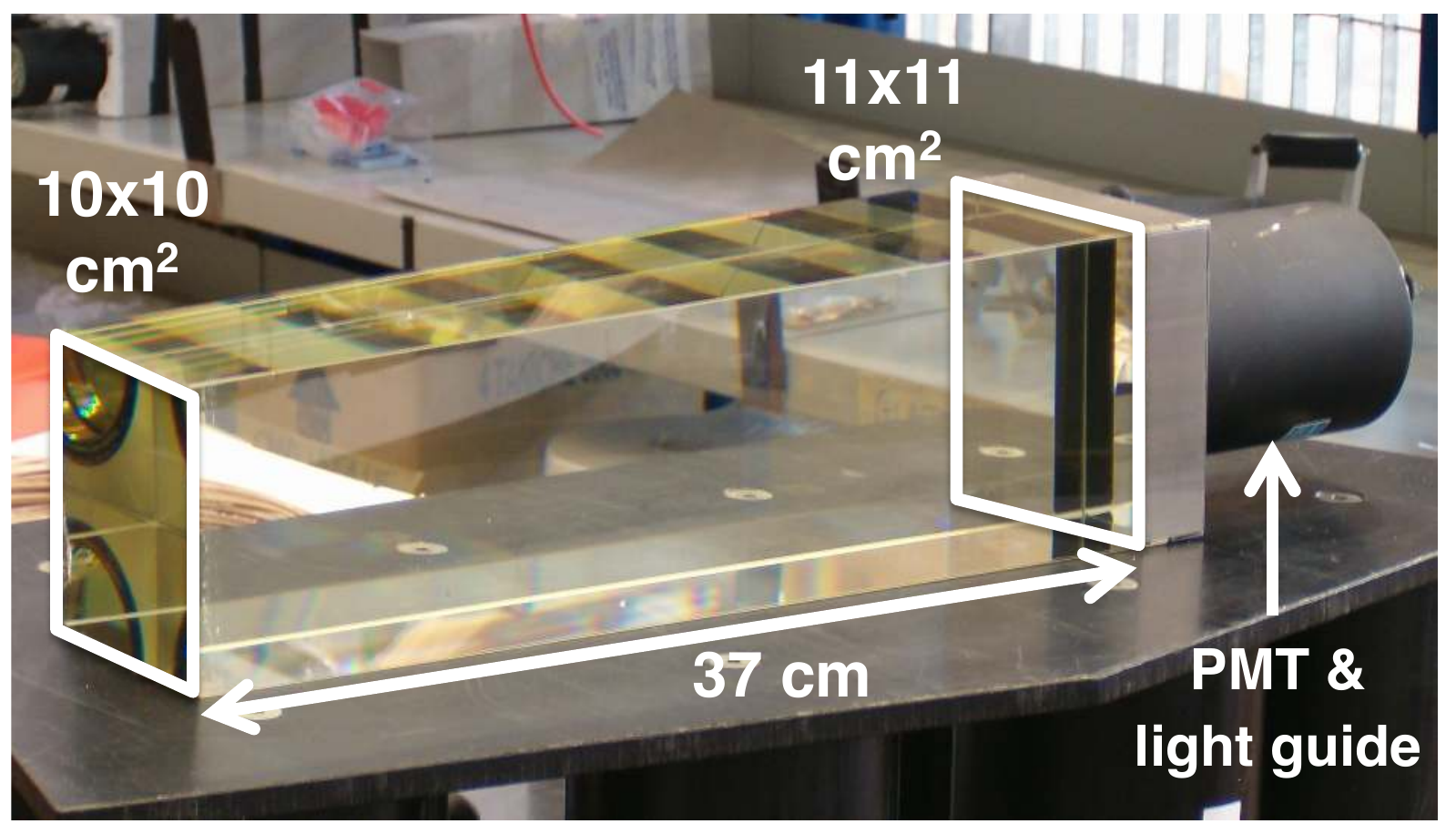}
\caption{A module from the OPAL calorimeter, without wrapping and with 
reinforcement plates at the interface between the glass and the steel flange.}
\label{fig:lavblock}
\end{figure}
The modules from the central part of the OPAL 
electromagnetic calorimeter barrel consist of blocks of Schott SF57 lead glass.
This material is about 75\% lead oxide (PbO) by weight and has a density
$\rho = 5.5~{\rm g}/{\rm cm}^3$ and a radiation length $X_0 = 1.50$~cm;
its index of refraction is $n \approx 1.85$ at $\lambda = 550~{\rm nm}$
and $n \approx 1.91$ at $\lambda = 400~{\rm nm}$. 
Electromagnetic showers in the lead glass are detected via  the Cherenkov light produced.
The front and rear faces of the blocks measure about 
$10\times10$ cm$^2$ and $11\times11$ cm$^2$, respectively. The blocks are 37~cm in length 
(the precise geometry depends on the ring of the OPAL calorimeter from which a block is extracted).
Each block is read out at the back side by a Hamamatsu R2238 76 mm diameter photomultiplier (PM),
 which is optically coupled via a 4 cm long cylindrical light  guide of SF57 of the same diameter 
as the PM. The rear face of the glass block is glued to a 1 cm thick stainless-steel mounting flange
with a circular cutout for the light guide. 
A complete module is the monolithic assembly of a block, PM, and light guide, as shown in \Fig{fig:lavblock}. 

A LAV station is made by arranging these modules around the perimeter of the sensitive volume of the 
experiment, with the blocks aligned radially to form an inward-facing ring. Multiple rings are used in each station in order to provide the depth required for the efficient detection of incident particles. The modules in successive rings are staggered in azimuth while the rings are spaced longitudinally by about 1~cm.

The LAV system consists of a total of 12 stations, the geometry of which is summarized in \Tab{tab:param}.
The 11 stations installed in the vacuum tank were manufactured in three different sizes, with the larger diameter stations
for installation further down the beam line. Apart from the different sizes and module configurations, the designs are conceptually similar.
The 12\textsuperscript{th} station is operated in air, requiring its design to be slightly different from that of the other stations. 

\begin{table}[ht]
\setlength{\tabcolsep}{3ex}
\renewcommand{\arraystretch}{1.1}
\caption{Parameters of LAV stations.}
\label{tab:param}
\vspace{2ex}
\centering
\begin{tabular}{@{}cccccc@{}}
\hline\hline
Stations & Diameter [mm] & \multicolumn{2}{c}{Ring radius [mm]} & Number of  & Number of \\
 & Outer wall & Inner & Outer & layers& modules\\
\hline
LAV1--LAV5  & 2168  &   537 &  907 & 5 & 160\\
LAV6--LAV8  & 2662  &   767 & 1137 & 5 & 240\\
LAV9--LAV11 & 3060  &   980 & 1350 & 4 & 240\\
LAV12       & 3320  &  1070 & 1440 & 4 & 256\\
\hline\hline
\end{tabular}
\end{table}

Since the spaces between the blocks are significantly smaller in the 
larger diameter vessels, fewer layers are necessary.  As a result of the 
staggering scheme, particles incident on any station are intercepted by
blocks in at least three rings, for a total minimum effective depth of 
21 $X_0$. Most incident particles are intercepted by four or more blocks (at least 27~$X_0$).

\begin{figure}
\centering
\begin{tabular}{cc}
\includegraphics[height=0.28\textheight]{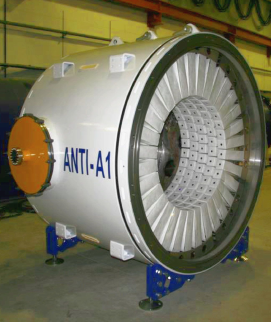} &
\includegraphics[height=0.28\textheight]{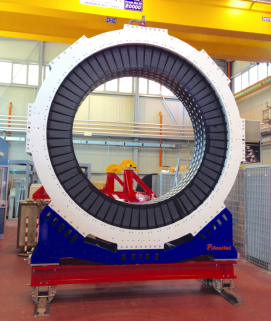} \\
\end{tabular}
\caption{Completed LAV1 (left) and LAV12 (right) stations before insertion in the beam line.}
\label{fig:station}
\end{figure}
In 2009, the LAV1 station was constructed as a prototype, installed in the
NA62 beam line, and tested with electrons and muons.
The completed detector is shown in \Fig{fig:station}-left.
After the beam test, various improvements were made, and the LAV2 station  
was constructed and tested with an unseparated, low-energy, positive beam
in the T9 beam line at the CERN PS in 2010. The results from this test confirmed that the  design 
performance had been achieved~\cite{A+11:LAV}.
Construction of the remaining stations operated in vacuum (LAV1--11)  took three years. 
The LAV12 design (\Fig{fig:station}-right) 
was finalized in 2013 and the completed detector was installed a year later. All 12 stations were fully
commissioned and operational during the  entire 2014 data-taking period.

\subsubsection{Construction details}

The OPAL modules were manufactured during the mid-1980s. Their recycling required substantial care throughout the assembly procedure.

The interface between the stainless-steel flange and lead-glass block is 
fragile, and was found to be critically damaged in a few percent of the modules 
upon first examination. This was attributed to thermally induced stress from
the differing expansion coefficients of the steel and glass.
The first step in the processing of the modules was therefore
to reinforce the interface. Using epoxy resin, 27~cm$^2 \times 0.3$~mm thick 
stainless-steel plates were attached across the glass-steel interface 
on all four sides of the block. Calculations indicated and static tests 
confirmed that the reinforced bond is several times stronger than the original
bond. In any case, to prevent breakage, a concerted effort was made to avoid
exposing the modules, even those in fully assembled detectors, to 
temperatures outside the range of 15--30~$^\circ$C.

The original OPAL wrapping for the lead-glass blocks
had to be removed for the reinforcement of the glass-steel interface and
subsequent cleaning.
For LAV stations 1--11, which are operated in the dark of the vacuum tank, 
the primary function of the wrapping is to increase the light collection 
efficiency as  light tightness is not necessary. The new wrapping
is made from Tyvek Brillon 4173D, a brilliant white grade of Tyvek. 
The whiteness of this material provides an ideal
diffusive reflector for light containment and collection. 
It has paper-like stiffness and is easily laser cut, folded
around the blocks, and heat welded.

For the LAV12 station, the wrapping must additionally exclude ambient light.
The new wrapping is similar to the original wrapping: it 
is a three-layer laminate commercially available for use as solar
backsheeting (Coveme dyMat T), consisting of 38~$\mu$m
black PVF (to block external light), 50~$\mu$m PET, and white 38~$\mu$m PVF (to
diffusively reflect light internally). The wrapping was laser cut,
folded around the blocks, and sealed with black PVC tape.

During the LAV1 beam test in 2009, ringing of the analogue signal was observed
to lead to errors in charge reconstruction using the time-over-threshold (ToT)
technique discussed in \Sec{ssec:lav-fee}.
This problem was traced to a small parasitic inductance in 
the PM dynodes and solved by replacing the original OPAL HV dividers soldered 
to the PMs with new, custom dividers.

After the dividers were replaced, the modules were tested and characterized 
12 at a time using a test station featuring an LED pulser and a cosmic-ray telescope. 
The PM gains were measured first, by varying the intensity of the light 
pulses from the LEDs as well as the PM HV settings and mapping out the 
response for each module. Using the gain curves  obtained, the PMs
were then set to a reference value of the gain ($9\times10^5$ or $1\times10^6$)
and the response to cosmic rays selected by the telescope was measured. The
photoelectron yield for each module (pe/MeV) was then obtained by 
assuming that cosmic rays at normal incidence on the front face of the horizontally arranged blocks deposit 77 MeV (the average MIP energy loss calculated using the Bethe-Bloch formula).
Photoelectron yields of 0.34~pe/MeV are typical.
Finally, using the gain curves 
and the measurements of photoelectron yield, the PM voltages were 
set to the values expected to produce a common output charge level of 4.5~pC
for cosmic-ray events. The response was measured and the HV setting was
validated. The 12-hour characterization cycle was fully automated  
and resulted in PM gain and photoelectron yield measurements
and operational HV settings for each of the 12 modules. Additional data 
(current-draw measurements, dark-count rates) were also collected 
for each module using the test station.

The OPAL design features an optical port at the base of each module. During construction, blue LEDs were installed in these optical ports for use with a planned calibration and monitoring system.
A low-capacitance LED was chosen to minimize the rise and fall times of the light pulse; this is important for use with the  ToT-based readout system discussed in the following
section. When completed (a custom LED pulser board is under development),
the system will allow monitoring of the operational status and
relative timing for each module. In principle, the system should also allow in-situ gain measurement.

\subsubsection{Front-End electronics}  \label{ssec:lav-fee}
The LAV detectors must detect incident photons with high efficiency and 
measure their times and energies. Due to multiple internal reflections of the 
Cherenkov light in the large lead-glass blocks, the time resolution of the
OPAL detector modules is about 1~ns. This is compatible with the required
performance for the LAV system and does not pose particular challenges 
in the design of the readout electronics. However, because of the large energy range 
of incident particles, from $\sim$100~MeV to 20~GeV, the 
energy measurement requires particular attention.  
For a photoelectron yield of 0.3 pe/MeV and a nominal gain of $10^6$, a MIP  produces 4.5 pC of charge 
at the PM, corresponding to a signal amplitude of 12 -- 15 mV.
Photons with energies at the very low end of the measurement range
($\sim$100~MeV) produce signals of this amplitude, while at the opposite end 
of the spectrum, a 20 GeV photon can give rise to signal amplitudes of 3~V or more.

For reasons of cost and simplicity, the readout scheme is based on the 
ToT technique. This scheme is implemented using a  
custom front-end ToT discriminator board, 
together with the TEL62 digital readout board used by various sub-detectors  (\Sec{ssec:TEL62}). 
The ToT discriminators convert the analogue signals from 
the detector to low-voltage differential signal (LVDS) pulses, with width 
equal to the duration of the analogue signal from the detector above a 
specified threshold. The signal from each PM is compared to two different
thresholds, a low threshold of about 5~mV and a high threshold of about 15~mV,
corresponding to two different LVDS outputs. For comparison, 
the noise level is less than 2~mV under usual operating conditions. 

\begin{figure}
\begin{center}
\includegraphics[width=0.8\linewidth]{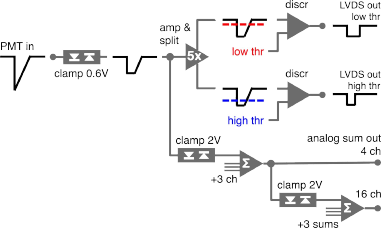}  
\caption{Conceptual schematic of the ToT discriminator board.}
\label{fig:fee}
\end{center}
\end{figure}
The ToT discriminator board is fully described in \cite{A+11:ToT}. 
A conceptual schematic is presented in 
\Fig{fig:fee} to illustrate the signal processing for a single channel.
Since each channel is discriminated against two different
thresholds, the board has 32 input channels and 64 output channels.

Before processing, the input signal is passively split.
One copy is summed with signals from adjacent channels to form front-panel 
diagnostic analogue outputs, as discussed below.
The other copy is discriminated for the ToT measurement.

Because signals from large showers may be several volts in amplitude, 
the input signal for the ToT measurement is first clamped for 
protection at 0.6~V by a circuit that maintains the timing of the
rising and falling edges of the pulse using a pair of fast, 
low-capacitance diodes. 

In order to maintain sensitivity to minimum ionizing photons and high 
detection efficiency for low-energy photons, the effective setting for 
the lower threshold for ToT discrimination must be about 5~mV. 
Amplification of the signal before discrimination accomplishes several 
purposes: it minimizes the effect of signal slope and overdrive on the
output response of the voltage comparator, improves the precision and
stability of the effective threshold setting, and improves the separation 
between signal and noise.  
A fast, low-noise, high-bandwidth (800 MHz) current-feedback amplifier 
(AD8001) is used to amplify the signal by a factor of 5.

The amplifier output is passively split into two copies, which are 
discriminated at the two different levels by high-speed comparators 
with LVDS output drivers (LMH7220).
A digital-to-analogue converter in the board-controller mezzanine supplies
programmable threshold levels; these can be adjusted over the range from
5 to 250 mV with 12-bit resolution.
To reduce double pulses from the comparator due to noise
in the input signal, 3 mV of hysteresis is also provided through a feedback 
resistor, so that the output signal is extended until the input signal falls
to 3 mV below the threshold. 
The LMH7220 comparator has a propagation delay of just 2.5~ns and rise and 
fall times for the LVDS signal of 0.6~ns. These features help to ensure 
good TDC performance.

The copy of the analogue input signal for diagnostic purposes is summed with 
the signals from 3 adjacent channels (and clamped at 2~V);
the resulting analogue sum is made available via a front-panel  connector.
These sums of 4 are in turn summed 4 at a time to produce sums of 16.
An easy way to perform single-channel diagnostics is to pulse the modules one at a time using the LED system and read out the analogue signal from these sums. 

\subsection{Liquid Krypton calorimeter (LKr) }\label{sec:LKR}
\begin{figure}
\begin{center}
\includegraphics[scale=0.75]{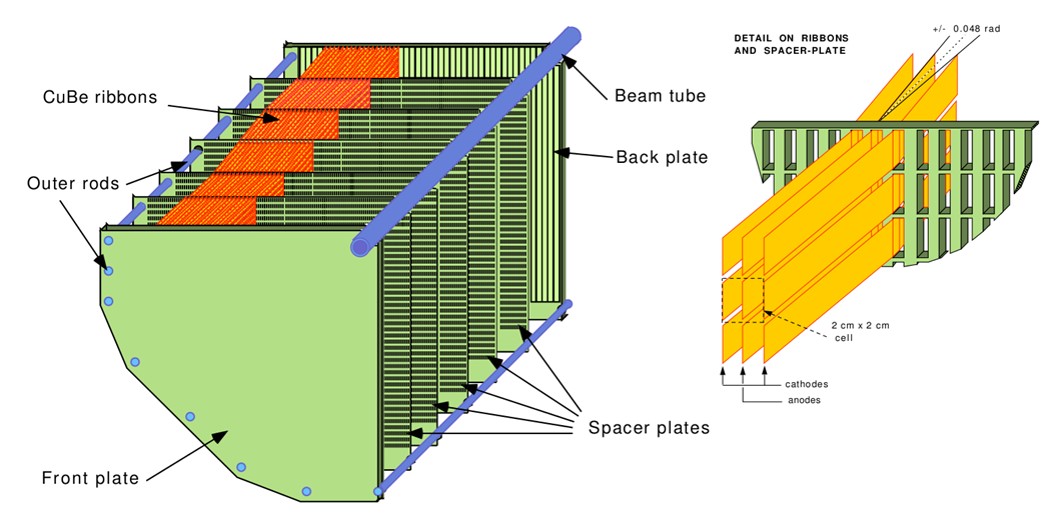}
\end{center}
\caption{\label{fig:CREAM-Picture0b} {Left: Schematic of the calorimeter structure (one quadrant). Right:  Detail of the calorimeter cells}}
\end{figure}

NA62 reuses the former NA48 liquid krypton calorimeter,  
which is fully described in \cite{NA48:2007}. The LKr is a quasi-homogeneous calorimeter filled with about 9000 litres of liquid krypton at 120 K, housed inside a cryostat. The calorimeter  extends from the beam pipe ($r\approx8$~cm) to a radius of 128~cm; its depth is 127~cm, corresponding to $27~X_0$. The sensitive area  is  divided into 13248 longitudinal cells  with a cross section of about $2 \times 2$ cm$^2$. The cells are formed by Cu-Be electrodes aligned along the longitudinal axis of the experiment, and have a zig-zag shape to avoid inefficiencies when a particle shower is very close to the anode (\Fig{fig:CREAM-Picture0b}).  The signal produced  by  a  particle  crossing  the  LKr  is  collected  by  preamplifiers inside the cryostat, directly attached to the calorimeter strips. The signal is sent out to the transceiver boards via 50~$\Omega$ coaxial cables and vacuum feedthrough connectors on top of the cryostat. The transceiver boards are mounted directly on the feedthroughs and are Faraday shielded by the cryostat.

The external components of the cryogenic system and the auxiliary parts of the readout system (power supplies, transceivers, HV, calibration system) were modified in order to prepare the detector for a new decade of data taking.
To satisfy the demanding rate requirements in NA62, the former  NA48 LKr readout system \cite{Gianoli:2000} based on gain-switching 10-bit Flash ADCs (FADC), was phased out and a new readout system was built and commissioned (\Sec{sec:TDAQ}).

\subsection{Small-Angle Veto system (SAV)}\label{ssec:SAV}

The small-angle veto system provides hermeticity for photons 
emitted at angles down to zero degrees with respect to the Z axis.
It consists of two detectors: the intermediate-ring ralorimeter (IRC) and
the small-angle calorimeter (SAC). Both are shashlyk
calorimeters, with lead and plastic-scintillator plates traversed by wavelength-shifting (WLS) fibres.
Photons from kaons decaying  in the decay volume that traverse the small-angle veto detectors 
have energies greater than 5 GeV; these photons must be detected with 
 a maximum inefficiency of $10^{-4}$.
For both detectors, the expected photon rates are of the  order of 1~MHz at the nominal beam intensity. 
The IRC is additionally exposed to muons from decays of beam particles, which are concentrated in 
a spot of a few cm in diameter to one side (towards negative X) of the beam line; the flux of muons 
from this spot increases the total rate of particles on the IRC to 10~MHz.

\subsubsection{Small-Angle Calorimeter (SAC)}\label{ssec:sac}

\begin{figure}[ht]
\begin{center}
\includegraphics[width=0.80\linewidth]{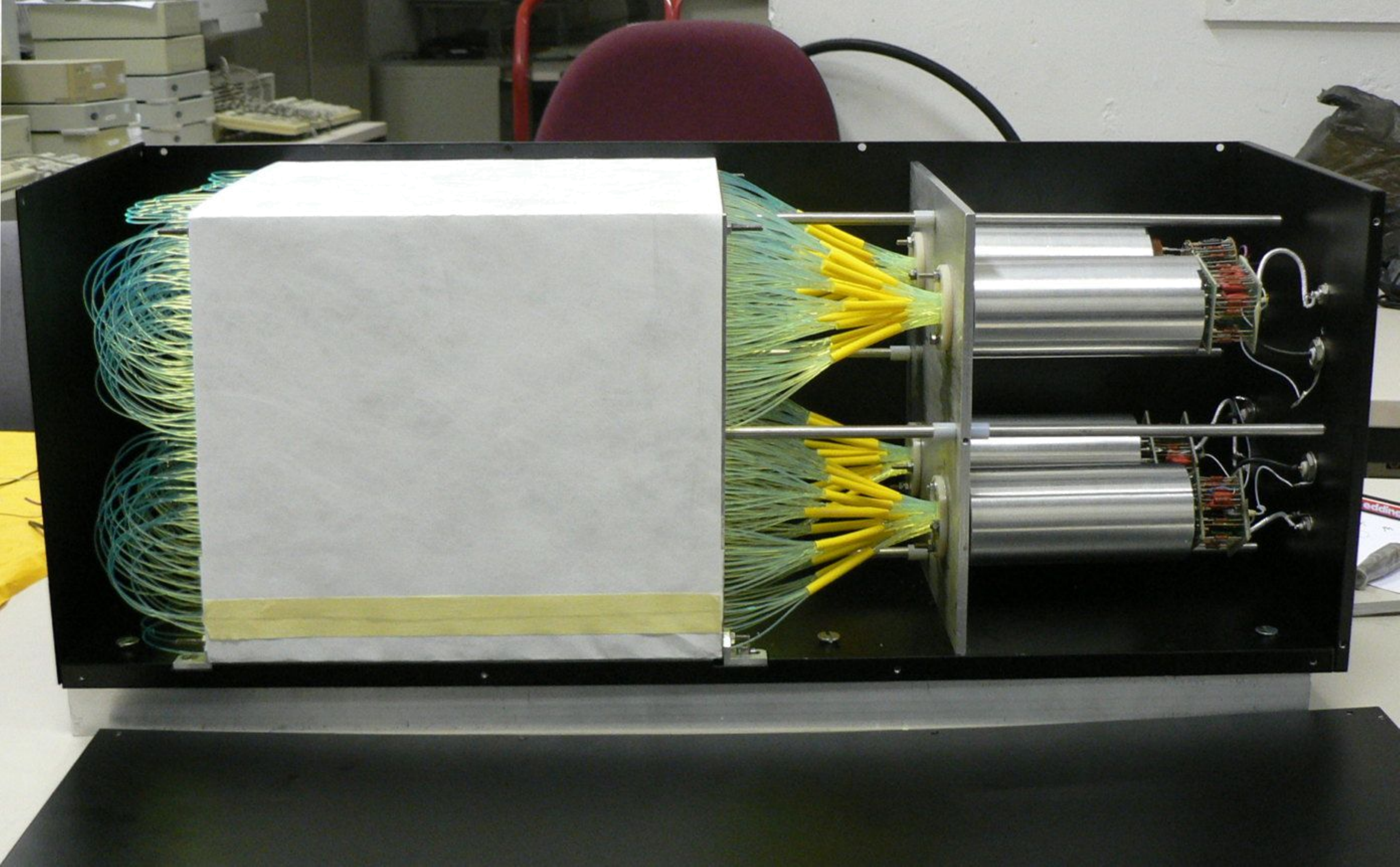}   
\end{center}
\caption{Picture of the  SAC after completion of assembly.} 
\label{fig:sac_assy}
\end{figure}
The small-angle calorimeter consists of 70 plates of lead and
70 plates of injection-molded plastic scintillator, both with
transverse dimensions of
$205 \times 205$ mm$^2$ and a thickness of $1.5$~mm,
for a total depth of $19~X_0$.
A rectangular grid of 1.5~mm diameter holes with
9.5~mm spacing is machined on each plate, for a total of 484
holes per plate. The plates are stacked with Tyvek sheets between the lead and scintillator layers 
to provide diffuse reflectivity, thereby increasing the light collection.
The scintillation light is read out by 240 1~mm diameter Kuraray Y11(250)MSJ WLS fibres. 
Each fibre is bent into a U shape and threaded through two holes,
allowing readout of both ends at the back of the detector.
The fibre ends are grouped into four bundles, each with 120 fibre ends
read out by  25~mm Hamamatsu R6427s PMs 
with a  rise time of 1.7~ns. To avoid anode current saturation and photocathode
evaporation, an absorptive neutral-density filter of
optical density~1.0 (Thorlabs NE10B)  is placed between the fibre bundle and the PM.
\Fig{fig:sac_assy} shows the SAC 
after completion of the assembly. Because the scintillator plates have no transverse
segmentation, the four readout channels are optically connected, and the
SAC is effectively a single-channel detector.

During a test beam at NA48 in 2006, the SAC was exposed to a 25~GeV
electron beam, with the trigger provided by the NA48 spectrometer
and hodoscope. The inefficiency was measured to be less than
$3 \times 10^{-5}$ \cite{NA62:2010}.
In 2012, after substitution of the PMs,\footnote{Originally, FEU-84 PMs,
with a  rise time of 5~ns for single photoelectrons, were used.}
the detector was re-tested at the Frascati Beam-Test Facility. 
The energy resolution was measured to be
\begin{equation}
 \frac{\sigma({\rm E})}{\rm E} = \frac{8.8\%}{\sqrt{\rm E}} \oplus \frac{7.1\%}{\rm E}~,
\end{equation}
where the energy is measured in GeV. 
The inefficiency for detecting 600~MeV electrons was
less than $5 \times 10^{-3}$ \cite{bib:sac-btf}. 

The detector is installed inside the beam vacuum towards the end of the beam pipe, just upstream of the 
beam dump (\Fig{fig:K12beam-downstream}).
To guarantee that photons incident on the SAC along the
Z axis do not traverse the detector along a WLS fibre
for more than 10~cm (half of the active length) without 
encountering any of the lead converters, the SAC is aligned 
at a 23 mrad angle to the Z axis in the horizontal plane.
For the positioning and alignment of the SAC, a support table
is used, which  allows fine movements in the horizontal and vertical directions
and a measurable rotation of up to $\pm$40~mrad in the horizontal plane. 
A precision on the SAC position of better than 500~$\mu$m was achieved.

\subsubsection{Intermediate-Ring Calorimeter (IRC)}\label{ssec:irc}

The intermediate-ring calorimeter is a lead/scintillator shashlyk calorimeter in the shape of an eccentric 
cylinder surrounding the beam pipe upstream  of the LKr. The detector has an outer diameter of 290 mm and is centered on the Z axis. The central bore has a diameter of 120~mm with an offset of 12~mm towards positive X to account for the beam deflection by the spectrometer magnet (\Fig{fig:K12beam-downstream}).
The IRC is divided into two longitudinal modules, with both the upstream and downstream modules measuring
89~and 154~mm in depth, respectively. The modules are spaced by 40~mm.
The inner diameter of the downstream module is 2.2~mm greater
than that of the upstream module, so that photons from kaon decays in
the decay volume do not hit the far downstream edge of the IRC, thus escaping detection.

To minimize the material in front of the sensitive volume, the IRC
incorporates the corresponding segment of the beam vacuum tube as its inner support
cylinder. The vacuum tube has three sections, with slightly different
outer diameters. The part that supports the IRC modules has a wall thickness
of 0.8~mm. The tube is made of 316L stainless steel, with a welded
longitudinal seam and a CF flange at the downstream end.
The upstream 40~mm  of the tube is machined to surface roughness
$\leq 1.6~\mu$m to allow the installation of a sliding vacuum flange sealed with an O-ring. 
The tube was extensively tested for structural
integrity: it was sealed, evacuated, and then placed for more than 18 hours inside a vessel at an 
overpressure of 1.5~bar. No substantial degradation of the vacuum was observed. A finite-element 
simulation indicated that the tube can withstand more than 2~bar of external pressure without buckling.

The upstream and downstream modules contain 25 and 45 ring-shaped layers, respectively. 
Each layer consists of a 1.5~mm thick lead absorber and a 1.5~mm thick scintillator plate.
The total depth for both modules is $19 ~X_0$.
The absorber plates are made of 97\% lead-3\% antimony alloy, 
which has a Brinell hardness of 9 HB (for comparison, pure lead has a
hardness of 5 HB). They are cut from a single piece of 
converter material and are not further divided into segments.
Each of the rings has 570 holes of diameter 1.5~mm.
To avoid alignment of  holes  at the same (X,Y) position in the two modules,  the grid of holes is shifted by 4.8~mm along the Y axis in the downstream module.
The scintillator rings are cut from plates of Saint Gobain BC-400 plastic
scintillator and coated with diffusely reflective Eljen EJ-510 paint.
The scintillator rings are divided into four optically isolated quadrants, which reduces the cross-talk between the channels.
The segmentation of the scintillator plates but not of the 
converter plates decreases the event rate for individual channels 
while avoiding overall detection inefficiencies from gaps in the converter. 
The scintillation light is read out by 1.2~mm, multi-clad Saint Gobain BCF-92 WLS fibres
traversing both the upstream and downstream modules.
Monte Carlo simulation shows that the light emitted towards the front of
the detector and reflected back along the fibre lengthens the signal width
from 15~ns to 25~ns, resulting in an increase of the probability
for events to overlap from 7\% to 12\% at a rate of 5~MHz per channel. 
The upstream ends of the fibres are therefore capped with black paper
to minimize reflection. At the downstream side, the fibres from each of the
four sectors are bundled and coupled to a Hamamatsu R6427 PM.
As in the case of the SAC, a neutral-density filter is placed between each fibre bundle and PM.
The detector is suspended by four aluminium ribbons to minimize 
stress on the beam tube. The entire detector is wrapped in black paper to ensure light tightness.

\begin{figure}
\begin{center}
\includegraphics[width=0.85\linewidth]{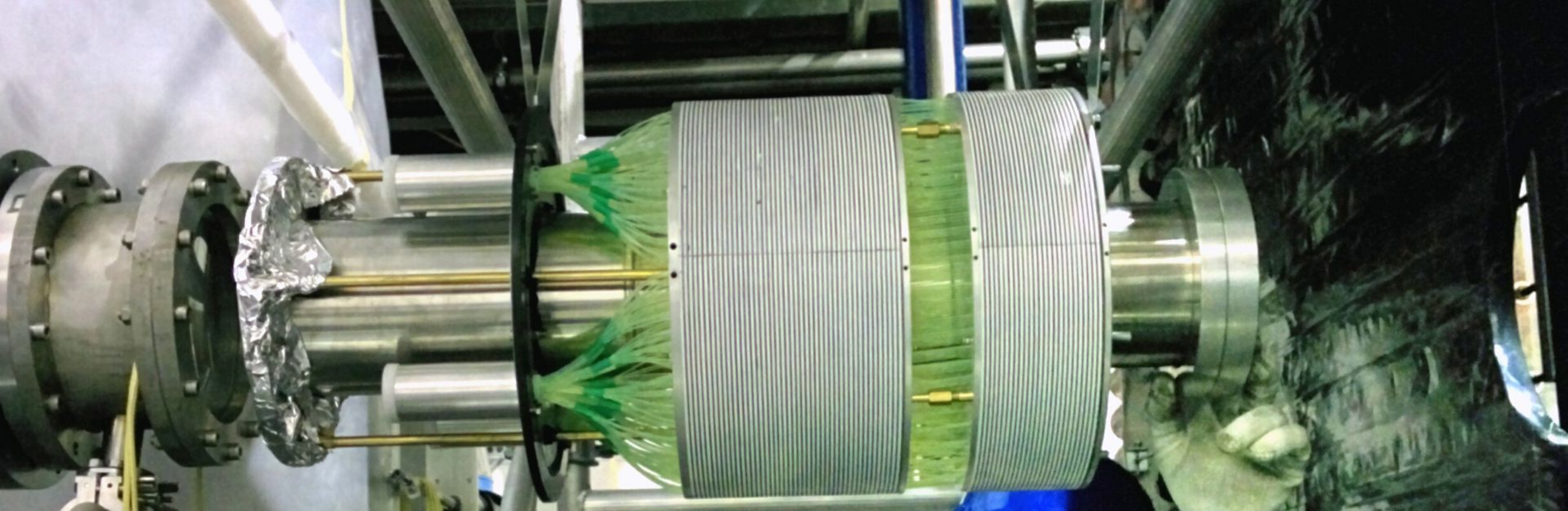}   
\end{center}
\caption{Picture of the  IRC during installation, prior to wrapping for light tightness. The upstream and downstream modules are clearly visible with the beam incident from the right.}
\label{fig:irc-installed}
\end{figure}
The detector, shown in \Fig{fig:irc-installed}, was installed in the beam line before the 2014 data-taking period.

\subsubsection{SAV readout}\label{ssec:savreadout}
The four channels of each calorimeter, IRC and SAC,  are read out using  both TEL62-based TDC  (\Sec{ssec:TEL62}) and CREAM FADC (\Sec{ssec:Cal-readout}) systems.
Using either system, the signals from the IRC and SAC can be used in the L0 trigger logic (\Sec{ssec:L0}). 
The redundant readout has proven to be useful for detector performance studies, as well as for the debugging and synchronization of the trigger and readout systems.

The signals from the detectors are split using a LeCroy 612AM two-stage, variable-gain linear amplifier with two outputs per channel.
The bandwidth of the amplifier is 140~MHz and the output signals have a rise time of 3.5~ns; 
the gain of the amplifier is kept at a value of 2.5 for the data taking.
For the TEL62-based readout, the signals from the amplifiers are input
into a LAV front-end board (\Sec{ssec:lav-fee}), where
each channel is discriminated against two adjustable thresholds,
nominally 10~mV and 30~mV, producing two channels of LVDS output
per input channel, in the same manner as for the LAV signals.
The leading- and trailing-edge times for the high- and 
low-threshold channels are digitized and read out using a 
TDCB mezzanine and TEL62 board (\Sec{ssec:TEL62}), 
providing ToT measurements.
For the CREAM-based readout, the signals from the amplifiers are shaped and
input to a CREAM module (\Sec{ssec:Cal-readout}) for full ADC readout and
inclusion in the calorimeter L0 trigger (\Sec{sssec:Cal-L0}). 

\subsection{Performance in 2015}
\subsubsection{LAV performance}\label{ssec:lav-perf}

The performance of the LAV system has been studied both with data
collected during standard  ``kaon runs'' and during dedicated
``muon runs''  as defined in \Sec{ssec:samples}.

\begin{figure}[ht]
\begin{minipage}{0.5\linewidth}
\centering
\includegraphics[width=1.\linewidth]{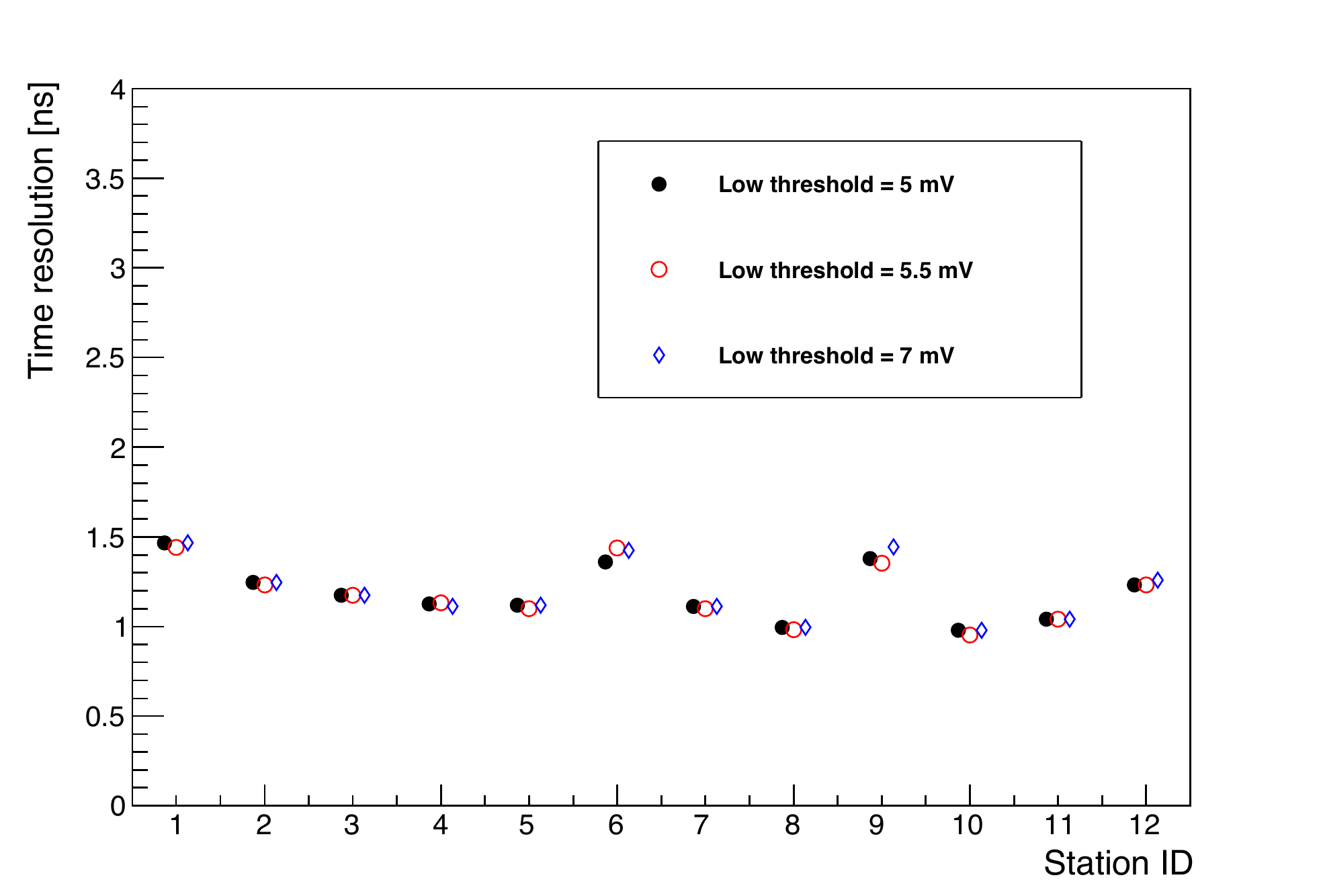}
\end{minipage}
\begin{minipage}{0.5\linewidth}
\centering
\includegraphics[width=1.\linewidth]{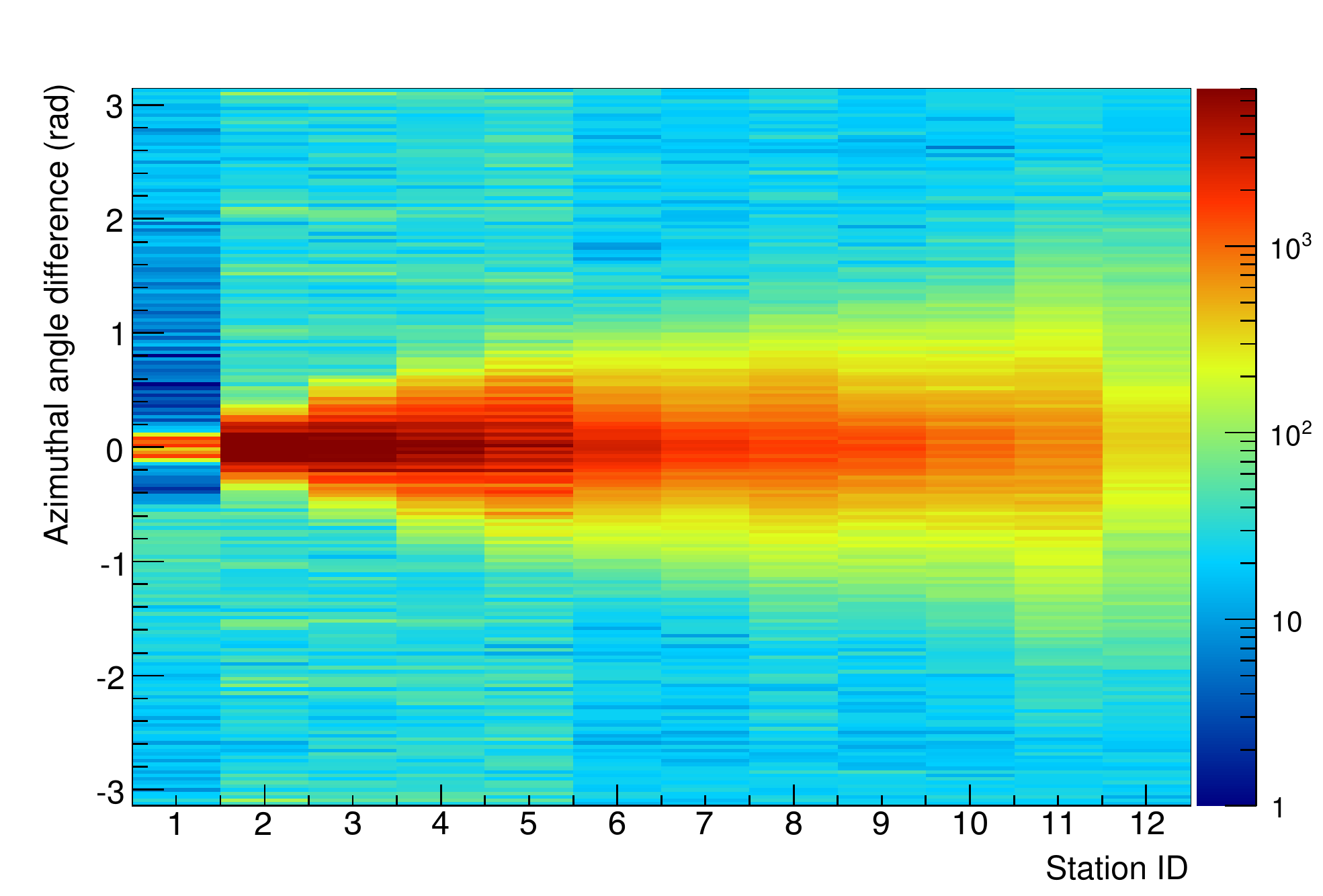}
\end{minipage}
\caption{\label{fig:lav-tres} Left: Time resolution for LAV stations for a high  threshold of 15 mV and three different values of the low threshold. 
Right: Difference in azimuthal angle between a cluster reconstructed in the most upstream layer of LAV station  1 (ID = 1) and clusters in successive layers, as a function of station ID, for data from a muon run.}
\end{figure}

Kaon runs are used to measure the time offsets for each channel with respect to the signals from the 
detectors that provide the event-time reference (NA48-CHOD or KTAG). Hit reconstruction
is then performed and slewing corrections are applied. A hit may be reconstructed from up to four time 
measurements, corresponding to the leading- and trailing-edge times on  the high and low thresholds.
The algorithm used to correct for slewing depends on how many and which
of the edges are used to reconstruct the hit. For example, if both
leading edges are present, the slewing correction is based on the
difference between the high- and low-threshold crossing times; if only
the low threshold is crossed, the slewing correction is based on a fit
to the measured distribution of leading-edge time as a function of ToT. After the application of slewing corrections,
time resolutions at the level of 1~ns are achieved for all LAV
stations, as shown in \Fig{fig:lav-tres}-left.  
Some difference from station to station is attributed to uncertainties in the determination of the channel-by-channel time offsets.
It should be noted that these results are obtained for samples that include all hit edge
configurations: not only complete hits built from leading/trailing-edge pairs for both thresholds, but also hits crossing only the low
threshold.

Muon runs are used to establish the threshold settings and to study the efficiency for the reconstruction of hits left by  MIPs. 
Hits on blocks adjacent in azimuth or layer and with compatible signal times are grouped into clusters. Penetrating muon ``tracks'' in the
LAV system are then identified by the correlation between clusters at the same azimuthal angle in different stations, as 
illustrated in \Fig{fig:lav-tres}-right.  
For certain configurations, it is possible to require hits on the blocks immediately
upstream and downstream of a given module to better determine its efficiency.
The efficiency for MIPs was found to saturate at about 97\%
for values of the low threshold below 6~mV, leading to a  value of 5~mV for the low-threshold working point.

A first, preliminary attempt has been made to study the photon detection efficiency with a sample of $\KTP$ decays in which the $\PIo$ decays to $\gamma\gamma$. One of the photons is required to be detected in the LKr, while the other may be extrapolated to one of the LAV stations. However, at present, the resolution on the extrapolated photon direction is not sufficient to allow  the efficiency  to be determined for individual LAV stations. 
This may be possible in the future by means of a complete kinematic fit
making use of all available information on the $\KPL$ trajectory from the GTK. 
Currently the focus is on estimating the global efficiency for the entire LAV system, so events are considered successfully matched if they contain a hit on at least one LAV block within 5~ns of the
$\KTP$ event time from the reference detector. MC studies demonstrate 
that the photon detection inefficiency determined by this method is dominated by geometrical inefficiencies and upstream photon conversions; the intrinsic inefficiency arising from the LAV detectors is less than about 15\% of the observed inefficiency. Relying on the MC estimate
for the contribution from the former effects, one finds the intrinsic inefficiency to be less than $10^{-3}$ with about 5\% of the detected
photons observed as a signal on an isolated block crossing only the low threshold. 
These results are preliminary; as noted above, further refinements to the method will be implemented. 

\subsubsection{LKr performance} 
The energy, space, and time resolutions of the LKr calorimeter have been quantified by NA48 \cite{NA48:2007}.  In the NA62 setup, the performance values are degraded, 
mainly because of the presence of a non-linear energy response and of extra material upstream of the calorimeter. A first estimation of the energy resolution as obtained from simulated data is :
\begin{equation}
\frac{\sigma({\rm E})} {\rm E} = \frac{4.8 \%} {\sqrt{\rm E}} \oplus \frac{11 \%} {\rm E} \oplus 0.9 \% ~, 
\end{equation}
where the energy is measured in GeV. 

Clusters of energy ${\rm E}_{\rm clus}$ are built from a seed --- any cell with energy  ${\rm E}_{\rm seed}$ above 250~MeV and larger than the sum of energies in the 8 neighbouring cells --- and all cells located within 11~cm of the cluster centre, defined as the energy-weighted position of the participating cells. A track is defined as an associated track if its impact point on the calorimeter front face coincides with the position of a cluster within 15~cm.  A photon candidate is a cluster of energy without  an associated track. 

\begin{figure}[ht]
\begin{minipage}{0.5\textwidth}
\centering
\includegraphics[width=1.05\textwidth]{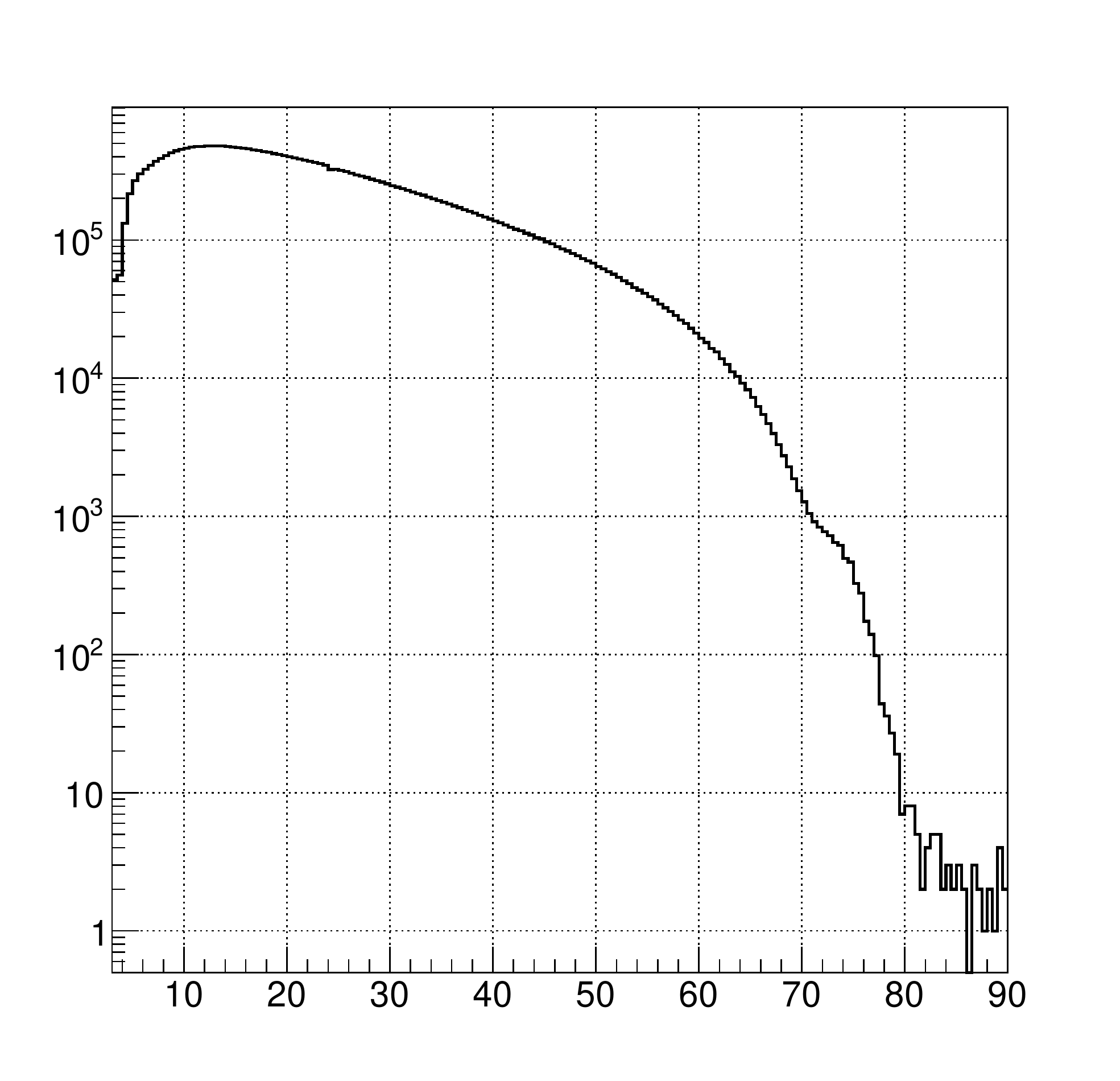}
\put(-75,0){ ${\rm E}_{\rm  clus}$ [GeV] }
\end{minipage}
\begin{minipage}{0.5\textwidth}
\centering
\includegraphics[width=1.05\textwidth]{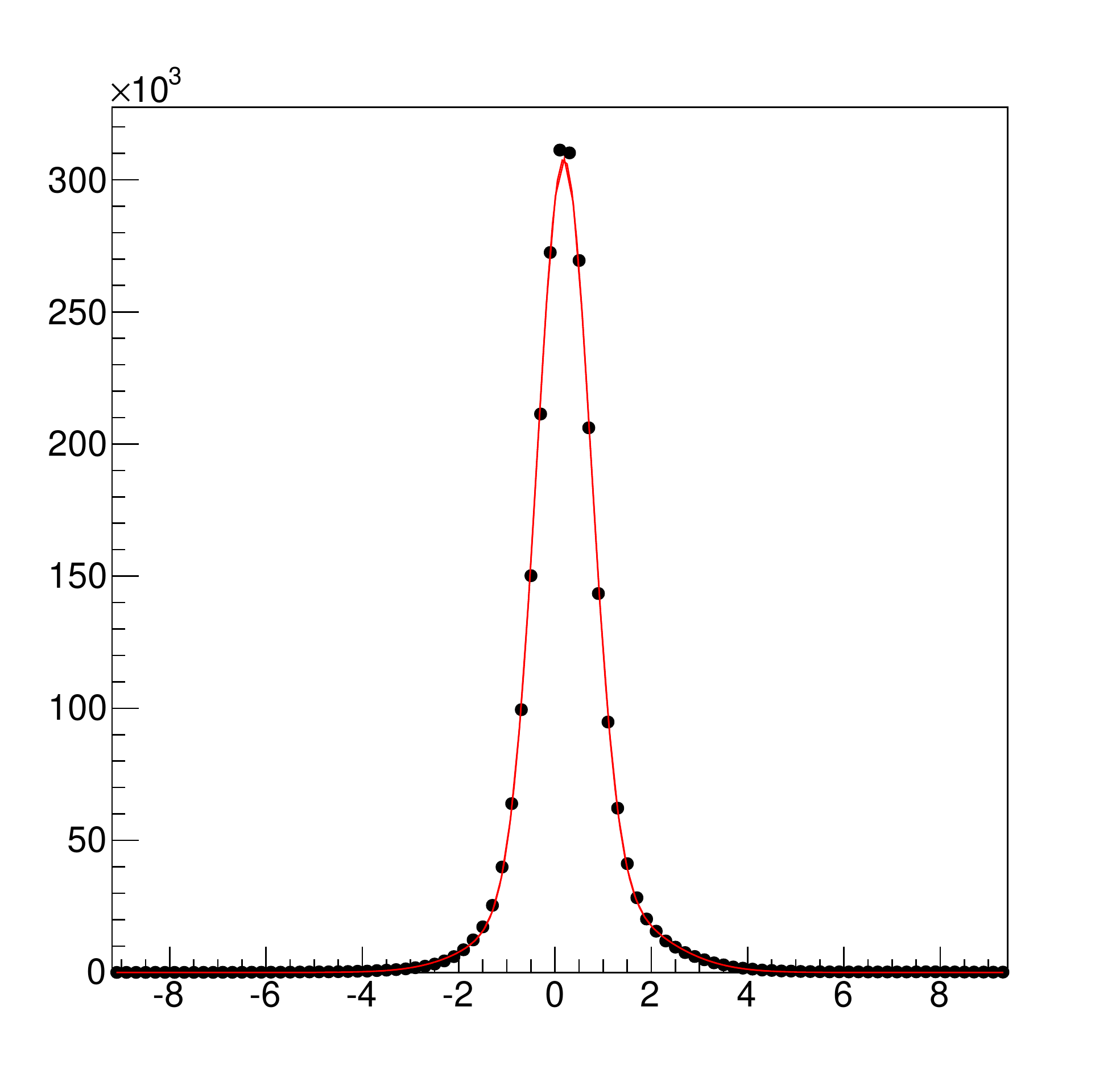}
\put(-100,150){$\sigma_{\rm tot}$ = 0.550 ns }
\put(-106,0){ ${\rm t}_{\rm photon} - {\rm t}_{\rm CHOD}$ [ns]}
\end{minipage}
\caption{\label{fig:CREAM_Picture11} Photon clusters detected in the LKr using selected $\KTP$ decays. Left: cluster energy distribution. Right: difference between the closest photon time and  the $\PPI$ candidate  time measured in the NA48-CHOD. The contribution of the NA48-CHOD to the quoted time resolution is negligible.}
\end{figure}

The energy distribution for photon candidates with ${\rm E}_{\rm clus} >3$~GeV, obtained with 2015 data, is shown in \Fig{fig:CREAM_Picture11}-left. 
Photon candidate clusters from selected $\KTP$ decays  
can be used to estimate the time resolution. The distribution of the difference between the NA48-CHOD time of the pion-candidate track and the LKr time of the closest in-time photon cluster with ${\rm E}_{\rm clus}>3$~GeV is shown in \Fig{fig:CREAM_Picture11}-right. The width of the distribution is 550~ps.

\begin{figure}[ht]
\begin{minipage}{0.5\textwidth}
\centering
\includegraphics[width=1.\textwidth]{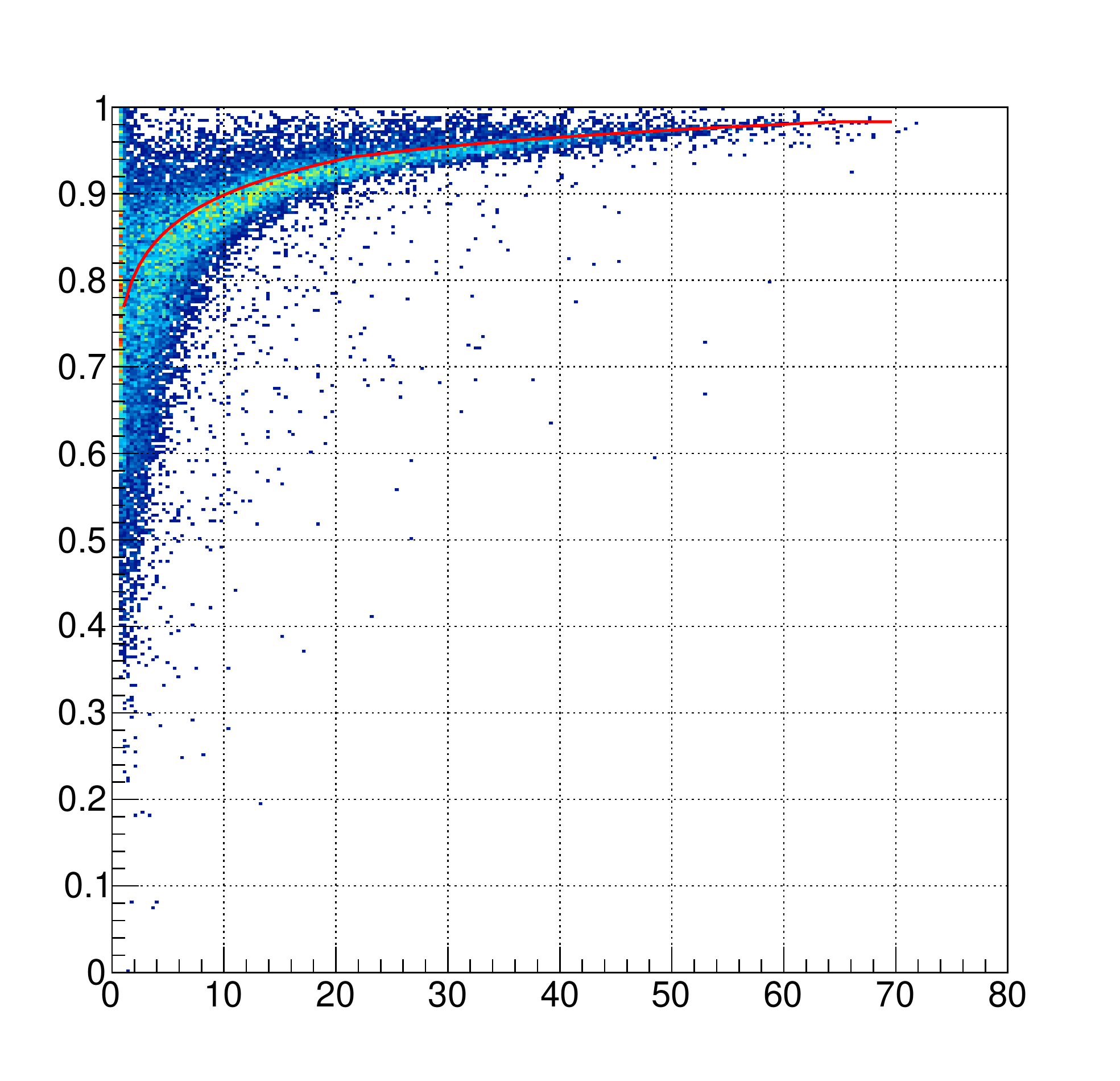}
\put(-105,0){ {\rm E}$_{\rm  clus}$, no ZS [GeV] }
\put(-215,85){\rotatebox{90}{${\rm E}_{\rm  clus}$, ZS / ${\rm E}_{\rm  clus}$, no ZS}}
\end{minipage}
\begin{minipage}{0.5\textwidth}
\centering
\includegraphics[width=1.\textwidth]{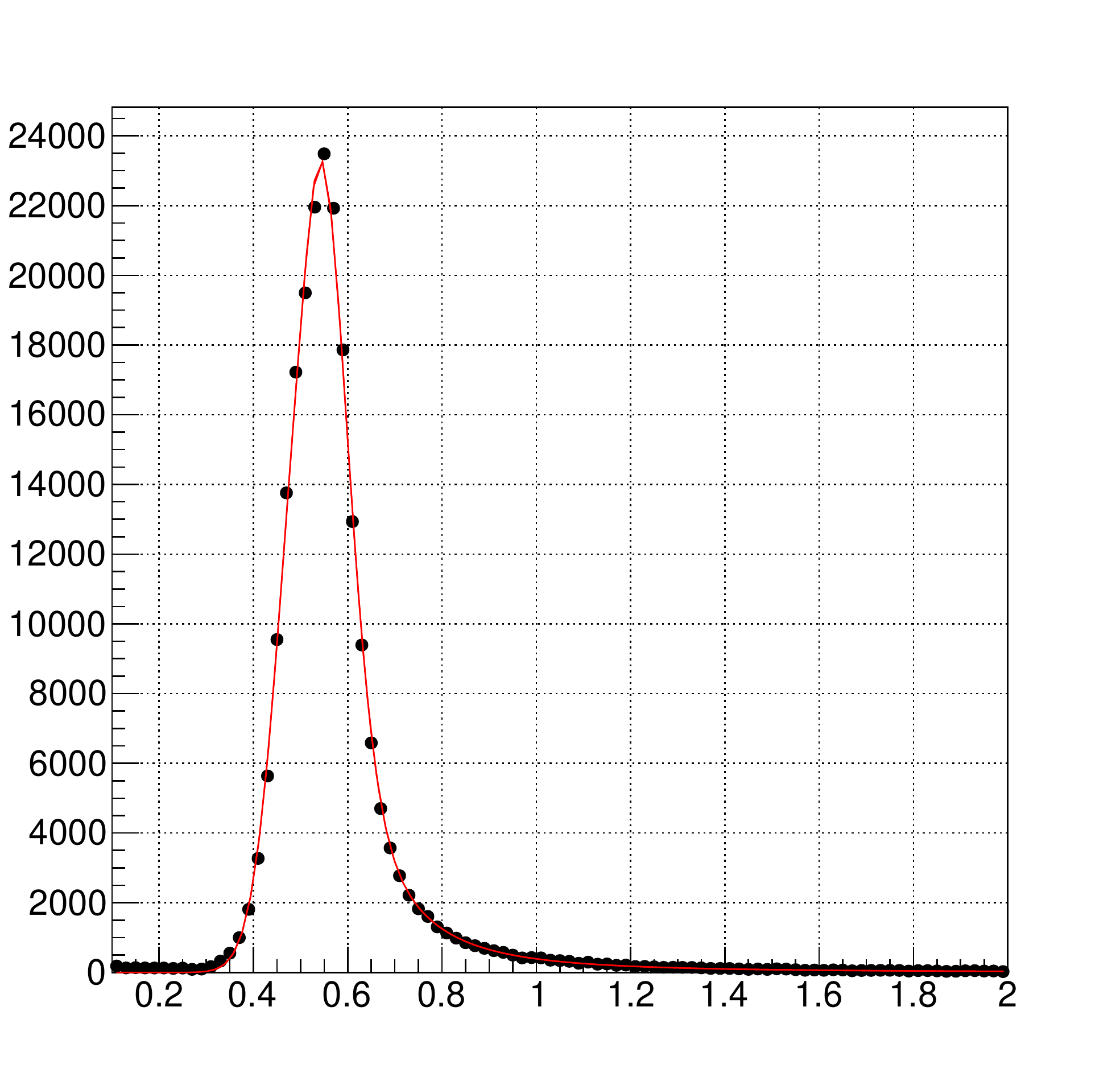}
\put(-75,0){ ${\rm E}_{\rm  clus}$ [GeV] }
\end{minipage}
\caption{\label{fig:LKr}Left: Non-linearity of the LKr cluster energy response. The red line represents the function used to correct for non-linearity of the energy response.
Right: Cluster energy distribution for muons from $\KMN$ decays.}
\end{figure}

The implementation of the zero-suppression (ZS) mechanism described in \Sec{ssec:Cal-readout} introduces a non-linearity in the energy response, due to the suppression of contributions  
from cells below the ZS threshold. The effect is larger at low energies.
\Fig{fig:LKr}-left shows the distribution of the ratio of the cluster energy with and without ZS as a function of the cluster energy without ZS, without any particular sample selection.
This distribution is used to correct the cluster energies obtained with ZS.

The ZS scheme allows the detection of MIPs  
with improved energy resolution and efficiency: in the NA48 scheme, a box of 121 cells was always read out around a seed and, in the case of MIPs, the noise contribution to the cluster energy was a large fraction of the deposited energy, while in the new scheme the 
noise level is strongly reduced and the energy resolution for MIPs is improved. As an example, \Fig{fig:LKr}-right shows the LKr energy distribution for muons from $\KMN$ decays, together with a fit by
a linear combination of Vavilov \cite{bib:vavilof} and Gaussian functions. The peak
value is~545 MeV, the average is 585~MeV, and the width (rms) of the Gaussian used in the fit
is 9~MeV. The detection efficiency for MIPs is $(99.0\pm0.5)$\% as measured from data.
The LKr information can be used together with other detector information for 
pion and muon identification (\Sec{sec:valid}).

The CREAM readout (\Sec{ssec:Cal-readout}) was fully commissioned and integrated with the NA62 data-acquisition system before the start of data taking.
Regular calibration runs are taken in standalone mode to verify the stability of the pedestals and the linearity of the electronics, which has been measured to be better than 1 per mille.

\subsubsection{SAV performance}

The response of the small-angle veto system has been studied
with muons. Both the spectrometer magnet (MNP33) and the beam-deflecting
magnet (BEND) in front of the SAC (\Fig{fig:K12beam-downstream}) are switched off.
A complete threshold scan\footnote{Threshold studies are conducted with the
TDC-based readout and without the linear amplifier.} is performed for each individual
channel by varying both the high and low thresholds from  
4~mV to 20~mV.
The observed rate as a function of the threshold is fitted with a cumulative Landau distribution function to 
obtain the most probable values of the signal amplitude $A_{\rm MIP}$ 
corresponding to the energy deposited by muons in the detectors. 
The method is sensitive to the inflection point of the rate as a function of the applied threshold. The value of $A_{\rm MIP}$,  about 5 mV for the IRC and 4 mV for the SAC, has been found to be stable throughout data taking.

Preliminary studies of the photon detection efficiency for
the SAC and IRC have been performed with tagged photons from $\KTP$ events in a manner similar to that used to estimate the LAV efficiency (\Sec{ssec:lav-perf}).
The spatial resolution of the extrapolation does not allow
the efficiency of the SAC and IRC to be studied separately; it is only
possible to estimate the overall inefficiency for the forward part of
the photon-veto system, including the SAC, IRC, and the inner region
of the LKr. The preliminary inefficiency for the forward photon veto,
including the mistagging probability of the method, has been found to be 
$7\times 10^{-4}$ and stable during data taking.

\begin{figure}[ht]
\begin{center}
\includegraphics[width=0.70\linewidth]{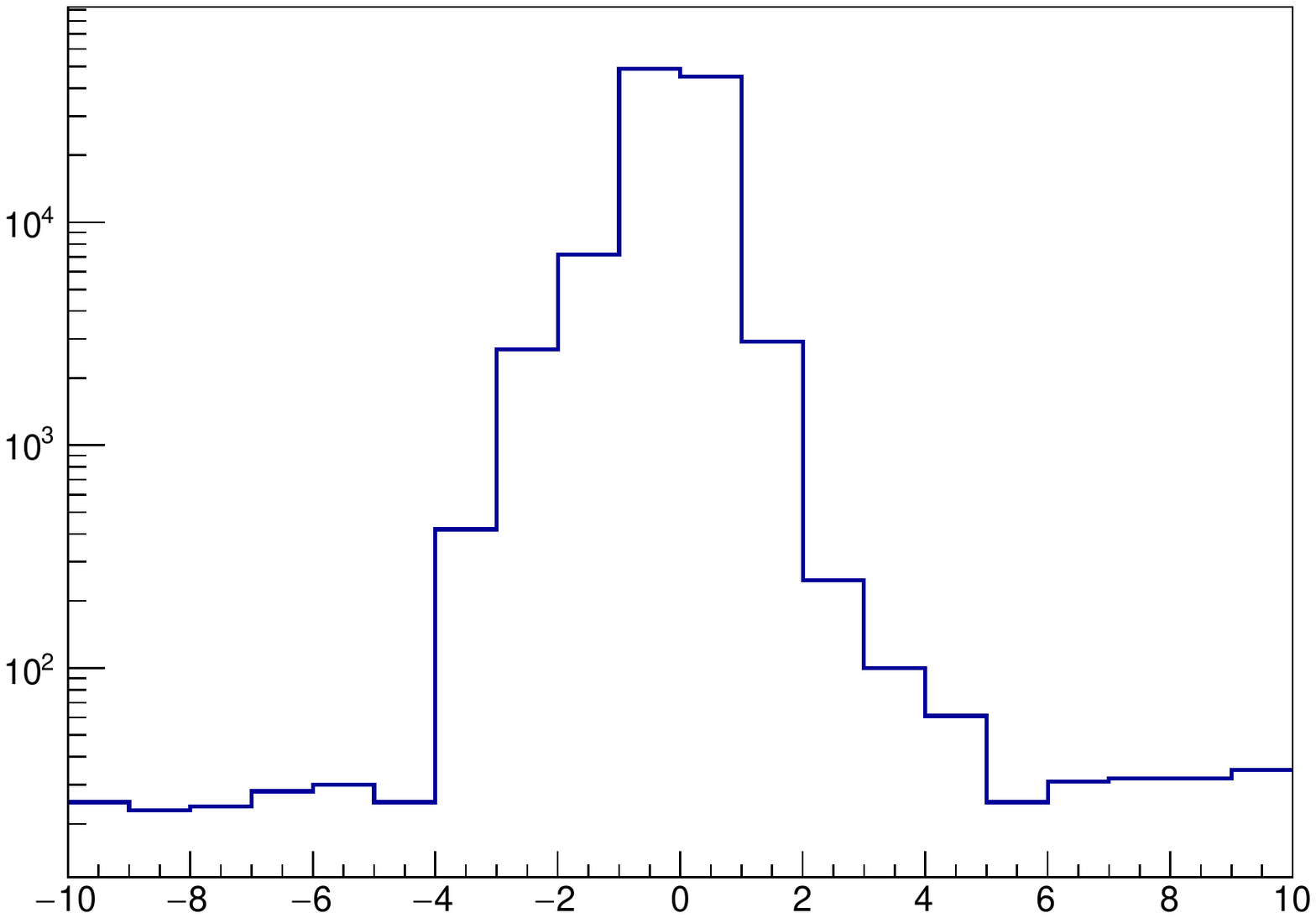}   
\put(-110,0){${\rm t}_{\rm SAV} - {\rm t}_{\rm CHOD}$ [ns]}
\put(-250,165){$\sigma_{\rm tot} = 0.89$ ns}
\end{center}
\caption{Time difference between the pion hit in the NA48-CHOD and the nearest
hit in time in the SAC or IRC for $\KTP$ events with a tagged photon.}
\label{fig:sav-chod-dt}
\end{figure}

The time resolution for muons is better than 2 ns for the SAC and
better than 1.6 ns for the IRC, while the time resolution for the SAC
and IRC together measured with tagged photons from $\KTP$ decays is
better than 1 ns (\Fig{fig:sav-chod-dt}).
The high rates in the SAC and IRC induce a non-negligible probability
for vetoing $\PPI \nu \bar\nu$ signal events due to accidental coincidence. The random veto
is estimated using the out-of-time sidebands of the recorded events and
applying the same veto criteria as for the signal.
With a 5~ns cut around the event time, at the nominal NA62
intensity the random veto probability is found to be $\sim$$5\%$.

%% file: Sec9-RICH_v2.tex
The RICH detector is designed to separate pions from muons between momenta of  15 
and 35 $\GEVc$ providing a muon suppression factor of at least 100 as part of the $\cal{O}$$(10^7 )$  overall rejection factor needed (\Sec{sec:intro}).
In order to have full efficiency for a 15 $\GEVc$ momentum pion, the
threshold should be about 20\% smaller or 12.5 $\GEVc$, corresponding to
$(n - 1) =  62 \times 10^{-6} $  matching almost exactly the
refractive index at atmospheric pressure and room temperature of neon gas that has been
chosen as radiator medium. The RICH measures the pion crossing time with
a resolution of about 100 ps, thus providing a possible reference time for charged tracks.

Two prototype detectors were built and tested in hadron beams to demonstrate the performance of 
the proposed layout. The results of these tests have been published \cite{Anzivino:2008, Angelucci:2010}.

\subsection{Radiator vessel description}\label{ssec:RICH-Vessel}

\begin{figure}[ht]
\begin{center}
\includegraphics[width=1.\linewidth]{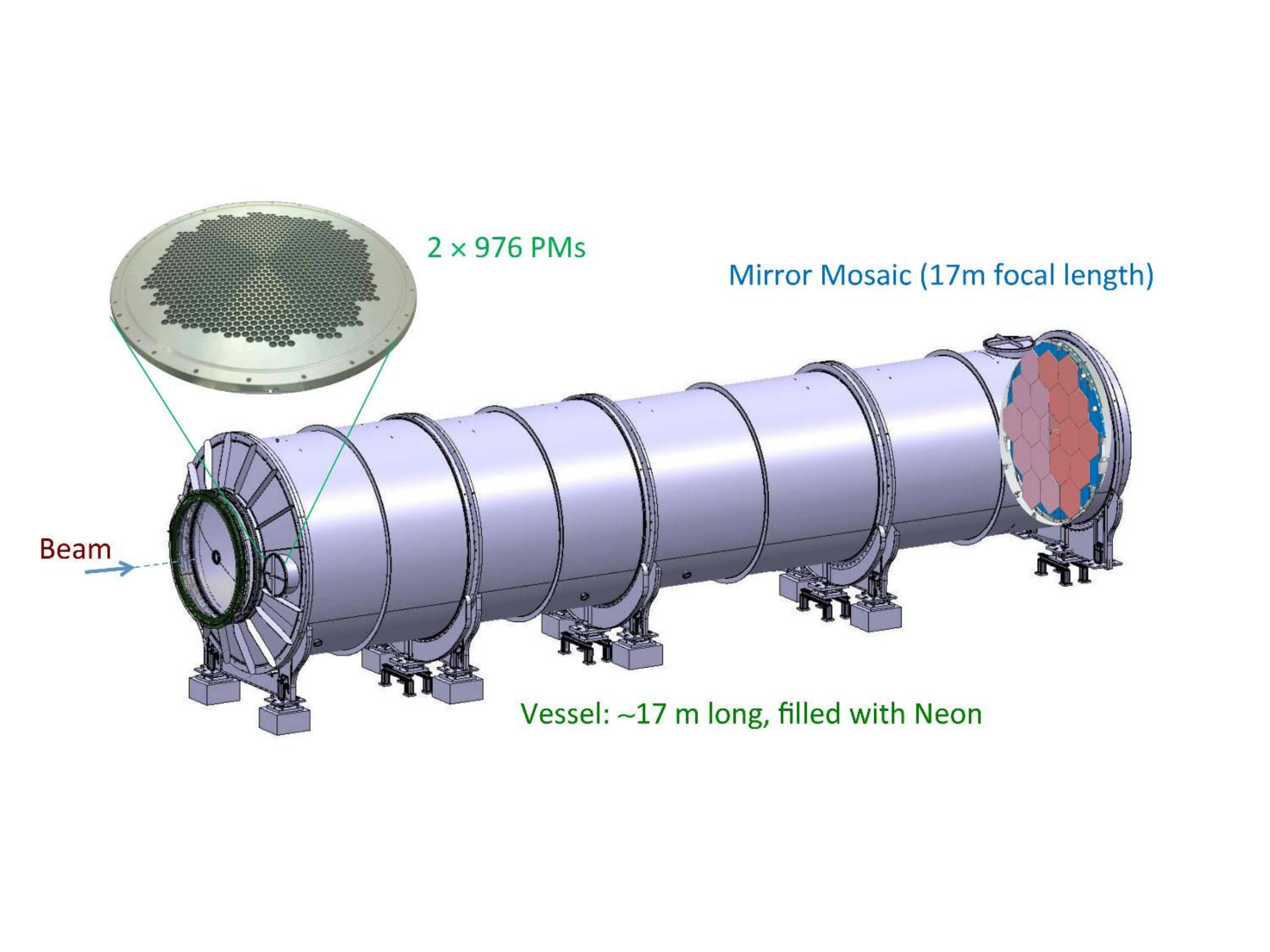}
\end{center}
\vspace{-3cm}
 \caption{\label{fig:RICH-Picture1} {
Schematic view of the RICH detector: The hadron beam enters from the left and travels throughout the length of the detector in  an evacuated beam pipe.  A zoom on one of the two disks accommodating the light sensors (PMs) is shown on the left; the mirror mosaic is made visible through the neon container (vessel) on the right.
}}
\end{figure}

The RICH radiator is a 17.5 m long cylindrical vessel made of
ferro-pearlitic steel (\Fig{fig:RICH-Picture1}) and filled with neon gas. The
vessel consists of four sections of gradually decreasing diameter and
different lengths. At the upstream end the vessel is about 4.2~m wide
 to accommodate the photomultiplier flanges outside the active
area of the detector. The diameter of the last vessel element is 3.2~m,
which is sufficient to house the mirrors and their support panel. The
integrity and cleanliness of the inner surfaces is achieved by a black
epoxy painting.

The active area of the detector extends  to a radial distance of 1.1~m from the beam axis at the RICH entrance and  to 1.4~m at the exit window.
The entrance and exit windows have a conical shape and are made of
aluminium with a thickness of 2 and 4 mm respectively. The entrance
window is the only separator between the decay vacuum volume and the
RICH radiator gas volume. 
 A lightweight aluminium tube, connected to the decay-tank vacuum, passes centrally through the vessel.

To minimize the material budget, especially at the upstream
detector side, the beam pipe is rigidly connected to a flange on the entrance window. The beam pipe
continues through the RICH exit window and is sealed with a sliding
O-ring which allows the beam pipe to move longitudinally.  This movement is
necessary to compensate for deformations of the entrance window.

The radiator vessel is evacuated before being filled with neon gas.
During operation, the neon pressure is then kept constant at about
990 mbar with the vessel sealed. Small gas losses due to
leaks are compensated by occasional top-ups. This concept has the advantage that
temperature variations do not influence the gas
density.

The photon detection sensitivity range starts at wavelength above 190 nm,
which makes the detector performance practically insensitive to
impurities like oxygen and H\textsubscript{2}O in the gas. Other
impurities, like for example CO\textsubscript{2} are  not present naturally and can be
kept sufficiently low \cite{Gersabeck:2013}.

The neon density influences the refractive index $n$ following the
relation
\begin{equation}
n = 1 + ( n_{0} - 1 )  \frac{\rho}{\rho_{0}}  ~,  
\end{equation}
where $n_{0}$ is the refractive index (1.000067) and $\rho_{0}$
is the density (0.9001 kg/m\textsuperscript{3}) of neon gas at NTP;
$\rho$ is the density at operating conditions ($\approx$ 0.814
kg/m\textsuperscript{3} for T= 25\textsuperscript{o}C and P $\simeq 1$ bar).

\subsection {Mirror layout}\label{ssec:RICH-mirrors}
\begin{figure}[ht]
\begin{minipage}{0.5\linewidth}
\begin{center}
\includegraphics[width=0.9\linewidth]{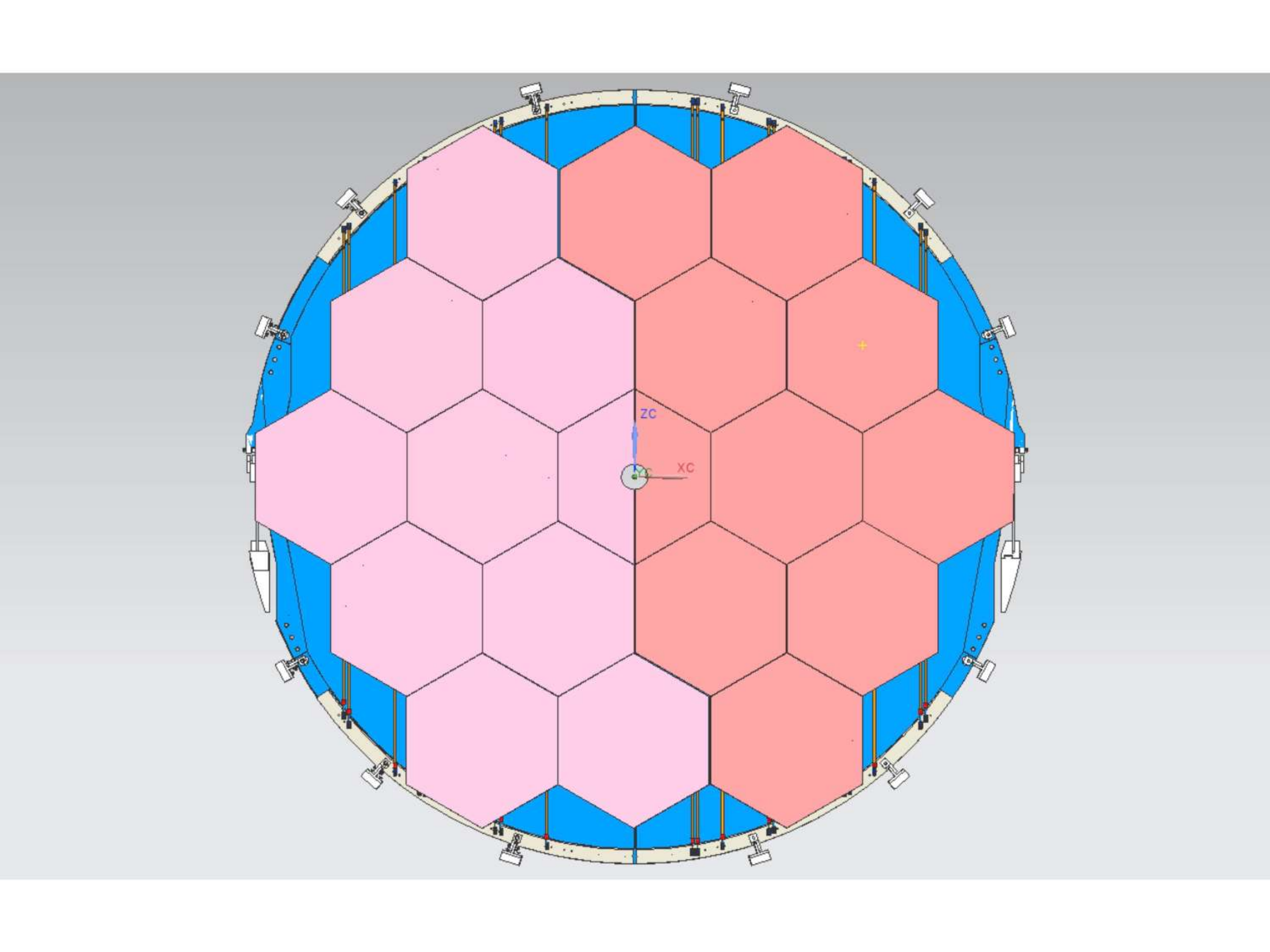}
\end{center}
\end{minipage}
\begin{minipage}{0.5\linewidth}
\begin{center}
\includegraphics[width=1.\linewidth]{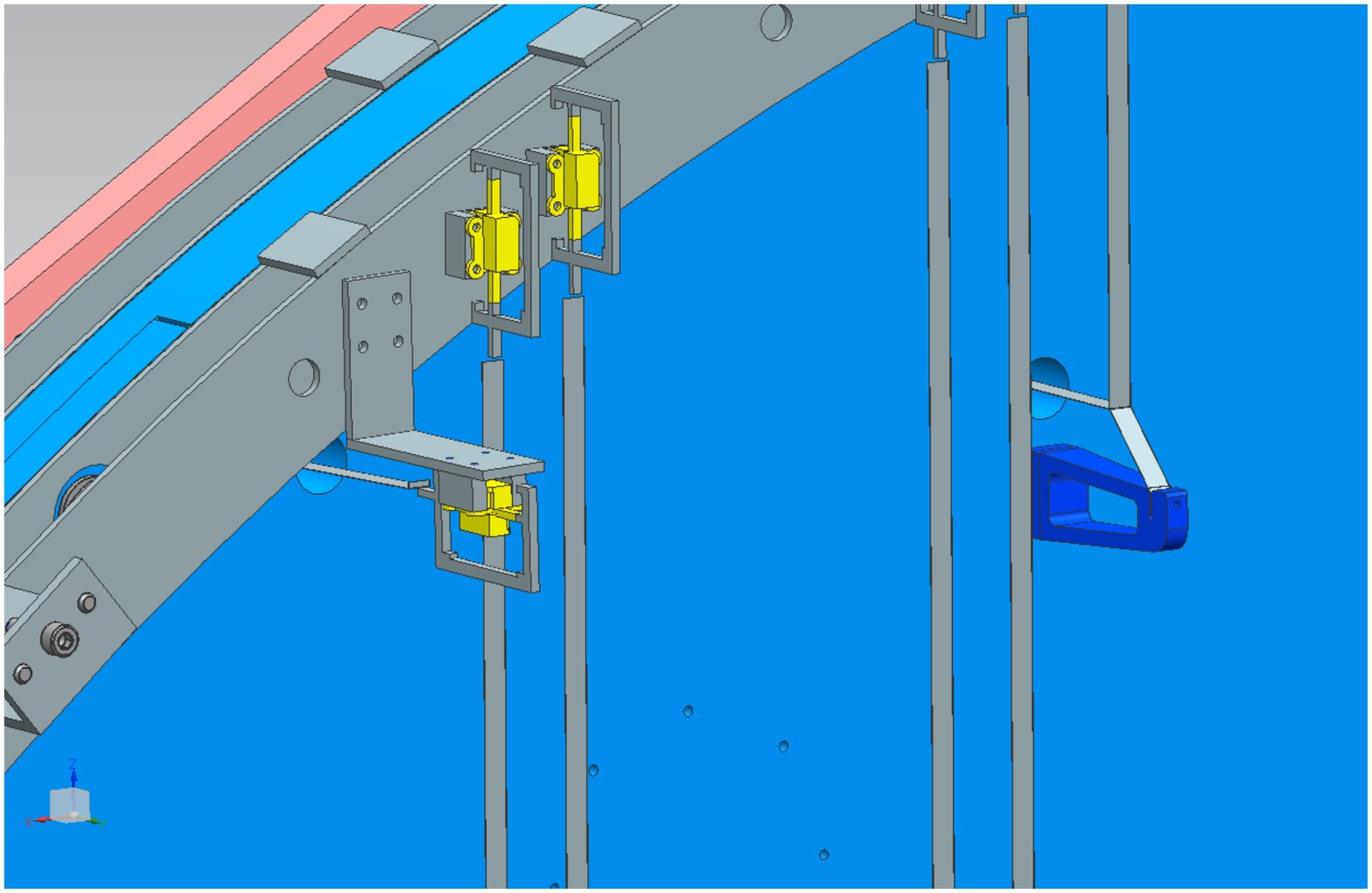}
\end{center}
\end{minipage}
\caption{\label{fig:RICH-Picture2} {Left: The mirror mosaic at the downstream end of the RICH contains 
18 hexagonal  (350~mm side) and  two
semi-hexagonal mirrors (adjacent to the beam pipe opening). Half of the
mirrors points to the left (right)  of the beam pipe. 
Right: Sketch of the mirror alignment system. Three piezoelectric motors (yellow boxes) can be seen; the leftmost motor is  pulling the mirror directly; on the right a transmission tool (blue object)  changes the ribbon direction from vertical to horizontal.
}}
\end{figure}

A mosaic of 20 spherical mirrors is used to reflect the Cherenkov light
cone into a ring on the PM array in the mirror focal plane (\Fig{fig:RICH-Picture2}-left). To avoid
absorption of reflected light by the beam pipe the mirrors are divided
into two spherical surfaces: one with the centre of curvature to the
left and one to the right of the beam pipe. The total reflective surface exceeds 6~m$^{2}$.

The mirrors have a nominal  radius of curvature of 34~m and hence a focal
length of 17~m. 
The mosaic includes 18 mirrors of regular hexagonal shape (350~mm side) and two half mirrors.
The two latter ones are used in
the centre and have a circular opening to accommodate the beam pipe. The
mirrors are made from 25~mm thick glass substrate coated with
aluminium. A thin dielectric film is added for protection and to improve
the reflectivity. The mirrors satisfy the following optical parameters:

\begin{itemize}
\item
 a diameter D$_0$ not larger than 4 mm (D$_0$ is the minimum diameter of a circle
 which collects 95\% of the light of a point-like source placed at the
 centre of curvature);
\item
  a radius of curvature within $\pm$20 cm from the nominal one;
\item
  an average reflectivity of about 90\% in the wavelength range (195, 650) nm.
\end{itemize}

The mirror support structure must sustain a total weight of
approximately 400 kg and must guarantee long-term stability of the
mirror positions. 
A 50 mm thick aluminum honeycomb structure was chosen as the mirror support panel to minimize interactions which deteriorate the performance of downstream detectors.

The supports for each mirror enable the fine adjustment in orientation necessary for its 
alignment  (\Fig{fig:RICH-Picture2}-right). 
A hole with a diameter of 12 mm was drilled on the back surface of each hexagonal
mirror close to its centre. A dowel with a spherical head was inserted
in each mirror hole and connected to the support panel. Two thin
aluminium ribbons, attached to the mirror rear surface at about 250 mm
from the hole and
at $\pm 45$ degrees with respect to the vertical direction,
keep the mirror in equilibrium and allow its orientation.
Each ribbon pulls the mirror in the horizontal direction, crosses the support panel 
in a dedicated hole and  is bent vertically by a proper transmission tool fixed 
on the panel.
The semi-hexagonal mirrors have two holes drilled on the rear surface and a single ribbon.

Two piezoelectric actuators, attached to the other end of the aluminium
ribbons, allow a remotely controlled two-axis orientation of each
hexagonal mirror. The two semi-hexagonal mirrors in the centre are
oriented only in the vertical direction by one actuator, the other axis
is fixed after the first alignment.

The piezoelectric actuators (type LEGS-LT02SV-10 produced by PiezoMotor)
have a $\pm$35 mm travel range with 1 nm resolution and can produce up to 20
N of push/pull force. These devices are self-locking and remain in the
same position if the supply voltage is turned off. The piezoelectric
actuator size is $22 \times 10.8 \times 21$ mm$^3$. All the piezoelectric actuators are located outside 
the particle acceptance.

After the installation, a laser mirror alignment was performed. The mirror alignment was checked with beam during data taking by selecting particles whose Cherenkov rings were completely contained in a single mirror and with a track angle measured by the magnetic spectrometer.

\subsection{Photon detection system }\label{sec:RICH-Pmt}

The granularity of the photon detection is an essential parameter to
optimize the angular resolution of the detector while the number of photodetectors 
has large impact on its cost. 
A reasonable compromise between the number of sensors, the photon
acceptance and the sensor dimensions has been pursued. According to the
detector simulation, a total of 2000 PMs is sufficient to
match the experiment requirements. This arrangement led to 976
PMs on each detector side and to a photosensor pitch of 18~mm. 

The Hamamatsu R7400 U-03 PM was chosen for its fast response, small dimensions and reasonable cost. 
This PM type has a single metal covered anode and a UV glass entrance window. The sensitivity starts at a wave length of 185~nm and peaks at a wave length of 420~nm. This PM has a cylindrical shape with a 16~mm wide base and an 8~mm diameter active area.

The HV divider is custom-made providing 8 dynode voltages (28 M$\Omega$ total
resistance) with a cylindrical shape of (17.0 $\pm$ 0.2)~mm diameter and
(15.0 $\pm$ 0.5)~mm length. Each HV divider is connected with three cables: a 2~m long RG-174/U cable for the signal output and two AWG22 cables for the negative high-voltage supply and ground.

The PM time response is of great importance and has been investigated in detail.
The transit time spread given by the vendor is expressed in FWHM.  The signal response has important tails and is far from a Gaussian shape. Several systematic contributions have been identified in the time distribution:

\begin{itemize}
\item
 a small early peak, attributed to the electron extraction from the first dynode instead of the photocathode;
\item
  a small, 1.2 ns late peak, attributed to electrons extracted from the photocathode, reflected from the first dynode towards the cathode and returning to the dynode.
\end{itemize}

The PM requires a low-noise 800 -- 1000 V negative-voltage supply. To reduce cost, four PMs are supplied by a single HV channel: 488 HV channels are required. The A1733N (12-channels) and A1535SN (24-channels) CAEN boards fulfil the requirements and 
are placed in four 16-slot SY1527 CAEN crates. The HV system is remotely controlled by the DCS system (\Sec{sec:online}). The A1733N board has a maximum of 4(3) kV output voltage (dual range) with 2(3) mA maximum output current. The voltage and current resolutions are 250~mV and 200~nA, respectively; the voltage ripple is smaller than 30 mV peak-to-peak. The A1535SN board has a maximum 3.5~kV output voltage, with 500~mV resolution, and 3~mA output current, with 500~nA resolution; the typical voltage ripple is smaller than 20~mV pp. All HV channels have a common floating return.

The PM output signal has a roughly triangular shape with an average rise-time of 0.78~ns and a
fall time about twice as long. Assuming 900~V PM
supply voltage (average gain of $1.5 \times 10^{6}$) the single photoelectron
output charge is about 240~fC, corresponding to a peak current of 200~$\mu$A and to
a negative peak voltage of 10~mV into 50~$\Omega$. There is also a large gain
variation among the PMs. In order to
exploit the fast PM response, the 8-channel NINO ASIC \cite{Anghinolfi:2004} was chosen as the discriminator. This chip has an intrinsic resolution of
50 ps and was developed for the output signal of multigap resistive
plate chambers. To match the optimal NINO performance region, the
PM output must be sent to a current amplifier with
differential output: a 32-channel customized printed circuit was 
used for this purpose, sending the output to a board containing 4 NINO
ASICs. The NINO chip is operated in time-over-threshold mode.  Its LVDS
output signal is sent to 512-channel TEL62 boards equipped with HPTDC
chips with 97.7 ps LSB producing 19 bits long words (corresponding to a
maximum of 51 $\mu$s). Both the leading and trailing edge of the LVDS signal
are recorded providing information on the original signal width useful
for offline
time slewing correction.
To simplify access, the PMs are mounted outside the
radiator gas volume. The assembly consists of two independent aluminium
flanges: a 23 mm thick radiator flange with quartz windows and an
independent 35~mm thick flange holding the PMs.

The light entrance holes on the radiator flange have the shape of a
truncated circular paraboloid (a ``Winston Cone'' \cite{Winston:1966})
covered with a highly reflective aluminized Polyethylenterephthalat 
(Mylar\textsuperscript{TM}) foil to funnel the light through the window aperture. The Winston Cone is 21.5 mm high, 18 mm wide at the entrance and 7.5 mm on the
opposite side. Outside each hole, on the PM side, a 1.5
mm deep, 14 mm wide cylindrical hole allows for positioning and gluing of the
quartz windows, using a 0.5 mm deep, 2 mm wide groove for the glue.
The PMs are mounted on an external aluminium flange in
front of the quartz windows. 
A cylindrical hole, 16.4~mm wide and 12.5~mm high has been drilled in the aluminium flange for each 
PM, followed by a 17.5~mm wide and 20 mm high hole for the HV divider. 
A 1 mm thick O-ring (17.5~mm outer and 13.5 mm inner diameter, positioned into a 1.5~mm thick groove in the hole, 1~mm above the end) has been placed in front of the PM and pressed against the quartz window to prevent external light from reaching the PM. A (5 $\pm$ 1)~mm thick O-ring
(with the same outer and inner diameter as the 1-mm O-ring) has been
placed on the back of the PM, after the end of the HV
divider, to close the hole and seal off external light. This latter O-ring
also guarantees good thermal contact between the PM and the aluminium flange and absorbs the tolerance in the PM total length. The PM HV divider dissipates about 30 mW per
tube or 30 W per side (1000 PM), hence a cooling system is not necessary.

\subsection{Performance in 2014 and 2015}\label{ssec:RICH-Performance}
\begin{figure}[ht]
\begin{minipage}{0.5\linewidth}
\begin{center}
\includegraphics[width=0.95\linewidth]{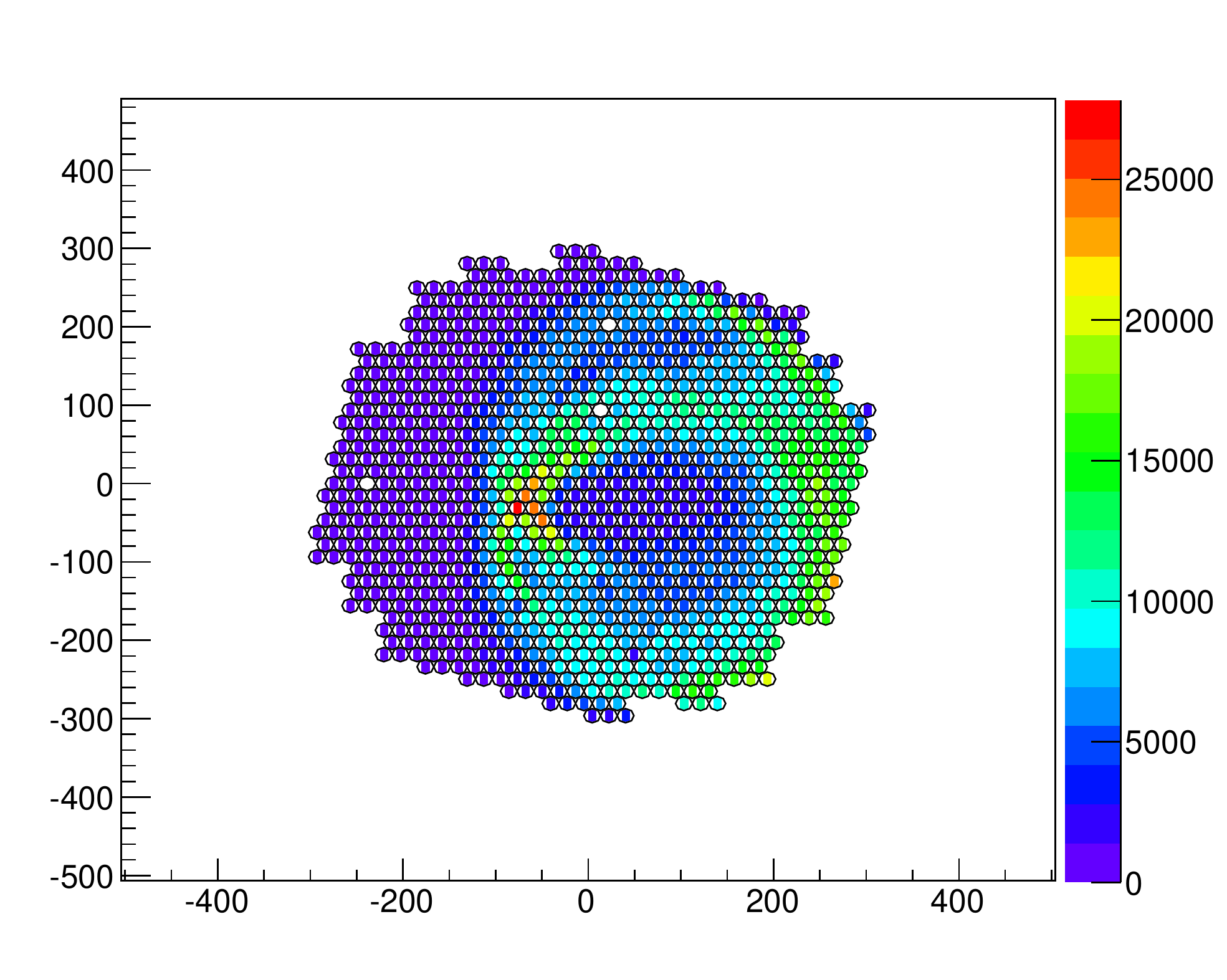}
\put(-60,0){x [mm]}
\put(-205,115){\rotatebox{90}{y [mm]}}
\end{center}
\end{minipage}
\begin{minipage}{0.5\linewidth}
\begin{center}
\includegraphics[width=0.95\linewidth]{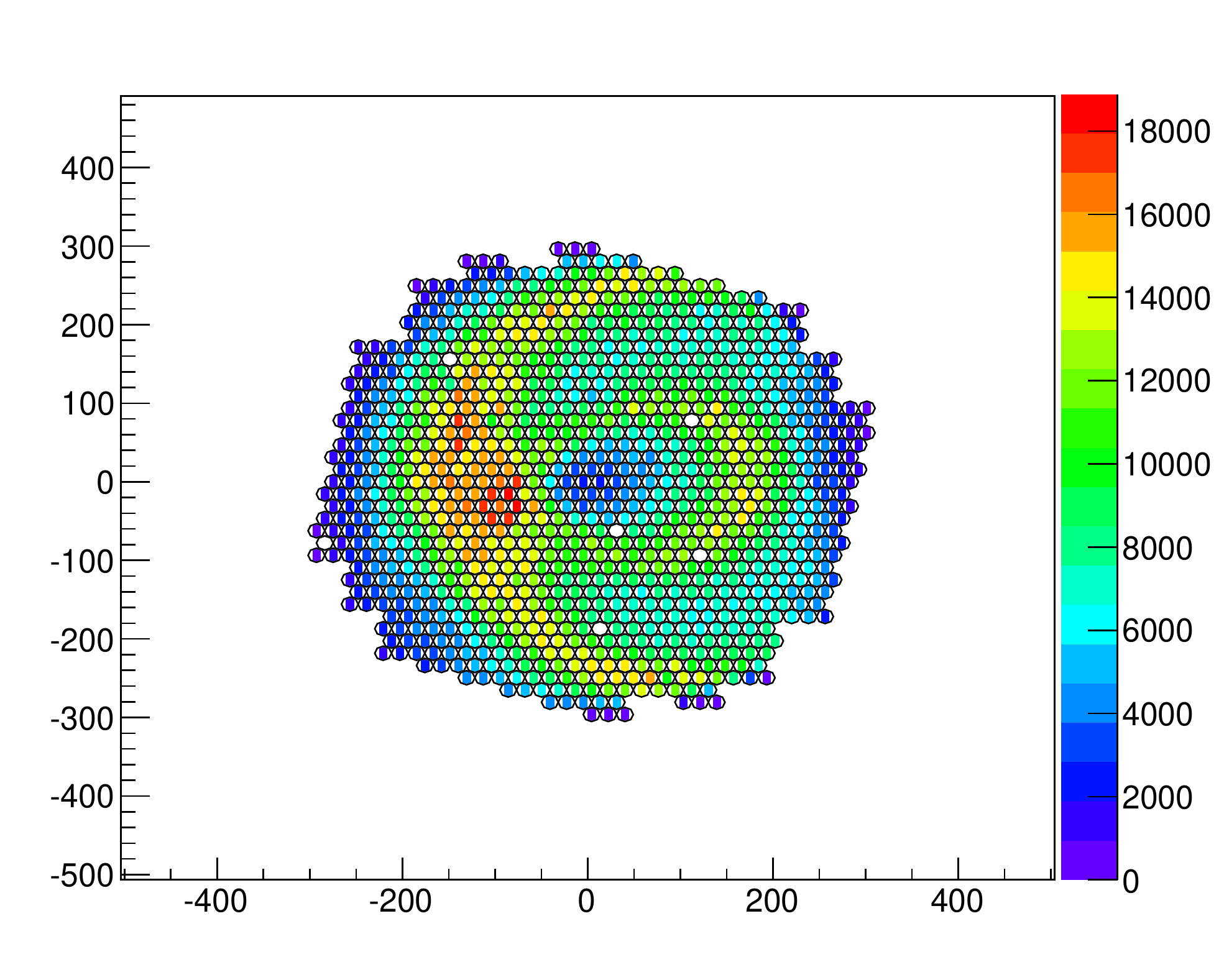}
\put(-60,0){x [mm]}
\put(-205,115){\rotatebox{90}{y [mm]}}
\end{center}
\end{minipage}
\caption{\label{fig:RICH-Picture3} {Cherenkov light illumination on the RICH PMs. The PM arrays on the left (negative X) and right (positive X) side of the beam are shown, using a  local coordinate system (x, y) centred in the middle of each array.}}
\end{figure}

The RICH installation was completed by the summer of 2014.  The detector has been commissioned with data in 2014 and operated throughout the 2015 data taking.
\Fig{fig:RICH-Picture3} shows the illumination of Cherenkov light on the two RICH PMs arrays on the left and the right side of the RICH.
\begin{figure}[ht]
\begin{minipage}{0.5\linewidth}
\begin{center}
\includegraphics[width=1.\linewidth]{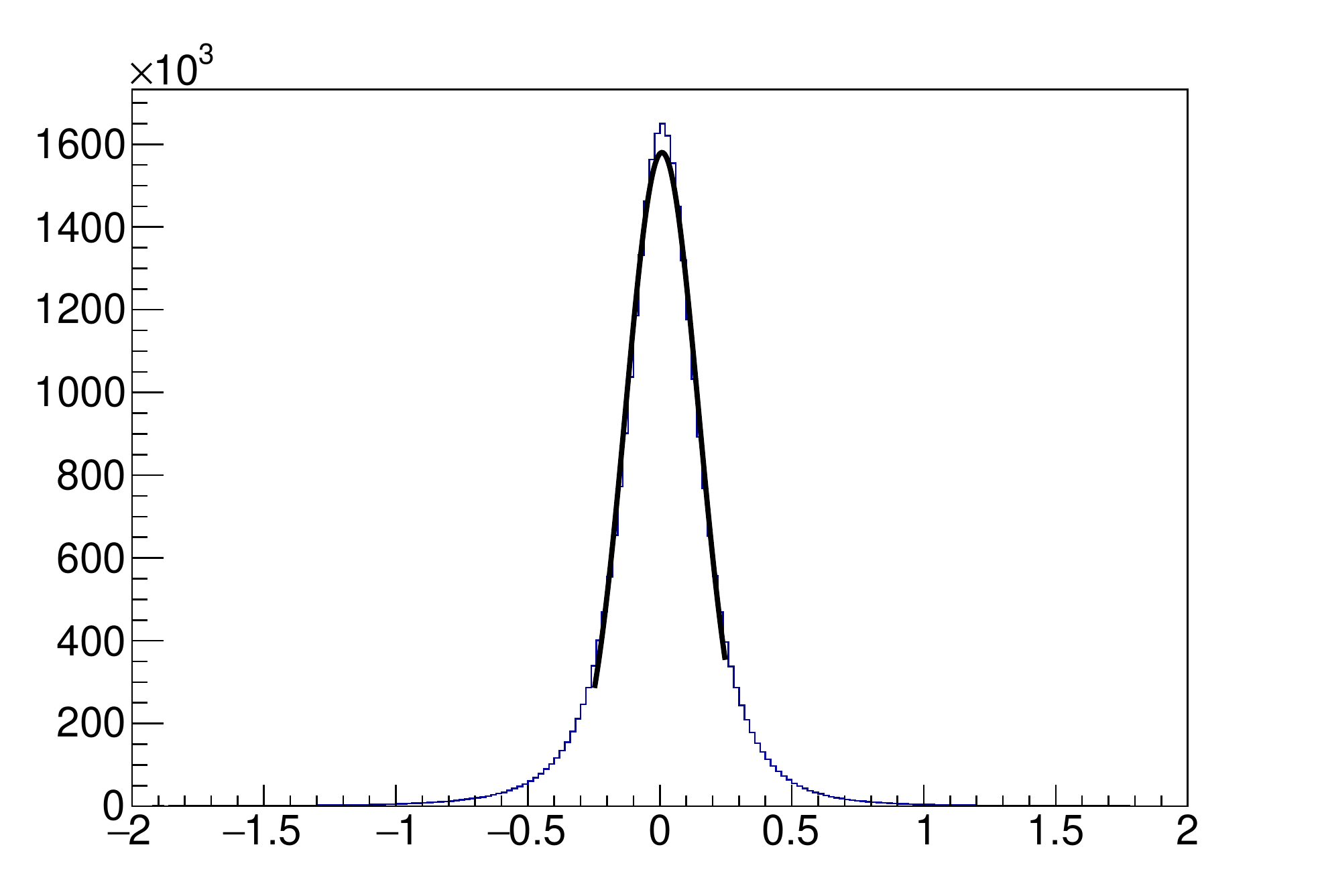}
\put(-90,-1){ ${\rm t}_{\rm Set 2} - {\rm t}_{\rm Set 1}$ [ns]}
\put(-80,95){$\sigma$ = 140 ps}
\end{center}
\end{minipage}
\begin{minipage}{0.5\linewidth}
\begin{center}
\includegraphics[width=1.\linewidth]{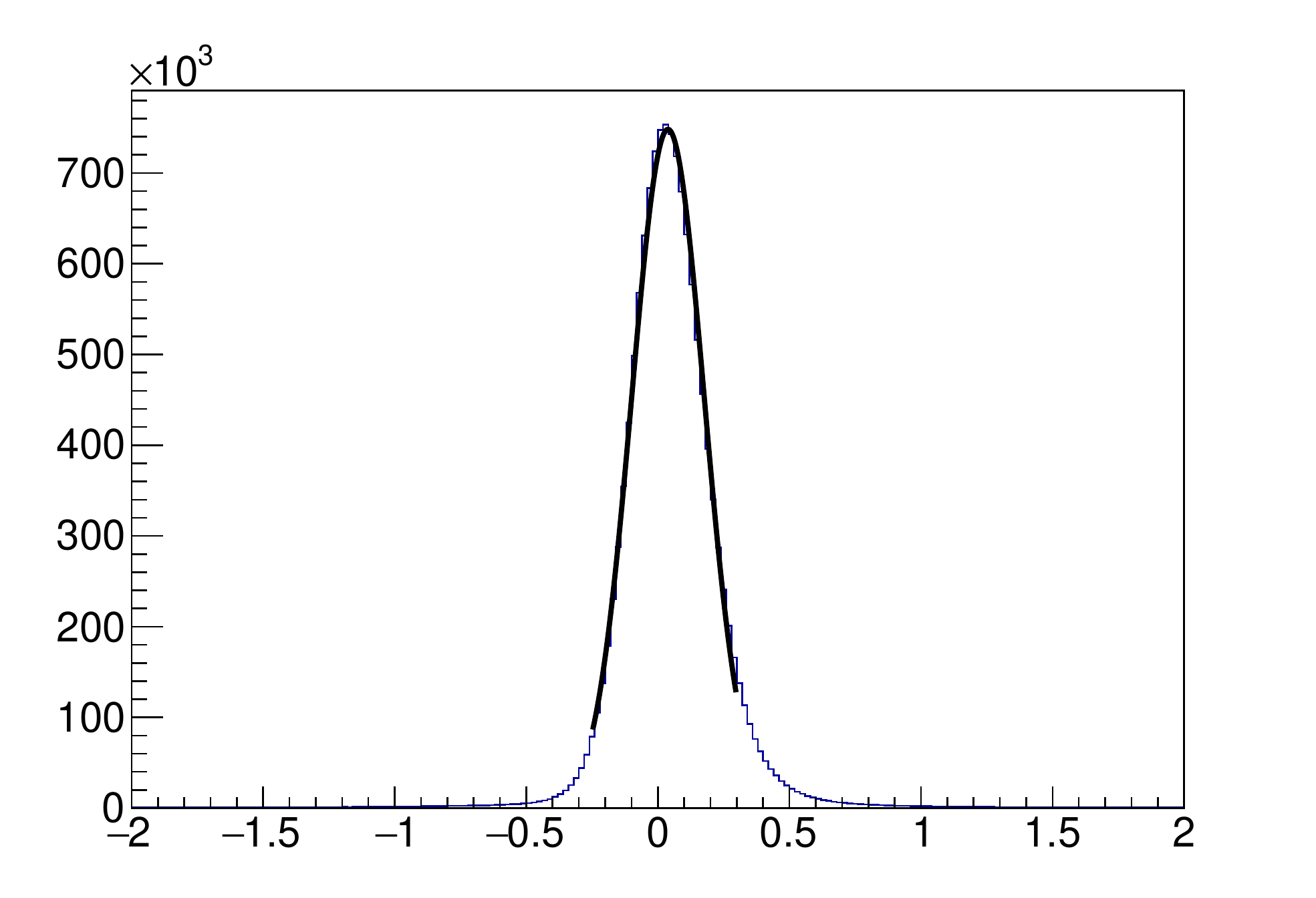}
\put(-101,-1){ ${\rm t}_{\rm RICH} - {\rm t}_{\rm KTAG}$ [ns]}
\put(-80,95){$\sigma$ = 140 ps}
\end{center}
\end{minipage}
\caption{\label{fig:RICH-Picture4}{Left: Intrinsic time resolution of the Cherenkov photons.
The detected photons of one Cherenkov ring are divided into two groups and the time difference is plotted.   
The width of the Gaussian fit is 140 ps. The time resolution of the full ring is one half of this width.
Right:  Time difference between the average time of a Cherenkov ring and the KTAG time. The width of the Gaussian fit is 140 ps. }}
\end{figure}

\Fig{fig:RICH-Picture4} demonstrates the RICH time resolution. On the left, the hits 
in a Cherenkov ring are divided into two halves and the difference of the average times is plotted yielding an intrinsic RICH event time resolution of about 70 ps. 
\Fig{fig:RICH-Picture4}-right shows the difference of the RICH and KTAG 
event times, which has a  width of 140 ps.  This resolution is about 40 ps larger than expected from the intrinsic  resolutions, but this increase could be due to remaining systematic uncertainties in the RICH time offsets.  

\begin{figure}[ht]
\begin{minipage}{0.5\linewidth}
\begin{center}
\includegraphics[width=1.0\linewidth]{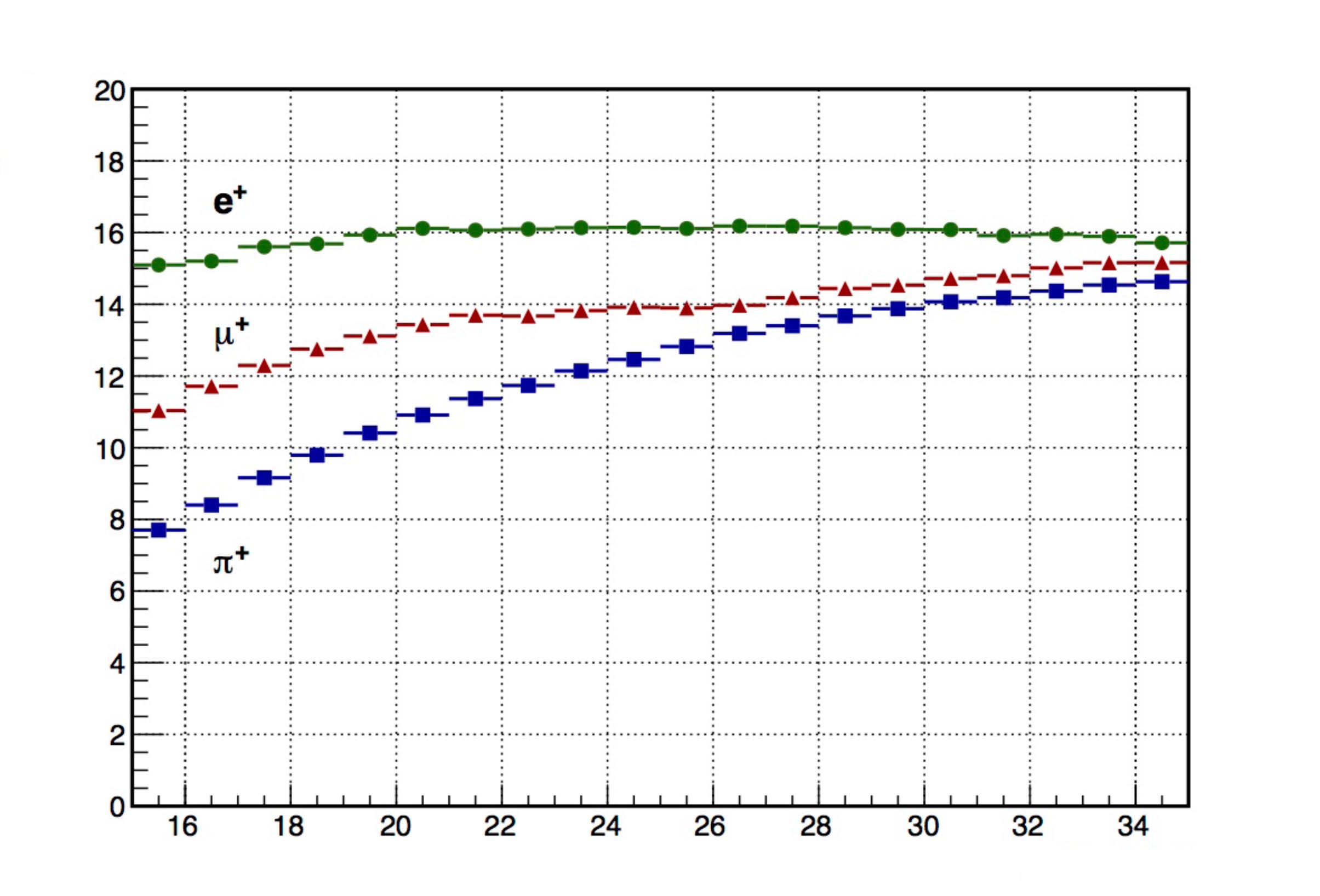}
\put(-65,0){p [$\GEVc$]}
\put(-212,100){\rotatebox{90}{$ \langle {\rm N}_{\rm hits} \rangle$}}
\end{center}
 \end{minipage}
 \begin{minipage}{0.5\linewidth}
\begin{center}
\includegraphics[width=1.06\linewidth]{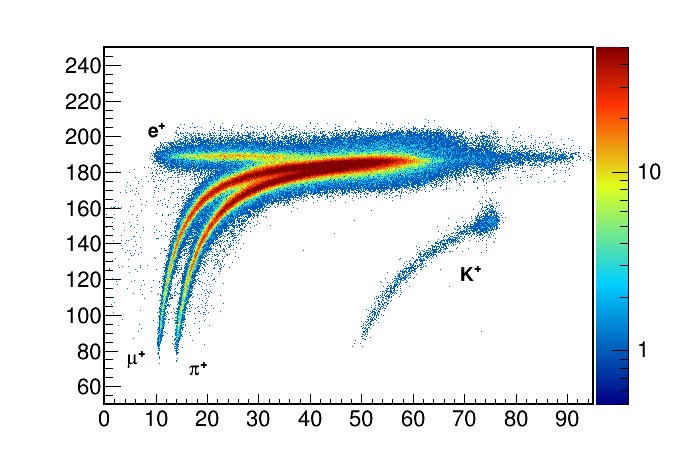}
\put(-75,0){p [$\GEVc$]}
\put(-222,60){\rotatebox{90}{Ring radius [mm]}}
\end{center}
 \end{minipage}
\caption{\label{fig:RICH-Picture5} {Left: Number of hits
per Cherenkov ring as a function of particle momentum; electrons, muons and 
charged pions were selected using spectrometer and calorimetric information. The number of hits in the electron case is not exactly constant due to the limited acceptance of the PMs. Right: Cherenkov ring radius as a function of particle momentum; electrons, muons, charged pions and scattered beam kaons can be clearly seen. 
Particles with momentum higher than 75 $\GEVc$ are due to halo muons.
}}
\end{figure}

In order to illustrate the performance of the RICH detector, samples of charged pions, muons and electrons were selected using calorimetric and spectrometer information. In \Fig{fig:RICH-Picture5}-left the number of hits per Cherenkov ring as a function of particle momentum (measured by the spectrometer) is shown for electrons, muons and charged pions. \Fig{fig:RICH-Picture5}-right shows the Cherenkov ring radius as a function of momentum with no selection on the particle type: electrons, muons, charged pions and scattered charged kaons can be clearly seen.

 Preliminary performance of the pion-muon separation is shown in \Sec{sec:valid}.

%% file: Sec10-CHOD_v2.tex
The NA62 setup includes a scintillator detector system called the charged particle hodoscopes. 
They cover the lateral acceptance downstream of the RICH and upstream of the LKr calorimeter defined by the LAV12 detector inner radius (1070~mm) and the IRC detector outer radius (145~mm). Their main function is to provide an 
 input for the L0 trigger (\Sec{sssec:TEL62L0}) when at least one charged particle crosses the annulus with the dimensions defined above. They are exposed to a nominal charged particle rate of 13~MHz, mainly due to beam kaon decays and, to a lesser extent, beam pion decays and muon halo.

The system consists of the NA48-CHOD detector from the former kaon experiment NA48~\cite{NA48:2007} and the newly constructed CHOD detector optimized for the high intensity conditions of NA62.
The NA48-CHOD and the CHOD are located, respectively, downstream  and upstream of the LAV12 detector, about 700~mm apart in the longitudinal direction; they are operated simultaneously and independently.
The NA48-CHOD exploits a high granularity design based on coincidence of signals in two planes of vertical and horizontal scintillator slabs, suitable to provide track timing with 200~ps precision. The detector was refurbished in 2012, and has been operated  since 2014.
The CHOD, with a single plane of scintillator tiles and a finer tile configuration in the high occupancy area near the beam axis, was installed after the 2015 run and commissioned in 2016. 

\subsection{NA48 Hodoscope (NA48-CHOD)}\label{ssec:NA48-CHOD}
\input{Sec10-NA48CHOD_v2.tex}

\subsection{Hodoscope (CHOD)}\label{ssec:CHOD}
\input{Sec10-NA62CHOD_v3.tex}

%% file: Sec10-NA48CHOD_v2.tex
\begin{figure}[ht]
\begin{center}
\includegraphics[scale=0.5]{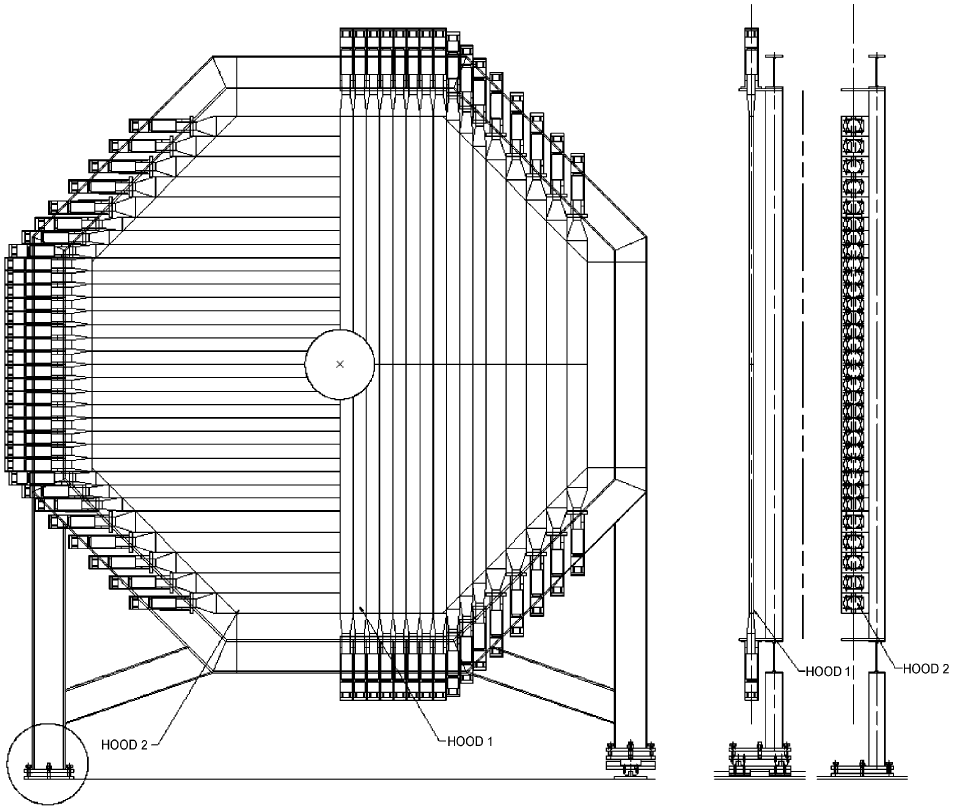}
\end{center}
\vspace{-6mm}
\caption{\label{fig:CHOD-Picture1} {Drawing of the NA48-CHOD detector. Half sections of the horizontal and vertical planes are shown, with the beam traversing at the centre of the detector.}}
\end{figure}

The NA48-CHOD consists of two consecutive planes made of 64 vertical and 64 horizontal BC408 plastic scintillator slabs of 20~mm ($0.10~X_0$) thickness. 
Each slab is read out at one end by a PHOTONIS\textsuperscript{TM} XP2262B photomultiplier through a fishtail-shaped Plexiglas light guide. 
The BC408 scintillator was chosen for its fast decay time (2.1~ns) and the large attenuation length (bulk value of 3800~mm). The fast 12-stage 51 mm XP2262 tube, with a bi-alkali photocathode, was chosen for its spectral characteristics matching the BC408 emission wavelength distribution, high gain (above $10^7$ for the NA48-CHOD HV operating values) and good timing properties (2.3~ns rise time and transit time spread of less than 0.5~ns). 
The 128 counters are assembled into 4 quadrants of 16 slabs in each plane. Slab lengths vary from 1210~mm (inner counters) to 600~mm (outer counters) 
forming an octagon  of 1210~mm apothem.
The slab widths are 65~mm in the central region close to the beam, where the particle flux is higher, and 99~mm in the outer region. The radius of the central hole crossed by the beam pipe is 128~mm.
The layout of the NA48-CHOD with its mechanical support is shown in \Fig{fig:CHOD-Picture1}.

Two independent time measurements are provided by the NA48-CHOD for each charged particle crossing a vertical and a horizontal slab, which reduces the 
possible tails in the event time distribution due to out-of-time accidental activity, interactions in the upstream material and back-scattering from the LKr calorimeter.

The front-end electronics consists of Time-over-Threshold (ToT) discriminators with double threshold setting availability, originally developed for the Large Angle Veto system (\Sec{ssec:lav-fee}). Four LAV front-end modules are used: each module processes 
16 analogue signals from a vertical quadrant and 16 signals from the corresponding horizontal quadrant, and produces 64 digitized output signals (32 for each threshold value).

The discrimination threshold values are chosen as a compromise between time resolution and noise suppression. 
The two-threshold feature of the front-end electronics is efficient in correcting LAV signal times over a wide amplitude range. However it does not provide significant improvement to the NA48-CHOD hit time resolution due to the relatively constant amplitude of the CHOD signals and is then not exploited.
The signal amplitude of a minimum ionizing particle has been adjusted to about 200~mV and a single value of the discrimination threshold has been set at 60 mV for all the channels.

The NA48-CHOD uses the common TDAQ system developed to read out most of the NA62 subsystems, based on TEL62 motherboards and TDCB mezzanines with HPTDC processors (\Sec{ssec:TEL62}). Both leading and trailing pulse edges are recorded in the TDCB and the corresponding times are sent to the TEL62 processors, which buffer data and produce trigger 
input in parallel with the readout. 

The nominal hit rate summed over the scintillator slabs, as obtained from simulation, is 35 MHz for each plane, well above the 13 MHz expected from  charged-particle events. The difference is due to high  multiplicity events  from interactions in the material of upstream detectors and back-scattering from the LKr calorimeter. 
This result has been confirmed by measurements performed  using 2014 data
extrapolated to the nominal beam intensity.

To keep input rates into HPTDCs and TDCBs within safe values with respect to the hardware  limits, 
a rarified scheme of the distribution of signals to TDCB channels is implemented. This arrangement is achieved by a splitter board 
between the LAV front-end modules and the TEL62 to match the 64 output signals from the front-end boards with the 128 input channels of the TDCB, thus reducing data feed into TDCB by a factor of two.  
In addition, thanks to a feature of the TEL62 board, a further 50\% reduction of input data is obtained by masking one of the two sets of output signals from the front-end modules.  
In this way, each TDCB processes 32 digitized pulses, i.e. 1 every 4 input channels, and can handle data at rates up to twice the nominal value. In total, one TEL62 board fully equipped with four TDCBs, one per LAV front-end module, is used.

The NA48-CHOD is used to provide L0 trigger signals and reference times for events with charged particles in the final state (\Sec{ssec:L0}). A loose trigger selection requires at least two hits within an adjustable time window. A more refined trigger selection could be provided by the coincidence between the signals of one vertical and one horizontal counter of consecutive quadrants within an adjustable time window. This coincidence allows particle times to be corrected for the hit impact point position on the detector slabs. 

The NA48-CHOD online time resolution has been estimated from the 2015 data.  
A Gaussian fit of the difference between the NA48-CHOD 
time and a reference time given by the RICH detector has a width of about 2~ns at nominal beam intensity, with negligible contribution from the reference time resolution. 

\begin{figure}[ht]
\begin{minipage}{0.5\linewidth}
\begin{center}
\includegraphics[width=1.\linewidth]{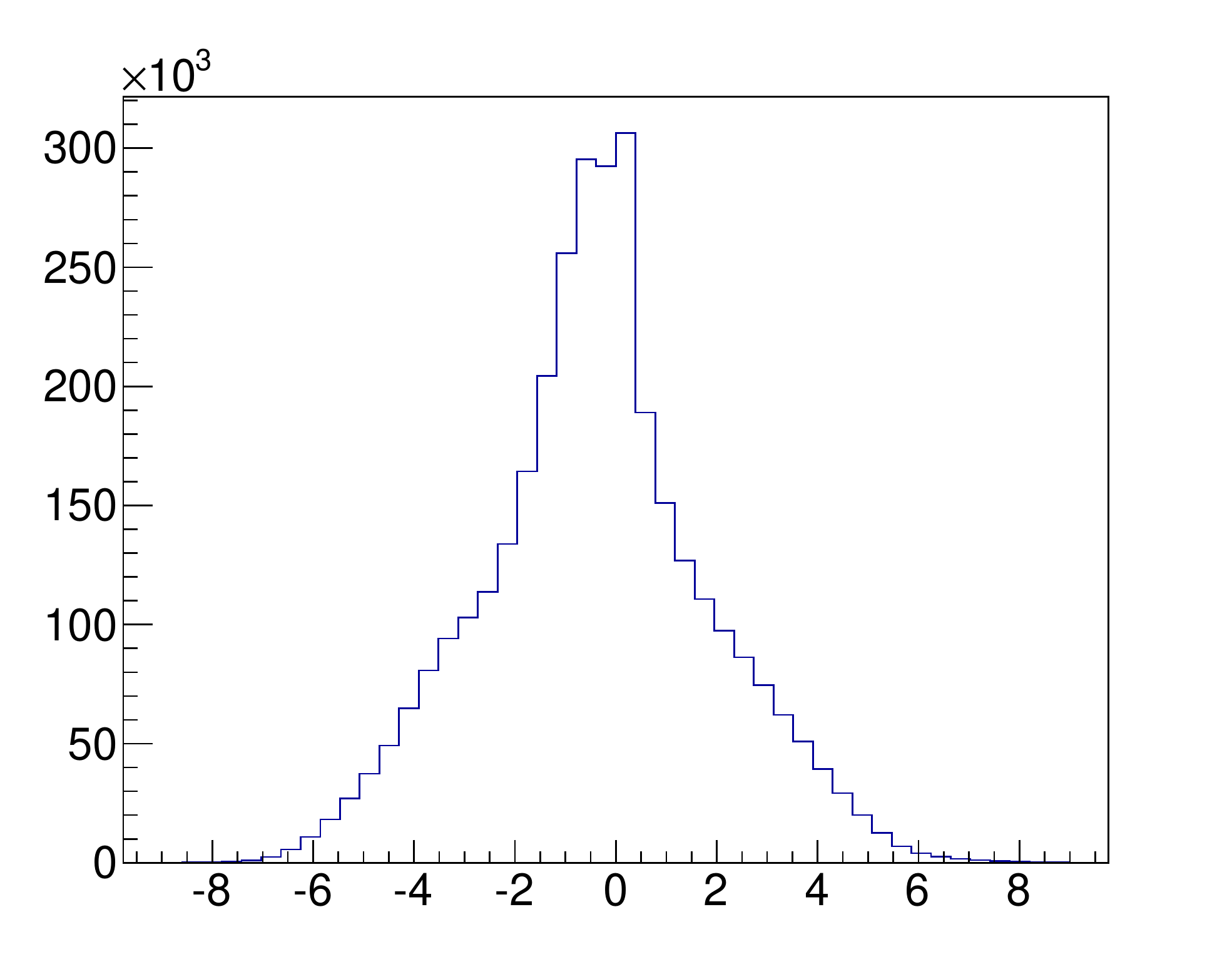}
\end{center}
\end{minipage}
\put(-180,50){\bf  \Large (a)}
\put(-90,50){rms = 2.5 ns}
\put(-70,-90){${\rm t}_{\rm V} - {\rm t}_{\rm H}$ [ns]}
\begin{minipage}{0.5\linewidth}
\begin{center}
\includegraphics[width=1.\linewidth]{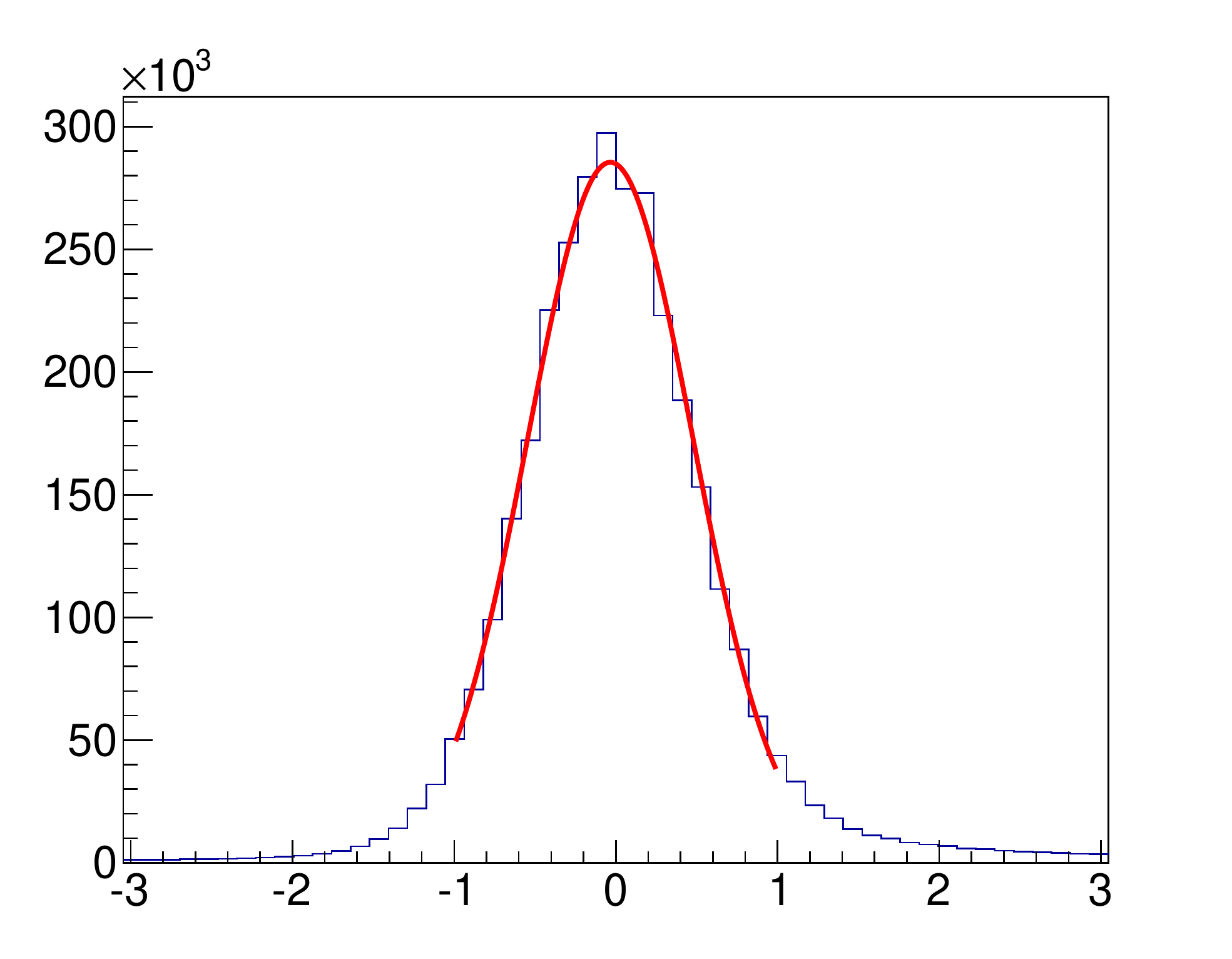}
\end{center}
\end{minipage}
\put(-85,50){$\sigma_{\rm t}$ = 0.51 ns}
\put(-180,50){\bf  \Large (b)}
\put(-70,-90){${\rm t}_{\rm V} - {\rm t}_{\rm H}$ [ns]}
\\
\begin{minipage}{0.5\linewidth}
\begin{center}
\includegraphics[width=1.\linewidth]{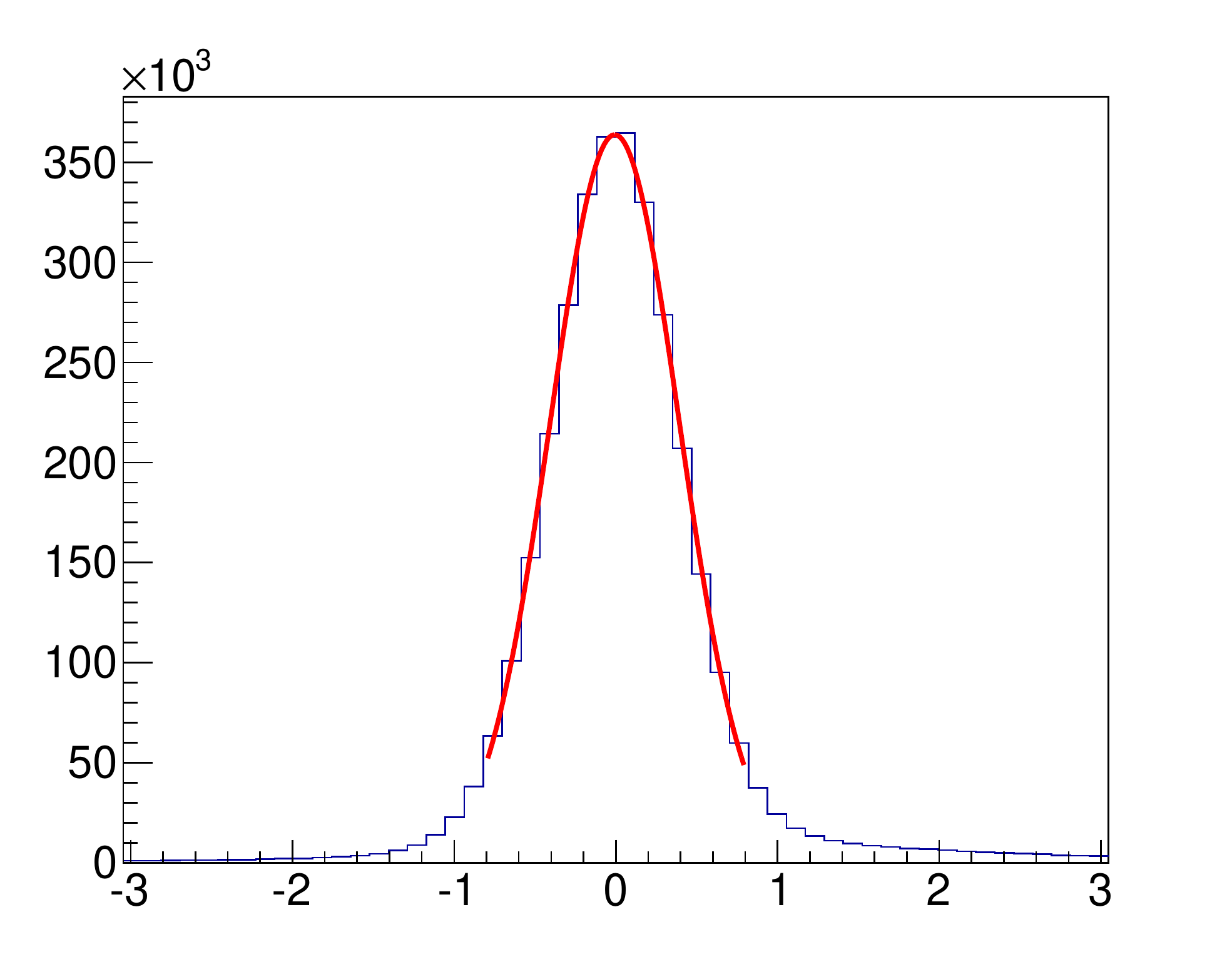}
\end{center}
\end{minipage}
\put(-180,50){\bf  \Large (c)}
\put(-85,50){$\sigma_{\rm t}$ = 0.40 ns}
\put(-70,-90){${\rm t}_{\rm V} - {\rm t}_{\rm H}$ [ns]}
\begin{minipage}{0.5\linewidth}
\begin{center}
\includegraphics[width=01.\linewidth]{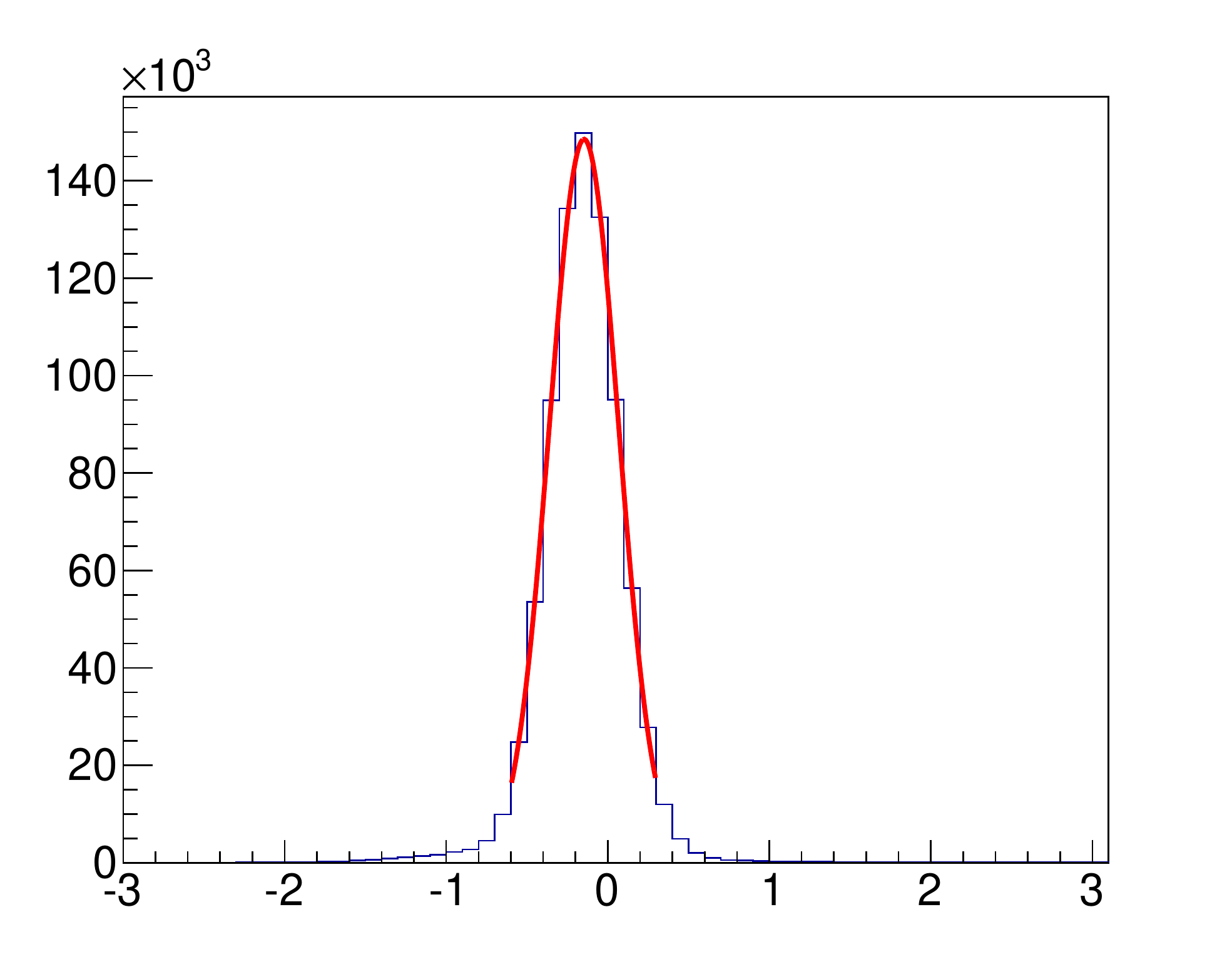}
\end{center}
\end{minipage}
\put(-85,50){$\sigma_{\rm t}$ = 0.21 ns}
\put(-180,50){\bf  \Large (d)}
\put(-105,-90){${\rm t}_{\rm CHOD} - {\rm t}_{\rm KTAG}$ [ns]}
\caption{\label{fig:CHOD-Picture2}  
(a) Difference of raw times of a  vertical  and a horizontal scintillator slab of the NA48-CHOD hit by a track;  
(b) difference after applying time corrections for the track impact point on the slabs; 
(c) difference after applying slewing time corrections;  
(d) Time difference between the NA48-CHOD candidate and KTAG kaon signal. 
All distributions have been obtained from data recorded at 1\% nominal beam intensity in 2015.}
\end{figure}
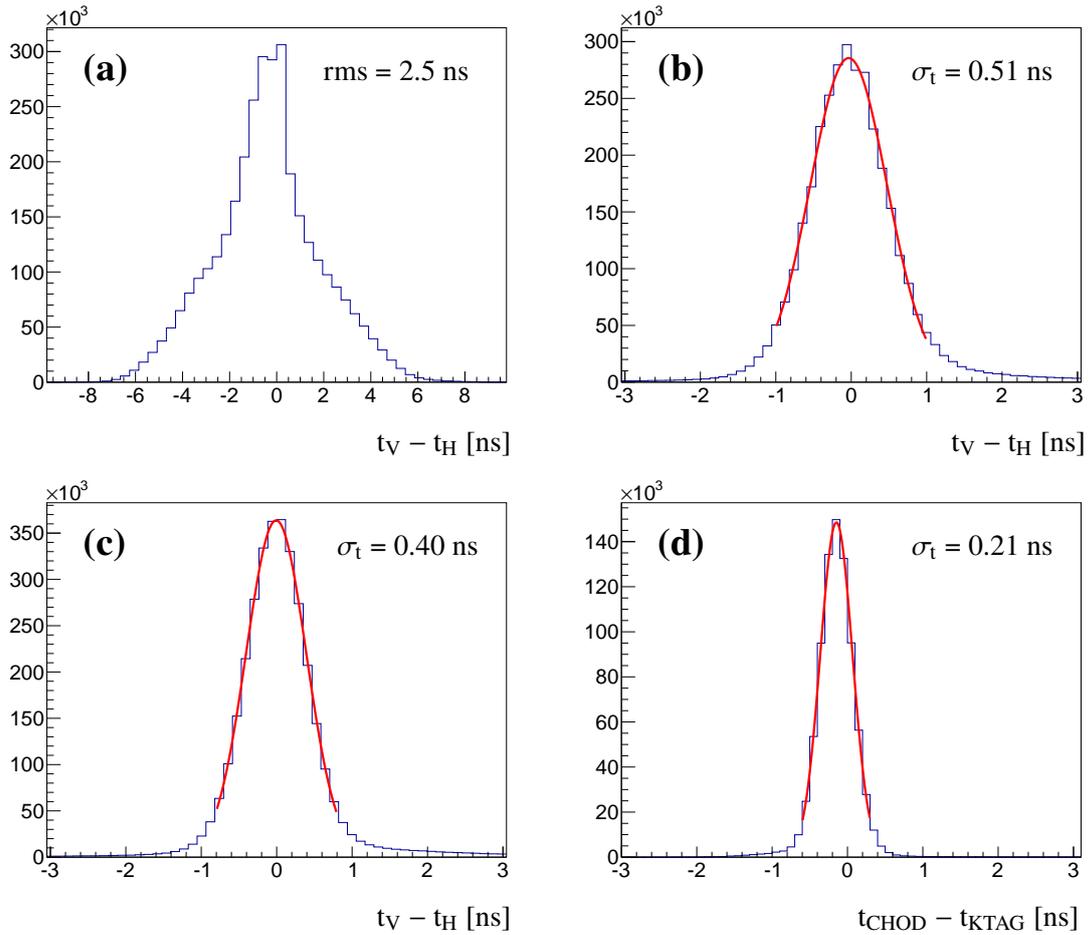

The distribution of the hit time difference between vertical and horizontal slabs is used to quantify the effect of each time correction.  The distribution of the raw hit time difference has a width (rms) of about 2.5~ns (\Fig{fig:CHOD-Picture2}-a).  A Gaussian fit to the distribution obtained after correction for the impact point position of the coincidence of one horizontal and one vertical slab gives a reduced rms width of 0.51~ns (\Fig{fig:CHOD-Picture2}-b).  After a further  hit time correction for slewing effects exploiting the time-over-threshold information, the  width decreases to 0.40~ns (\Fig{fig:CHOD-Picture2}-c).
This value results in a single counter time resolution better than 0.30~ns. The NA48-CHOD track signal is defined as the coincidence of hits in one vertical and one horizontal counter within a predefined time window. The signal time is the mean time of two hits.
The distribution of the time difference between a NA48-CHOD track signal and the KTAG beam kaon candidate is shown in \Fig{fig:CHOD-Picture2}-d. 
The width of the Gaussian fit superimposed to the data is 0.21~ns, the time resolution of the KTAG being 0.07~ns (\Sec{sec:KTAG}). 
The values of the time resolution obtained at 1\% nominal beam intensity are confirmed by data at higher beam intensities (up to the nominal one).

A detector inefficiency at the per mille level  has been measured using 2015 data by extrapolating tracks reconstructed by the spectrometer to the NA48-CHOD planes. This inefficiency is due to signals with amplitudes below discrimination threshold and due to tracks impinging into counter edges.

These results are consistent with the performance measured in the NA48 experiment \cite{NA48:2007} and match the NA62 requirements.

%% file: Sec10-NA62CHOD_v3.tex
\begin{figure}[ht]
\begin{center}
\includegraphics[width=0.85\textwidth]{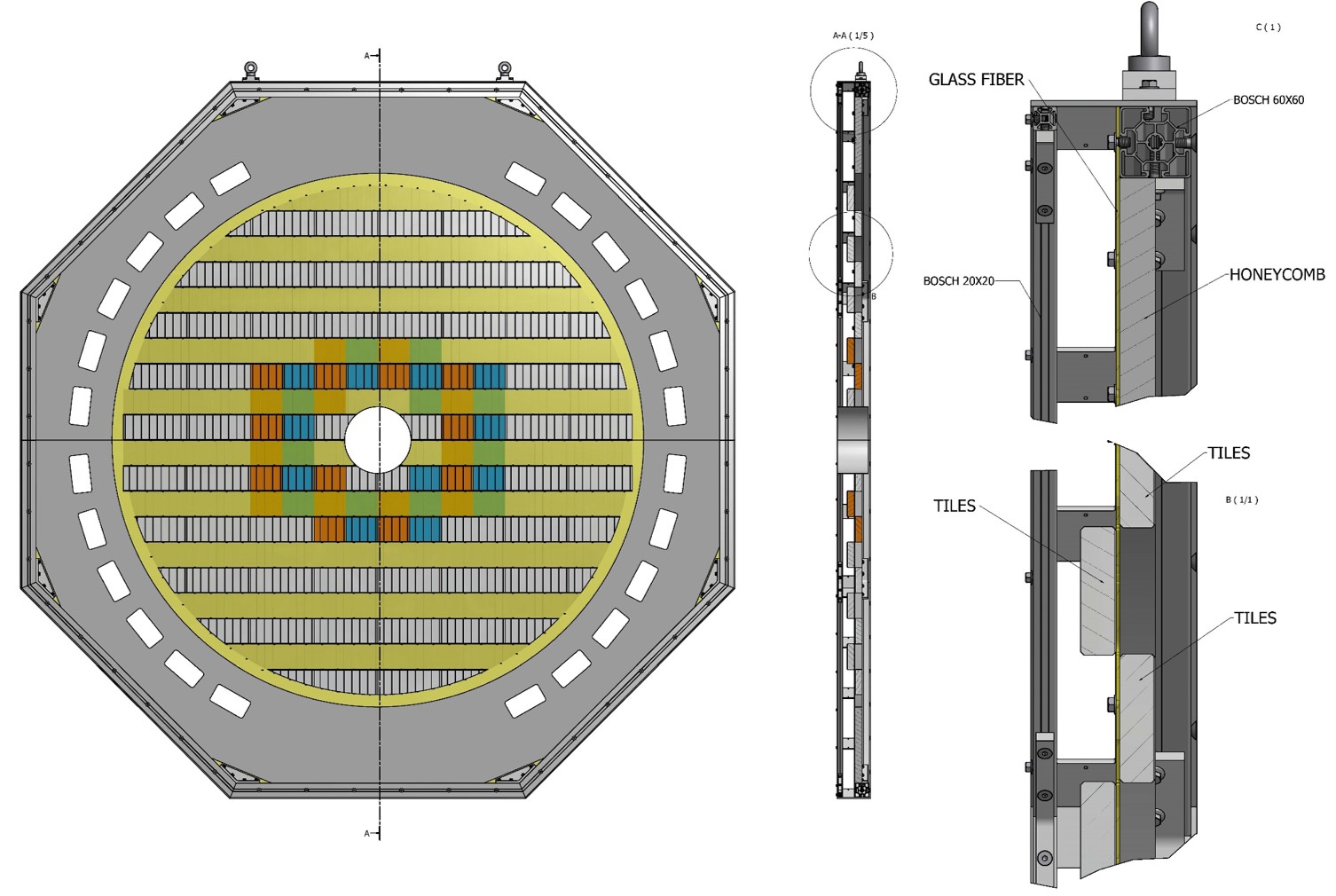}
\end{center}
\vspace{-7mm}
\caption{\label{fig:CHOD-Picture5} {The CHOD detector is mounted on the front face of LAV12 (left). The 152 scintillator tiles are mounted front and back of thin G10 support panel (bottom right). The structure is stiffened at the periphery with honeycomb and aluminium construction profiles (top right).}}
\end{figure}

The CHOD active area is an array of 152 plastic scintillator tiles of 30~mm thickness covering  
an annulus with inner (outer) radii of 140~mm (1070~mm). Subdivision of the acceptance surface into tiles leads to an optimized distribution of hit rates, and different groups of tiles can be selected to contribute to specific trigger requirements. The tiles are 108~mm high (except for 12 tiles near the external edge); laterally most tiles are either 134~mm or 268~mm wide. The tile centres are spaced vertically by 107~mm, resulting in a 1~mm overlap. This is possible by placing rows of tiles alternatively on the upstream and downstream sides of a 3~mm thick central support foil (G10 with 35~$\mu$m Cu lining on both sides) suitably perforated for the passage of 4.4~mm wide, 0.25~mm thick,   
steel panduits (two per tile) to secure the tiles firmly in their positions (\Fig{fig:CHOD-Picture5}). The total thickness of the detector in the active area is $0.13~X_0$.

The overall lateral dimensions of the CHOD are those of an octagonal box with 1550~mm apothem. The structure is divided horizontally into two  halves at the height of the beam. The thickness of the box, including the 0.5~mm thick foils which seal it longitudinally, is 140~mm. The structure is built with aluminum     
profiles with additional lateral 140~mm wide, 5~mm thick Al plates, with suitable openings for 20 mother boards located on the vertical and $45^\circ$ inclined sides of the octagon.

The scintillation light is collected and transmitted by 1~mm diameter Kuraray\textsuperscript{TM} Y11 S wavelength shifting fibres, and is detected by  $3\times 3~\rm{mm}^2$ SensL\textsuperscript{TM} SiPMs pairs mounted on the motherboards (each SiPM  hosts four fibres). The fibres are glued with 
optical epoxy in 1.5~mm deep and 1.1~mm wide grooves milled on the tiles; their lengths range from 135~cm to 200~cm. The odd and even numbered fibres of each tile are connected to different SiPM pairs. In total, there are 304 SiPM pairs and corresponding pre-amplifiers plugged on the mother boards in direct connection with the SiPMs. Their  
outputs can be chosen to correspond to either the individual SiPM pairs detecting the light from the two interleaved sets of fibres of each tile, or to their linear OR. The signals are shaped using constant fraction discriminators to improve the trigger time resolution, and read out by a 512-channel TEL62 board equipped with four TDCB mezzanines. The SiPMs can be accessed directly without opening the front or back of the octagonal box. Before mounting, each tile has been tested with cosmic rays, and it has been checked that the efficiency of the coincidence between the pre-amplifier outputs of the two sets of fibres exceeds 99\%. The expected time resolution of the CHOD signals is of $\cal O$(1~ns).

\begin{figure}
\begin{center}
\includegraphics[width=\textwidth]{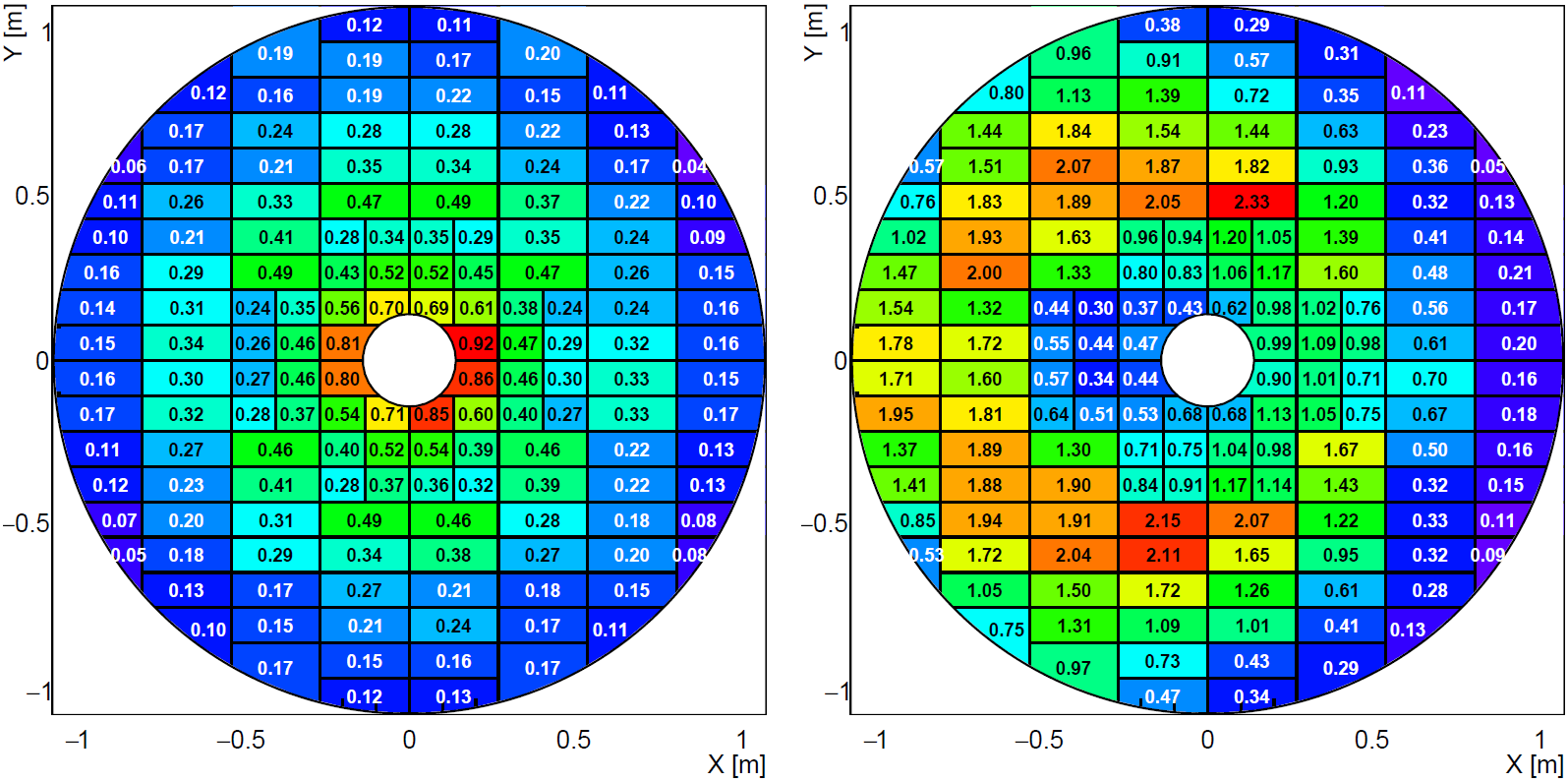}
\end{center}
\vspace{-7mm}
\caption{\label{fig:CHOD-Picture4} {Left: expected rates in CHOD tiles at nominal beam intensity (in MHz). Right: probability of detecting a signal in each CHOD tile for $\PNNP$ decays satisfying the signal selection conditions $105~{\rm m}< {\rm Z}_{\rm vertex}<165~{\rm m}$ and $15~\GEVc < p_\pi < 35~\GEVc$ in each tile (in percent). Both are calculated with MC simulations.
}}
\end{figure}

The nominal hit rate summed over the tiles obtained with Monte Carlo simulations is 45~MHz, which is higher than the nominal charged-particle rate of 13~MHz due to high multiplicity events produced by interactions of photons and beam pions upstream of the detector. Tile layout details and expected signal rates in tiles at nominal beam intensity are shown in~\Fig{fig:CHOD-Picture4}-left. 
The probability of detecting a signal in each CHOD tile for $\PNNP$ 
events satisfying the standard momentum and decay vertex position selection conditions is shown in~\Fig{fig:CHOD-Picture4}-right. The rate pattern suggests the possibility of optimising the trigger algorithms by masking specific tiles with high background rate and low signal acceptance.

%% file: Sec11-MUV12_v5.tex
The Muon Veto System (MUV)  complements the RICH (\Sec{sec:RICH}) and 
the LKr (\Sec{sec:PV}) detectors in the discrimination between muons and pions.
The MUV system consists of a 
hadron calorimeter comprising two detectors MUV1 and MUV2, an 80~cm filtering iron wall
and a fast Muon Veto detector (MUV3). The characteristics of these detectors are displayed in \Tab{tab:MUV-table1}.

 The energy released in the calorimeter  can be used as an input to the L0 trigger  to reject muons.
This trigger logic is complemented by a tight veto requirement on MUV3 signals. 

\begin{table}[!h]
\setlength{\tabcolsep}{3ex}
\caption{Characteristics of the Muon Veto system elements.}
\label{tab:MUV-table1}
\vspace{2ex}
\centering
\begin{tabular}{l >{\centering\arraybackslash}m{2.5cm} >{\centering\arraybackslash}m{2.cm} >{\centering\arraybackslash}m{1.5cm}>{\centering\arraybackslash}m{1.5cm} }
\hline\hline
Detector & Description &   Thickness  (interaction length)   & No. of strips per plane&  No. of readout  channels\\
\hline
MUV1  &  iron/scintillator & 4.1 & 44 &176\\
MUV2  &  iron/scintillator & 3.7  & 22 & 88\\
Muon filter  & 80 cm iron & 4.8 & -- & -- \\
MUV3  & scintillator tiles & -- & -- & $2 \times 148$ \\
\hline\hline
\end{tabular}
\end{table}

\subsection{Hadron calorimeter (MUV1, MUV2)}
The hadron calorimeter is a sampling calorimeter made from alternate layers of iron and scintillator
corresponding to about 8 interaction lengths. 
The calorimeter is divided into two independent detectors: 
the front detector (MUV1),  
newly built for the NA62 experiment with a fine transversal segmentation to better disentangle hadronic and electromagnetic shower components, and the back detector (MUV2) inherited from  the NA48 experiment \cite{NA48:2007}.  

\subsubsection{Design and construction}
The MUV1 detector consists of 24~layers of 26.8~mm thick SE35 steel plates, corresponding to a total thickness of 4.1~interaction lengths (including the scintillating material). The innermost
22~layers have outer dimensions of $2700  \times 2600$ mm$^2$ while the first and the last layer 
are  larger to cover the readout fibres and to serve as
support of the whole structure. The distance between consecutive iron layers is 12~mm.  

The iron plates are interleaved with 23~layers of scintillator strips of 9~mm thickness and 60~mm width (54~mm width for the four~inner strips which end at the beam pipe). 
Consecutive layers of scintillators are alternately aligned in the horizontal and vertical direction, resulting in  12~layers with horizontal and  11~layers with vertical strip direction.
Most of the strips have a length of 2620~mm and span the whole transverse extension of the detector. 
The four outer strips  are  shorter by up to 240~mm to accommodate the support
structure. The six horizontal and eight vertical  strips at or close to the beam pipe are split in
two strips of half length to accommodate the pipe and the high particle rate close to the beam, respectively. 

The strip width of 60~mm was chosen to optimize the muon-pion separation.
Simulation studies showed that a smaller strip width of 40~mm would increase the muon rejection only on a percent level.
The scintillating strips were produced 
by melting polystyrene (Styron 143E)  with additions of p-terphenyl and POPOP in a vacuum of $10^{-4}$~bar at about 250~$^\circ$C. All
strips were diamond polished and wrapped in aluminized mylar foils.

The  strips are read out by wave-length-shifting (WLS)
scintillating fibres connected to PMs. 
Two different fibre types are used (about 30\% GC~Technology BCF-92\textsuperscript{TM} and
70\%  Kuraray Y-11\textsuperscript{TM}), both with 1.2~mm diameter and
multi-cladded. Each scintillating strip has two grooves in its longitudinal direction separated by 30~mm,
which contain the WLS fibres and optical glue. Except for the shorter central
strips, which are read out at one side only,\footnote{For theses strips the fibres on the non-readout side 
are terminated with reflective foil.} the fibres extend on each side of the strips by about 90~cm 
over the strip end and are routed to 1-inch Hamamatsu R6095S\textsuperscript{TM} PMs. All finished strips  
underwent a selection procedure and only those  providing sufficient and uniform light yield were 
installed. Each  PM receives all fibres 
from similarly aligned strips having the same transverse position,
i.e.\ 24~fibres for 12~consecutive horizontal and 22~fibres for 11~consecutive
vertical strips, respectively. The  detector in total has 176~channels. 
The average signal from a minimum ionizing particle 
crossing the whole detector in the longitudinal direction was measured to vary from  35 to 50 
detected photoelectrons in one PM, depending on the impact point of the particle on the front face.

The MUV2 detector is the refurbished front module of the NA48 hadron calorimeter. 
Similarly to the MUV1 detector, it is built as a sandwich calorimeter and consists of 24~iron plates of
25~mm thickness, each followed by a layer of ELJEN NE~110\textsuperscript{TM}  plastic scintillators. 

Each scintillator plane 
consists of 44~scintillating strips with each strip spanning half the calorimeter. The strips are 1300~mm  long, 119~mm wide, and 4.5~mm  thick. The two central strips of
each half-plane are shaped on one end to enclose  
the central hole for the beam pipe.  The strips are alternately aligned in the horizontal and vertical directions in consecutive planes.
As in the MUV1 detector, the strips with identical alignment are coupled
to the same PM but in this case using Plexiglas light guides
read out by 3-inch Electron Tubes 9265KA\textsuperscript{TM}  PMs. 
 In total the MUV2 has  88~readout channels.

\subsubsection{Readout and calibration}
The  MUV1 and MUV2 detectors are read out using FADCs and CREAM boards developed for the LKr readout (\Sec{ssec:Cal-readout}).
The fast pulses from the photomultipliers are shaped to differential LKr-like pulse shapes by active 
NIM  modules originally used for the NA48-HAC readout \cite{NA48:2007}.  Full  details about the CREAM readout and the L0  hardware trigger  using the calorimeter  information can be found in \Sec{sec:TDAQ}.

The required hadron calorimeter  performance can be achieved only after precise calibration of the detector response.
The calibration procedure consists of four steps: 
\begin{enumerate}
\item{\bf PM gain equalization.}
The PM gains are adjusted by changing the applied high voltage (HV). In special runs with a muon-based trigger, HV scans for each channel are performed to obtain an operating HV value corresponding to a charge peak of 1500~pC  (MUV1) and 1000~pC (MUV2)  for minimum ionizing particles.
The charge reference values are chosen as  a compromise between the dynamic range of the CREAM boards and the separation of signal from background. 

\item{\bf Light attenuation and propagation in scintillators.}
The correction for the impact point of the particle plays a major role in the calorimeter energy and time resolution.
Dedicated muon runs (\Sec{ssec:samples}) are used to precisely determine the signal attenuation and delay as a function of the impact point.
Tracks reconstructed in the STRAW spectrometer are 
matched with the NA48-CHOD (used as time reference) and the calorimeters.
The distribution of the charge collected as a function of the position is parameterized with a double exponential function to take into account reflections inside the scintillators. 
Following a similar procedure,  the time differences between the MUV1(MUV2)  channels and the NA48-CHOD signals are parameterized  with a power function.  
\Fig{fig:MuResult} shows the effect of the impact point corrections  in muon events.

\begin{figure}[!h]
\centering
\includegraphics[scale=0.85]{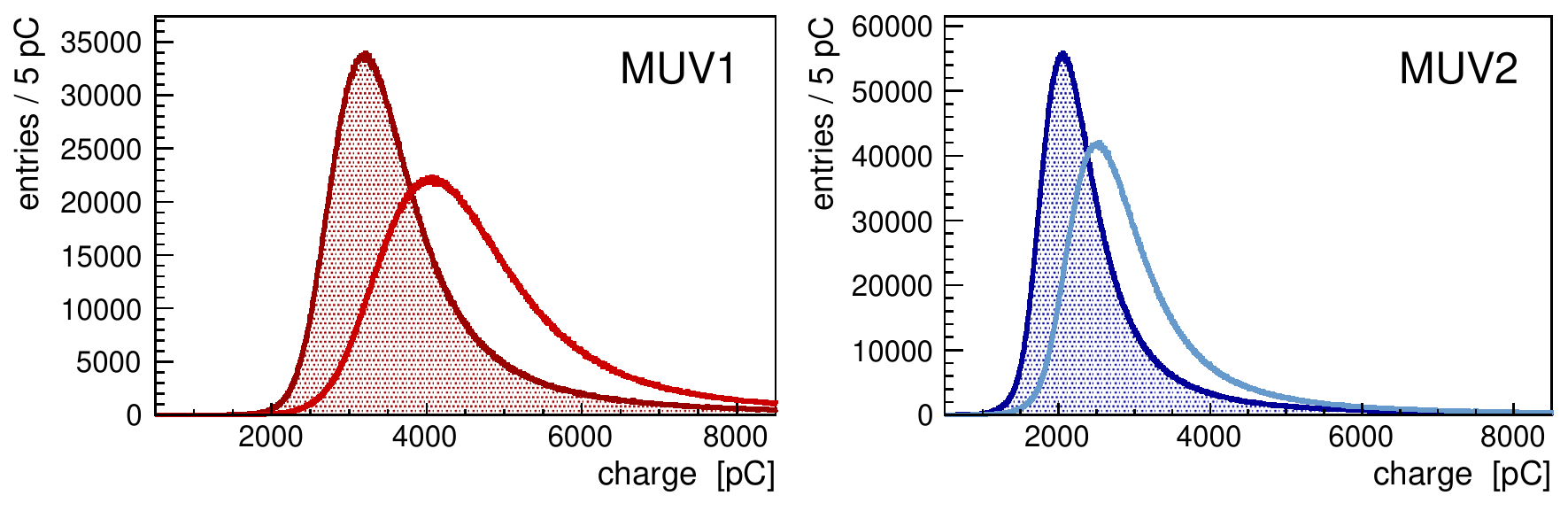}
\caption{\label{fig:MuResult}Charge distributions in the MUV1 (left) and MUV2 (right) detectors when crossed by muons. The raw data distributions of the total collected charge (lines), i.e. the sum of the collected charges of the vertical and horizontal strips, and the distributions after the impact point correction (shaded) are shown.}
\end{figure}
\item{\bf Energy Scale.}
The absolute energy scale of the calorimeter is determined independently for each detector by fitting the collected charge distribution using a Landau function convoluted with a Gaussian.  
The Landau distribution describes the energy deposited by minimum ionizing particles, while the Gaussian term parameterizes the smearing effect from the resolution. The energy scale is derived by dividing the value at the peak of the energy deposition from simulated data by the most probable charge value of the fitted Landau distribution.
For the 2015 data,  after fixing the charge equalization values and correcting for the effect of the impact position,
 the energy scale factors were estimated to be  317 keV/pC for MUV1 and 450 keV/pC for MUV2, respectively.

\item{\bf Pion Calibration.}
The two calorimeters have slightly different structure and therefore different responses to impinging pions.
In particular the sampling fraction is almost twice larger for MUV1  than for MUV2. 
As a consequence,  the two detectors are calibrated separately. 
The calibration is derived from standalone Monte Carlo simulations of the MUV1 and MUV2 detectors.
The ratio between the energy deposited in the MUV1,2 detectors (E$_{\rm visible}$) and the pion momentum is studied as a function of the momentum as shown in \Fig{fig:EnVis}.

\begin{figure}[!h]
\centering
\includegraphics[scale=0.85]{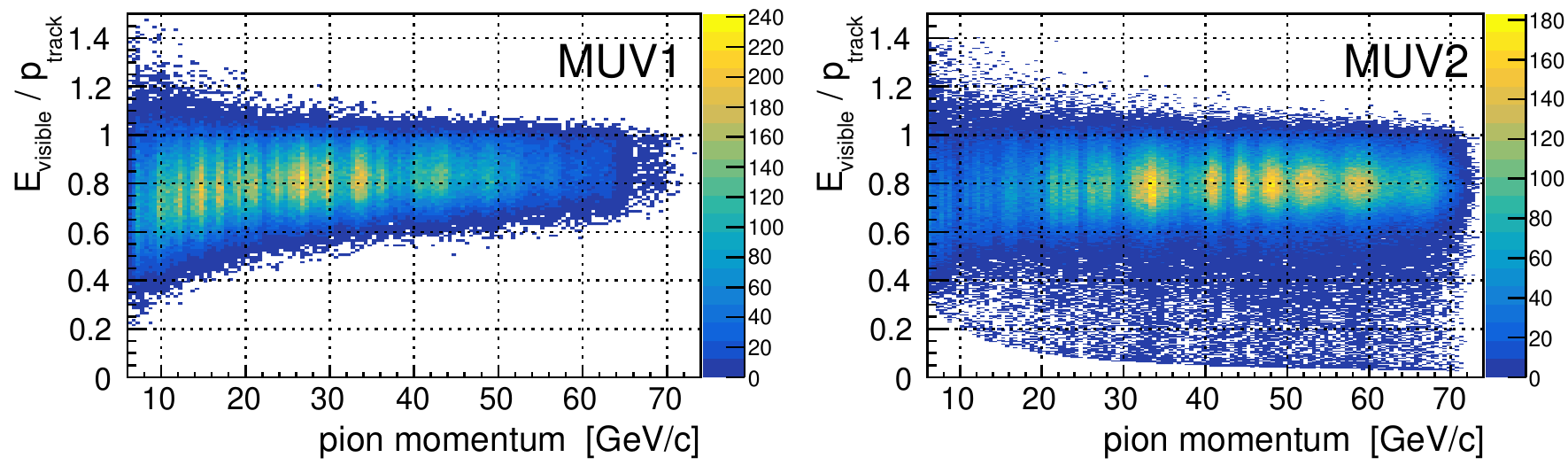}
\caption{Fraction of the visible energy in the MUV1 (left) and MUV2 (right) detectors for simulated pions as a function of their momentum. }
\label{fig:EnVis}
\end{figure}

The distinction of hadronic and electromagnetic components improves the energy resolution of the calorimeter. A parameter $w=\sum_i E_i^2 / E_\text{tot}^2$ is defined as the ratio between the sum of the squared energies collected in each channel $i$ and the squared total energy collected in the detector.\footnote{To avoid double-counting, MUV1 channels from strips with two-sided readout obtain half of the sum of the measured energies of each side.} This parameter $w$ can take values between 0 and 1 and is sensitive to the shower composition: electromagnetic showers with only few contributing channels have $w$ close to 1 while for pure hadronic showers $w$ is close to 0. The visible energy $E_{\rm visible}$ is then corrected to ${\rm E}_{\rm corr} = {\rm E}_{\rm visible} + F(w)  \times  {\rm E}_{\rm invisible}$ with a Fermi-Dirac-like function $F(w)$ which ranges from full correction for $w=0$ to zero at $w=1$. 
The correction ${\rm E}_{\rm invisible}$ is the difference between the mean visible energy and the track momentum, and is taken from simulated pion showers, as shown in \Fig{fig:EnVis}.  
\end{enumerate}
\subsubsection{Performance in 2015}
\begin{figure}[!h]
\centering
\includegraphics[scale=0.9]{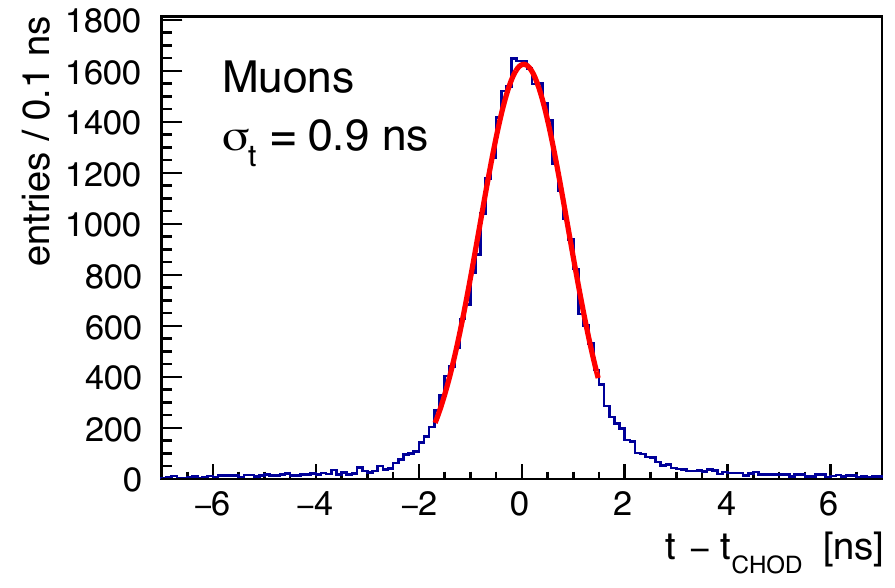}
\caption{Calorimeter time resolution for muons from $\KMN$ decays with respect to the NA48-CHOD reference time.} 
\label{fig:TRes}
\end{figure}

 The  detector was operated during the whole 2015 data taking period. 
Samples of $\KMN$ and $\KTP$  decays were used to 
validate the functionality and performance of the calorimeter.
With the detector time defined as the energy-weighted mean of the channel times, the time resolution 
was measured to be 0.9~ns for muons   
with respect to the NA48-CHOD detector (\Fig{fig:TRes}).

\begin{figure}[!h]
\includegraphics[scale=0.75]{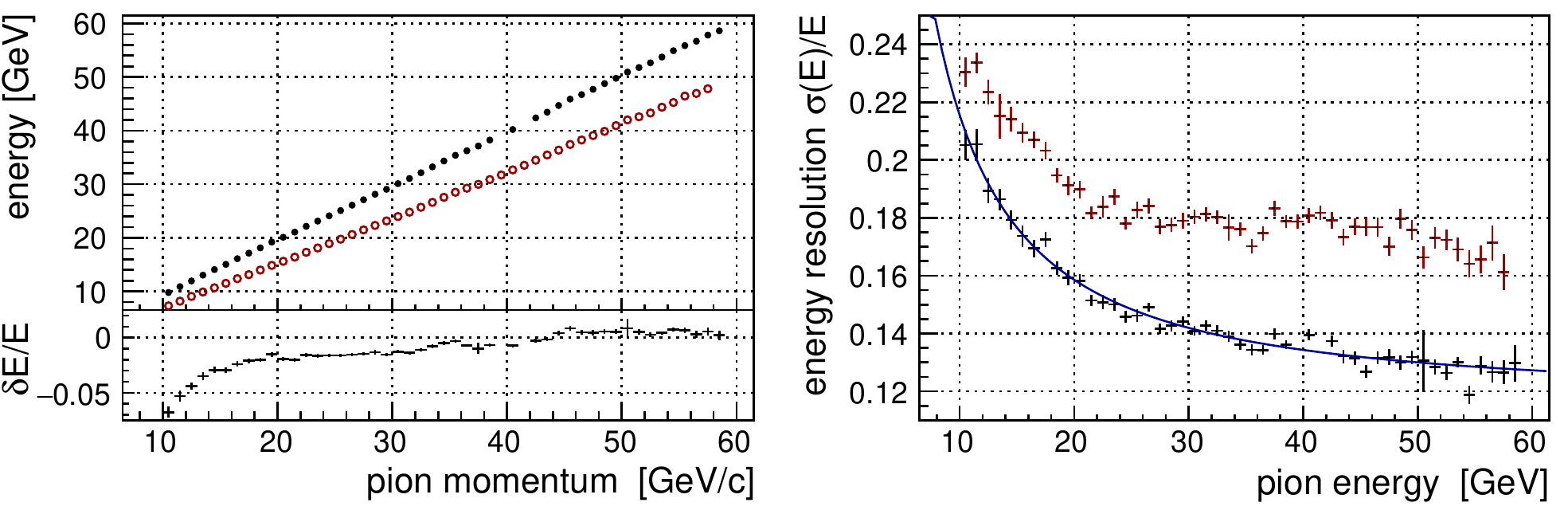}
\caption{\label{fig:EnRes}Calorimeter energy response for charged pions from selected $\KTP$ decays with minimum energy deposition in the LKr calorimeter. Left: Linearity of the response before correction (red  open circles)  and after all corrections (black filled dots). The lower plot shows the relative deviation $\delta$E / E from an exact linear energy response. Right:  Energy resolution before correction (red  crosses) and after all corrections (black crosses).  The  line corresponds to the parameterization of the energy resolution given in \Equ{eq:hacresol}.}
\end{figure}

 The calibration procedure was validated using the data by applying each correction in turn
and checking the quantitative improvement in terms of resolution and linearity of the energy response.
The energy calibration of the calorimeter was tested by requiring  the pion in the $\KTP$  selected sample to behave as a minimum ionizing particle in the LKr. 
The energy collected in the hadron calorimeter was then  fitted in bins of pion momentum with a Crystal Ball function  (a Gaussian function describing the core and a power-law function describing the low-end tail below a certain threshold) to take into account the energy lost in the previous interactions. 
It can be noted that the correction for ${\rm E}_{\rm invisible}$ has no effect on the linearity of the energy response but  improves the energy resolution.
 \Fig{fig:EnRes} shows the improvement for the calorimeter energy response linearity  and resolution 
 after introducing all corrections.

The obtained resolution, measured with 2015 data, is:

\begin{equation}{\label{eq:hacresol}}
\frac{\sigma({\rm E})}{\rm E} = 0.115  \oplus \frac{0.38}{\sqrt{\rm E}} \oplus \frac{1.37}{\rm E},  
\end{equation}
where E is in $\GEV$. 
The total resolution on the measured hadronic energy may be improved with respect to \Equ{eq:hacresol} by adding information from the preceding LKr calorimeter.

%% file: Sec11-MUV3_v2.tex
\subsection{Fast Muon Veto (MUV3)}\label{ssec:muv3}
\begin{figure}[ht]
\begin{center}
\vspace{-35mm}
\includegraphics[width=0.55\linewidth]{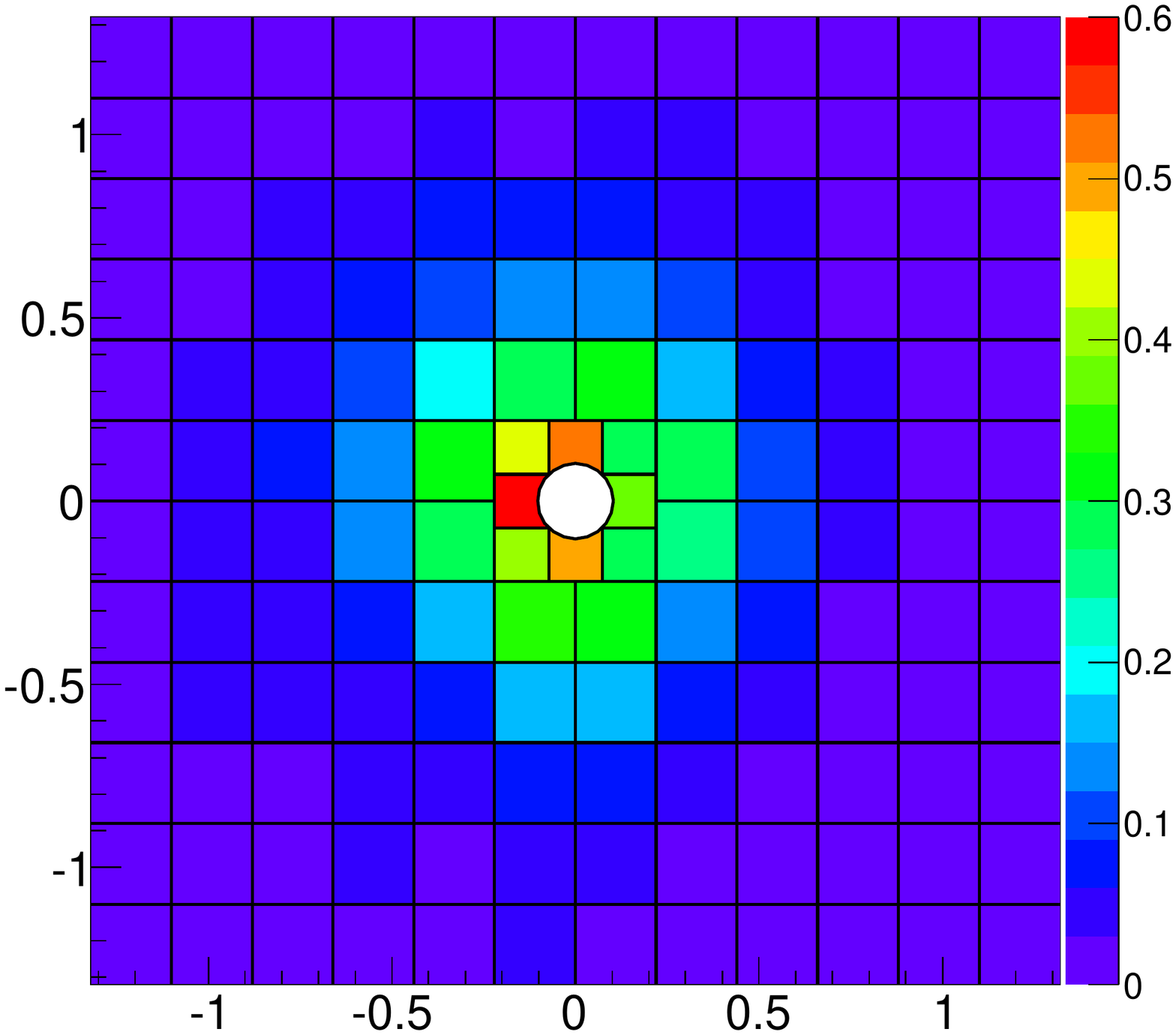}
\put(-112,-1){X coordinate [m]}
\put(-243,120){\rotatebox{90}{Y coordinate [m]}}
\put(-50,202){[MHz]}
\includegraphics[width=0.40\linewidth]{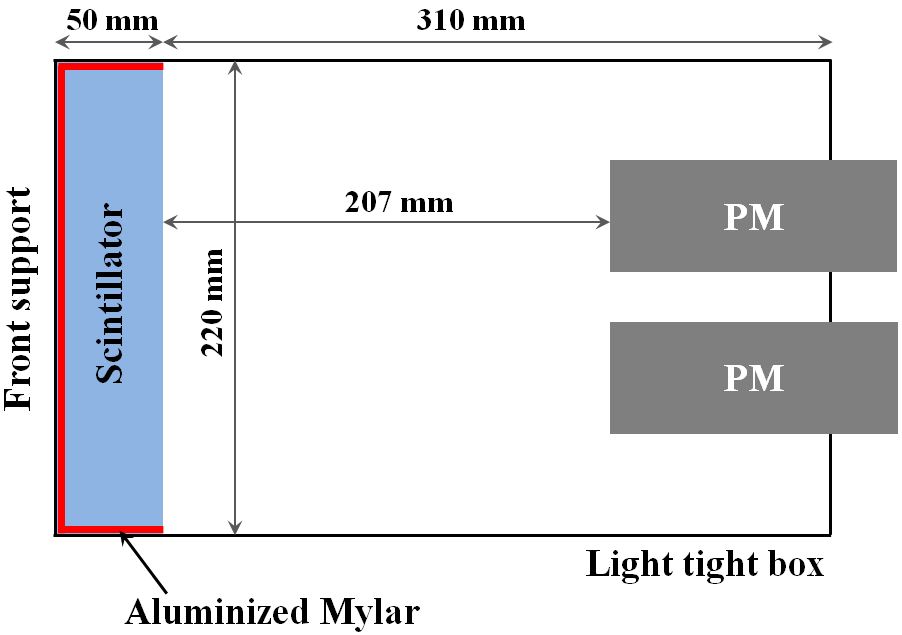}
\caption{Left: MUV3 tile geometry and expected signal rates in tiles at the nominal beam rate (in MHz). The rate in the ``hot tile'' on the negative X  side of the beam pipe is 3.2~MHz (outside the colour scale), dominated by muons from beam pion decays. Right: Schematic drawing of a regular MUV3 detector cell, including a scintillator tile, a light tight box and two PMs. 
\label{fig:MUV3-rate}}
\end{center}
\end{figure}

The MUV3 detector, located downstream of the hadron calorimeter behind a 80 cm thick iron wall, provides fast L0 trigger signals and is used for muon identification; it detects charged particles traversing the whole calorimeter system (LKr, MUV1,2 and the iron wall) with a total thickness of over 14 interaction lengths.
The MUV3 has a transverse size of $2640 \times 2640$~mm$^2$ and is built from 50~mm thick scintillator tiles, 
including 140~regular tiles of $220 \times 220$~mm$^2$ transverse dimensions and 8~smaller tiles adjacent to the beam pipe, as required by the high particle rate near the beam (\Fig{fig:MUV3-rate}-left).

The front and lateral surfaces of each tile are covered with aluminized Mylar\textsuperscript{TM} foil, while the back surface faces a light-tight box to avoid cross talk between tiles. Two 2-inch PMs  facing towards the tile are placed 
behind each tile; the distance between the scintillator and PM windows is 207~mm
(\Fig{fig:MUV3-rate}-right). Out of the 296 PMs used, 280~are of type EMI 9814B and 16 (serving 8 peripheral tiles) are Philips XP 2262. The average number of photo-electrons released at the PM photocathode for a minimum ionizing particle traversing the tile is 35, as measured in a test beam.

The anode signal from each PM is fed to a constant fraction discriminator (CFD) which reduces the time jitter associated with the amplitude range~\cite{Gedcke:1967}, and the CFD output signal is sent to 
a TEL62 readout board (\Sec{ssec:TEL62}) equipped with 3 TDCB mezzanines to accommodate the 296 MUV3 readout channels.

\begin{figure}[h]
\begin{center}
\includegraphics[width=0.5\linewidth]{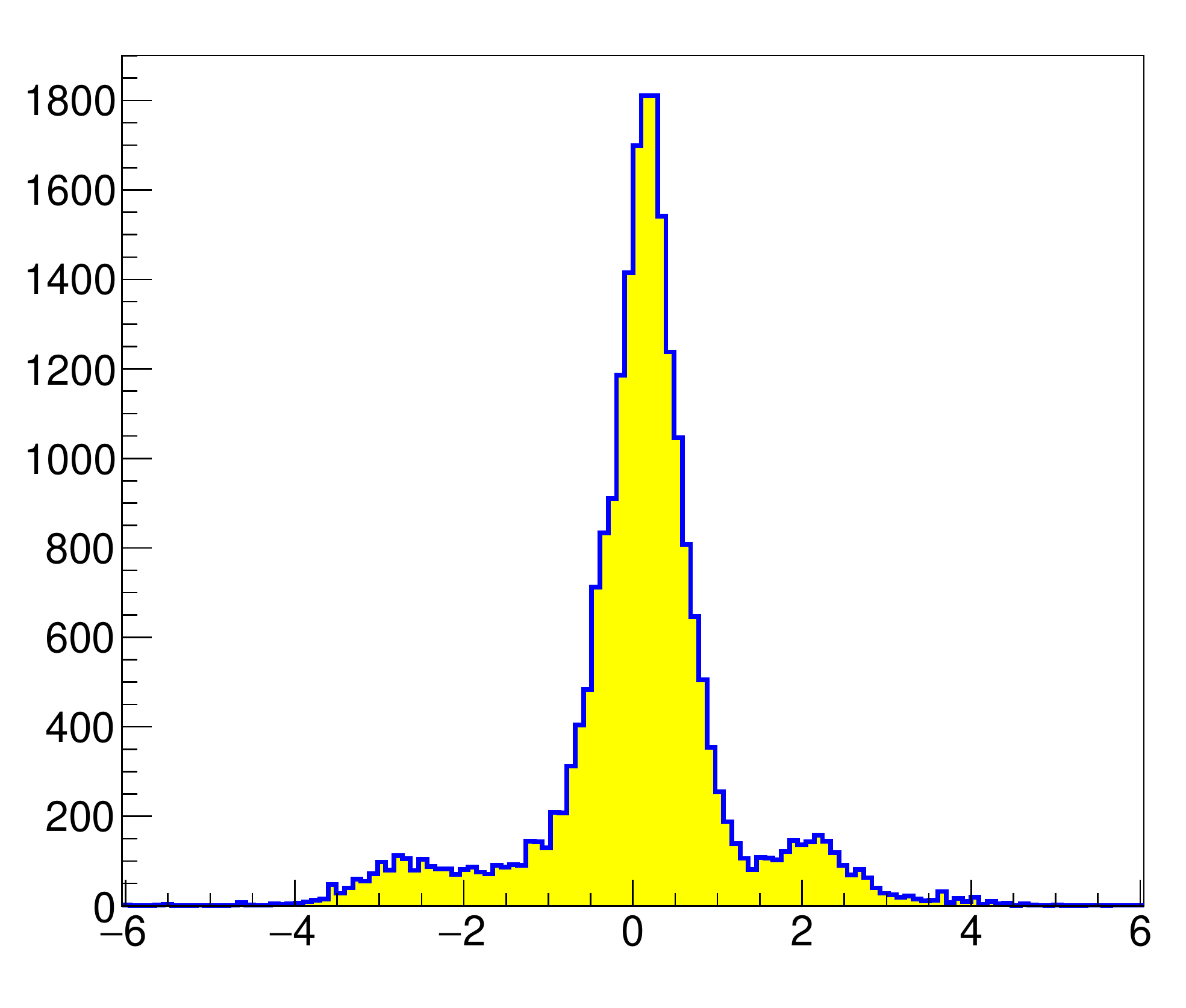}%
\includegraphics[width=0.5\linewidth]{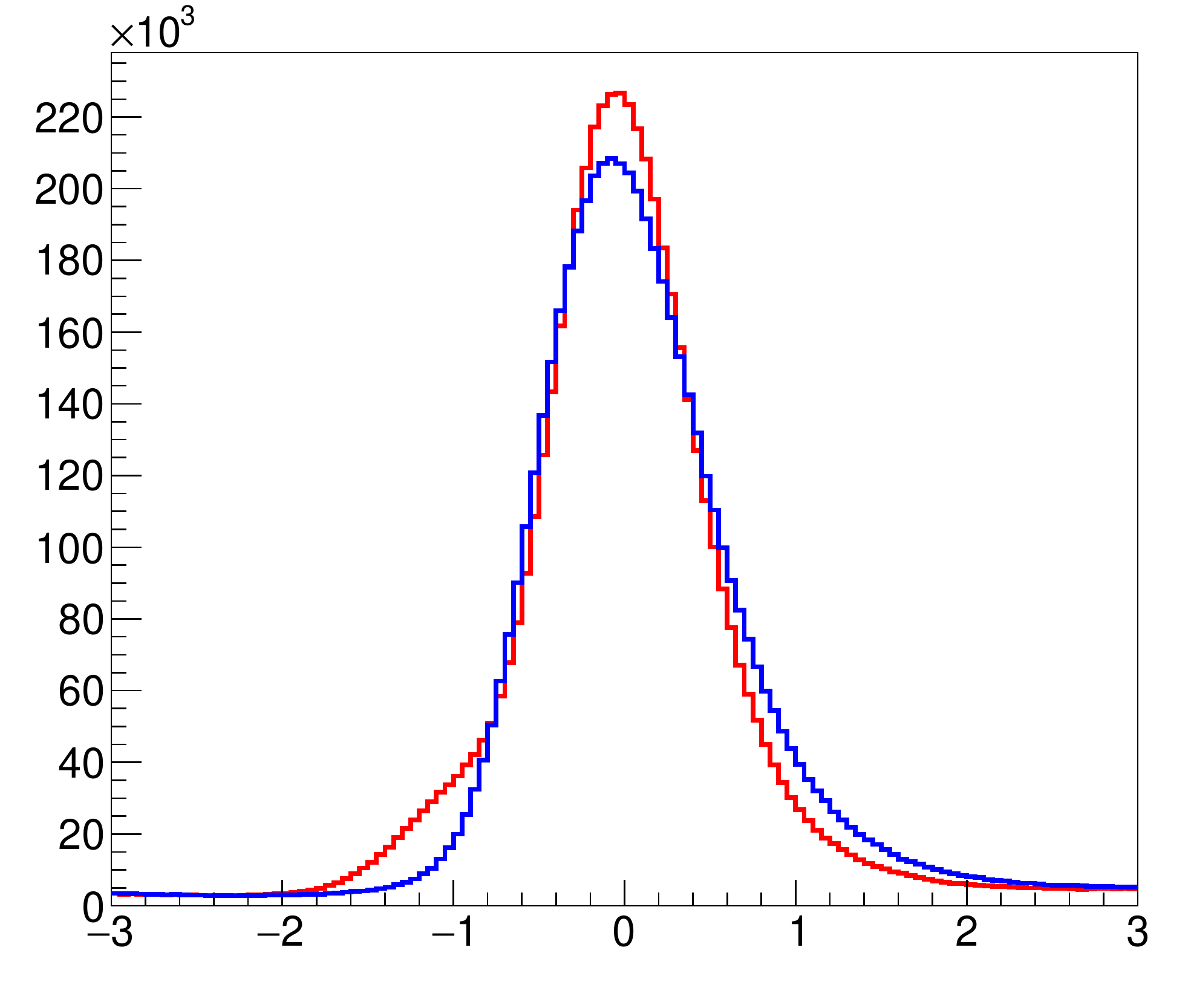}%
\put(-185,0){MUV3 muon time $-$ reference time [ns]}
\put(-404,0){Time difference between two signals [ns]}
\put(-92,150){\color{red}Average time}
\put(-65,45){\color{blue}Latest time}
\caption{Left: Distribution of the time difference between the signals from the two PMs of a single MUV3 cell. The effect of muons traversing a PM window is clearly visible on both sides of the central peak.  A Gaussian fit to the central peak gives a resolution of 0.5~ns. Right: MUV3 muon time defined as the average and the latest of the two signal times in a cell (the latter definition reduces the bias due to early Cherenkov photons) with respect to the KTAG reference time (which has a resolution of 70~ps). The distributions are integrated over all MUV3 cells, with appropriate time offsets applied to channel times. The time resolutions for the two muon time definitions with respect to the KTAG reference time are 0.41~ns and 0.48~ns, respectively. The effect of early Cherenkov photons is visible in the average time distribution. 
\label{fig:MUV3-timing}}
\end{center}
\end{figure}

The total rate of muons traversing MUV3 at the nominal beam intensity is 13~MHz. The corresponding signal rates in tiles computed with Monte Carlo simulations are shown in \Fig{fig:MUV3-rate}-left. The signal time resolution in individual channels measured with the 2015 data is in the 0.4 -- 0.6~ns range (rms). Muons traversing a PM window generate Cherenkov radiation;  the probability of that happening for a muon traversing a regular tile is 8\%. These Cherenkov photons arrive on average about 2.5~ns before scintillation photons, which affects the muon time reconstruction. The time resolution measurements are illustrated in \Fig{fig:MUV3-timing}.

\begin{figure}[ht]
\begin{center}
\includegraphics[width=0.6\linewidth]{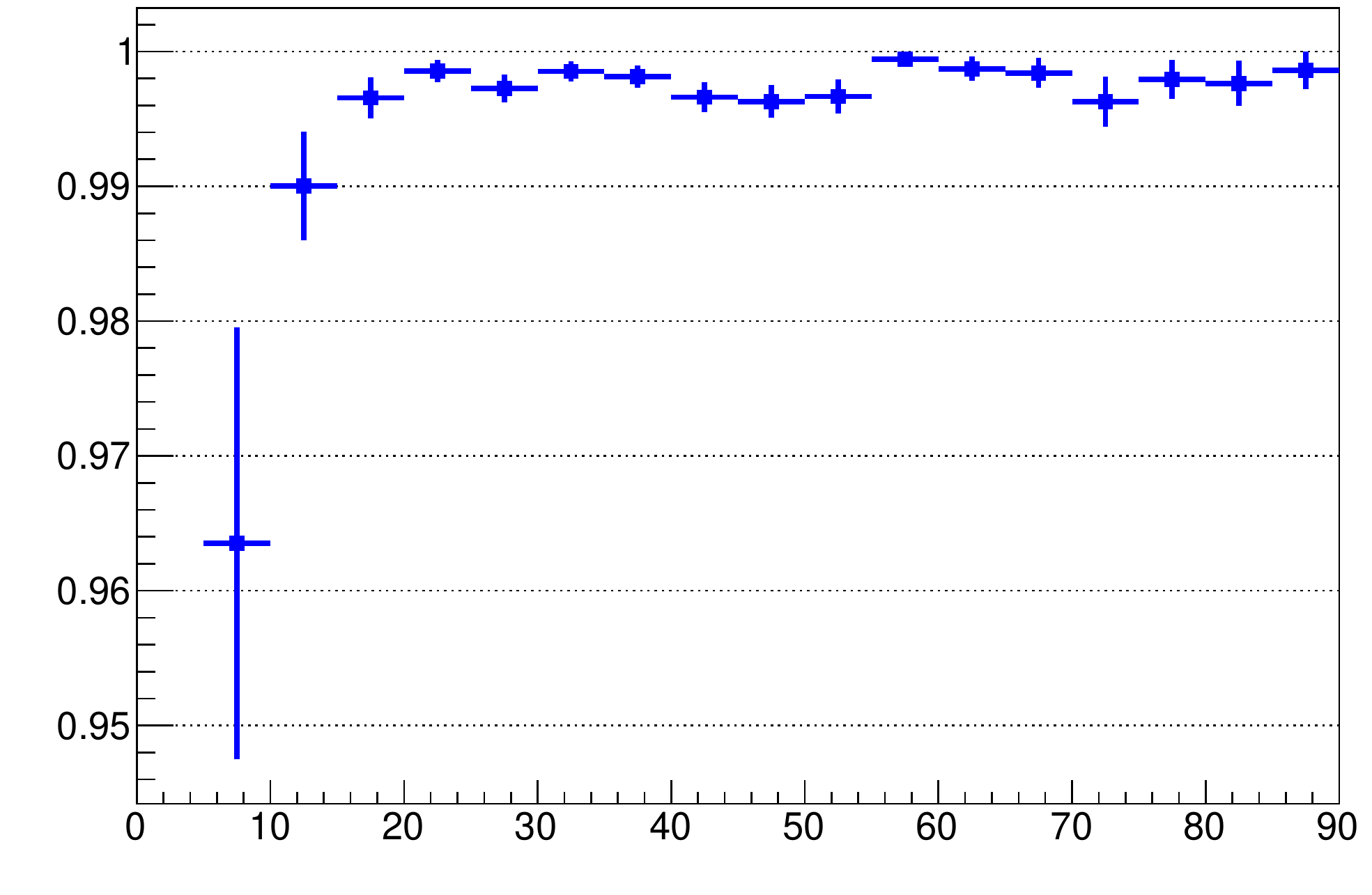}
\put(-124,0){Muon momentum [$\GEVc$]}
\put(-262,58){\rotatebox{90}{Identification efficiency}}
\caption{\label{fig:MUV3-efficiency}Muon identification efficiency of the MUV3 detector measured with beam halo muons as a function of momentum reconstructed by the straw spectrometer. }
\end{center}
\end{figure}

MUV3 muon identification efficiency was measured in 2015 using a sample of beam halo muons triggered by the NA48-CHOD detector and reconstructed by the spectrometer. It exceeds 99.5\% for muon momenta above 15~$\GEVc$, as shown in \Fig{fig:MUV3-efficiency}.

%% file: Sec12-Vetoes_v4.tex
Two complementary veto detectors are used for the detection of pions from
$\KTAU$ decays escaping the lateral acceptance of the STRAW chambers.
Both detectors were installed after the 2015 data-taking period, and integrated within the NA62 trigger and data acquisition system during the 2016 physics period. Their performance will be evaluated with the 2016 data.

\subsection{Peripheral Muon Veto (MUV0)}

The MUV0 detector is a scintillator hodoscope designed to detect $\MPI$ emitted in $\KTAU$ decays with momenta below 10~$\GEVc$, deflected towards positive X by the spectrometer magnet (which adds to the deflection of the $\KPL$ beam by TRIM5 magnet), and  leaving the lateral acceptance near the RICH.

\begin{figure}[h]
\begin{minipage}{0.5\linewidth}
\begin{center}
\includegraphics[width=0.9\linewidth]{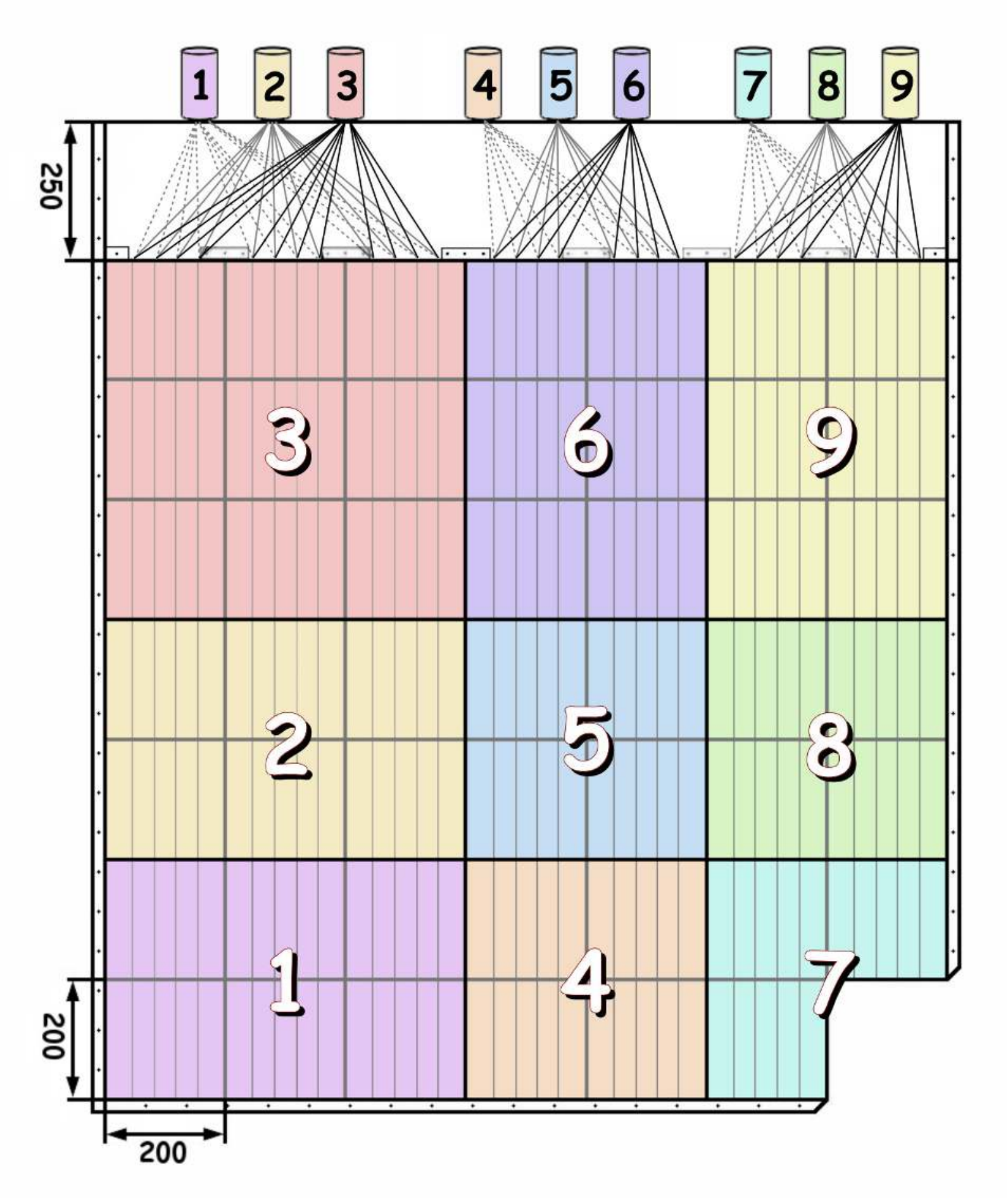}
\end{center}
\end{minipage}
\begin{minipage}{0.5\linewidth}
\begin{center}
\includegraphics[width=1.\linewidth]{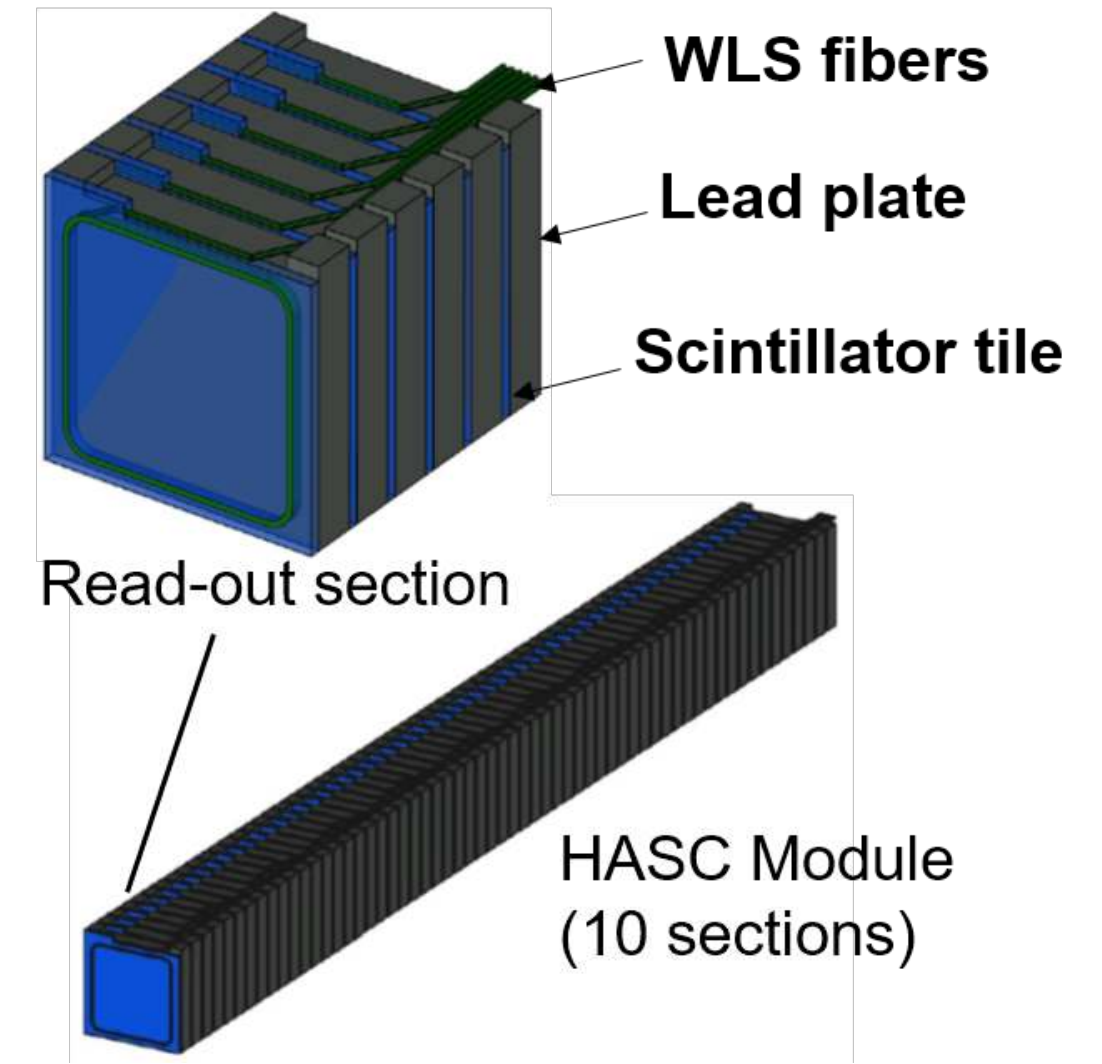}
\end{center}
\end{minipage}
\caption{Left: Sketch of the MUV0 detector showing the pattern of scintillator tiles  and super-tiles. Right: 
A HASC module with $100 \times 100$~mm$^2$ cross section; Six consecutive scintillators are connected to one optical fibre, leading to 10 fibres per module. Each fibre is read out with a SiPM.
\label{fig:HASC-Picture1}}
\end{figure}

The MUV0 is mounted on the downstream flange of the RICH. 
Its active scintillator area of $1.4\times1.4$~m$^2$ covers the periphery of the lateral acceptance (1.545 m $<$ X $< 2.945 ~{\rm m}, |$Y$| <0.7$~m), and consists of two layers of 48~plastic scintillator tiles with dimensions of $200\times 200\times 10$~mm$^3$. The tiles are grouped in 9 super-tiles forming a pattern shown in~\Fig{fig:HASC-Picture1}-left. Each super-tile is read out with wavelength shifting fibres and a Hamamtsu R7400 PM using the TEL62/TDC readout: 
the analog output signals are sampled with two thresholds in a time-over-threshold board (Section~\ref{ssec:lav-fee}) whose LVDS signal output length is digitized by a TDCB mezzanine (Section~\ref{ssec:TEL62}) to give a total of 18 readout channels.

\subsection{Hadronic Sampling Calorimeter (HASC)}

The hadronic sampling calorimeter (HASC) is used for the detection of $\PPI$ emitted in
$\KTAU$ decays with momentum above 50~GeV/$c$ and propagating through the beam holes in the centres of the straw chambers. The detector is located downstream of MUV3 and the BEND dipole magnet (Figure~\ref{fig:K12beam-downstream}) which sweeps these pions out of the $\KPL$ beam towards negative X. The detector covers the lateral acceptance of $-0.48$~m $<$ X $<-0.18~{\rm m}, |$Y$| <0.15~{\rm m}$. 

The HASC is constructed of 9~identical modules (\Fig{fig:HASC-Picture1}-right) re-used from a prototype of the Projectile Spectator Detector~\cite{Golubeva:2012} developed by the NA61 collaboration. The active element of a module is a sandwich of 60 lead plates (16~mm thick) interleaved with 60 plates of scintillator (4~mm thick) of $100\times 100$~mm$^2$ transverse dimensions.

Each module is organized in 10~longitudinal readout sections. Each scintillator tile is optically coupled to WLS optical fibres. In the rear side of the module there are 10~optical connectors, originally designed to be coupled with $3 \times 3$~mm$^2$ green-sensitive, Micro-pixel Avalanche Photodiodes (MAPD). For the NA62 application, the photodetectors have been replaced with $3 \times 3$~mm$^2$ SiPM sensors (Hamamatsu S12572-015C) coupled to an amplifier and a bias board. The analogue output signals are sampled with 4 thresholds in a time-over-threshold board (Section~\ref{ssec:lav-fee}) whose LVDS signal output length is digitized by a TDCB mezzanine (Section~\ref{ssec:TEL62}). 
This results in a total of 360 readout channels.

%% file: Sec13_TDAQ_v3.tex
%********************************  TDAQ ********************************************************
The intense flux of NA62 dictates the need for a high-performance triggering and data acquisition 
system, which must minimize dead time while maximizing data collection reliability.
A unified trigger and data acquisition (TDAQ) system was designed in order to address such 
requirements in NA62 in a simple and cost-effective manner.

The focus on a design with a large integration, in terms of channel density and functionality, aims at optimizing 
the implementation and maintenance effort, while at the same time allowing an excellent control and reproducibility 
of the trigger conditions. The other guiding principle was that of scalability and versatility, as desirable for 
an experiment adopting a new technique and one in which triggering conditions cannot be fully spelled out 
\emph{a priori}, both due to unavoidable uncertainties in the detector and beam performances, and the possibility of 
later expanding the physics scope and reach of the experiment.

Cost and performance issues prevent a fully trigger-less (i.e. software trigger only) system to be implemented for 
a medium-scale experiment such as NA62, which deployed a system with a single hardware trigger level (called L0), 
initially involving only a small set of fast sub-detectors, but potentially extensible to most of them in a 
straightforward way.
An early decision was made to allow a very long (by HEP standards) maximum latency time of 1 ms for the hardware L0 
trigger: this choice allows the possibility to explore the innovative concept of introducing massively parallel 
commercial processors (GPUs) at the lowest trigger level, by using such devices in hard-real time 
\cite{Collazuol:2012}, a side project which is being actively pursued within NA62.

With an estimated 10 MHz rate of decays in the detector, the L0 hardware trigger maximum output rate was chosen to be 1 MHz.  
Further data reduction  down to about 10 kHz, dictated by the available bandwidth for data-storage to tape, should be achieved by L1 and L2 software trigger levels.
In this context, three key design ingredients were identified to push towards a high integration: 
the full unification of the trigger and readout, the use of a single unified path for trigger and control of 
individual system boards, and the use of commodity output-data links.

The first point led to a design in which the hardware L0 trigger is performed by evaluating conditions on the very same complete set of digitized data which are transmitted from the detectors to L0.  This is somewhat contrary to the traditional approach in which a reduced sub-set of information is processed by a dedicated sub-system trigger.
The higher data throughput required by the chosen
approach is nowadays sustainable, and provides several advantages: elimination of a dedicated trigger data path, 
resulting in reduced hardware, full control and offline reproducibility of the trigger algorithms with no uncertainty 
related to independent flows, as well as the possibility of a fully accurate simulation, no design limitation on the 
kind of trigger elaboration which can be done on the detector data, ease of integrating any sub-detector in the 
trigger chain and of re-defining the hardware trigger at will. 
The NA62 TDAQ system is rather unique in allowing such a fully-digital flexibility on this scale, in which any 
information available from the detector can be used in the trigger. 

Moreover, the extraction of some ``slow-control'' and monitoring information from the TDAQ system is also fully 
integrated within the data flow path: at times during a burst (most notably at the end of burst) special 
triggers are delivered to the system, to which all boards react by sending special monitoring data along the 
standard data links. This approach has the advantage of not requiring additional slow control data paths, and also 
ensures by design the availability of monitoring information together with the main data, as required for offline 
analysis, without any additional effort.

The second point, the use of a single link for clock, trigger and control distribution, fully pursues to its logical conclusion 
an established trend in HEP DAQ systems: each part of the TDAQ system receives the same synchronization, 
L0 trigger timing, and slow timing information, and also L0 trigger type, calibration control and monitoring commands, 
as well as data related to flow integrity, reach them along the same common and unique path.

Finally, the choice of using standard commodity output links (Gigabit Ethernet) was not only dictated by 
considerations on cost and simplicity, but resulted in a large flexibility in reconfiguring the downstream part of 
the TDAQ system, which has been exploited in modifying the design of the L1/L2 computing farm architecture to adapt
to emerging challenges, different needs for sharing bandwidth between L0 trigger and main readout information, and 
the availability of cheaper and more powerful machines, in an easy and cost-effective way without requiring 
changes in the upstream part of the system itself. The data are transferred using UDP \footnote{User Datagram 
Protocol (UDP) is a minimal and connectionless network protocol.}, thus with no reliability built-in in the protocol, 
because of data performance requirements: this was not an issue in the original TDAQ concept in which each 
sub-detector used dedicated PCs and only point-to-point links were present. In order to optimize PC usage, a paradigm 
shift was introduced in which data from the acquisition boards go directly onto a switched network, and
possible data losses in the switching fabric have to be monitored.

Broadly, the NA62 experiment comprises about 15 sub-detector systems, most of them sharing similar requirements in 
terms of precision timing response and readout capabilities; this led to the design of a common, unified and 
versatile trigger and data acquisition system based on high-precision TDCs, suited for use in most of NA62 
sub-detectors, described in \Sec{ssec:TEL62}. 
For the straw spectrometer, with the largest channel count among the TDC-based sub-detectors and an 
intrinsically poorer time resolution for which less precise TDCs are sufficient, it was later decided to implement 
a dedicated solution for an even higher integration and reduced cost (\Sec{ssec:Straw-readout}).

The GigaTracker (GTK) has the largest number of channels in a highly-integrated miniaturized system, which required a dedicated system (\Sec{ssec:GTK-readout}).
Calorimeters instead use continuous pulse sampling via FADCs to extract information, with the system 
described in \Sec{ssec:Cal-readout}.

The L0 hardware trigger is described in \Sec{ssec:L0}, and the software triggers in \Sec{ssec:HLT}.

\begin{figure}
\begin{center}
\includegraphics[width=0.9\linewidth]{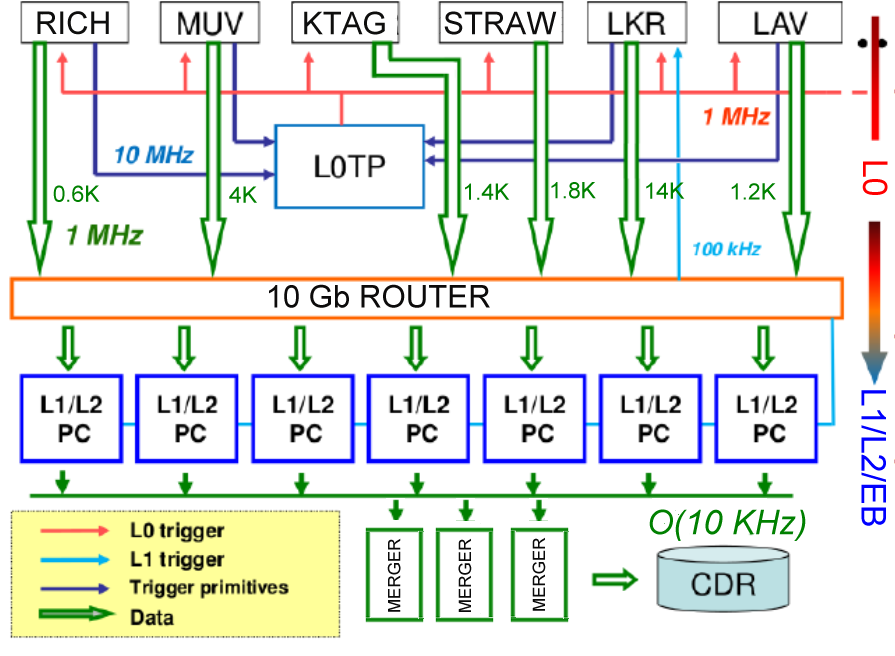}
\end{center}
\caption{\label{fig:TDAQ-scheme} {Overview of the NA62 TDAQ Trigger and Data Acquisition system. The average data size in byte is reported  close to the data arrow. For sake of simplicity, only few detectors are displayed, in particular GTK is not represented.}}
\end{figure}

\Fig{fig:TDAQ-scheme} illustrates the overall TDAQ scheme of NA62 and the relevant trigger rates 
and connections. 

\subsection{Common signal distribution}\label{ssec:TTC}

Practical considerations for an experiment of the size and time scale of NA62 suggested evaluating the adoption of 
existing systems and infrastructures. This led to the choice of using the Timing, Trigger and Control (TTC)  
\cite{Taylor:2002} system, developed at CERN and adopted by all LHC experiments for clock and trigger distribution, 
based on time-multiplexed transmission of synchronous and asynchronous data with 25 ns precision over a low-jitter
clock.
The TTC is a unidirectional optical fibre based transmission system where two channels are multiplexed and encoded using a 40 MHz clock and transmitted at 160 MHz rate. One channel is used to carry the L0 trigger signal only, 
while the other carries encoded information concerning resets and trigger types. 

Two important differences in the use of the timing system with respect to LHC are related to the fixed-target SPS 
environment: in NA62 the main clock is centrally generated by a high-quality clock generator\footnote{Hewlett-Packard
8656B Signal Generator.} in the experimental area; its frequency of 40.079 MHz is chosen to fall within the locking 
range of the QPLL (Quartz Crystal Phase-Locked Loop) jitter-cleaning system \cite{Moreira:2003}, used  
to guarantee the required high accuracy and stability, but is free-running and 
otherwise unrelated to any clock from the accelerator system. Moreover, the SPS beam structure is very different
from that of the collider, and expected to be roughly uniform in rate for a few-seconds long machine spill, 
separated from the next one by a variable time which can be as long as a minute. 
This leads to the machine spill (or burst) being the unit of data-taking in NA62, with all electronics running in 
a fully synchronized way while it lasts, and performing clean-up tasks in an independent way outside the spill.

All the synchronous elements of the NA62 TDAQ system run on the centrally distributed TTC clock and are synchronously 
reset by a Start Of Burst (SOB) command delivered through the same links and generated by aligning to 25 ns 
precision the SPS Warning of Warning of Ejection (WWE), roughly 1 second before the first beam particles are 
delivered; each system also synchronously stops on a similar common End Of Burst (EOB) signal, after a number of 
25 ns clock periods (which can vary from burst to burst but is common to all systems for any given burst). 
By resetting all coarse time counters on SOB through the same link which delivers the clock, any relative delay 
between sub-systems due to differences in propagation time from the clock generator is intrinsically corrected for.

The actual clock and L0 trigger signal distribution occurs through the use of a NA62-modified version of ALICE 
Local Trigger Unit (LTU) boards \cite{ALICE:Trigger}, of which 12 are deployed (roughly one per sub-system); these 
modules take care of time multiplexing the synchronous L0 trigger signal received by the central L0 Trigger Processor 
(L0TP) and the asynchronous 8-bit Trigger Type word which accompanies each L0 trigger to allow for different 
processing. 
Trigger type coding allows for up to 31 physics triggers with possibly different readout options, and the same 
number of service triggers (calibration, monitoring, etc.), some being used by the L0TP to acknowledge errors or 
requests for pausing the L0 trigger due to data congestion in some system. 
Also the command to inject calibration pulses in selected sub-detector front-end boards is delivered through the 
TTC link. Strictly sticking to the use of a unique and universal TDAQ link to all sub-system electronics to 
basically funnel all kinds of real-time,  the TDAQ controls was an early design choice which greatly simplified 
inter-communication both in terms of required hardware and state logic.

The parts of the TDAQ system which receive data (through Ethernet links) after L0 trigger are all asynchronous; 
the online computing farm, on which software trigger levels run,  is also loosely synchronized to the burst time 
structure through a DIM \cite{Gaspar:2000} software service which makes available the in burst/out of burst status 
driven by the SOB and EOB signals, with coarse (software) time precision.

\subsection{Common TDC-based trigger and readout system (TEL62 and TDCB)}\label{ssec:TEL62}

In this section the common TDC-based TDAQ system currently used for KTAG, CHANTI, LAV, NA48-CHOD, CHOD, RICH, MUV3, 
SAC and IRC sub-detectors is discussed.

Specific requirements for the common TDC-based system were an electronic time resolution of 100~ps, comparable to the signal jitter,  
the possibility to deliver some pulse-height information, a large channel integration and flexibility to implement different L0 trigger conditions. 
Crucial for an apparatus expected to be running for 10 years, programmability, scalability and flexibility were also key requirements  
to be able to meet such changing and varying requirements as the understanding of the 
detector and backgrounds progresses, so that an FPGA-based solution was the natural choice.

A good starting point for this common DAQ system was identified in the TELL1 board \cite{Haefeli:2006} developed 
for the LHCb experiment. A major redesign of the board was undertaken  
to significantly increase its performance in terms of computational and storage power, as well as inter-communication capabilities, resulting in the TEL62 board \cite{Angelucci:2006}. 
This is a 9U-size board which hosts 4 identical  FPGA-controlled\footnote{ALTERA Stratix III EP3SL200 devices with 200K logic elements.} 640 MB/s input data channels, 
with up to 2 GB of fast dynamic RAM (DDR2) storage each, which are then merged into an identical fifth FPGA which eventually drives a custom quad-Gigabit Ethernet board for output. TTC interfacing, auxiliary buses for board interconnection and on-board PC for slow control are also present on the board.

The TEL62 board is mechanically and electrically compatible with the TELL1 board and could house the input cards 
(ADC and optical receiver) developed by LHCb, but a high-precision TDC board was required for NA62,
and this was also developed within the collaboration.
The TDC Board (TDCB) \cite{Pedreschi:2015} is a mezzanine card for the TEL62 which houses 4 HPTDC chips \cite{Christiansen:2004}, under the
control of an FPGA, and can digitize leading and trailing edge times of 128 detector channels presented as LVDS 
signals on 4 VHDCI connectors. The time measurement is obtained by using Delay Locked Loops on high-precision 40 MHz 
clock, resulting in a 100 ps resolution. 

The TDCB on-board FPGA\footnote{Altera Cyclone III EP3C120 with 120K logic elements.} is configured via a 
I\textsuperscript{2}C interface\footnote{Inter-Integrated Circuit is an industry-standard serial bus developed by 
Philips Semiconductors (now NXP Semiconductors).} from the carrier (TEL62) board, and takes care of configuring 
HPTDC chips via JTAG bus.\footnote{An industry-standard serial bus developed by a consortium of hardware vendors and 
widely used for integrated-circuit diagnostics.} 
The trigger windowing logic of the HPTDCs is not used to store data waiting for a L0 trigger due to its short maximum 
latency, but is conveniently used to format the digitized data in 6.4 $\mu$s long frames, to simplify its processing 
and storage within the TEL62 board. The 32-bit data from each HPTDC is read out on a separate parallel bus at the 
maximum speed at which it can be delivered by the chips themselves (one 32-bit word every 25 ns), monitored on the 
fly within the FPGA and delivered to the carrier board at the same speed. 
When equipped with 4 TDCBs, a TEL62 can therefore readout (and trigger on) 512 detector channels.

The following is a brief outline of the data processing performed on the TEL62 board.
Within each of the FPGAs (called Pre-Processing FPGA, or PP) handling one TDCB card, the continuous streams of 
6.4 $\mu$s long TDC data frames from each TDC chip are time matched and merged, recording the possible presence of 
error words generated on the TDCB. Data (leading and trailing edge hit times) in each merged frame is rearranged 
in 25 ns wide time slots corresponding to the granularity at which it will be requested for readout after a L0 
trigger, and then such data is compacted into a block of adjacent data words and a structure of data pointers, 
both stored into a large time-slot indexed memory, allowing several ms of maximum latency.

When a L0 trigger signal, with 25 ns time resolution and an accompanying trigger type word, is received by the central 
FPGA (called Sync-Link FPGA, or SL) via the TTC system, data from a programmable and trigger-type dependent number 
of 25 ns time slots around the trigger time is extracted from memory in each PP, delivered to the SL which merges 
it into an event fragment. Event fragments are further patched together in multi-event UDP packets to optimize the 
bandwidth, and sent to the online PC farm on one or more Gigabit Ethernet links.

In parallel with the processing described above, two more identical copies of the complete input data are made 
available to the firmware in both the PP and SL FPGAs, for monitoring purposes and for generating L0 trigger 
primitives respectively.   
Sub-detector-specific firmware is developed for L0 trigger primitive generation, working on 128-channels data in 
each single PP FPGA and on 512-channels within the SL. For sub-detectors with more channels participating in L0 
trigger generation, TEL62 boards can be made to inter-communicate in either a star or daisy chain configuration by 
using a pair of custom buses connected to the SL FPGA: a daughter card with twin 2 Gb/s serial links called 
Inter-TEL was developed and built for this purpose and can be integrated in the firmware where needed.

The TEL62 can host a four-port Quad Gigabit Ethernet card (QGBE) \cite{Muller:2005}, originally designed for the LHCb 
experiment TELL1 board, which is used to send L0 trigger primitives to the L0TP for the boards involved in L0 trigger 
generation, as well as detector data to the PC farm after a L0 trigger is received.

\subsection{GTK readout system}\label{ssec:GTK-readout}
The GTK off-detector readout (GTK-RO) system acts as an interface between the on-detector ReadOut-Chips (ROC, 
also known as TDCpix) on the detector side, and the clock and trigger distribution performed by the TTC system 
\cite{Taylor:2002} on the experiment side.
The GTK-RO system is highly modular and is composed of ten custom cards per station, each card serving one TDCpix chip.

\begin{figure}
\begin{center}
\includegraphics[width=0.95\linewidth]{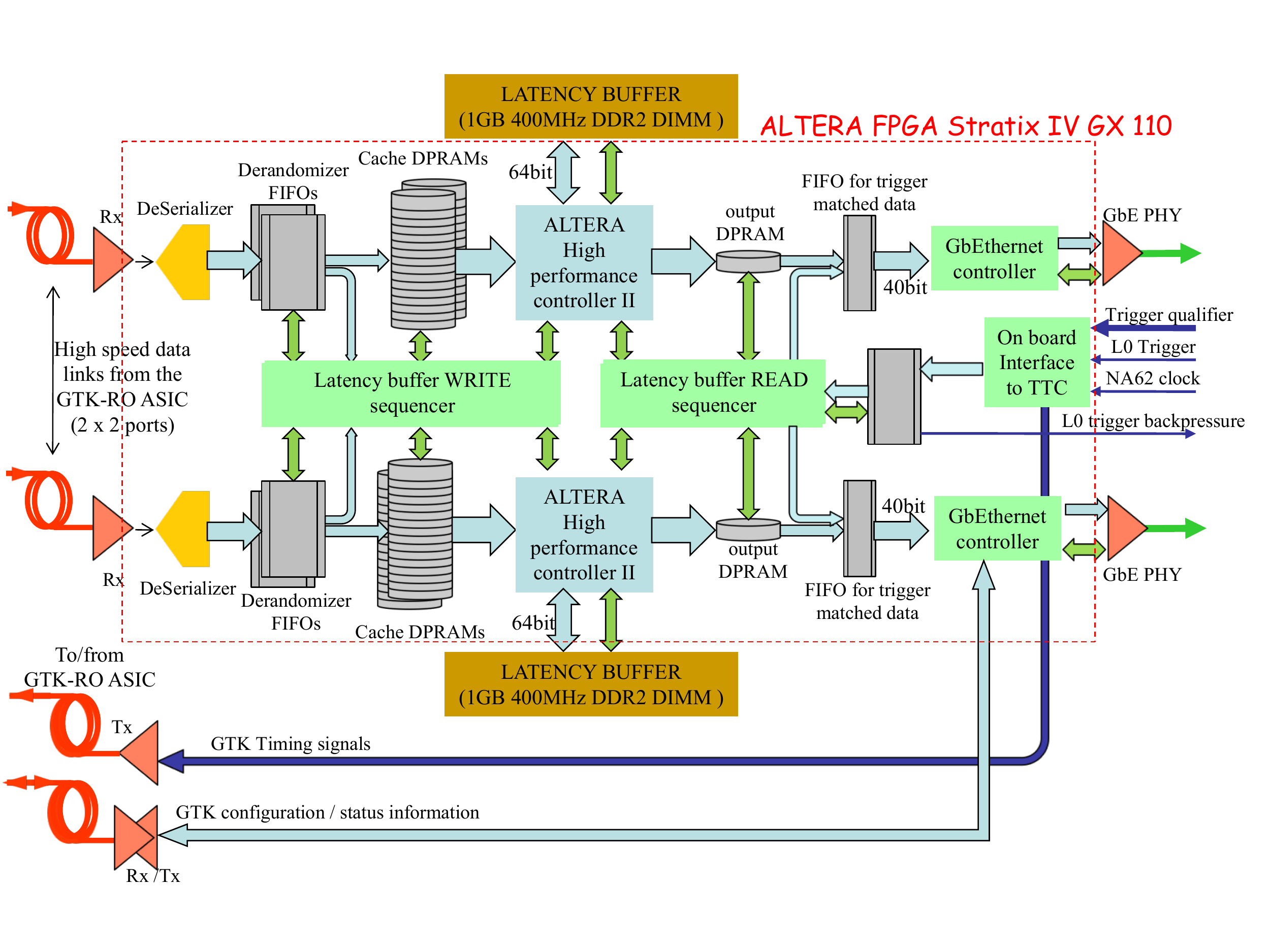}
\end{center}
 \caption{\label{fig:GTK-blockdiagram} Block diagram of a GTK-RO motherboard. One GTK-RO serves one TDCpix ASIC.}
\end{figure}

The GTK-RO board is a custom built module made of two decked cards centered around an FPGA\footnote{Altera Stratix 
GX110.} with 2~GB DDR2 SDRAM. 
Each main board has four 3.2 Gb/s optical links, one low speed (320 Mbit/s) optical link and two Gigabit Ethernet copper links. 
The block diagram of a GTK-RO card is shown in \Fig{fig:GTK-blockdiagram}. The interface of the GTK-RO with the TTC is implemented on a dedicated daughter card, which adds one spare 
Gigabit Ethernet copper link, and hosts a TTCrq mezzanine card with a TTCrx receiver \cite{TTCrq} for interfacing the system with the TTC. 

Each TDCPix sends the data via four 200~m long optical fibres to the corresponding GTK-RO board located in the 
PC-farm room outside of the experimental area. Each optical fibre transmits the output of a 3.2~Gb/s serializer, 
which serves a quarter of a TDCpix.
Two additional optical fibres per TDCpix are used to transmit service information such as the 320~MHz 
clock and  chip configuration signals.
The total data bandwidth is 12.8~Gb/s for each chip, corresponding to 128~Gb/s per station. 

The transmission is driven by the data which are sent by the TDCpix chips. The data are temporarily stored in the 
large on-board memory addressed by the hit time slot. 
The arrival of a L0 trigger with a fixed latency identifies the time slots of interest which are then 
extracted. Three time slots (corresponding to a 75~ns window) centred around the L0 time are read from the memory.
Extracted data are then transmitted to the six sub-detector PCs, one for each group of 5 chips, through Gigabit Ethernet switches.

The purpose of the six identical sub-detector PCs is to collect and organize the data identified by the L0 triggers 
and to send them to the NA62 PC farm. To balance the amount of data, each PC serves five GTK-RO cards corresponding 
to a row of chips in a station. The PCs are dual-CPU servers with at least 24~GB of RAM and 10~Gb/s Ethernet ports, 
running Linux as operating system.
Since each GTK-RO sends the data to a PC using two Gigabit Ethernet links, and since the foreseen data rate for 
each GTK-RO is of the order of one Gb/s, an Ethernet switch with 24 1~Gb/s ports and 2 10~Gb/s ports is used as a multiplexer. 
In the sub-detector PCs two 10~Gb/s Ethernet ports are used, one to receive the data and one to send them to 
the NA62 PC farm. To achieve the full 10~Gb/s throughput from the Ethernet ports,   
a custom multi-threaded program was written
 that uses the \emph{zero copy} module of the PF\_RING\textsuperscript{TM} libraries \cite{pfring} to avoid  unnecessary memory to memory copy inside the PC.
The modularity of the software and hardware design allows the GTK data to be read either at the L0 or at the L1 trigger
level; since the GTK data is not expected to be used in the L1 software trigger, the choice of reading at the lower 
L1 trigger rate was taken, which puts much lower stress on the GTK readout system.

\subsection{STRAW  tracker readout system}\label{ssec:Straw-readout}
\input{Sec13_STRAW_v2.tex}

\subsection{Calorimeter readout system}\label{ssec:Cal-readout}

\paragraph{Calorimeters readout:}
The CREAM ({\bf CREAM}: Calorimeter REadout Module \cite{Ceccucci:2011}) readout 
was developed specifically for the LKr  
and is also used for all other calorimeters, i.e. the  hadronic calorimeters (MUV1 and MUV2) and the small-angle vetoes (IRC and SAC),  which however account for   
only 2\% of the total number of channels (\Tab{tab:CREAM-table1}). 
 The system uses 14-bit, serial-output 40 MHz FADCs, FPGAs for the handling of the data and trigger requests, and large DDR3 memories to store the data for an entire SPS burst. 
\paragraph{CREAM module:}
The CREAM is a 6U VME 64 board developed by CAEN, under specifications 
provided by NA62. It consists of a daughter-board, where the analogue input signals are shaped and digitized, and a motherboard, where the data are processed and sent out if the required trigger conditions are met.  

\renewcommand{\arraystretch}{1.2}
\begin{table}[ht]
\setlength{\tabcolsep}{3ex}
\caption{Number of channels in the five NA62 Calorimeters readout with the CREAM system.}
\label{tab:CREAM-table1}
\vspace{2ex}
\centering
\begin{tabular*}{0.95\textwidth}{lccc}
\hline\hline
Detector &  Type & Channels & Trigger channels \\
\hline
LKr & electromagnetic & 13248 & 864 \\
IRC & electromagnetic & 4 & 1 \\
SAC & electromagnetic & 4 & 1 \\
MUV1 & hadronic  & 176 & 12 \\
MUV2 & hadronic  & 88 & 6 \\
\hline
Total &  & 13520 & 884 \\
\hline
\hline
\textbf{}
\end{tabular*}
\end{table}
\renewcommand{\arraystretch}{1.0}

\begin{figure}
\begin{center}
\includegraphics[width=0.95\linewidth]{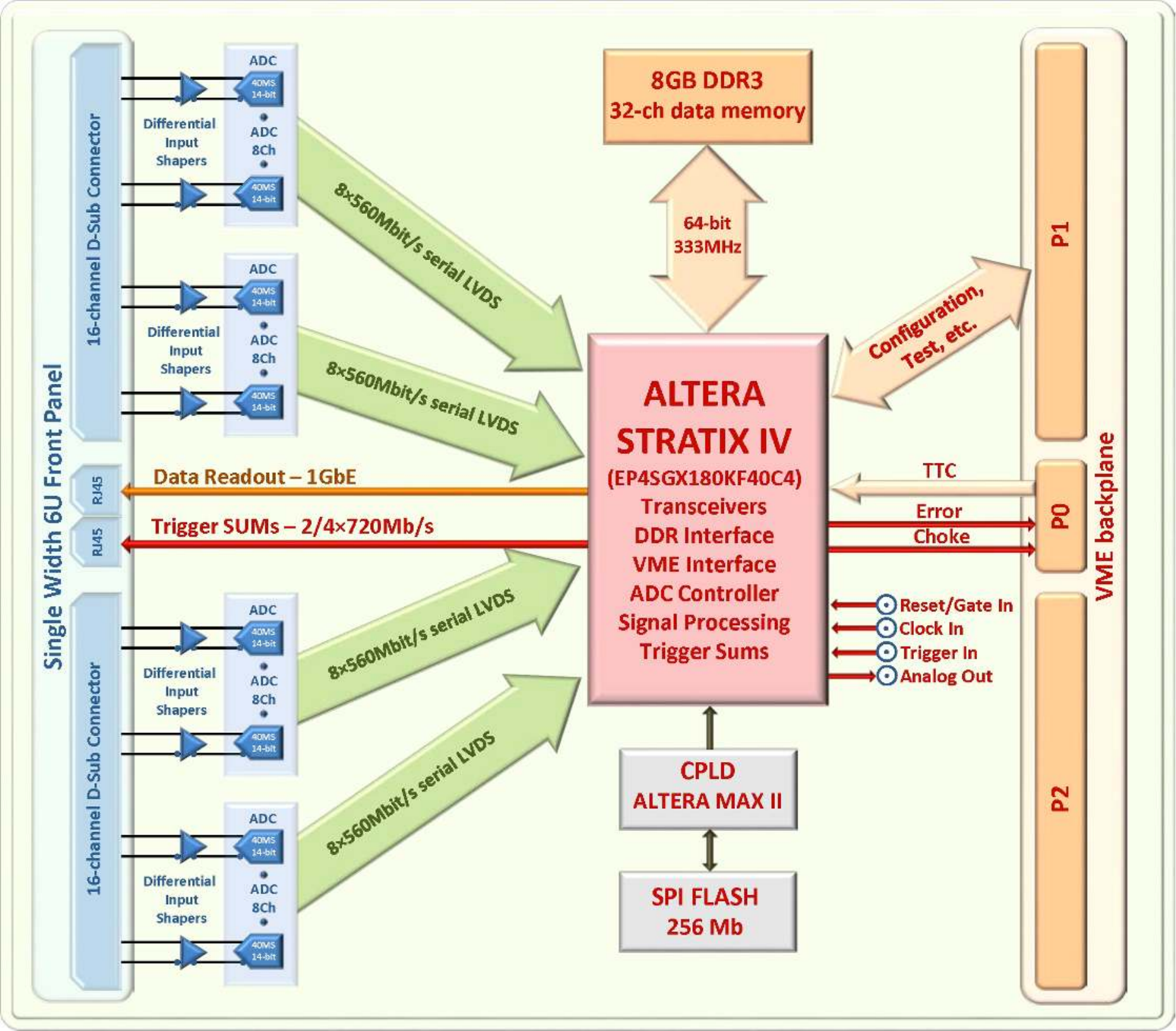}
\end{center}
\caption{\label{fig:CREAM-Picture2} {CREAM module block diagram.}}
\end{figure}

The CREAM block diagram is shown in \Fig{fig:CREAM-Picture2}.  A crate can host up to 19 CREAMs housed in 8 racks located above the LKr cryostat; a total of 432 CREAMs are used to 
read all LKr cells, requiring 28 VME crates.

The CREAM configuration is provided via the VME bus. A PC equipped with a four-channel optical link PCI express 
card can control up to 8 daisy-chained VME bridges per link, thus handling the configuration of all CREAMs. 
The board firmware can be updated through either the VME backplane or JTAG.

\begin{figure}
\begin{center}
\includegraphics[width=1.0\linewidth]{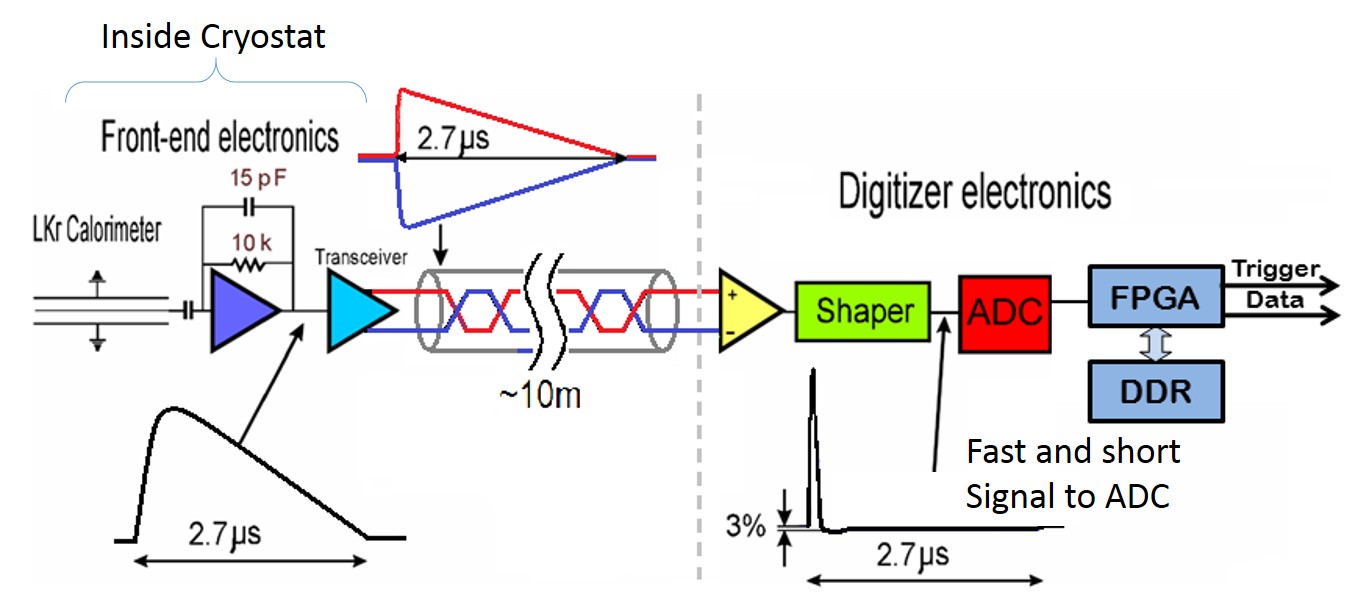}
\end{center}
\caption{\label{fig:CREAM-Picture3} {Schematics of the front-end electronics inside the cryostat, unchanged with respect to NA48. 
The shaping and digitizing is done in the CREAM.}}
\end{figure}

\paragraph{Analogue signal processing:}
Signals from the calorimeters arrive at the CREAM on differential lines, a total of 32 channels per CREAM module.
The signals are AC-coupled and have a maximum amplitude of $\pm$1~V; their shape is triangular, with 20 ns rise-time 
and 2.7 $\mu$s total duration \cite{Gianoli:2000, DeLaTaille:1996}. \\
Before being sent to the ADCs, the signal is shaped by first differentiating it with a 20~ns time constant and then 
by feeding it into a  Bessel filter (\Fig{fig:CREAM-Picture3}). The result is a 70~ns FWHM 
pseudo-Gaussian shaped signal, followed by an undershoot at about 3\% of the pulse amplitude which lasts for a time  equal to the drift time of the electrons across the calorimeter cell.

\paragraph{Signal digitization:} 
The shaped signals are digitized by a FADC using the 40 MHz clock distributed by the TTC-LKr 
module as sampling clock. The FADC\footnote{AD9257 by Analog Devices \cite{adc}.} is a 14-bit, 2~V peak range, 8-channel serial-output device; 4 such chips, placed on the daughter-board, are required to digitize all the input signals of a CREAM 
module. The baseline of each channel can be set by a 16 bit DAC to fully exploit the available 
dynamic range; during normal data-taking the baseline of each channel will be set at about 400~ADC counts. 
Samples from each chip are then serialised and sent to a FPGA\footnote{Altera Stratix IV EP4SGX180 \cite{altera}.} 
mounted on the motherboard. Here data are continuously copied on a circular 256~Mbit buffer implemented in a 8~GB 
storage capacity DDR3 SODIMM module. The maximum latency allowed by this buffer is 12.5~ms, well above the maximum 
L0 trigger latency.
 
\paragraph{Trigger signal processing and output data:}
The CREAM data flow is shown in \Fig{fig:CREAM-Picture2}. When a CREAM receives a L0 trigger signal through 
the custom backplane, a configurable number of samples (up to 256) is extracted from the circular buffer at fixed, 
but configurable, latency and copied into the 255 $\times$ 256~Mbit wide L0 buffer, implemented in the DDR3 module, which 
can contain up to 16 seconds data-taking (longer than the SPS spill duration) at the nominal L0 trigger rate of 
1 MHz for 8 extracted samples per channel. 
Each event is uniquely identified by an event number and a timestamp, associated with each L0 trigger. 
The CREAM firmware allows the management of several trigger types according to the trigger type word received after each 
L0 trigger through the backplane; different actions could be performed for different trigger types.

A L1 trigger decision is notified by the PC farm to the CREAM through a UDP Multi-Request Packet (MRP) sent to the 
Data Link (DL), a 1 Gb/s Ethernet connection on the front panel. Each MRP can contain up to 100 L1 trigger requests, 
each of them specifying the event number of the requested event. CREAM data corresponding to the requested event 
number are sent as a Sub-Detector Event (SDE) UDP packet to the PC farm through the same link, being the IP address of the 
requesting PC  automatically retrieved from the MRP by the CREAM firmware. 
Data from the DLs of the 16 CREAMs in the same crate are collected by a switch which routes all the data packets to 
the PC farm through a 10~Gb/s optical link. 

MRPs are sent to the CREAMs in multicast mode, provided that the CREAMs have previously joined a proper multi-cast 
group by issuing a IGMP packet.\footnote{Internet Group Management Protocol is a standard network protocol for the 
management of multicast groups.} This feature reduces the network traffic (request packets are distributed 
at switch level to all CREAMs) and also allows an easy way to define LKr regions of interest: different 
calorimeter regions can join different multicast groups, thus requiring only a part of the LKr data on the basis of 
data used for the L1 trigger decision.

Besides the normal data acquisition mode, in which data from all channels are sent to the PC farm when a L1 request 
reaches the CREAMs, other acquisition modes can optionally be activated:
\begin{itemize}
\item Continuous mode: 65536 samples (corresponding to about 1.6 ms data-taking) per channel are sent to the DL 
when the acquisition is activated; 
\item L0 readout mode: Packets are sent to the DL when a L0 signal is received through the custom backplane; 
\end{itemize}

In both the standard and the L0 readout modes, it is possible to activate a channel-based zero suppression 
algorithm, to reduce the amount of data to be read. The minimal condition for keeping channel data 
is to have at least one sample above a fixed threshold; since this is sensitive to pedestal variations, an improved 
algorithm was implemented, requiring the difference of the larger and smaller samples to be higher than a fixed value.

In addition to the lines used to deliver the TTC information to the CREAMs, the custom backplane also hosts 20 CHOKE and 20 ERROR single-ended CMOS lines, one for each CREAM slot. 
Each line can be driven high to indicate the presence of  a condition: 
the CREAM module memory is overloaded with data and approaching a situation in which triggers cannot be served without losing data (CHOKE) or  data are lost (ERROR). 

\paragraph{Trigger Sum Link (TSL):}
The information from CREAMs cannot be fully exploited to take a L0 trigger decision, because the bandwidth required 
to move the corresponding data would be too large; even sending all LKr data to the PC farm after a L0 trigger would require about 2~Tbit/s bandwidth. 

On the other hand, the information from the calorimeters is fundamental to obtain the rejection power required by 
the NA62 trigger system especially at the L0 trigger level. For this reason the sums of the digitized signals 
from sets of 4 $\times$ 4 cells (super-cells) are computed by the CREAM firmware every 25~ns and sent to the Cal-L0 
processor, described in \Sec{sssec:Cal-L0}.  

The sample value from each channel is first individually baseline-subtracted and gain-adjusted to take into account the differences between channels (mainly due to different pre-amplification factors and different gains of the ADC channels). The sum is then computed and the result serialized  and sent out through a connector on the front panel  on the TSL. Two such sums are produced by each CREAM and serialized on two individually shielded pairs of an Ethernet cable. 
The 16-channels sums use a 12/16~bit encoding scheme in the serialization, allowing a resolution of 56~MeV on the super-cell energy.

\begin{figure}
\begin{center}
\includegraphics[width=1.0\linewidth]{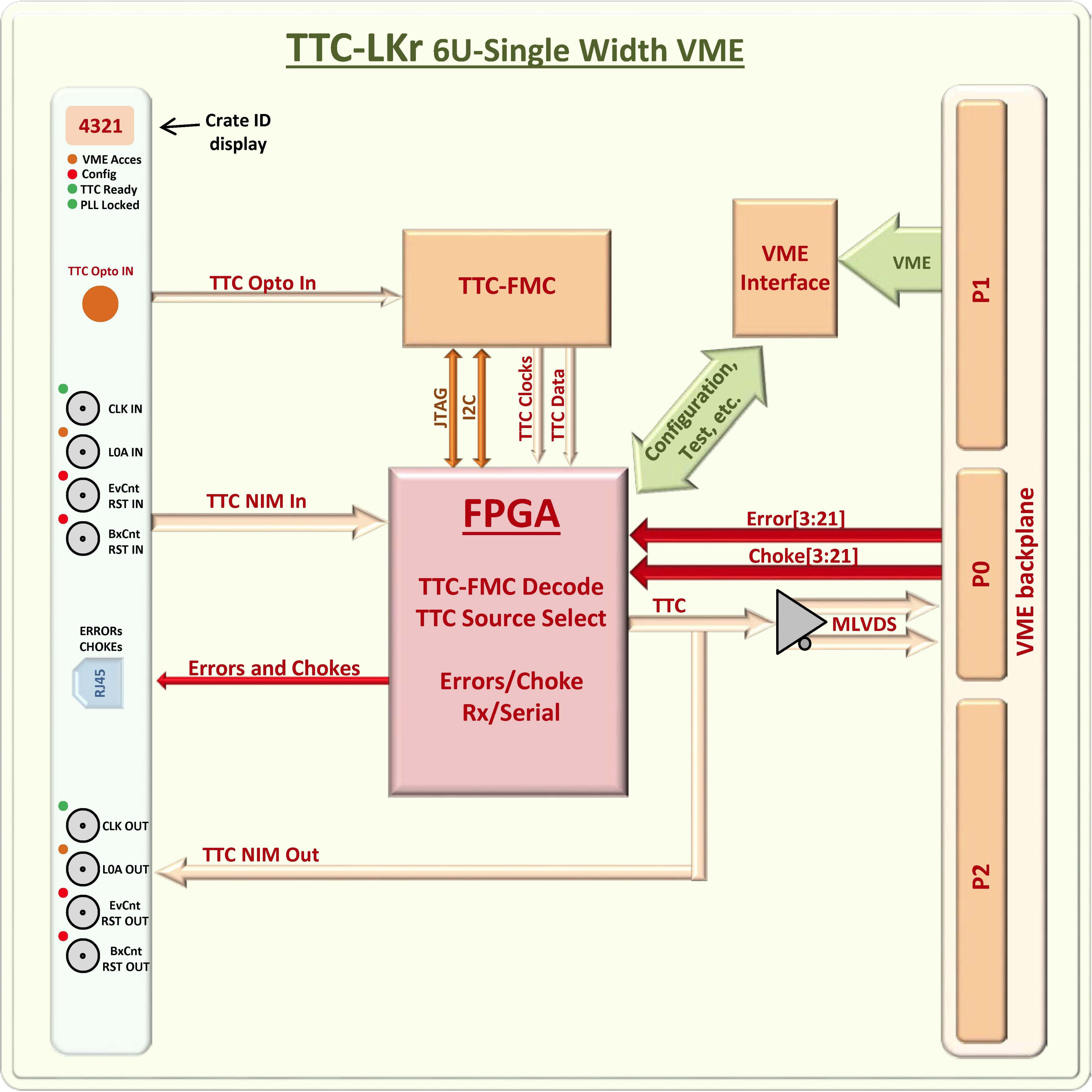}
\end{center}
\caption{\label{fig:CREAM-Picture6} {Front face and block diagram of the TTC-Cal  board.}}
\end{figure}

\paragraph{TTC-Cal board:}
The TTC-Cal board (\Fig{fig:CREAM-Picture6}) receives the clock and the L0 trigger signal 
information through the optical TTC link \cite{TTCrq}, decodes  and distributes it to all CREAMs in a crate. 
The TTC-Cal is a 6U VME 64 board  mounting a FPGA\footnote{Xilinx Spartan-6.} and hosting a 
TTC-FMC mezzanine with a data recovery integrated circuit\footnote{Analog Devices ADN2814 \cite{ADN2814}.}. 
The board is placed in the 11th slot of each VME crate. \\
The TTC-Cal board turns the TTC information into multipoint LVDS signals and distributes them to all CREAMs using 
a custom backplane, which is also used by each of the 16 CREAMs to signal the presence of CHOKE or ERROR conditions. 
Apart from the optical source, the TTC-Cal board can be configured to provide the clock and L0 trigger signal through 
a front panel input, the VME bus or an internal generator. 

\subsection{Level 0 hardware trigger}\label{ssec:L0}
\input{Sec13_L0MUV3_v1.tex}

%%%%%%%%%%%%%%%%%%  L0 calorimeter %%%%%%%%%%%%%%%%%%%%%%%%%%
\subsubsection{Calorimeter Level 0 trigger (Cal-L0)}\label{sssec:Cal-L0}
The Level 0 Calorimeter Trigger (Cal-L0) identifies clusters from the electromagnetic (LKr, IRC and SAC) and hadronic (MUV1 and MUV2) calorimeters, 
readout via the CREAM system. 
The Cal-L0 prepares time-ordered lists of reconstructed-clusters together with their times, positions, and energies. 
Cluster search is performed in parallel for each of the five detectors, to allow 
trigger primitive generation based on complex energy and cluster multiplicity combinations 
\cite{Fucci:2011,Bonaiuto:2013}.

\begin{figure}[h]
\begin{center}
\includegraphics[width=1.\linewidth]{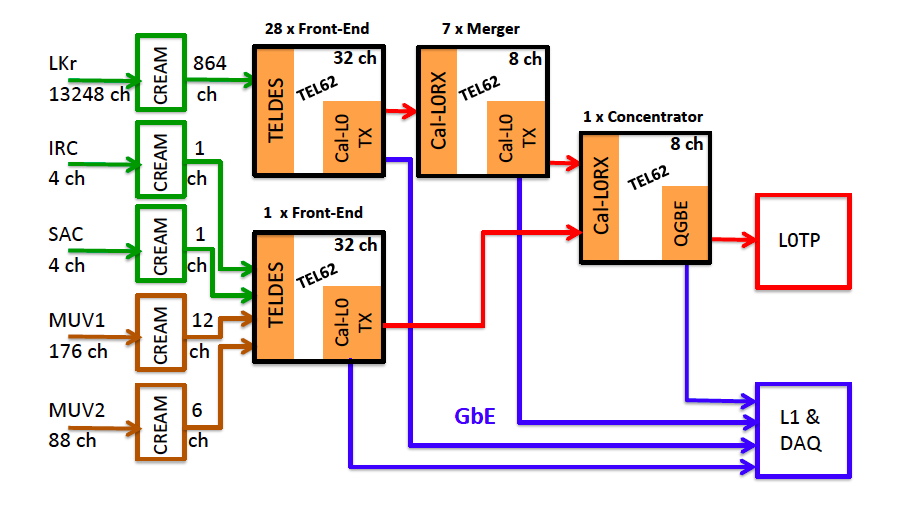}
\end{center}
\caption{\label{fig:CalL0-combi} {Schematic diagram of the Cal-L0 trigger system implementation.}}
\end{figure}

The Cal-L0 is composed of Front-End, Merger and Concentrator boards 
(Figure \ref{fig:CalL0-combi}), all based on 
the TEL62 boards (\Sec{ssec:TEL62}) complemented with custom mezzanine cards. 
The basic functions of the three stages are:
\begin {itemize}
\item Front-End boards receive trigger sums from CREAM modules (\Sec{ssec:Cal-readout}), perform 
peak searches and compute time, position and energy for each detected peak.
\item Merger boards (only for LKr) receive trigger data from multiple Front-End boards and merge peaks into single clusters.
\item A single Concentrator board receives reconstructed-clusters from the five calorimeters, performs cluster counting, 
computes sums of electromagnetic and hadronic energies and generates trigger primitives.
\end {itemize}

Cluster search in the LKr is performed in two steps, acting on vertical slices of the calorimeter. 
In the first step peaks in space and time are searched independently in each slice with a uni-dimensional algorithm;
in the second step peaks which are close in time and space are merged and assigned to the same electromagnetic 
cluster. \\
Cluster search in the other calorimeters is performed in a single step with a one-dimensional algorithm: the MUV1 
detector is divided in 6 vertical and 6 horizontal slices, and MUV2 in 3 vertical and 3 horizontal slices, while 
IRC and SAC are read out by a single channel each (\Tab{tab:CREAM-table1}).

The system also provides a coarse-grained readout of the calorimeters that might be used in software trigger 
levels: two different kinds of data can be readout from the Cal-L0 upon reception of a L0 trigger: raw data from 
Front-End boards, and reconstructed clusters information (time, position and energy) from Merger and Concentrator 
boards. The available output bandwidth is sufficient to read out reconstructed clusters data, while reading of raw 
data requires some data compression stage.

The system is composed of 37 TEL62 boards and 111 mezzanine cards used for data transfer, housed in three 9U crates.
A more detailed description of the individual boards follows.

\begin{figure}
\begin{center}
\includegraphics[width=0.8\linewidth]{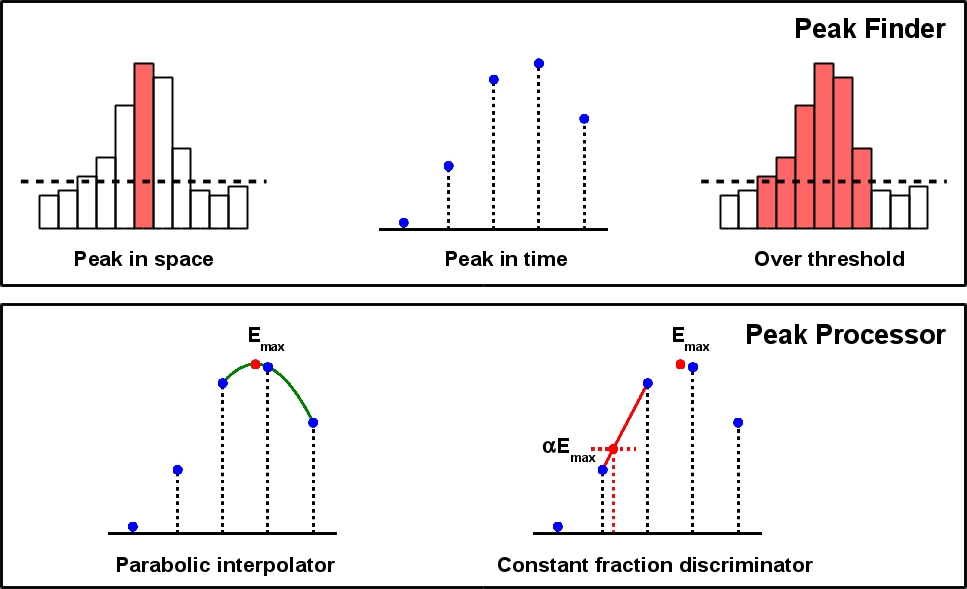}
\end{center}
\caption{\label{fig:CalL0-Picture3} {
Five step peak-finding and reconstruction in the Front-End board.}}
\end{figure}

\paragraph{Front-End board:} 
The Front-End board receives up to 32 trigger sums from CREAMs, aligns them in time and performs the peak search 
algorithm \cite{Badoni:2014}, executed in parallel 
in the following steps (\Fig{fig:CalL0-Picture3}):
\begin{itemize}
\item
peak search in space, as defined by the following condition: \\ 
\mbox{$ E_{i-1} [n] \le E_i [n] $ AND $ E_{i} [n] \ge E_{i+1} [n]$}, where $E$ indicates the ADC count, $i$ the 
super-cell number and $n$ the sample number (timestamp);
\item
peak search in time, as defined by \mbox{$E_{i} [n-2] < E_i [n-1] \le E_{i} [n] $ AND  $ E_{i} [n] \ge E_i [n+1]$};
\item
threshold check: $E_{i} [n] > E_{th}$;
\item
parabolic energy interpolation using samples $n-1$, $ n$ and $n+1$ to obtain the peak value 
$E_{max}$;
\item
linear time interpolation in the interval between subsequent samples $ n-2$ and $n-1$ to determine the crossing time of a 
threshold corresponding to a programmable fraction of $E_{max}$.
\end{itemize}

Times, positions and energies of reconstructed peaks are transferred to the Merger boards (for the LKr) or to the
Concentrator board (for IRC, SAC, MUV1 and MUV2 calorimeters).
Raw data received by the CREAMs are also stored in memories, so that they can be read out after a L0 trigger is 
received. 

The 28 Front-End boards of the whole LKr calorimeter and the single board of all other calorimeters were installed,  each equipped with two TELDES and one TX mezzanines (see next paragraphs).

\begin{figure}
\begin{center}
\includegraphics[width=0.5\linewidth]{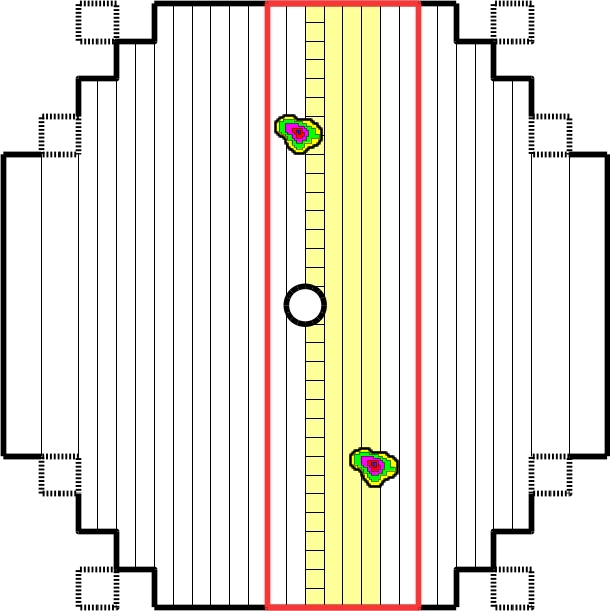}
\end{center}
\caption{\label{fig:CalL0-Picture4} {Cluster reconstruction in the LKr: each Merger board receives reconstructed peaks from a vertical section in the calorimeter and merges 
peaks which are close in time and in space. To avoid double counting, clusters with maxima in the inner region 
(yellow) are reconstructed by the Merger itself (e.g. the cluster in the right bottom part of the figure) while 
other clusters are reconstructed by neighbouring Mergers (e.g. the cluster in the upper part of the figure).}}
\end{figure}

\paragraph{Merger board:}
Each Merger board (LKr only) receives data from up to 8 Front-End boards, covering a region of 8 super-cells along 
the horizontal axis and 32 super-cells along the vertical axis. 
The region covered by a Merger board is divided in an inner and an outer region: only clusters with a maximum along 
the vertical axis in the inner region are managed by the Merger board, to avoid cluster double-counting (\Fig{fig:CalL0-Picture4}).
Reconstructed clusters are also stored in memories for readout after a L0 trigger.

Seven Merger boards are installed for the LKr, each equipped with two RX and one TX mezzanine (next paragraphs).

\paragraph{Concentrator board:} 
The Concentrator board receives time-ordered lists of reconstructed clusters from 7 Merger boards (for LKr) and from one Front-End board (for other calorimeters), performs the trigger algorithm and delivers time-ordered trigger primitives to the L0TP via Gigabit Ethernet.

\begin{figure}
\begin{center}
\includegraphics[width=\linewidth]{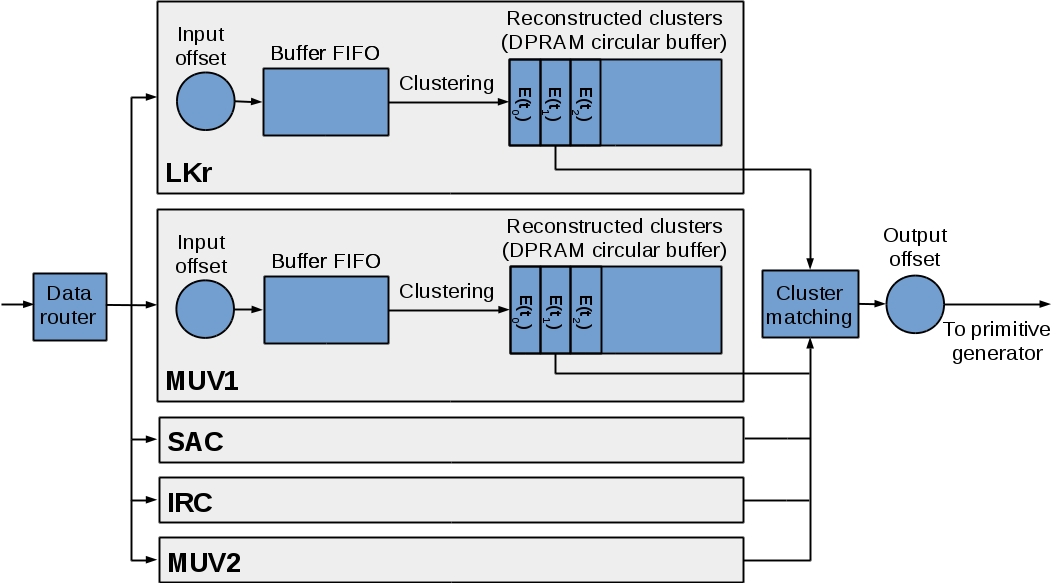}
\end{center}
\caption{\label{fig:CalL0-Picture5} {Implementation of the  Concentrator board FPGA including five Dual Port RAMs (one per calorimeter) each containing time-binned reconstructed clusters.}}
\end{figure}

The main building block of the Concentrator board is a Dual Port RAM (DPRAM) used as a circular buffer \cite{Ammendola:2015}: it contains  a time-binned reconstructed cluster histogram, with each memory cell corresponding to a time interval and the stored 
value representing a reconstructed cluster (with its associated energy, position and time). The bin size can be 
adjusted, with a minimum value of 6.25 ns. Each memory has a depth of 16384 locations and can thus store data 
related to a time interval of  102.4 $\mu$s or more. The timestamp and the fine-time of the data are 
used to address the DPRAM (\Fig{fig:CalL0-Picture5}).
The Concentrator hosts eight DPRAMs: seven for the LKr boards and one for the other calorimeters. 
The first received data packet of the burst is written in the middle of the corresponding memory, and this sets the 
reference time for the memory locations of all buffers. Half of the memory (51 $\mu$s or more) is enough to order 
incoming data packets in time. All eight buffers are read simultaneously starting from the beginning, and the output 
data are sent to the primitive generation logic, based on programmable energy sums and cluster counting.

\paragraph{Input/Output mezzanines and data transfer:}
In order to allow point-to-point, low latency, high-throughput data transfer between different parts of the Cal-L0, 
four custom mezzanine cards are used: the TELDES, the Cal-L0-TX, the Cal-L0-RX and the QGBE.

The TELDES (TEL62 DESerializer) \cite{Checcucci:2015} receives serialized trigger sum signals (16~bits at 40~MHz) 
from up to 8 CREAM modules, equalizes and de-serializes them, making them available to the Front-End board FPGAs.
All 58 TELDES mezzanines were installed.

The Cal-L0-TX transmits trigger data between different parts of the system on custom low-latency links. 
The mezzanine is based on an high-end FPGA\footnote{Altera Stratix II EP2S60.} connected to two 48-bit serializers. 
Each TEL62 equipped with such a mezzanine can transmit trigger data to two other boards at a rate of 3.5~Gb/s. 
A single-error correction, double-error detection protocol and a standard 23-bit pseudo-random binary sequence 
test system have been implemented in the Cal-L0-TX FPGA. 
An additional connector mounted on the Cal-L0-TX mezzanine will allow the connection of other cards for future use, 
such as a Gigabit Ethernet card foreseen for the readout. In total 36 Cal-L0-TX mezzanines were installed.

The Cal-L0-RX mezzanine receives trigger data from the Cal-L0-TX card, de-serializes them and delivers 48-bit word trigger data to the Merger or Concentrator board hosting it.
The mezzanine is equipped with 4 48-bit deserializers. Each TEL62 equipped with two Cal-L0-RX mezzanines can
receive trigger data from 8 different Cal-L0-TX mezzanines for a total of 16 mezzanine cards.

The standard QGBE mezzanine (\Sec{ssec:TEL62}) is mounted on the Concentrator board and used to send L0 trigger primitives to the L0TP.

\subsubsection{Level 0 Trigger Processor (L0TP)}\label{sssec:LTP}
The main function of the L0TP is to acquire trigger primitives, to sort them in time and to 
search time aligned matches with any of the active trigger masks. 
The time alignment is based on information contained in the primitive data, namely a 25~ns precision time-stamp 
(32~bit) and a 100~ps precision fine time. The primitives are sent to the L0TP as UDP packets over Gigabit Ethernet links, in periodic bunches every 6.4~$\mu$s, thus providing a coarse synchronisation.
The trigger processor verifies the relevant matching conditions taking into account the time resolution of each 
source, as measured at the trigger level. When a set of primitives matches one of the allowed masks, a L0 trigger 
signal is generated by sending a  synchronised TTC signal with 25~ns time precision and a fixed specified latency with respect to the event occurrence time.
Due to the requirements of the TTC trigger transmission protocol, two consecutive L0 triggers must be separated by at least 75~ns. 

Two L0TP versions were developed for NA62: one purely FPGA-based, and a second one where trigger primitive processing 
is performed in a PC. Both versions share a common multiple Gigabit Ethernet receiver board, which is based on a 
commercial FPGA evaluation board.\footnote{Terasic DE-4, with an Altera Stratix IV FPGA on board.}
The purely FPGA-based L0TP was used in data taking for triggering, while the PC-based version was used so far in 
parasitic mode, receiving primitive data from the same sources for monitoring and control purpose.  

The FPGA-based L0TP performs trigger primitive processing entirely in hardware. 
It receives packets with multiple primitives via Ethernet links, unpacks them and then stores primitives in the 
FPGA memory for the time alignment using the time information to generate the memory address. 
This results in primitives from different sub-detectors within a given time window written to the same memory address.
The memory is circular, with programmable time binning depending on the number of time bits used. 
The smallest possible time bin is 3.125~ns, achieved by using 11~bits for the time-stamp and 
the 3 most significant bits of the fine time; the available depth of the memory in this case is 51.2~$\mu$s, 
corresponding to the maximum possible time difference between the latencies of different primitive generators. 

After time alignment, the algorithm determines the trigger conditions (primitive identifiers) which were 
simultaneously satisfied within the programmable time window: matching primitives must appear in the same or adjacent 
memory slots, but during the reading process a more accurate time-matching cut, different for each source, can be 
performed using the entire information of the fine time.
The time-matching primitives are compared with those of a number of programmable reference masks to take a trigger 
decision. If a trigger mask is satisfied, a programmable downscaling can be applied to the generation of a L0 trigger. 

While the checking of the time matching condition always starts from a seed (positive primitive) determined by a 
single pre-defined "reference" detector, a secondary system allows the generation of triggers which do not require the 
presence of such detector, thus allowing the collection of  sets of independently-triggered events for efficiency 
measurements: this is achieved by issuing triggers according to a downscaled stream of selected incoming primitives 
from one sub-detector. 

The fixed L0 trigger latency is obtained by storing each trigger into a dual-port 800~$\mu$s deep circular buffer which is read in a sequential way starting after such fixed delay. 
The time-matched primitives which match a trigger mask are stored in a data buffer and eventually sent to the PC farm 
for recording together with the other event data, and an 8-bit trigger word is delivered to all sub-detectors 
together with the L0 trigger signal, thus allowing for selective readout according to the trigger type. 

The FPGA-based L0TP was used throughout the 2015 run for providing L0 triggers originated by different sub-detector primitives.

In the PC-based solution, the 8xPCIexpress (PCIe) Gen.2 bus available on the FPGA evaluation board is used to 
deliver primitive data into the RAM of a PC running a standard Linux OS.\footnote{Intel Core i7-4930K with 3.40~GHz clock running Scientific Linux 6.} The system can exploit the entire PC memory, thus allowing the storage of all  primitive information for the whole length of the spill.
The primitive receiving part is the same as in the previously described implementation, and on the software side, 
a process running in user space on the CPU reads primitives from memory and, by using a single-threaded algorithm, 
aligns them in real time, by looking for overlapping time windows.
After time alignment, primitive IDs are compared with pre-set masks and selections are applied.
Positive trigger decisions are then shipped back to the FPGA via PCIe, and the fixed latency delay is applied before transmission to the TTC system for broadcasting.
Since all primitive information is accessible to the matching algorithm, it can use all time-stamp and fine-time 
bits, resulting in a time matching granularity up to the maximum precision of 100~ps. 
Studies were carried out to measure the time needed to take the trigger decision with realistic primitive 
generators and input rates up to 14~MHz, showing that the trigger processing time can be kept at all times well below the maximum available latency time of 1~ms.

Back-pressure and system integrity control are achieved by using two dedicated lines per sub-system, called CHOKE and ERROR, used by each system to signal anomalous conditions to the L0 Trigger Processor. 
Such signals actually reflect the status of all boards in a sub-system through appropriate multiplexing.
The assertion of a CHOKE line indicates that a sub-system is approaching a critical situation, due to a data flow above its capabilities, in which it would not be able to handle properly the L0 triggers.  
The L0 Trigger Processor responds to such condition by stopping the delivery of L0 triggers and signalling such occurrence to all systems through a special L0 trigger, which must be acknowledged. 
When the overload condition disappears the CHOKE line is driven low and the L0TP restarts trigger dispatching, again after notifying all sub-systems through another  special L0 trigger. 
The CHOKE mechanism is not  meant as a flow control system, but as a protection against  undetected TDAQ inefficiencies. The ERROR line is similarly driven by any sub-system when a condition was reached  in which data was actually lost; also in this case L0 triggers are paused and all sub-systems are notified.

\subsubsection{GPU-based Level 0 trigger}\label{sssec:GPU}

The possible use of GPUs (Graphics Processing Units) at L0 trigger has been under investigation in NA62 since its conception, and actually drove the choice of a large L0 trigger maximum latency \cite{Collazuol:2012}. 
This requires the RICH TEL62 boards to continuously dispatch the entire set of hits, arranged in a highly compressed format and packed together in periodic packets, to one or more GPU cards through a communication channel with low and highly predictable latency. GPUs can run much more complex and flexible trigger algorithms than FPGAs, and thus produce more elaborate L0 trigger primitives. 

In the present case under study  the GPU-based RICH trigger collects all hits related to the RICH detector for the GPUs to perform ring identification and determination of its centre and radius.  
This trigger was tested parasitically on half of the RICH channels (corresponding to a single PM array), receiving event data from dedicated Gigabit Ethernet ports of the main RICH TEL62 boards using a custom FPGA-based network interface card called NaNet.
NaNet implements a direct data transport mechanism from the network channel to the GPU memory (Remote Direct Memory Access and GPUDirect peer-to-peer capabilities), which avoids any intermediate data buffering and host PC intervention on the time critical data path \cite{Lonardo:2016}. 
The fitter stage, running on a graphics processor card\footnote{Currently an nVIDIA K20c GPU with 2500 cores providing 3.5 Teraflops of computing power.} is based on an algorithm specifically developed for this purpose. 
The measured average computing time per event (on a single GPU) is  
of $\cal{O}$(350~ns), matching the throughput required for a 3~MHz event-rate on the RICH \cite{GAP:2016}.
A new board featuring four 10 Gb/s links, called NaNet-10, will be used in the  
next data taking to overcome the input bandwidth limitation of the NaNet board; this will allow the handling of  all RICH channels, with a further reduction of the latency and the possibility to offload the serial part of the data pre-processing (data merging and decompression) on the more performant  FPGA \cite{Ammendola:2016}.

\begin{figure}[ht]
\begin{center}
\includegraphics[width=0.85\linewidth]{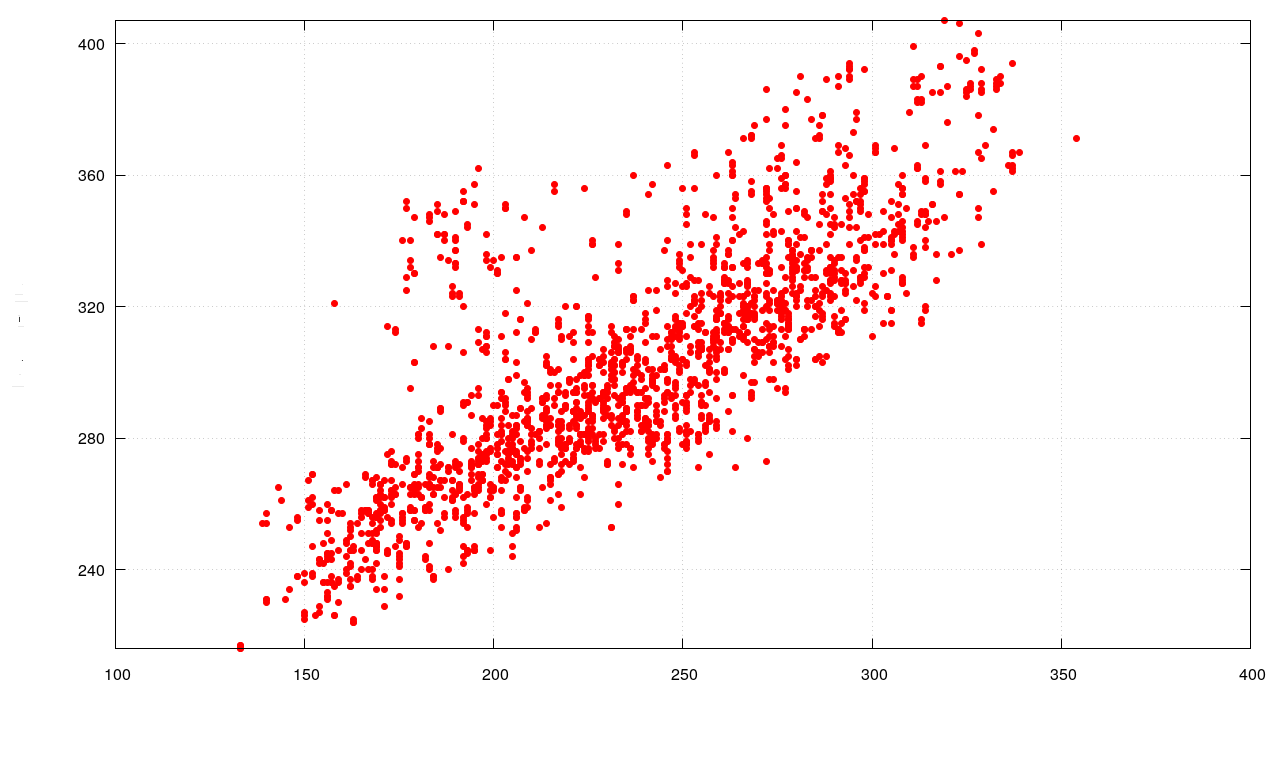}
\put(-195,0){Events collected per GPU receiving buffer}
\put(-360,145){\rotatebox{90}{Total time [ $\mu$s]}}
\end{center}
\caption{\label{fig:TDAQ-Picture2} Latency (total processing time) as a function of the number of events collected in about 400 $\mu$s in a GPU receiving buffer and then sent to the K20c nVIDIA GPU.}
\end{figure}

Figure \ref{fig:TDAQ-Picture2} shows results obtained with a beam intensity of $4 \times 10^{11}$ protons 
per spill during the 2015 beam period. The test showed that high quality primitives for the RICH can be produced within 
the latency required by the trigger system. Highly selective trigger conditions would be implemented to increase 
the purity of the online selection and optimize the readout bandwidth. 

\input{Sec13_HLT_v1.tex}

%% file: Sec13_STRAW_v2.tex
The on-detector electronics for the straw tracker consists of an 8-channel analogue front-end chip containing a 
fast pre-amplifier, semi-Gaussian shaper, ions-tail cancellation circuitry, base line restorer and discriminator. 
Ions-tail cancellation is of great importance for straws at high particle rates, as pile-up would cause loss of both 
efficiency and time resolution. The CARIOCA  readout chip \cite{Bonivento:2003}, developed for the
LHCb muon chambers, is used. 
Both leading and trailing edges from the straw signals provide useful information. 
The leading edge time depends on the particle track distance from the wire and provides precise radial crossing 
position of the particle in the straw. The trailing edge occurs at a fixed delayed time with respect to the particle 
crossing time, independently of the crossing distance from the wire, corresponding to the arrival of the last 
primary ionization cluster close to the straw wall; the trailing edge time can therefore be used as a validation of 
hits belonging to the same track, thus reducing false hits and improving track reconstruction. 
The straws are equipped with TDCs measuring both leading and trailing edges. The front-end electronics modularity 
follows the arrangement of straw tubes. 
There are 16 straws in one basic cell, which share common gas, high voltage and readout connectivities.
A front-end board thus contains two CARIOCA chips, each serving 8 straws; one I\textsuperscript{2}C-controlled chip 
with 16 DACs for setting the discriminator thresholds, and one FPGA which contains 32 TDCs with a time bin of 0.78 ns.
Standard halogen-free category 6 Ethernet cables are used for transmitting the data and controlling the front-end boards.

The front-end board is also used as a cover for the gas operation, so it was built gas tight, always using multiple blind vias for passing the signals from inside to outside.  
HV power is supplied using pass-through vacuum-tight HV connectors. A picture of a cover is shown 
in \Fig{fig:STRAW-Picture5}.

The Back-End of the readout manages up to 16 FE-cover boards in a custom-made Straw Readout 
Board (SRB), 
which is configured, controlled and monitored via a VME-based Single Board Computer (SBC).

One chamber (four views) is served by 8 SRBs housed in one VME 9U crate, positioned about 10-15 meters from the detector.  
The Back-End boards receive precise system clock, timing information 
and control from the common NA62 TTC system \cite{Taylor:1998}, and time-align the data from the straws by
attaching the required time-stamps. The SRB board provides fine tuning of clock delays for timing of the detector. 
Adjustable delays are needed due to the spread of propagation time through components and cables, and
time-of-flight of particles along the beam.

The incoming 8/10 bit-encoded, 400 Mb/s links are received in the SRB Front-End Interface (FE\_INTF) FPGA,   
which handles 4 cover links, each containing a clock line, a control line and two data read-back lines. 
One data line is used for parameter read back, but in case a higher data bandwidth were needed it can be 
reconfigured for use in data transmission.
FE\_INTF formats and reorganizes received data and sends them to VME Board Manager FPGA (VME\_BM) and Event Manager 
Interface FPGA (EM\_INTF). Serial 2.5 Gb/s links are used for data transmission. The control and parameter loading 
commands are received from VME\_BM and passed to the corresponding cover board.

The EM\_INTF FPGA receives four 2.5 Gb/s links, one from each FE\_INTF. The data are de-randomized in FIFOs and 
multiplexed to time-reordering and trigger-matching circular buffers. 
The buffers can accommodate data for up to 1.5 ms, exceeding the maximum L0 trigger latency. 
Trigger selected data are sent via Ethernet interface to the NA62 PC farm, while data out of the time window are discarded.

The VME\_BM FPGA takes care of communications through the VME bus, 
the loading of configurations and parameters both for SRB and connected covers, and the storage
of raw and trigger selected monitoring data. It receives four 2.5 Gb/s links, 
one from each FE\_INTF, and stores data in DDR3 based buffers. The readout is performed via VME bus and controlled 
by the VME SBC.

%% file: Sec13_L0MUV3_v1.tex
The L0 hardware trigger is meant to filter the events based on inputs from a small set of fast sub-detectors and 
has a maximum design output rate of 1 MHz and a maximum latency of 1~ms. 
In its starting implementation, the participating sub-detectors are:

\begin{itemize}
\item
  \textbf{CHOD}, providing positive primitives for any charged track based on hit multiplicity and pattern and a  high quality reference time;
\item
  \textbf{RICH}, providing positive primitives for any charged track above Cherenkov threshold, based on hit multiplicity;
\item
  \textbf{LAV}, providing photon (and halo muon) vetoing primitives based on multiplicities of adjacent blocks hit;
\item
  \textbf{MUV3}, 
  providing muon primitives based on tile multiplicities, used both in positive and vetoing logic;
  \item
  \textbf{Calorimeters} (LKr electromagnetic calorimeter and MUV1, MUV2 hadron calorimeters), providing positive identification 
  for pions based on energy deposits and vetoing primitives based on LKr calorimeter cluster multiplicity.
\end{itemize}

For CHOD, RICH, LAV, and MUV3 the L0 trigger primitives are generated by TEL62 boards equipped with TDCs, as described in \Sec{sssec:TEL62L0}. For the calorimeters the primitives are produced by processing the continuous ADC sampling from the CREAM readout system, as described in \Sec{sssec:Cal-L0}.
Trigger primitives are generated asynchronously in a variable time, currently not exceeding 100 $\mu$s. 
Each L0 trigger primitive consists of a 64-bit data block\footnote{To be possibly reduced to 32-bit in the 
future to allow for higher primitive rates without exceeding bandwidth limitations.} containing sub-25 ns time information and an identifier which indicates which conditions are satisfied at that time.
If different conditions are satisfied at times with a difference comparable to the sub-detector time resolution, 
the primitives  are merged into a single one. The dispatching of L0 trigger primitives also occurs asynchronously over dedicated 
Gigabit Ethernet links from the TEL62 boards.

All systems must necessarily reply with some data to every L0 trigger received, thus providing a tight coupling of 
all readout systems which allows a strict checking of data acquisition integrity.
Exceptions are the GTK readout system (\Sec{ssec:GTK-readout}) and the CREAM system used for calorimeters (\Sec{ssec:Cal-readout}) whose large data fluxes do not allow readout at the L0 trigger rate; 
these systems save data into local temporary buffers upon reception of a L0 trigger, but actual readout of such data is postponed to when a L1 trigger is dispatched, at a ten times lower rate.

\subsubsection{TDC-based Level 0 Trigger primitives}\label{sssec:TEL62L0}

All TDC-based L0 triggers are based on the identification of groups of hits belonging to the same event, which 
require time clustering to produce the trigger primitive. Such an operation requires hit sorting, which is a 
time-consuming operation whose time requirement scales with the number of hits.
While the first generation of L0 trigger algorithms worked by processing 6.4 $\mu$s long data frames, as they are 
produced by the TDCs themselves, recent implementations exploit the time sorting performed in the common 
TEL62 firmware for formatting the data into the 25 ns long time slots used for storage.
The trigger algorithms are significantly simpler if fed with already sorted data, with a significant gain on their sustainable input rates.

\paragraph{CHOD L0 trigger:} 
The CHOD L0 trigger is based on sets of hit-multiplicity ranges  
with the goal of providing primitives compatible with either single-track or three-track events. 
The limited number of channels allows processing all of them within a few TDC boards belonging to the same TEL62. The clustering of hits based on time 
is performed; when using the NA48-CHOD with long scintillator bars, 
in order to exploit the good intrinsic time resolution, an impact point time correction due to the light propagation 
delay is required; for a single track the impact point is determined directly by the two elements, horizontal and 
vertical, of the hodoscope, but the correction becomes increasingly complex and time demanding in case of high hit 
rates. Different versions of the algorithm were developed to cope with large multiplicities at the price of a coarser 
time-correction. 

The CHOD based on scintillator tiles does not require such correction and   
enables the exploitation of 
geometric patterns of neighbouring tiles for L0 trigger primitive generation, thereby providing more selective conditions. 

\paragraph{RICH L0 trigger:} 
The RICH L0 trigger is based on hit multiplicity after time clustering. 
The large number of channels is distributed over four TEL62 boards.  Therefore,  primitive evaluation requires inter-communication among these boards using Inter-TEL cards.
An alternative solution was adopted which exploits the digital OR of discriminated PM signals 
available for each group of eight channels at the output of the pre-amplifiers in the front-end boards. 
By connecting such signals to an additional TDC-equipped TEL62 board, 
the full information from the whole detector is available on a single board, which can then evaluate multiplicities in terms of the above OR signals. 

Investigations on the possibility of generating more complex trigger primitives for the RICH are discussed in  \Sec{sssec:GPU}.

\paragraph{LAV L0 trigger:}
The LAV L0 trigger is meant to reject events with photons or muons within the acceptance; this requires using all 
LAV stations, and since each one is handled by an individual TEL62 board, inter-communication among the 12 boards 
in a daisy chain using the Inter-TEL cards is needed. 
A first implementation exploits only the LAV12 station (and therefore a single TEL62 board). 

A LAV trigger primitive is generated for events containing signals from one or more blocks that cross both 
discriminator thresholds. Signals arriving close in time from different blocks are assumed to be originated 
from the same event, so they generate a single primitive. 

As the first step, in the PP FPGAs, events are distributed among different FIFOs, one for each high- and 
low-threshold channel. While the FIFOs are filled, a finite state machine (FSM) searches for a signal crossing 
both the high and low threshold for the same block within a programmable time window, and with a 100~ps binning over a 
range of 0 to 25~ns. 
Once the association has been made, the slewing-corrected event time is computed as:
\begin{displaymath}                                                                                                                
  {\rm t} = {\rm t}_{\rm low} - \frac{({\rm t}_{\rm high} - {\rm t}_{\rm low}) \cdot {\rm t}_{\rm low}}{V_{\rm high} -V_{\rm low}},
\end{displaymath}
where $V_{\rm high}$, $V_{\rm low}$ are the high and low threshold voltages and ${\rm t}_{\rm high}$, ${\rm t}_{\rm low}$ are 
the respective crossing times. This computation is performed in the FPGA by using a computing block generated by High-Level-Synthesis technology.

The data are then transferred from the four PP FPGAs to the SL, where they are merged into a single stream. 
Hits occurring within a given programmable time window are grouped into clusters, whose times are defined as the 
average of hit times. Finally, the time values of the clusters are sorted and used to build the L0 trigger primitives 
which indicate when the LAV hit multiplicity exceeds any of four programmable thresholds.

\paragraph{MUV3 L0 trigger:}
The MUV3 L0 trigger is derived from fast signals produced by muons crossing the detector plane.  The primitives based on tile multiplicities can be used either as a positive or as a veto condition in the trigger logic.

At nominal beam intensity, the expected hit rate in the whole plane (148 tiles, \Sec{ssec:muv3}) is 13~MHz while the eight inner tiles contribute 6~MHz. In the same beam conditions, the estimated multi-muon event rate is 1.1~MHz for a 5~ns coincidence time window: 55\% of this rate is due to accidental coincidences between unrelated muons and 45\% is due to signals in multiple MUV3 tiles from single muons.  The contribution of single kaon decays to muons represents more than half of the latter rate. The expected trigger rate reduces to about 0.5~MHz if considering only the outer tiles in the  multi-muon trigger logic.

The 296 input channels are accommodated on a single TEL62 board equipped with three TDCB mezzanines.
Only the signal leading edge time measurements are used. A tight hit is produced when two signals from the same tile (one from each PM) arrive within a defined coincidence time window. 
To ensure that only the measurements of scintillation light (and not the earlier Cherenkov light due to a particle crossing the PMT window) are considered, the time of 
the tight hit is defined as the later time of the two signals. If the two signals are farther apart than the 
coincidence window, the signals are considered to be isolated and each forms a loose hit. 
Loose hits are typically due to inefficiencies and/or noise in the detector or the front-end electronics. 
The efficiency of the trigger can be optimized by using both tight and loose hits.
The hits produced using 3 PP FPGAs (corresponding to the three TDC boards) are read into the SL FPGA, where clustering of hits within a defined matching window and time sorting are performed. 
The time of each cluster is the average time of all the hits in the cluster, with tight and  loose hits treated equally. 
Along with time information, hit clusters contain metadata comprised of four counters of the number of hits, for  each of the four hit categories: loose (tight) hits in inner (outer) tiles. Each cluster is converted to a MUV3 trigger primitive by encoding the metadata in the 16-bits allocated for the primitive ID.

%% file: Sec13_HLT_v1.tex
\subsection{High Level Triggers (HLT)}\label{ssec:HLT}
The maximum L0 trigger rate is 1 MHz.  Significant subsequent rate and data reduction is required to match 
the available bandwidth for permanent data storage of $\cal{O}$(10~kHz). The NA62 TDAQ system uses two software trigger 
levels to achieve the necessary reduction:

\begin{itemize}
\item The L1 trigger reduces the data rate by a factor of 10 to a maximum of 100 kHz, with algorithms 
using standalone information from individual sub-detectors. The calorimeters (LKr, MUV1 and MUV2) 
cannot be used for the L1 decision as they are only read out following a positive L1 trigger decision.  
\item The L2 trigger reduces the data rate by another factor of $\sim$ 10, down to the allowed storage 
rate of $\cal{O}$(10~kHz). The L2 event filter is based on partially reconstructed events and exploits  correlated information from several sub-detectors.
\end{itemize}

Both trigger levels run sequentially on the same processors in the online PC farm (\Sec{sec:online}). 

\paragraph{L1 Trigger:}  The following L1 trigger algorithms were deployed  
in 2015 and were operated at 10\% of the nominal beam intensity. They were applied sequentially on top of each other. Below, ``efficiency'' denotes the number of triggers after L1 with respect to the input considered which maybe of different types, while ``data reduction factor'' corresponds to the inverse of the efficiency with respect to  all L0 triggers presented as input.

\begin{figure}[ht]
\begin{minipage}{0.5\linewidth}
\begin{center}
\includegraphics[width=1.02\linewidth]{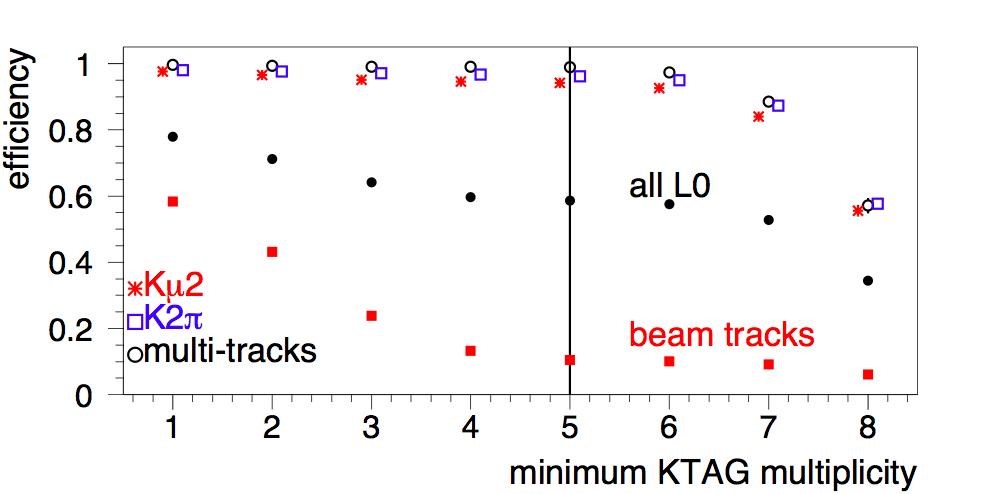}
\put(-32,87){\bf  \large (a)}
\end{center}
\end{minipage}
\begin{minipage}{0.5\linewidth}
\begin{center}
\includegraphics[width=1.02\linewidth]{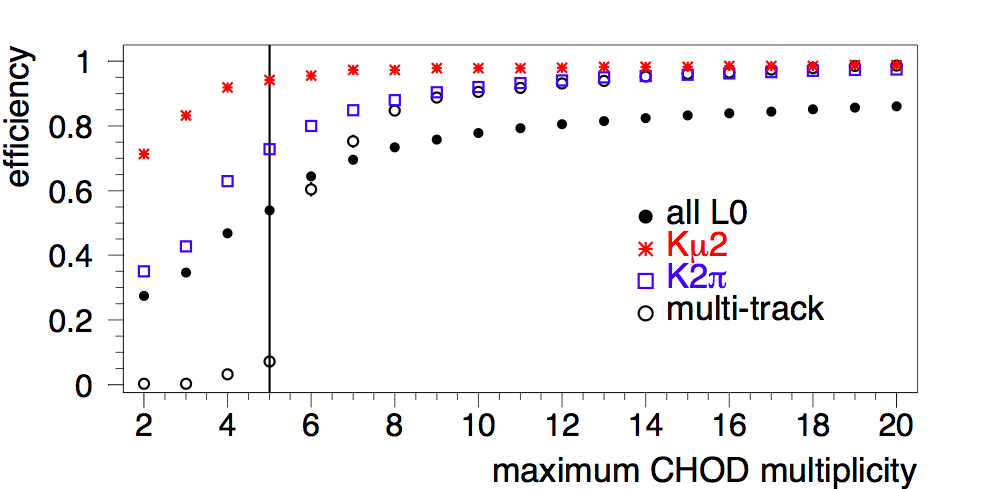}
\put(-190,87){\bf  \large (b)}
\end{center}
\end{minipage}
\\
\begin{minipage}{0.5\linewidth}
\begin{center}
\includegraphics[width=1.02\linewidth]{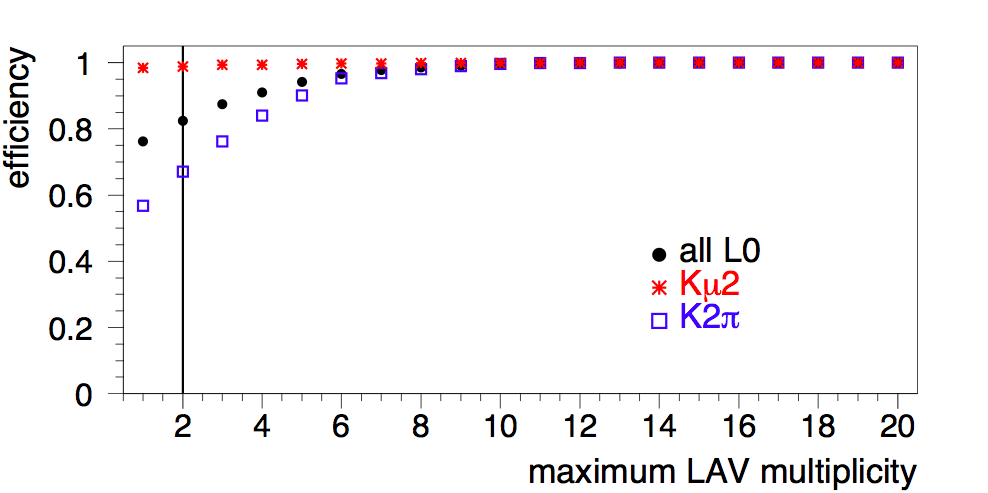}
\put(-32,30){\bf  \large (c)}
\end{center}
\end{minipage}
\begin{minipage}{0.5\linewidth}
\begin{center}
\includegraphics[width=1.02\linewidth]{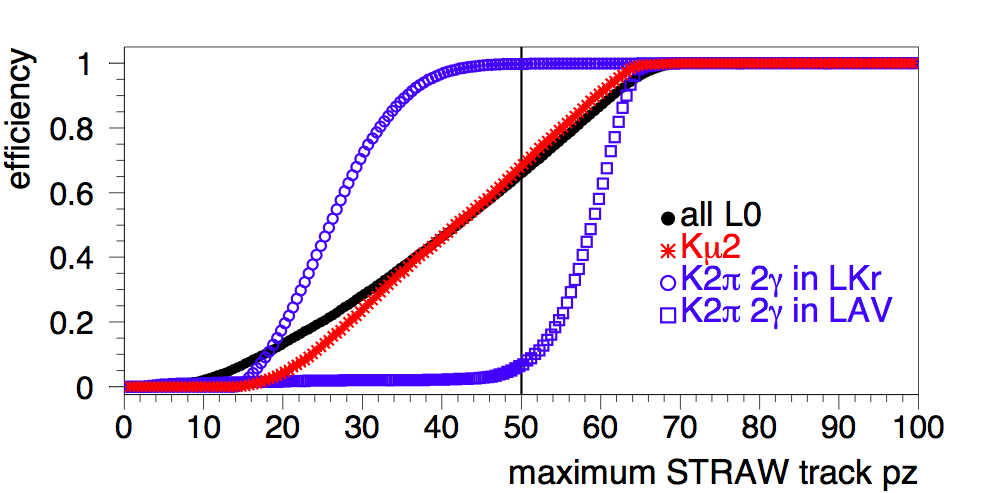}
\put(-190,30){\bf  \large (d)}
\end{center}
\end{minipage}
\caption{\label{fig:l1-eff}Fraction of L1 selected events for different types of L0 input events. The vertical lines correspond to the chosen cuts. (a) The minimum KTAG sector-multiplicity rejects events on the left of the line. (b) The maximum CHOD slab-multiplicity rejects events on the right of the line. (c)  The maximum LAV hit-multiplicity rejects events on the right of the line.  (d) The maximum STRAW track longitudinal momentum ${\rm p}_{\rm Z}$ ($\GEVc$) rejects events on the right of the line.}
\end{figure}

\begin{itemize}
\item The {\bf KTAG L1} trigger uses the KTAG sector-multiplicity 
 to positively identify a beam kaon and reject non-related accidental L0 triggers. The selection requires a minimum of five out of eight KTAG sectors in coincidence and in time with the L0 trigger.
Dedicated studies 
show an efficiency of more than 95\% for kaon decays and of 10\% for the accidental component (beam scattered tracks). 
\Fig{fig:l1-eff}a shows the efficiency  of this trigger for various input types, the overall 
data reduction factor being 1.7 for the standard cut at five sectors.  
\item The {\bf CHOD L1} trigger relies on the slab-multiplicity of the NA48-CHOD to reject multi-track events as well as
$\KTP$  events followed by $\PIo$ decay to photons and photon conversion in the upstream material.   
Requiring a maximum of five CHOD slabs hit 
in time with the
L0 trigger can be used to reject such events. The performed studies  
yield an inefficiency  of a few percent on genuine single-track events ($\KMN$ in \Fig{fig:l1-eff}b) and a data reduction factor of 1.8. . 
In the future, the tile-based CHOD detector (\Sec{sec:CHOD}) 
should  improve the rejection power of this trigger because of its geometrical segmentation in the XY plane. 
\item The {\bf LAV L1} trigger uses a hit-multiplicity cut in the twelve LAV stations aiming at further 
reduction of the $\KTP$  background by identifying photons emitted at large angles. A cut requiring no more than two
 hits in the whole LAV station system was applied to the events satisfying the L0 trigger (\Fig{fig:l1-eff}c). 
 Efficiency studies performed on selected $\KTP$  decays show an efficiency of 67\% while keeping a single-track event efficiency greater than 95\% and a global data reduction factor of 1.2. 
\end {itemize}

The above studies 
indicate that the 2015 L1 trigger algorithms performed as expected, resulting in a combined (KTAG + CHOD + LAV) L1 trigger efficiency of more than 68\% for  
genuine single-track events and an overall data reduction factor of 3.7.   

The L1 trigger is modular by design and its performance can be improved by adding requirements based on other sub-detectors. For example, a L1 trigger exploiting information from the STRAW spectrometer was studied offline with 2015 data
aiming at rejecting  multi-track events and one-track events 
reconstructed either outside the decay region or outside the track momentum range
required in the $\PNNP$  selection. 
The STRAW L1 trigger reconstructs particle tracks by performing a two-dimensional Hough transform and a crude momentum evaluation. The algorithm  applies the following loose requirements:  1) longitudinal track momentum ${\rm p}_{\rm Z}$ smaller than 50~$\GEVc$; 2) track closest distance of approach ($CDA$) to the beam axis smaller than 200~mm; 3)  longitudinal position at the $CDA$ smaller than 180~m that is upstream of Straw chamber 1.
To minimize the reconstruction of fake tracks by the L1 algorithm, a minimal ${\rm p}_{\rm Z}$
of 3~$\GEVc$ is required and the absolute values of the track slopes in the XZ and YZ planes have to be both smaller than 20~mrad. Multi-track events are identified by the reconstruction of a pair of tracks with a $CDA$ smaller than  30~mm. 
Trigger efficiency studies show a signal efficiency better than 99\%,  evaluated with $\KTP$  
decays with both $\PIo$ photons reconstructed in the LKr and the $\PPI$ track momentum in the range $(15-35)~\GEVc$. \Fig{fig:l1-eff}d shows the efficiency of this trigger for various event types with at least one reconstructed track. A data reduction factor of 1.7 is obtained. 
Applying this L1 STRAW trigger after the L1 KTAG, CHOD and LAV triggers would result in a reduction factor of  6.3. 
The online implementation of the L1 STRAW trigger started towards the end of 2015. The deployment and online tests being performed using later data.
\paragraph{L2 Trigger:} 
The L2 software trigger architecture was implemented in 2015, although no algorithms 
were applied. Trigger algorithms combining information from several sub-detectors are envisaged to be 
deployed in the future, such as a partial track reconstruction combining information from the GTK and from the straw tracker to allow cuts on the vertex position. 

High-level trigger software will optimize processing time versus required bandwidth for data recording. 
The available time budget is determined by the hardware, the data distribution among PC farm nodes, 
and the overall duty cycle of the beam. Features such as global and local downscaling, 
event by-passing and flagging provide the required degree of flexibility for testing and implementing these trigger algorithms.

%% file: Sec14_PCfarm_v2.tex
\subsection{PC Farm and data handling}\label{ssec:PCfarm}
Digitized data are received from detector electronics through optical links, and are transmitted to the receiving computers in the online farm on the network in packets, using  UDP as internet protocol (IP): it has been chosen over TCP (the other possible IP protocol) for achieving the highest possible transmission band-width, reducing the overhead due to the much simpler implementation, the smaller header size, 
and also for the possibility of shipping multi-event packets (MEPs). This choice has influenced the entire hardware and software architecture of the system, since UDP (contrary to TCP) does not implement a flow control and just needs a source and destination port for the packets, in addition to the length and checksum data, without any acknowledgement or handshake.

\begin{figure}[ht]
\centering
\includegraphics[width=1.0\linewidth]{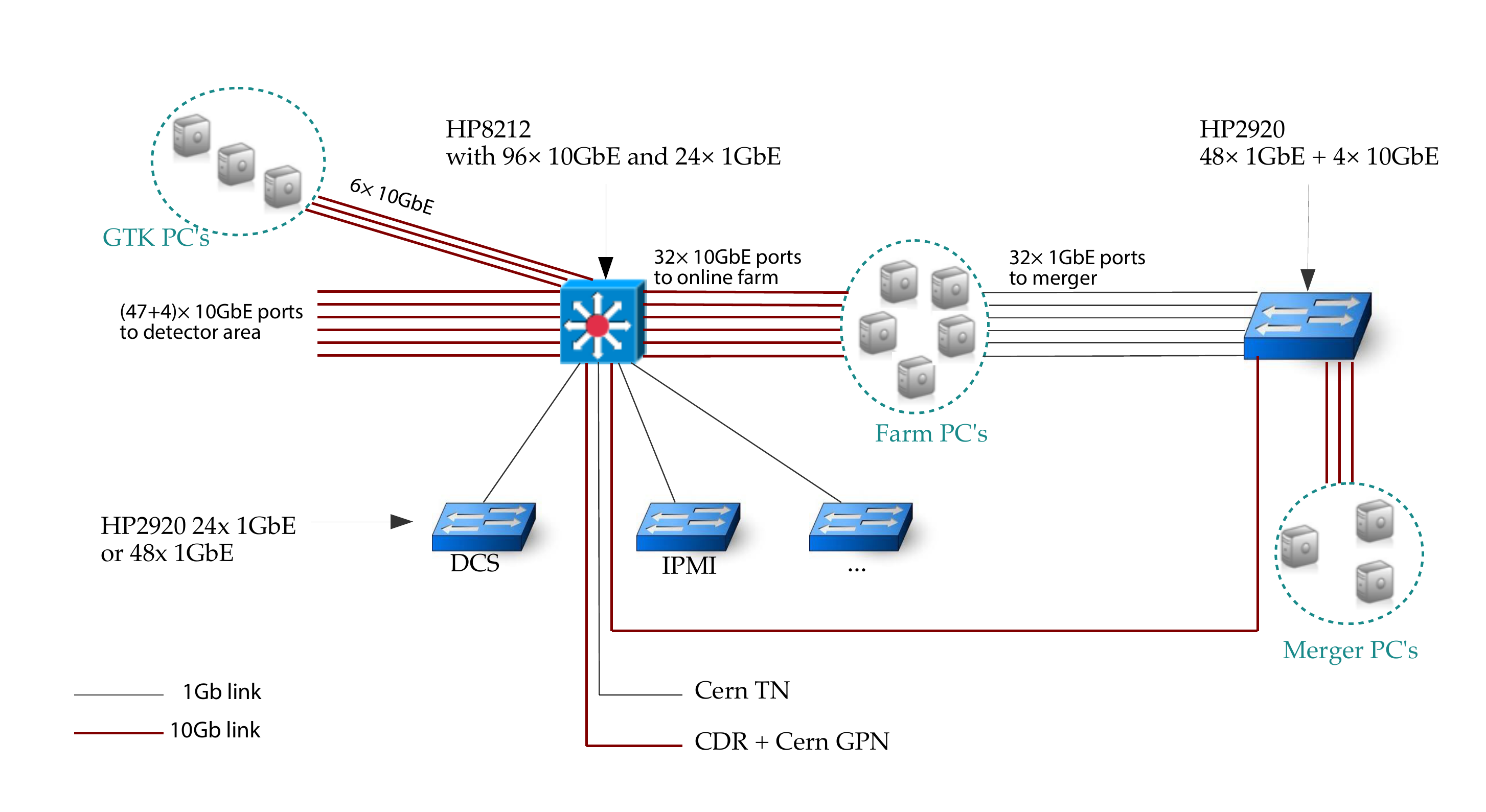}
\caption{\label{fig:Online-Picture1} Schematic diagram of the data link between the detector and the online farm.}
\end{figure}

A block diagram of the data link between the detector and the online farm is shown in Figure \ref{fig:Online-Picture1}. Front-end boards in the underground experimental area 
are aggregated via 1GbE links to network switches, and then through 10 GbE links to the main network router (HP8212) and the PC-farm, both housed  in the same room of the surface building.
The number of PC-farm nodes depends on the processing time of the software triggers and I/O capacity.  In 2015, the farm consisted of 30 PCs, but this number will increase with future demands.

Events from the same burst passing all three trigger levels are transferred to one of three merger PCs  and written in a common raw data file.  A single PC merging all the events from the farm would be sufficient, but 
to be safe with respect to rate variations depending on the trigger or beam conditions, the event merging is spread over three dedicated PC's. 

Assuming the maximum L0 and L1 rates (1 MHz and 100 kHz, respectively) and a total input bandwidth to L1 of about 6 GB/s, the total bandwidth to L2 is about 3.5 GB/s,  
dominated by 2 GB from calorimetric detectors and 0.6 GB/s from GTK data. 
A SPS burst with an effective duration of approximately 3.5 s therefore produces 
$\cal O$(10 Gb) to be written on tape.

The merger PCs also houses the main farm storage (approximately 42 TB) which provides:
\begin{itemize}
	\item a buffer for data before the transfer to the Central Data Recording system (CDR), that takes care of logging to tape;
	\item disk space for services, data quality histograms, software, etc.
 \end{itemize}
A buffer of at least 48h of continuous data-taking is usually required to decouple from CASTOR,\footnote{CERN Advanced STORage manager.}  the CERN tape system, requiring at least 20 TB of disk. In order to have sufficient bandwidth for CERN services, such as EOS,\footnote{EOS is a disk-based service providing a low latency storage for physics experiments.} CASTOR, etc., the CDR is connected with a 1 GbE dedicated link to the CERN general purpose network (GPN). 
The main task of the CDR system is to transfer the RAW files produced by the farm and stored in the merger PC storage areas, to CASTOR for permanent logging to tape and subsequent reconstruction. An automatic script-based system transfers files and logs and cleans up after recording to tape, at a sustained speed up to 120 MB/s. The CDR also produces meta-data for keeping track of the file until it is migrated to tape.

The NA62 reconstruction software is run on the raw data files, and the output is generally made available for analysis using the disk-based EOS system. 
Part of the raw data is reconstructed on the PC farm, directly from the merger PC buffer, for online monitoring of sub-detector performance and beam parameters. An automatic, continuous task produces a set of reference histograms for display.

%% file: Sec14_RunC_v3.tex
\subsection{Run Control}

Run Control is the central control and monitoring software for trigger and data acquisition systems (TDAQ). 
The information collected from various sources is summarized in synthetic states presented to the operator and redistributed to the experiment sub-systems.  An example is the number of events satisfying  different trigger levels  (L0TP and HLT in the farm nodes) used  by other systems (KTAG during a pressure scan).  Another example is the set of values (intensity on the T10 target and beam line scalers) received from the SPS and reformatted by the PC farm  before inclusion in the experiment data stream. Run Control is itself a  producer of meta-data information used by other systems (run number, burst number, list of participating detectors).
In addition, Run Control manages the configuration and running state of the various sub-systems. 
A summary of the controlled elements is given in \Tab{tab:rc_equipment}.

\begin{table*}[h]

\caption{List of devices integrated in the Run Control.}
\label{tab:rc_equipment}
\vspace{2ex}
\centering
\begin{tabular*}{0.85\textwidth}{ lc}
\hline\hline
\vspace{1 mm}
Equipment type & Number of elements \\
\hline
\vspace{1 mm}
 TEL62 boards  & KTAG(6), CHANTI(2), LAV(12), RICH(5) \\
 & CHOD(1), IRC/SAC(1), MUV3 (1) \\
Level 0 Trigger Processor & 2\\
GTK DAQ controller & 3 \\
STRAW DAQ controller & 4 \\
CREAM controller & 3 \\
L0-Cal Trigger &  1 \\
LTU & 11 \\
Raspberry Pi (thresholds) & 79 \\
PC nodes & 33 \\
\hline 
\vspace{1 mm}
Total & 164 \\
\hline
\hline
\end{tabular*}
\end{table*}

\begin{figure}
\centering
\includegraphics[width=0.85\linewidth]{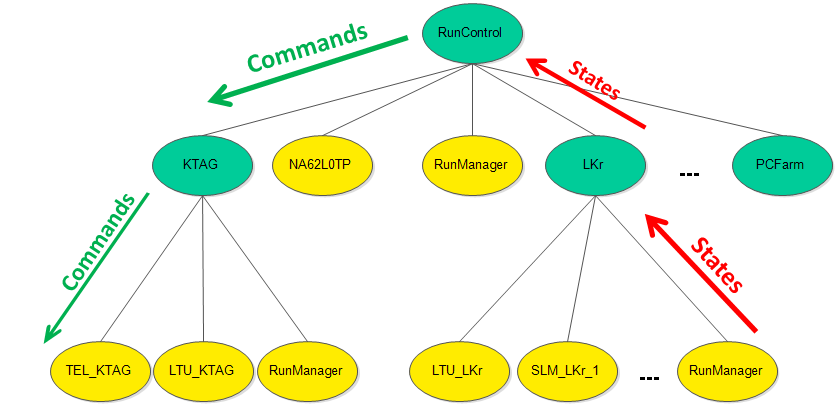}
\caption{Tree-like hierarchy of units. Green shapes are logical units and yellow shapes are device units. Commands are
propagated downwards and states are propagated upwards and summarized in the logical units.}
\label{fig:FSMTree}
\end{figure}

\paragraph*{Main functionalities:}
Run Control is based on  a hierarchy of finite state machines (FSM). Each device connected to the system is modelled
internally as a list of states and transitions between them. The change from one state to another can be triggered either 
by a device reporting the modification of some conditions or by a command issued at the interface. 
Each FSM is a node in a hierarchical structure. The outer nodes are device FSMs (modelling hardware or software devices), while the inner nodes are logical FSMs. The latter collect the states of the device
nodes and compute their own states according to logical rules with respect to inputs from the device FSMs.  Nodes are thus assembled and summarised in
logical sub-systems.  Run Control is itself a FSM representing the global state of the TDAQ of the
experiment (\Fig{fig:FSMTree}).

The commands issued on a node are propagated through logical FSMs to a device FSM. The command is then
sent through the network to the corresponding device, according to a standard interface. 
The difference between the various devices and their relative complexity  
are thus hidden at the Run Control level. Device details and interfacing protocols are contained in separate control programs directly connected to the hardware.

A flexible configuration scheme has been devised. XML files are stored in a database (recipe database) and retrieved at
configuration time by Run Control. The content of the file is transmitted to the device controller through a common
network file system. To ensure that the configuration has been loaded correctly, a snapshot of the current configuration
is requested after the start of run command.

A second database is used to store a wide range of values acquired by Run Control, for
example: configuration reports, included detectors, included devices, enabled trigger, and run information. All the data contained in this database are exported offline and can be used at a later analysis stage.

Run Control includes scripts which automate some checks and critical actions: readings and set points of beam line magnets are monitored;  an alarm is immediately issued if an anomaly is detected. The PC farm is also under constant surveillance, and an automatic recovery mechanism is
implemented in case of both software and hardware problems.

\paragraph{Technologies and Architecture:} 
 Operation of the experiment involves several control systems related to hardware equipment  like DCS, Detector Control System  (\Sec{ssec:DCS}) and 

DSS, Detector Safety System, gas, vacuum and cryogenic controls. The development of all these systems is
based on a widely used industrial software WinCC Open Architecture \cite{Simatic:0000}.   This software is the central part of two frameworks  developed at CERN, JCOP \cite{Gonzalez-Berges:2003} and UNICOS  \cite{Milcent:2009}. 
 The hardware connected to Run Control is mostly custom made and the communication
is exclusively handled through the network with the DIM protocol \cite{Gaspar:2001}. 
 Dedicated control software has been written to receive and execute commands on the electronics of the sub-detectors, on the electronics of the trigger system and on the PC Farm.

\begin{figure}
\includegraphics[width=0.85\linewidth]{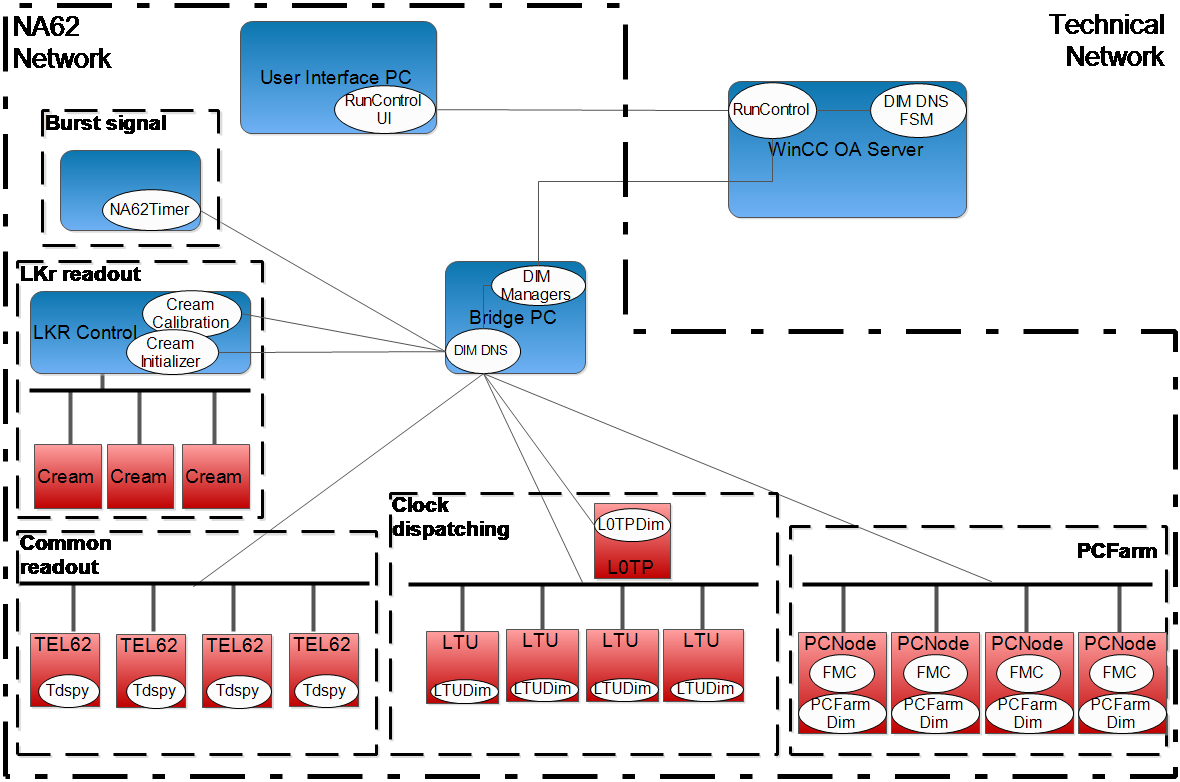}
\caption{Schematic of the infrastructure of the DAQ and Run Control of the NA62 experiment.}
\label{fig:RCInfra}
\end{figure}

 Run Control is built on a distributed architecture, as sketched in Figure \ref{fig:RCInfra}. 
The core of the system is  hosted on a dedicated server on the Technical Network. The  hardware elements are spread across the underground experimental hall 
and the server room on the NA62 Network. A central machine contains the DIM managers and is loosely binding both networks. In case of network problem or unavailability of this central machine, the DAQ will continue taking data waiting for the communication to be re-established.
The user interface is running on a different computer in the control room and is remotely connected to the main system.

%% file: Sec14_DCS_v3.tex
\subsection{Detector Control System}\label{ssec:DCS}

The Detector Control System (DCS) is dedicated to  detector monitoring and operation~\cite{Golonka15}. The system is built on robust infrastructure and runs on hardware guaranteed to provide 24/7/365 operation with minimal human supervision.

\paragraph*{Main Functionalities:} 
The system operates and monitors a variety of devices, reacts to changes of measured values, compares them to settings and tolerances  and issues warnings and alarms.  Besides an alarm screen, the system provides synoptic views of the state of the experiment as a whole as well as in parts.  It allows operators to drill down to quickly identified individual devices.  

Trend plots of historical data are available for monitored values. The history of acquired data (conditions) is recorded in an Oracle database and can be retrieved  later in offline analysis programs. 
Remote access and expert operation are possible through dedicated terminal servers and/or secondary consoles, which run the same user-interface application as the main control room console. Partitioned operation and sub-tree ownership prevent two operators from sending conflicting commands.

Different levels of access rights can be given to the central operator  and sub-detector experts according to their need.
Advanced functionalities are implemented to perform, for example, a pressure scan of the KTAG gas radiator. This tool allows the sequencing of parameter settings that would otherwise require manual setting on the Gas Control System console.

\begin{figure}[ht]
\centering
\includegraphics[width=0.85\linewidth]{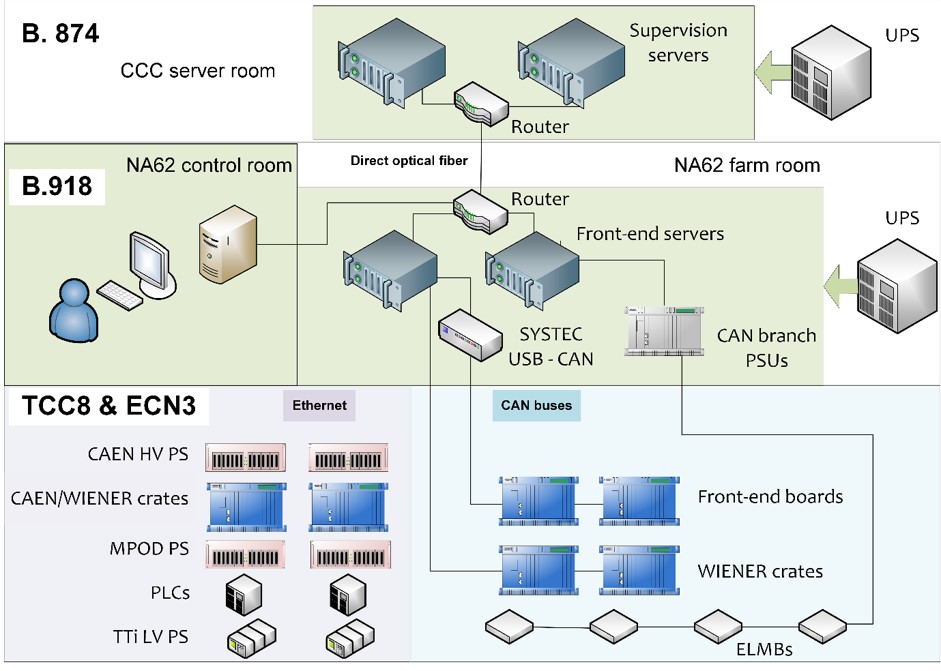}
\caption{\label{fig:DCS-Picture1} Block diagram of the DCS system. TCC8 and ECN3  are located  in the underground area. Building 918, at the surface level, hosts the NA62 control room  where the operator supervises the experiment and the PC-farm room while  Building 874 hosts infrastructure for the Cern Control Center (CCC).}
\end{figure}

\paragraph{Technologies and Architecture:}
The system has a layered architecture with supervision, process/front end and field/equipment layers, shown in \Fig{fig:DCS-Picture1}, together with the geographical 
location of the components.

The supervision layer consists of 9 supervision applications, one per sub-detector (or group of sub-detectors) and a central DCS application.  

The front-end layer is composed of dedicated computers, hosting hardware-interface cards (for CAN bus) and interface hardware. The front-end computers aggregate the traffic from hardware, translate the necessary middleware technologies eliminating platform-dependent factors and interface with the supervision layer over a dedicated network. Two Programmable Logic Controllers (PLCs) are used for detector specific tasks: one in LKr low voltage control for digital I/O handling and another in KTAG to drive detector alignment motors.

The high and low voltage systems ($\sim$ 4 000 channels), electronic crates, and I/O boards with sensors populate the field and equipment layer (\Tab{tab:DCS-param}). A significant number of custom front-end-electronics boards are configured and monitored.

User interface programs run on a set of consoles located in the NA62 control room. A direct optical fibre link connects the CCC server room and the experiment facilities.

\begin{table*}[h]
\caption{List of hardware equipment controlled by DCS.}
\label{tab:DCS-param}
\vspace{2ex}
\centering
\begin{tabular*}{0.85\textwidth}{lcr}
\hline\hline
Equipment Type & Devices & Channels \\
\hline
CAEN HV power supply mainframes  & 25 &  3 690 \\
CAEN VME crates & 28 &  \\
WIENER electronic crates    & 34 & \\
WIENER MPOD power supply & 4 & 32 \\
Aim TTi LV power supplies  & 7 & \\
Front-end electronic  & & 2 916  \\
Analogue  measurements  & & 4 446 \\
Temperature, pressure and humidity sensors & & 30  \\
\hline
Total   & & $\approx$~11 100  \\
\hline
\hline
\textbf{}
\end{tabular*}
\end{table*}
The supervision layer is built using the same standard CERN control software as Run Control.
CAEN high voltage power supplies, WIENER electronic crates and ELMBs \cite{Burckhart:2001} are integrated within the JCOP framework. 

Legacy hardware associated with the liquid krypton calorimeter LKr is interfaced through a mixture of control frameworks and technologies. For the low voltage system, a PLC powered by UNICOS provides control of digital I/O, and ELMB I/O boards, integrated through the JCOP framework, acquire analogue data.  For the high voltage system, CAEN power supplies are integrated by the JCOP framework , while custom electronics provide channel-by-channel monitoring interfaced through DIM \cite{Gaspar:2001}.

Front-end boards are controlled by two similar technologies. The ELMBs embedded in KTAG and RICH electronics are controlled by a CANopen OPC server and corresponding JCOP framework component. CHANTI front-end board control is also based on CANopen, however, in this case, a custom board firmware was developed. Consequently, the supervision integration is also custom made.

The DIM protocol is used to integrate WIENER MPOD and Aim TTi power supplies.  
DIP \cite{DIPDIM:0000} allows communication 
with external control systems like the gas control, in particular to implement the KTAG pressure-scan tool.

The front-end computers, network hardware, consoles and data servers are powered through UPS units providing up to an hour of autonomy in case of a power-cut. The mandatory computer security is ensured by running the system over an insulated network with remote access through gateways and terminal servers.

%% file: Sec15-perf2015_v3.tex
Data collected in 2015 with a minimum bias trigger at 1\% of the nominal beam intensity offer the possibility to validate the full detector performance in terms of kinematic resolution, particle identification, and photon rejection  by combining information from several detectors.  The results can be compared with the requirements of the $\PNNP$ measurement (\Sec{ssec:req}). 
\subsection{Advanced single track selection}
A single-track selection that is tighter than but similar to the ``single-track event''  definition of  \Sec{ssec:samples}, has been set up as a preliminary step towards the  $\PNNP$  measurement. Tracks reconstructed in the straw spectrometer with matching energy depositions in the calorimeters and  in the NA48-CHOD are selected. 
The matched NA48-CHOD signal defines the time of the track with 200~ps resolution. 
A single track event is defined by the presence of a 
track not forming a common vertex within the decay region with any other in-time track. To refine the selection of single-track events originating from a kaon decay, a GTK track is required to match in time and space the straw track, to form a vertex with this track in the decay region ($CDA < 1.5$ cm), and  to match in time a kaon signal in KTAG. 
This sample of single-track events originating from kaon decays  
is used to study the kinematic resolution, particle identification, and photon rejection.  

\begin{figure}[hbt]
\begin{minipage}{0.5\linewidth}
\begin{center}
\includegraphics[width=1.0\linewidth]{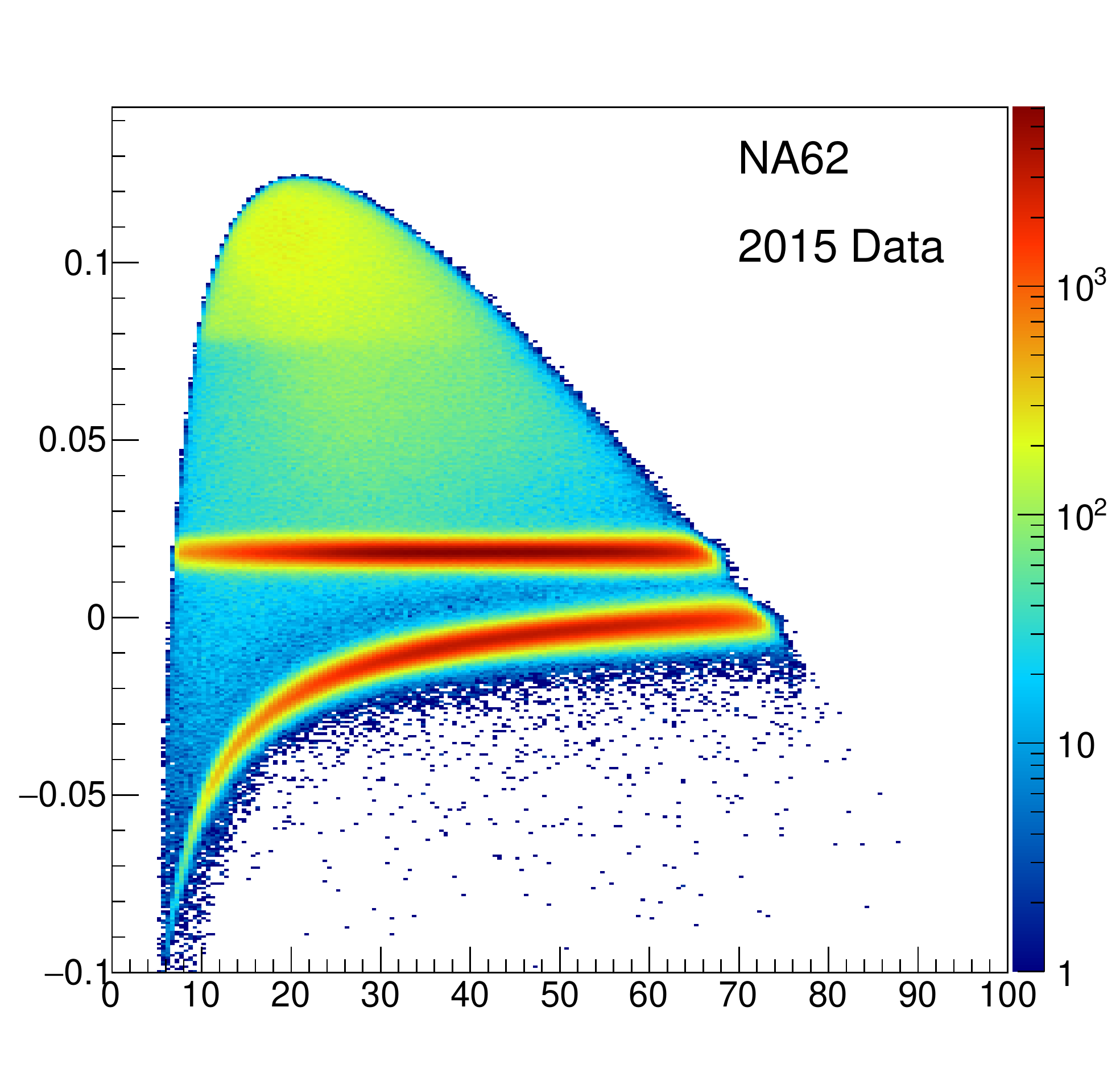}
\put(-75,0){${\rm p}_{\pi^+}$  [$\GEVc$] }
\put(-220,117){\rotatebox{90}{$m^2_{\rm miss}$  [GeV$^2 /c^4$]}}
\put(-160,150){$K_{3\pi}$}
\put(-140,108){$K_{2\pi}$}
\put(-60,88){$K_{\mu2}$}
\end{center}
\end{minipage}
\hfill
\begin{minipage}{0.5\linewidth}
\begin{center}
\includegraphics[width=1.0\linewidth]{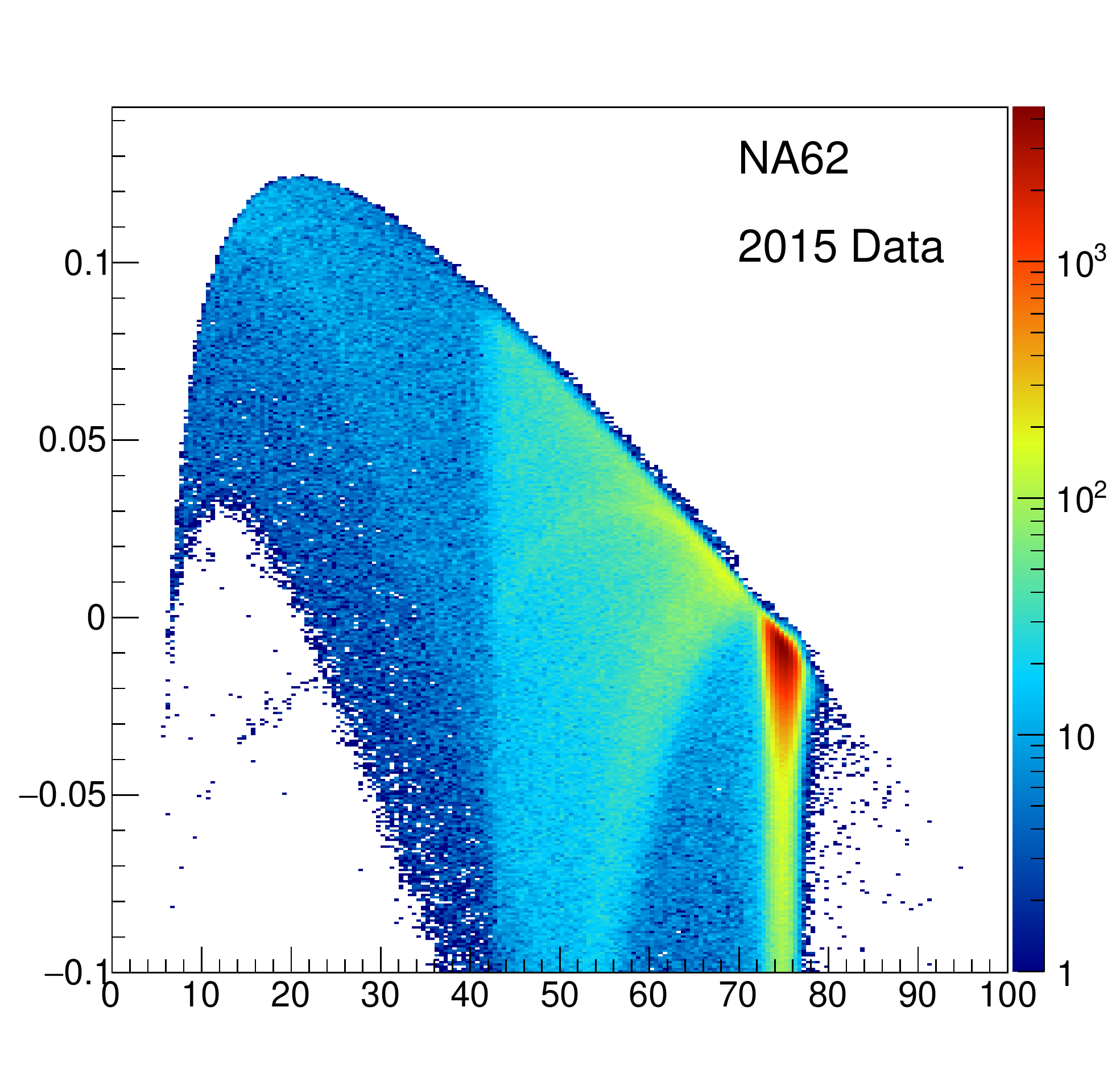}
\put(-75,0){p$_{\pi^+}$ [$\GEVc$]}
\put(-220,117){\rotatebox{90}{$m^2_{\rm miss}$  [GeV$^2 /c^4$]}}
\end{center}
\end{minipage}
\caption[]{Left: Distribution of $m^2_{\rm miss}$  under $\KPL$ and $\PPI$ mass hypotheses as a function of the pion momentum measured in the straw spectrometer
 for selected single-track events originating from kaon decays. Right: Corresponding distribution when using KTAG in anti-coincidence with NA48-CHOD and GTK times to select single-track events not related to kaon decays.}
\label{fig:dataplot}
\end{figure}

\Fig{fig:dataplot}-left shows the $m^2_{\rm miss}$ distribution computed under the $\KPL$ and $\PPI$ mass hypotheses (\Equ{eq:mmis}) as a function of the $\PPI$  momentum.
The $m^2_{\rm miss}$ range accessible depends on the pion momentum 
because the kaon beam is nearly monochromatic. Regions corresponding to the main $\KPL$ decay modes, 
$\KMN$ ($K_{\mu2}$),  $\KTP$  ($K_{2\pi}$), $\KTAU$ and $\KTAUo$ ($K_{3\pi}$)
are clearly seen. The $m^2_{\rm miss}$ distribution for  $K_{\mu2}$  
decays varies as a function of the particle momentum because the pion mass hypothesis is used. 
Semileptonic decays like $K^+\rightarrow\pi^0 e^+\nu$ and 
$K^+\rightarrow\pi^0 \mu^+\nu$ are not kinematically constrained and hence populate the whole plane. 

The KTAG can also be used in anti-coincidence with respect
to the NA48-CHOD and the GTK times to select single-track events not related to kaon decays. The corresponding $m^2_{\rm miss}$ distribution as a function of the  
pion momentum is displayed in \Fig{fig:dataplot}-right. The geometrical acceptance of the straw spectrometer and the maximum available $m^2_{\rm miss}$ define the boundaries of the distribution. Events in the region above the 42.5~$\GEVc$  track momentum threshold come from the decay of  75~$\GEVc$ beam pions ($\PPI \to \mu^+\nu$). 
Events in the 75~$\GEVc$ band correspond to beam particles entering the 
STRAW acceptance after elastic scattering 
in the material along the beam line (KTAG and GTK), while events with tracks  produced 
in  inelastic interactions  span the whole plane.

\subsection{Kinematic resolution } 
\label{ssec:kine}
\begin{figure}[ht]
\begin{center}
\includegraphics[width=0.6\linewidth]{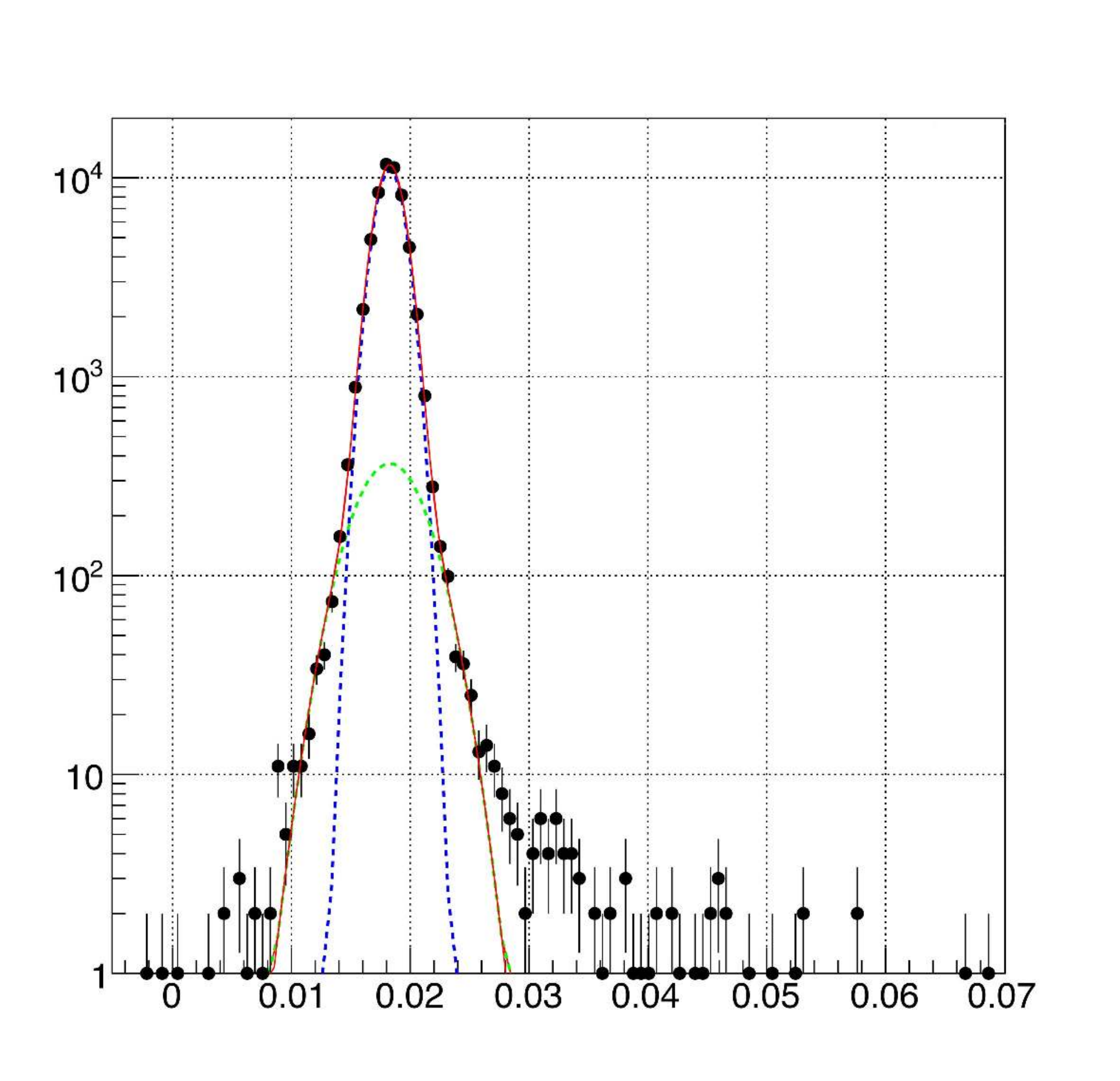}
\put(-100,0){$m^2 _{\rm miss}$  [GeV$^2 /c^4$]}
\end{center}
\caption{\label{fig:pip0peak} Distribution of $m^2_{\rm miss}$ for $\KTP$ events to measure the kinematic performance. Dots correspond to data, the solid curve to the two-Gaussian fit result and the dotted lines represent the two individual Gaussian contributions.}
\end{figure}

Reconstructed $\KTAU$ decays are used to set the momentum scale of the straw spectrometer (\Sec{sec:STRAWSpec}) and to calibrate and align the GTK (\Sec{sec:GTK}). The global performance of the kinematic reconstruction is investigated using $\KTP$ events 
selected from the sample of  single-track events described above by requiring the additional presence of two photons in the LKr calorimeter compatible with the decay of a $\PIo$ within the decay region.  However,  to avoid possible bias in the distribution of the far tails of the $m^{2}_{\rm miss}$ distribution, no matching requirement is applied to the $\ZVT$ positions of the charged track and reconstructed $\PIo$ .

The resulting $m^{2}_{\rm miss}$ spectrum is shown in \Fig{fig:pip0peak}. The mean value of the distribution $\langle m^{2}_{\rm miss}\rangle = m^{2}_{\pi^{0}}$ 
has been extracted from a two-Gaussian fit 
constrained to a single mean parameter. The standard deviation of the first Gaussian 
depends on the intrinsic resolution of the tracking system when the pion track matches the track of the parent kaon in the GTK. The second Gaussian corresponds to 
 the kinematic resolution in case of mismatching between the pion and kaon tracks.

\begin{figure}[ht]
\begin{minipage}{0.5\linewidth}
 \centering
 \includegraphics[width=1.05\linewidth]{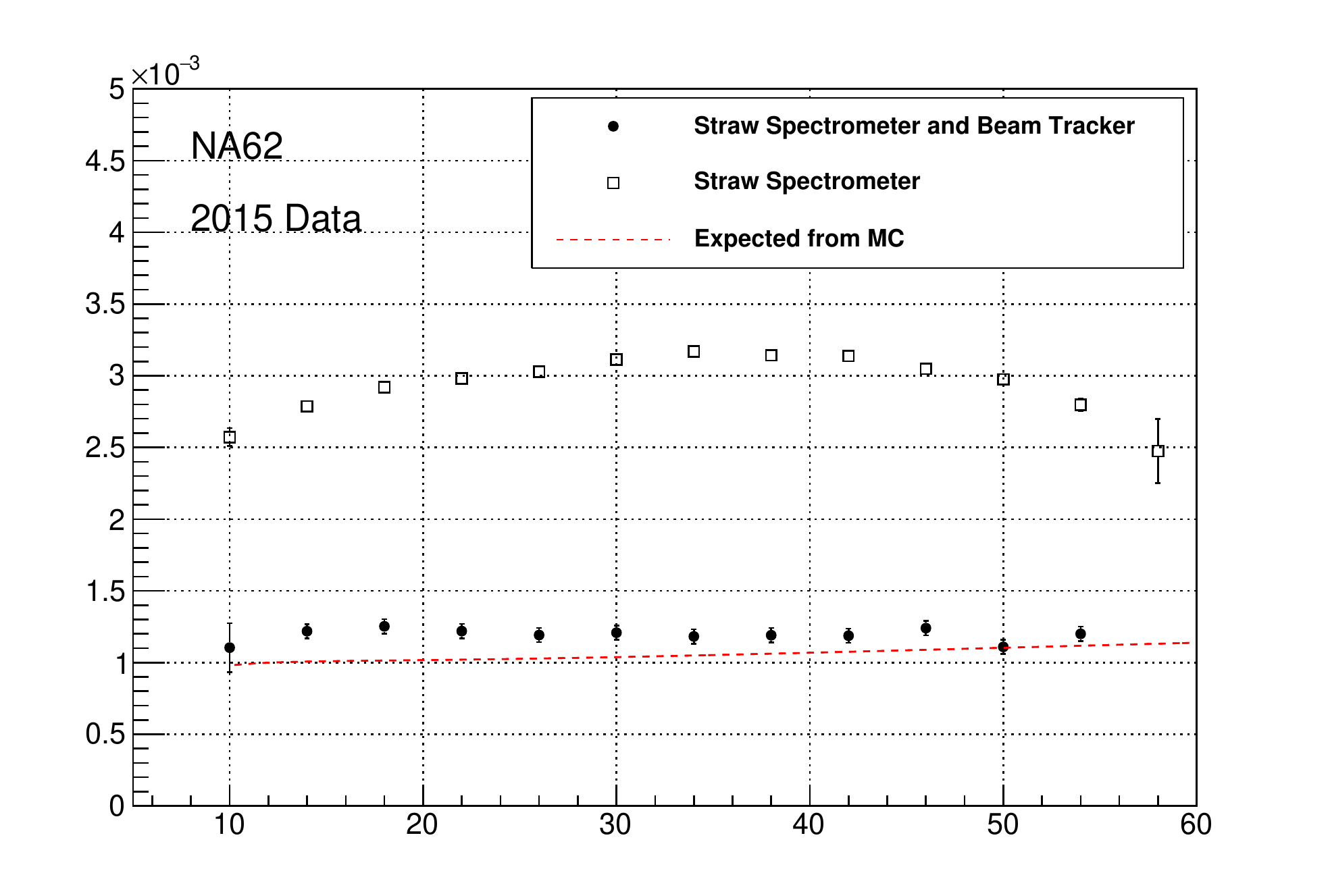}
\put(-75,0){p$_{\pi^+}$ [$\GEVc$]}
\put(-230,51){\rotatebox{90}{$\sigma(m^2_{\rm miss})$  [GeV$^2 /c^4$]}}
\end{minipage}
\begin{minipage}{0.5\linewidth}
 \centering
 \includegraphics[width=1.05\linewidth]{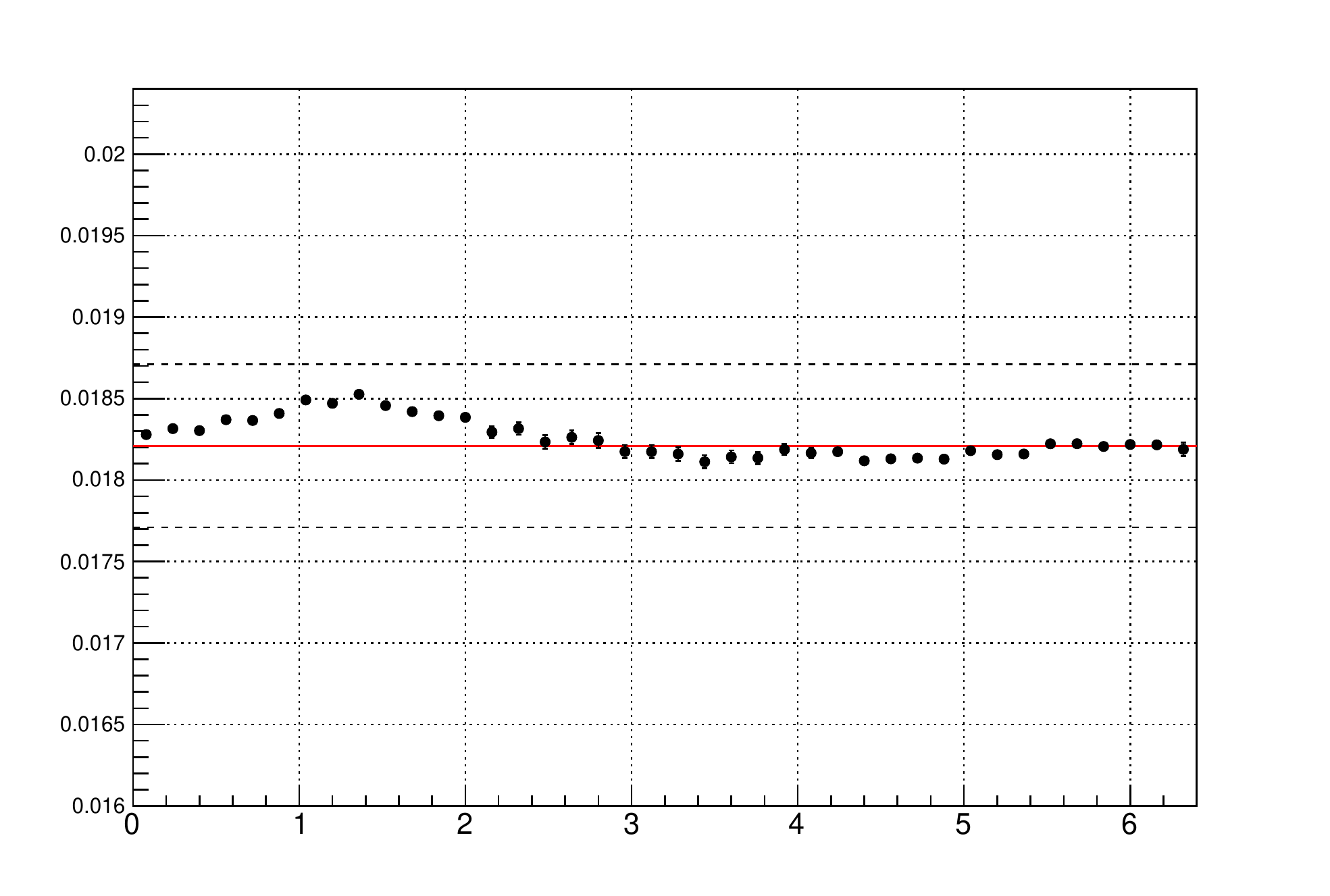}
\put(-73,0){$\phi(\PPI)$ [rad]}
\put(-230,58){\rotatebox{90}{$\langle m^2_{\rm miss} \rangle$ [GeV$^2 /c^4$]}}
\end{minipage}
\caption{\label{fig:datakine} Left: Standard deviation of the $m^{2}_{\rm miss}$ distribution of $\KTP$ events as a function of the $\PPI$ momentum. Full dots correspond to the $m^{2}_{\rm miss}$ computed using the $\KPL$ momentum and direction as measured by GTK,  open squares to the resolution of the $m^{2}_{\rm miss}$ computed using the nominal $\KPL$ momentum and direction. Right: Mean value of the $m^{2}_{\rm miss}$ distribution of $\KTP$ events as a function of the azimuthal angle of the $\PPI$.  The azimuthal angle runs anti-clock-wise with its origin on the X axis. The red solid line corresponds to  the expected $m^{2}_{\pi^0}$ value and the dotted lines define a spread of  $\pm$ half the $m^{2}_{\rm miss}$ resolution. }
\end{figure}

\Fig{fig:datakine}-left illustrates the measured $m^{2}_{\rm miss}$ intrinsic resolution as a function of pion momentum. The resolution  approaches that
expected, the difference being due to the partial hardware configuration of GTK in 2015. 
The resolution is mostly dominated by multiple scattering in the last GTK station and  the first straw chamber.
The measured resolution is compatible with an angular resolution for the kaon and
pion tracks below 20 and 60~$\mu$rad, respectively. 
The kinematic resolution obtained when using the nominal $\KPL$ momentum and direction instead of the event-by-event GTK measurements is also shown for comparison.
In the later case,  the beam divergence and momentum bite dominate the resolution. The GTK improves the resolution by a factor of about three. The effect of the relative geometrical alignment of the straw spectrometer and GTK on the intrinsic resolution 
becomes negligible  once the $\KTAU$ alignment procedure has been applied.
The effect of residual misalignments  is checked by looking at the  
variation of  the mean $\langle m^{2}_{\rm miss}\rangle$ value as a function of the azimuthal angle of the pion track,  which is found to stay within half  the $m^{2}_{\rm miss}$ resolution (\Fig{fig:datakine}-right).

The tracking system is designed to provide a rejection factor  of $\mathcal{O}(10^4)$ for $\KTP$
decays on the basis of the  $m^{2}_{\rm miss}$ cuts. 
The previously described $\KTP$ sample is used to measure the kinematic suppression factor. 
 Considering for illustration only the $m^{2}_{\rm miss}$ intervals $(0,0.01)$ and $(0.026,0.068)$ GeV$^2/$c$^4$,
the measured $\KTP$ suppression factor is  $0.5\times10^3$. 
The suppression power is mainly limited by the partial hardware arrangement of the GTK used in 2015 , because of the $m^{2}_{\rm miss}$ tails due to beam track mis-reconstruction. 
\subsection{Particle Identification}
The particle identification (particle-ID) system is designed to separate $\pi^+$ from $\mu^+$ 
to obtain at least  seven orders of magnitude of suppression of the $\KMN$ decay, in addition to the kinematic rejection, for the track momentum range  (15, 35)~$\GEVc$.  
The RICH detector, all calorimeters and MUV3 are employed  for this purpose. 

The pion-muon separation in the RICH is studied with the sample of $\KTP$ decays used for kinematic studies and a sample of $\KMN$ decays selected 
among single-track events from kaon decays by requiring signals to be present in MUV3.
Samples with higher pion  purity are obtained by requiring  the absence of signals compatible with photons in the LAVs, IRC and SAC in time with KTAG.
The absence of photons in the LKr improves the purity of the muon sample.
Samples of positrons  can also be obtained  by requiring the ratio of the energy deposition in LKr to the track momentum (E/p) to be close to 1.

\begin{figure}[ht]
\begin{minipage}{0.5\linewidth}
\begin{center}
\includegraphics[width=1.0\linewidth]{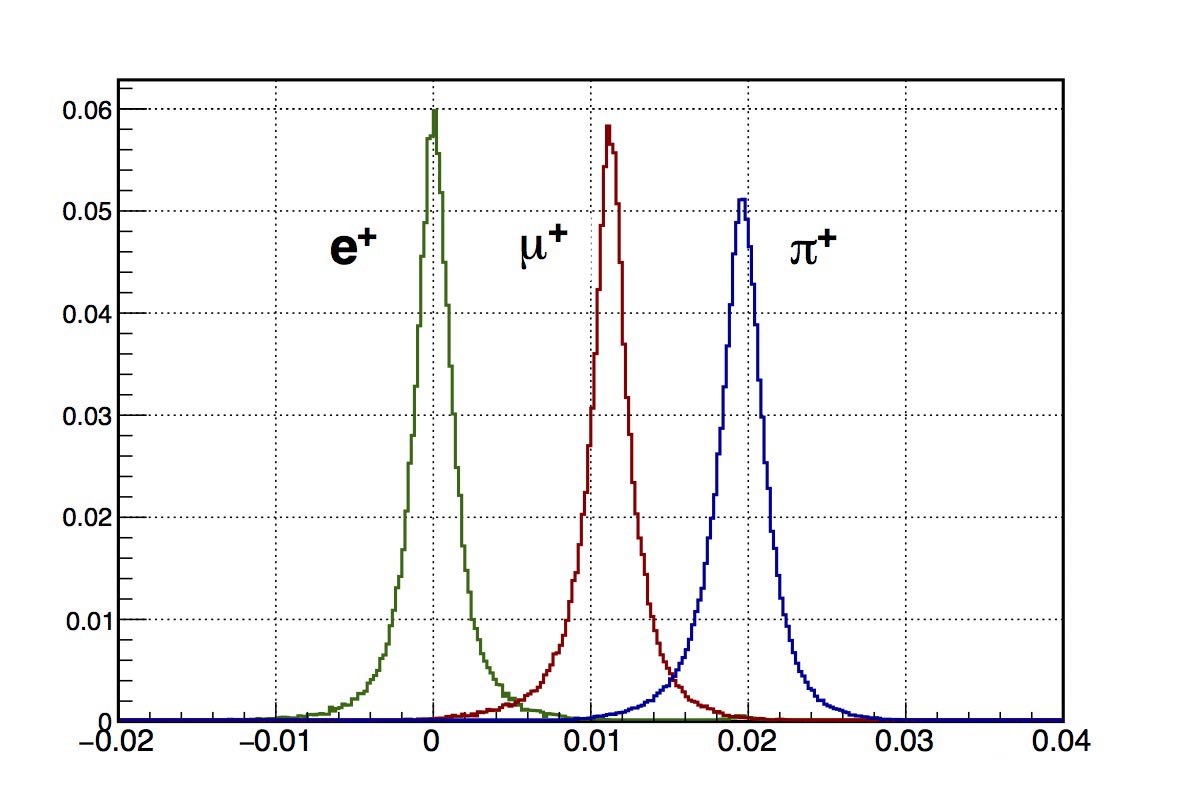}
\put(-85,-5){${\rm M}^2$  [GeV$^2 /c^4$]}
\put(-220,65){\rotatebox{90}{arbitrary units}}
\end{center}
\end{minipage}
 \begin{minipage}{0.5\linewidth}
\begin{center}
\includegraphics[width=1.0\linewidth]{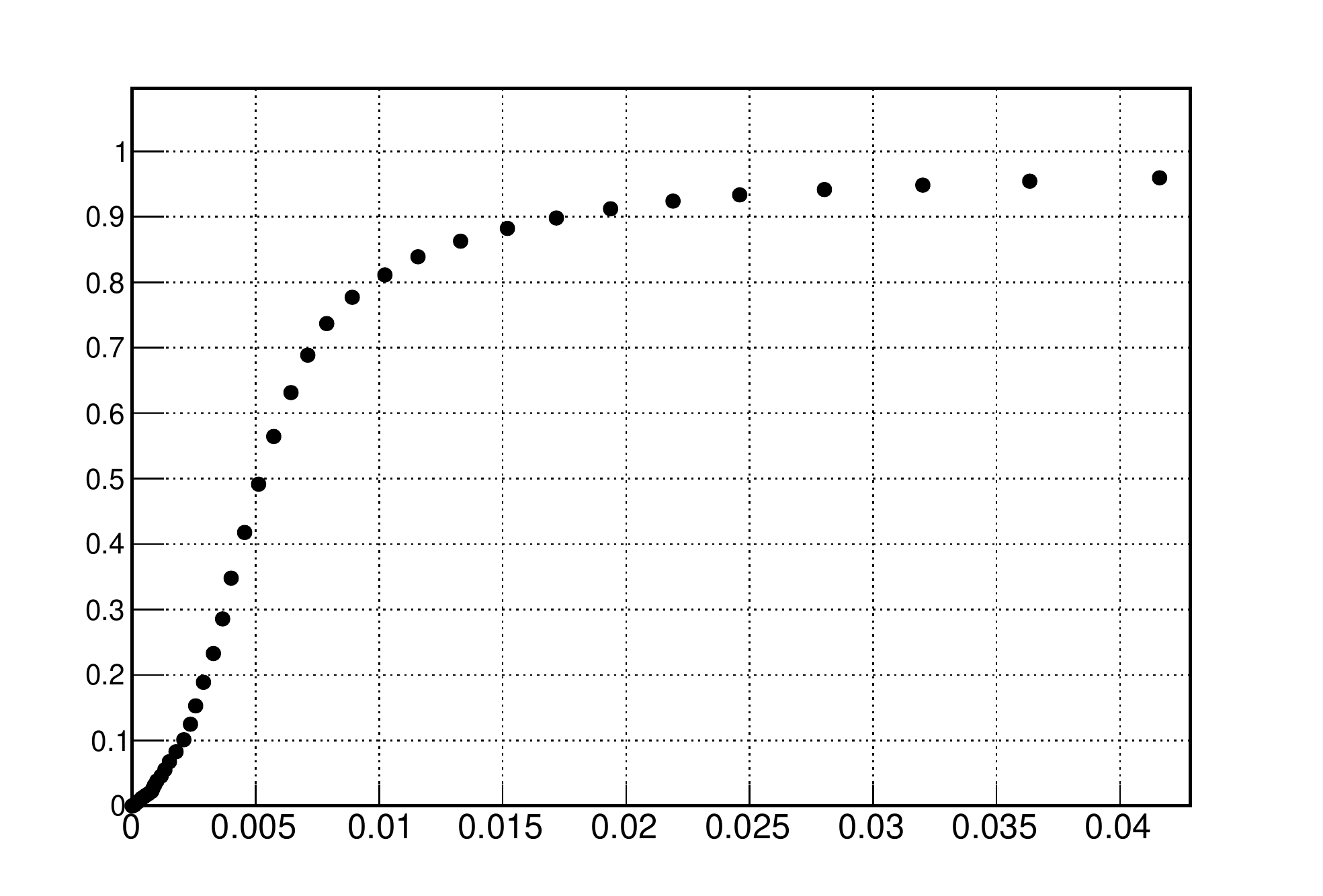}
\put(-113,-5){efficiency (muon-ID)}
\put(-220,45){\rotatebox{90}{efficiency (pion-ID)}}
\end{center}
 \end{minipage}
 \caption{\label{fig:RICH-Picture6} {RICH particle identification for track momenta between 15 and 35 $\GEVc$. Left: Squared particle mass reconstructed using the velocity estimated by the Cherenkov angle measured by the RICH and the momentum measured by the spectrometer, for positrons, muons, and charged pions selected using spectrometer and calorimetric information. Right: Pion-ID efficiency versus muon-ID efficiency for different choices of the reconstructed mass range.}}
\end{figure}

\paragraph{RICH particle identification:}
The particle mass can be reconstructed using the velocity estimated from the Cherenkov angle measured by the RICH and the momentum measured by the spectrometer.
\Fig{fig:RICH-Picture6}-left shows the squared reconstructed particle mass after selecting the  
momentum region between 15 and 35 $\GEVc$ 
for samples of positrons, muons, and charged pions as defined previously. By imposing a reconstructed mass range, charged pions can be selected and muons rejected with known efficiencies. \Fig{fig:RICH-Picture6}-right shows the pion-ID efficiency as a function of the muon-ID efficiency for different choices of the selected reconstructed mass range.
A muon suppression factor of $10^2$ (i.e.\  a muon-ID efficiency of 0.01) corresponding to 80\% pion-ID efficiency is measured. 
Above 35 $\GEVc$,  the separation degrades quickly as expected from the Cherenkov threshold curves for $\pi^+$ and $\mu^+$ in neon.  The pion-muon separation is expected  to improve in 2016 
after mirror alignment. The RICH provides even better separation between $\pi^+$ and $e^+$ (\Fig{fig:RICH-Picture6}-left).  

\begin{figure}[h]
\centering
\includegraphics[width=0.65\linewidth]{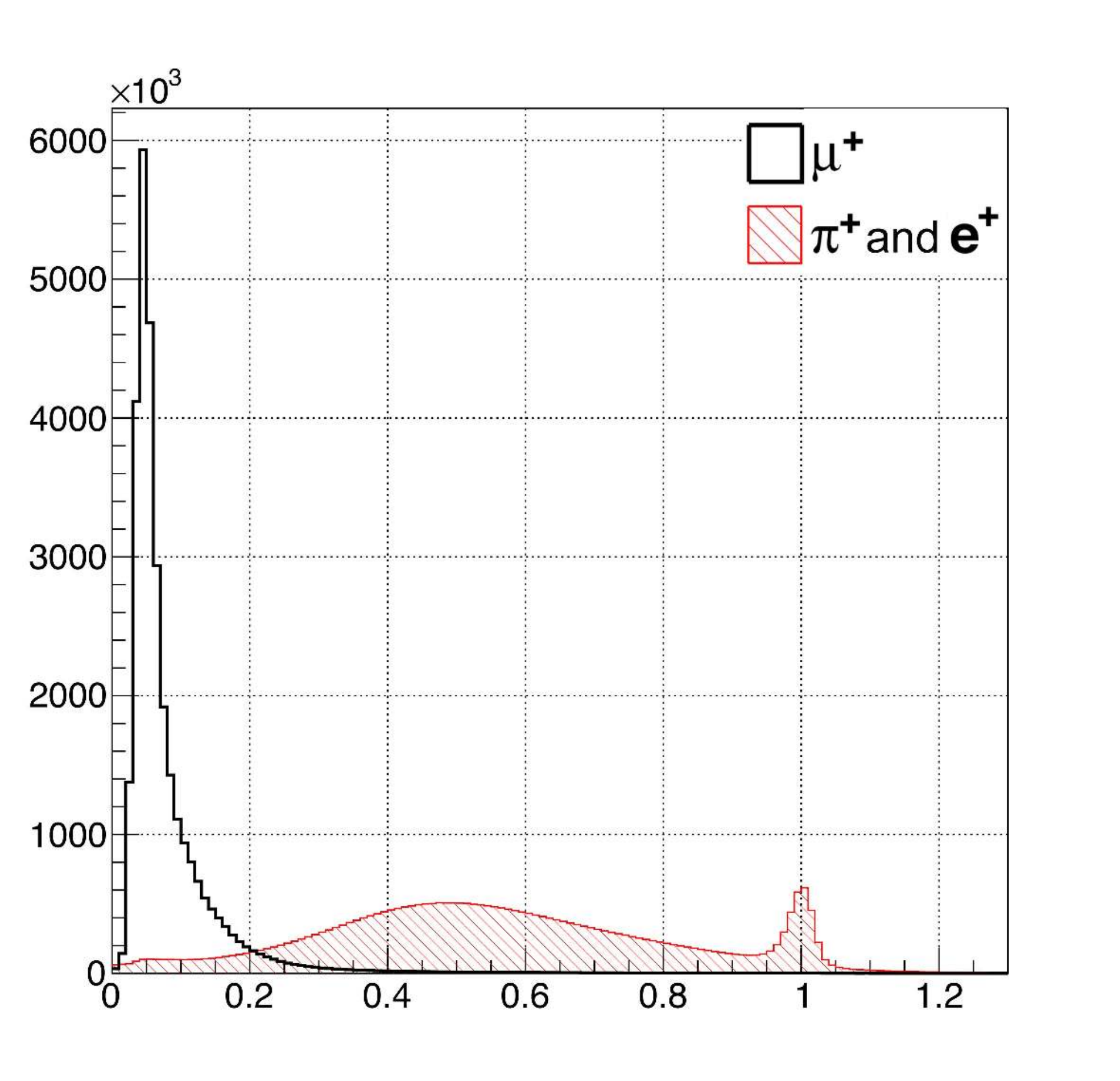}
\put(-60,5){${\rm E}_{\rm total}$/ p }
\caption{\label{fig:calopimu} Ratio of the total energy measured in the calorimeters and the momentum of the particle measured by the straw
spectrometer. The solid (black) distribution corresponds to muons, the (red) shaded distribution to pions and positrons. Positrons are under the peak around E$_{\rm total}$/p = 1 and their energy is released in LKr only.}
\end{figure}

\paragraph{Calorimetric particle identification:}
Samples of $\pi^+$ and $\mu^+$ selected as described previously, but without applying the energy requirements in the calorimeters allow the calorimetric pion-muon
separation to be investigated. Pions, muons, and positrons are characterized by different responses of the LKr/MUV1/MUV2/MUV3 system. These particles can be distinguished  
by the energy sharing between LKr, MUV1 and MUV2, the type of energy release in the calorimeters, the total energy measured, and the presence of signals in MUV3. For illustration, Figure \ref{fig:calopimu} 
shows the ratio between the total energy measured in the calorimeters and the momentum measured with the straw spectrometer for pions, muons, and positrons.   
A MIP releases on average 550 MeV in the LKr  and 1.5 GeV in MUV1 and MUV2. A pion, instead, is stopped by the the calorimeters, with a punch-through probability below the per-mille level. 
A  cut-based analysis using 
the previously defined characteristic variables provides a muon suppression factor of 
$10^{5} ~(10^{6})$ for a pion-ID efficiency of 70\% (40\%), respectively. A multivariate analysis is under development to optimize 
the separation. Preliminary studies show that a  pion-ID efficiency of up to 85\% can be obtained while keeping the muon suppression factor at the level of $10^{5}$.  
\subsection{Photon rejection}
NA62 is designed to suppress the $\KTP$ decay by eight orders of magnitude by detecting at least one of the photons from the $\pi^0$ decay in one of the 
electromagnetic calorimeters (LAV, LKr, IRC, SAC) spanning the angular region 0--50~mrad. 
The $\PIo$ suppression benefits from the angle-energy 
correlation of the two decay photons and from the maximum $\PPI$ momentum cut at 35~$\GEVc$. Under these conditions, the above suppression can be reached even in the presence of a single-photon 
detection inefficiency of $10^{-5}$  above 10 GeV and of even higher inefficiency 
at lower energy (\Tab{tab:phveto}). The suppression of $\pi^0$ from $\KTP$ decay is measured from data 
as the fraction of events remaining after photon rejection in the $m^2_{\rm miss}$ control region  (0.011, 0.026)~GeV$^2 / c^4$.

A photon is defined by the reconstruction of one the following energy signatures: 
a signal in any LAV block within $\pm$5~ns of the NA48-CHOD time associated to the track; 
an energy cluster in the LKr within 25~ns around the NA48-CHOD time if the cluster energy is above 2~GeV or within $\pm$8~ns if the energy is below 2~GeV; 
a signal in IRC or SAC within $\pm$5 ns of the NA48-CHOD time. 
After applying photon rejection, no $\pi^+\pi^0$ event remains out of $1.5 \times10^6$ events in the $m^2_{\rm miss}$ control region, inclusive of the extended $\PPI$ momentum range
between 15 and 55~$\GEVc$. This sets an upper limit on the $\pi^0$ rejection inefficiency of $2 \times10^{-6}$ at 90\% CL.
The data sample collected in  2015, corresponding to $\sim 3  \times 10^8$ kaon decays,  is not large enough to address  the design $\pi^0$ rejection.

The random rejection of signal-like ($\pi\nu\bar\nu$) events induced by the presence of accidental photons has been measured to be  2\% using a $\KMN$ sample. 
Finally, the loss of signal induced by photon rejection because of $\pi^+$ interactions in the RICH material producing extra clusters in LKr, LAVs and IRC has been quantified to be 
about 5\% using a sample of 75~$\GEVc$ beam pions scattered elastically in the last station of the beam tracker.

\subsection{Further opportunities for NA62}
Although the NA62 setup is optimized for the measurement  of the $\PNNP$ decay, several other physics opportunities can also be addressed. The improved apparatus with respect to the previous generation of kaon decay in flight experiments and the high-intensity beam 
allow precise measurements of rare decays and searches for possible dark interactions. 
\paragraph{Rare and forbidden \boldmath{$\KPL$} and  \boldmath{$\PIo$ }decays:}
For rare decays accepted by the trigger, the expected NA62 sensitivities to branching ratios are 
$\sim$10$^{-12}$ for $\KPL$ decays and $\sim$10$^{-11}$ for $\PIo$ decays (the primary source of neutral pions being the $\KTP$ decay). The advanced particle identification and background suppression capabilities  open the way to a programme of rare decay measurements at improved precision, including $\Pll$, $\KPL\to\PPI\gamma\ell^+ \ell^- $, $\KPL\to\ell^+\nu_\ell\gamma$ ($\ell=e,\mu$), $\KPL\to\PPI\gamma\gamma$, $\KPL\to\PPI \PIo e^+e^-$, and $\PIo\to e^+e^-$. 
Exploration of the semileptonic decays $\KPL\to\PPI\MPI\mu^+\nu$ and $\KPL\to\PIo\PIo\mu^+\nu$,
for which there is little existing data, also seems possible.

Furthermore, searches for lepton-number-violating decays $\KPL\to\MPI\ell_1^+\ell_2^+$, $\KPL\to\ell_1^-\bar\nu\ell_2^+\ell_2^+$ ($\ell_{1,2}=e,\mu$), 
lepton-flavour-violating decays $K^+\to\pi^+\mu^\pm e^\mp$, $\PIo\to\mu^\pm e^\mp$,  and the 
$C$-violating decay $\PIo\to3\gamma$  with sensitivities below 
the current experimental upper limits are within reach.
\paragraph{The dark sector:}
The hermetic photon coverage  and the good missing-mass resolution provide an opportunity to 
search for the decays 
$\KPL\to\PPI X$, $\KPL\to\ell^+X$, $\KPL\to\ell^+\nu X$, and $\PIo\to\gamma X$, where $X$ is a new 
particle which is either invisible or decays into SM particles in the  decay region of the detector.  These searches can be interpreted in 
terms of phase-space limitations for the dark photon~\cite{ho86, po09}, axion~\cite{Izaguirre:2016dfi}, 
inflaton~\cite{Bezrukov:2009yw,Bezrukov:2014nza} and heavy neutral leptons~\cite{atre2009} . 
A by-product of the $\PNNP$ measurement is an improved limit on the $\PIo\to {\rm invisible}$ decay.

Furthermore, NA62 is capable of searching for decays of the dark photon, heavy neutral leptons and axion-like particles~\cite{dobrich2016} produced by the interactions of protons from the  SPS in the kaon production target or a collimator located downstream in a beam-dump-like experiment. This particular setup may be proposed for a later phase of NA62.

%% file: conclusion_v1.tex
The NA62 detector and its dedicated beam line have been described in detail,  together with the physics performance required to measure the ultra-rare $\PNNP$ decay, the main goal of the experiment.
At the time of this publication,  the experimental setup is essentially complete and operational.

Several novel techniques, specifically developed for this experiment, have been exploited in the domain of low-mass trackers. Three stations of silicon pixel assemblies operating in vacuum, sustaining a 750 MHz particle rate, associated to  micro-channel cooling  and a specific ASIC chip (TDCPix) make up the kaon tracker,  accounting for less than $0.5\% ~X_0$ per station. Four large  straw chamber stations, arranged in an inventive layout to avoid any flange material close to the beam passage, operating in vacuum  and read out by custom-made electronics, form the downstream spectrometer measuring the kaon charged decay products, accounting for $1.8\% ~X_0$  total material.  Both  trackers have precise timing capabilities.

 Hermeticity and redundancy  are key features to identify and veto other kaon decays and accidental coincidences. A variety of experimental techniques have been used by Cherenkov counters, electromagnetic and hadronic calorimeters, and several scintillation counters, allowing for charged-particle and photon identification.  Again, precise timing capabilities  are of prime importance and are used at trigger level.
  
The performance presented here has been obtained from data recorded during 2014 and 2015 using limited beam intensity  (1\% of nominal intensity)  and a minimum bias trigger. The kinematic rejection,  the charged-particle identification and the photon-rejection capabilities  match the design performance closely enough to ensure that NA62 will reach its goal within the programme approved to the end of 2018. 

 In the next few years, commissioning of the trigger and data acquisition system  will be performed up to the nominal beam intensity. High-level triggers, already partially deployed, will permit 
the study of the main decay mode to be complemented
with other exotic searches. Performances will be continuously improved both by accumulating  data samples with larger statistics and by developing elaborate software algorithms.

%% file: acknow.tex
The design, construction, installation and commissioning of the NA62 beam line and detector would 
have been impossible without the devoted efforts of our numerous technical colleagues from the 
CERN departments and from the NA62 Collaboration Institutes.
 We are particularly thankful to: \\
J. Bauche, K. Cornelis,  M. Calviani, M. Donze, M. Dumas, A. Ebn Rahmoun, Y. Gaillard, X. Genillon, R. Gorbonosov, D. Grenier, V. Kain, J. Lendaro, A. Masi and Ph. Schwarz for the beamline, target and controls;\\
P.~Carrie  and F.~Merlet for the detector gas systems;\\
J.~Bremer for the cryogenics;\\
A. Honma, F. Manolescu and I. Mcgill for the GTK wire bonding;\\
G. Bisogni, R. Di Raddo, V. Lollo and V. Chimenti for the work on the LAV detectors;\\
S. Casenove and M. Collignon  for the network installations;\\
C. David and M. Van Stenis  for the RICH mirror coating;\\
P.-A. Giudici, and F. Garnier for the B-field measurements;\\
M. Arnaud, M. Jeckel,  D. Parchet and E. Perez-Duenas for the surface building, civil engineeering and infrastructure;\\
S. Blanchard, V. de Jesus, A. Gutierrez, and T.~Mongelluzzo for the vacuum system;\\
A. Beynel, M.K. Boruchowski, B. Cumer, A. Froton, D. Mergelkuhl and M. Troillet  for the  
numerous survey measurements;\\
G.~Roche for the careful and safe transport of detectors;\\
Ph. Carri\`{e}re, S. Deschamps and J. Spanggaard for help on the CEDAR counter;\\
C. Crombez and M.B. Lonjon for the electrical installation;\\
M. Brugger, S. Danzeca and E. Nowak for radiation monitoring;\\
V. Martins de Sousa dos Rios and D. Vaxelaire for the access system;\\
W.T. Bannister, O. Crespo-Lopez, J. Lehtinen, G. Peon, X. Pons, D. Lefils and P. Valente for cooling  and ventilation;\\
J. Gulley and H. Wilkens for safety;\\
M. Gonzalez Berges for DCS.\\
The contributions of all these persons are sincerely  appreciated and gratefully acknowledged.\\
The experiment is efficiently and warmly supported by our secretary Veronique Wedlake. \\

The cost of the experiment and of its auxiliary systems were supported by the funding agencies of 
the Collaboration Institutes. We are particularly indebted to: \\
F.R.S.-FNRS (Fonds de la Recherche Scientifique - FNRS), Belgium;\\
BMES (Ministry of Education, Youth and Science), Bulgaria;\\
MEYS (Ministry of Education, Youth and Sports),  Czech Republic;\\
BMBF (Bundesministerium f\"{u}r Bildung und Forschung) contracts 05H09UM5, 05H12UM5 and 05H15UMCNA, Germany;\\
INFN  (Istituto Nazionale di Fisica Nucleare),  Italy;\\
MIUR (Ministero dell'Istruzione, dell' Universit\`a e della Ricerca),  Italy;\\
CONACyT  (Consejo Nacional de Ciencia y Tecnolog\'{i}a),  Mexico;\\
IFA (Institute of Atomic Physics),  Romania;\\
INR-RAS (Institute for Nuclear Research of the Russian Academy of Sciences), Moscow, Russia; \\
JINR (Joint Institute for Nuclear Research), Dubna, Russia; \\
NRC (National Research Center)  ``Kurchatov Institute'' and MESRF (Ministry of Education and Science of the Russian Federation), Russia; \\
MESRS  (Ministry of Education, Science, Research and Sport), Slovakia; \\
CERN (European Organization for Nuclear Research), Switzerland; \\
STFC (Science and Technology Facilities Council), United Kingdom;\\
NSF (National Science Foundation) Award Number 1506088,   U.S.A.;\\
ERC (The European Research Council)  ``UniversaLepto" advanced grant 268062, ``KaonLepton" starting grant 336581, Europe.\\

Individuals have received support from:\\
Charles University (project GA UK number 404716), Czech Republic;\\
Ministry of Education, Universities and Research (MIUR  ``Futuro in ricerca 2012''  grant RBFR12JF2Z, Project GAP), Italy;\\
The Royal Society  (grants UF100308, UF0758946), United Kingdom;\\
The Science and Technology Facilities Council (Rutherford fellowships ST/J00412X/1, ST/M005798/1), United Kingdom;\\
The European Research Council (ERC ``UniversaLepto'' advanced grant 268062,  ERC ``KaonLepton'' starting grant 336581).